%% file: tifrthesis.tex
\documentclass[12pt,tightenlines]{tifrthesis}
\usepackage{hyperref}
\usepackage[pagewise, mathlines]{lineno}
\usepackage[nottoc]{tocbibind}
\usepackage{booktabs}
\usepackage{setspace}
\usepackage{cite}

\usepackage{fancyhdr}
\usepackage{tocloft}
\usepackage{type1cm}
\usepackage{courier}
\usepackage{url}
\makeatletter
\def\url@leostyle{%
  \@ifundefined{selectfont}{\def\UrlFont{\sf}}{\def\UrlFont{\small\ttfamily}}}
\makeatother
\usepackage{graphicx}
\usepackage{relsize}
\usepackage[latin1]{inputenc}
\usepackage{amsfonts}
\usepackage{amsmath}
\usepackage{amssymb}
\usepackage{amsthm}
\usepackage{mathrsfs}
\usepackage{bm}
\usepackage{caption}
\usepackage{xcolor}
\usepackage{xspace}

\newcommand{\La}{\langle}
\newcommand{\Ra}{\rangle}

\newcommand{\nc}{\newcommand}
\newcommand{\mean}[1]{\left\langle #1 \right\rangle}
\nc{\pt}{p_\mathrm{T}} \nc{\kt}{k_\mathrm{T}} \nc{\mt}{m_\mathrm{T}}
\nc{\Kt}{K_\mathrm{T}} \nc{\Mt}{M_\mathrm{T}} \nc{\pL}{p_\mathrm{L}}

\newcommand{\btu}{\bigtriangleup}
\newcommand{\eps}{\varepsilon}

\begin{document}
%% you need to comment this if you want to supress line numbering for drafts
% \pagewiselinenumbers
% First, declare the parts of your title page
%

\author{Amaresh Jaiswal}
\regno{PHYS166}
\title{Formulation of relativistic dissipative \\[0.10in] fluid dynamics and its applications in \\[0.10in] heavy-ion collisions}
\dept{Department of Nuclear and Atomic Physics}
\subtime{February, 2014}
\gradtime{June, 2014}
\subject{Physics}

%
% The following creates the title page
%

\maketitle

\advisorname{Prof. Subrata Pal}
%\coadvisorname{Prof. Rajeev S. Bhalerao}
%\newpage\mbox{}\newpage
%
% The following \bigotimescreates a page used to copyright your thesis
% (This is optional and depends on the traditions of your dept.)
% You need either a copyright or a blankpage
\disscopyright
%
%Blank Page
%\newpage\mbox{}\newpage
%
%  My Dedication
%
%\dedication{\it\Large Dedicated to the people of India \\ for their
%generous support to research in basic sciences}
%\newpage\mbox{}\newpage
%
% Declaration goes here
%
\include{Declaration}
%\newpage\mbox{}\newpage
%
% Bring in Acknowledgement from separate files named ``Acknowledgments.tex''
%
\include{Acknowledgments}

%\newpage\mbox{}\newpage
%
% Abstract goes here.
%
\include{Abstract}
%\newpage\mbox{}\newpage
%
% Synopsis of your thesis goes here
%
\include{Synopsis}

%
% Make the Table of Contents and other good stuff
%

% The graduate school requires that the headings are centered for the toc, lof, lot

\begin{center}
\startcontentspace
\tableofcontents
\startdoublespace
\end{center}
\pagebreak
\begin{center}
\listoffigures
\end{center}
\pagebreak
%\begin{center}
%\listoftables
%\end{center}

%
%% This package can behave badly, so we put it after the TOC and other lists
%\pagewiselinenumbers
%

%
% The following is a lis\bigotimest of chapters.  Each is brought in from a\bigotimes
% separate file using the \include{} command.
%

\include{Chapter1}	% Introuduction
\include{Chapter2}	% Relativistic Viscous Hydrodynamics-the formulism
\include{Chapter3}	% Causal Viscous Hydrodynamics in 2+1-dimension
\include{Chapter4}	% Shear Viscosity Effects\biguplus
\include{Chapter5}	% Shear Viscosity Effects\biguplus
\include{Chapter6}	% Shear Viscosity Effects\biguplus
\include{Chapter7}	% Shear Viscosity Effects\biguplus
\include{Chapter8}	% Shear Viscosity Effects\biguplus
\include{Chapter9}	% Other estimation of Viscosity

%
% If you have appendices in your thesis, you will need the
% following, else keep it commented. 
%
\appendix
\include{Appendex}

%\include{AppendixB}
%\include{Glossary}
%

%
% The all important bibliography file at the end of your document!! Use
% the bibstyle you (your department) like in the \bibliographystyle{}
% % statement and list the name of your bibliography database file in
% the \bibliography{} statement.  In this example, ``bibfile.bib'' is
% the name of the database.  See the LaTeX manual appendix B for details
% about the bibliography database and BibTeX.
%

% single spacing (with double spacing between entries) is required
% by the graduate school, so leave \singlespacing here
\singlespacing
\bibliographystyle{unsrt}
\bibliography{}

\end{document}

%% file: Declaration.tex
\vspace*{0.6in}
\begin{center}
{\Large\bf DECLARATION}
\end{center}

\vskip 1cm

\noindent 
\startonehalfspace{\large This thesis is a presentation of my 
original research work and has not been submitted earlier as a whole 
or in part for a degree/diploma at this or any other 
Institution/University. Wherever contributions of others are 
involved, every effort is made to indicate this clearly, with due 
reference to the literature, and acknowledgement of collaborative 
research and discussions.}\\

\noindent 
\startonehalfspace{\large The work was done under the guidance of 
Prof. Subrata Pal at the Tata Institute of Fundamental Research, 
Mumbai.}

\vskip 2.0cm

\begin{flushright}
\startonehalfspace{\large Amaresh Jaiswal}
\end{flushright}
\begin{flushright}
\startonehalfspace{\large {\bf [Candidate's name and signature]}}
\end{flushright}

\vskip 1.5cm

\noindent 
\startonehalfspace{\large In my capacity as supervisor of the 
candidate's thesis, I certify that the above statements are true to 
the best of my knowledge.}

\vskip 2.0cm

\begin{flushleft}
\startonehalfspace{\large Prof. Subrata Pal}
\end{flushleft}
\begin{flushleft}
\startonehalfspace{\large {\bf [Supervisor's name and signature]}}
\end{flushleft}

\vskip 0.5cm
\noindent\startonehalfspace{\large Date:}

%% file: Acknowledgments.tex
\begin{acknowledgements}

\startonehalfspace{\large First and foremost, I would like to 
express my deepest gratitude to my advisor Prof. Subrata Pal, not 
only for his scientific guidance and valuable advice during my 
graduate study, but also for giving me the freedom to develop my own 
ideas independently. I appreciate very much his concern for my 
future and consider myself very fortunate to have had him as my 
thesis advisor. I am also greatly indebted to Prof. Rajeev S. 
Bhalerao for his supervision, brilliant insights, helpful comments 
and suggestions, without which my early research work would not have 
found its proper direction.}

\startonehalfspace{\large  I thank Dr. Sreekanth V. for the fruitful 
collaboration and all the discussions we have had, scientific or 
not, during the last couple of years. For work covered in this 
thesis, I gratefully acknowledge the enlightening interactions with 
Prof. J. P. Blaizot, Dr. M. Luzum, Prof. S. Minwalla and Prof. J. Y. 
Ollitrault. I am also grateful to Dr. G. S. Denicol, Dr. A. El and 
Prof. M. Strickland for several helpful correspondences. I thank 
Prof. N. Mathur and Prof. R. Palit for keeping a keen interest in my 
progress.}

\startonehalfspace{\large My sincerest thanks to the TIFR staff 
members of computer centre, DNAP office, library, photography 
section and university cell for their help.}

\startonehalfspace{\large Special thanks are due to Sayantan Sharma 
for his comradely affection and countless study sessions from which 
I learned a lot. I thank my friends and colleagues Amitava, Anjani, 
Bharat, Chandrodoy, Deepika, Gourab, Jasmine, Nikhil, Nilay, 
Padmanath, Pankaj, Prashant, Purnima, Rahul, Ravitej, Saikat, 
Sayani, Sudipta, Tarakeshwar and Vivek for making my stay at TIFR a 
memorable experience.}

\startonehalfspace{\large Last but not least, I thank my wife 
Pallavi and my family for their unconditional love and support.}

\end{acknowledgements}

%% file: Abstract.tex
%  The dissertation abstract can only be 350 words.
\begin{abstract}

\startonehalfspace {\large Relativistic fluid dynamics finds 
application in astrophysics, cosmology and the physics of 
high-energy heavy-ion collisions. In this thesis, we present our 
work on the formulation of relativistic dissipative fluid dynamics 
within the framework of relativistic kinetic theory. We employ the 
second law of thermodynamics as well as the relativistic Boltzmann 
equation to obtain the dissipative evolution equations.}

\startonehalfspace{\large We present a new derivation of the 
dissipative hydrodynamic equations using the second law of 
thermodynamics wherein all the second-order transport coefficients 
get determined uniquely within a single theoretical framework. An 
alternate derivation of the dissipative equations which does not 
make use of the two major approximations/assumptions namely, Grad's 
14-moment approximation and second moment of Boltzmann equation, 
inherent in the Israel-Stewart theory, is also presented. Moreover, 
by solving the Boltzmann equation iteratively in a Chapman-Enskog 
like expansion, we have derived the form of second-order viscous 
corrections to the distribution function. Furthermore, a novel 
third-order evolution equation for shear stress tensor is derived. 
Finally, we generalize the collision term in the Boltzmann equation 
to include non-local effects.  We find that the second-order 
dissipative equations derived using this modified Boltzmann equation 
contains all possible terms allowed by symmetry.}

\startonehalfspace{\large In the case of one-dimensional scaling 
expansion, we demonstrate the numerical significance of these 
formulations on the evolution of the hot and dense matter created in 
ultra-relativistic heavy-ion collisions. We also study the effect of 
these new formulations on particle (hadron and thermal dilepton) 
spectra and femtoscopic radii.}

\end{abstract}

%% file: Synopsis.tex
\begin{synopsis}
\vspace{-1cm}
%%%%%%%%%%%%%%%%%%%%%%%%%%%%%%%%%%%%%%%%%%%%%%%%%%%%%%%%%%%%%%%%%%%%%%%%

\section*{\Large{1. Introduction and motivation}}

Fluid dynamics is an effective theory describing the 
long-wavelength, low frequency limit of the microscopic dynamics of 
a system. It is an elegant framework to study the effects of the 
equation of state on the evolution of the system. Relativistic fluid 
dynamics has been quite successful in explaining the various 
collective phenomena observed in astrophysics, cosmology and the 
physics of high-energy heavy-ion collisions. The collective behaviour 
of the hot and dense matter created in ultra-relativistic heavy-ion 
collisions has been studied quite extensively within the framework 
of relativistic fluid dynamics. 

In application of fluid dynamics, it is natural to first employ the 
zeroth-order (gradient expansion for dissipative quantities) or 
ideal fluid dynamics. However, as all fluids are dissipative in 
nature due to the uncertainty principle \cite{Danielewicz:1984wwS}, 
the ideal fluid results serve only as a benchmark when dissipative 
effects become important. The earliest theoretical formulation of 
relativistic dissipative hydrodynamics also known as first-order 
theories, are due to Eckart \cite{Eckart:1940zzS} and Landau-Lifshitz 
\cite{LandauS}. However these formulations, collectively called 
relativistic Navier-Stokes (NS) theory, involve parabolic 
differential equations and suffer from acausality and numerical 
instability. The second-order Israel-Stewart (IS) theory \cite 
{Israel:1979wpS}, with its hyperbolic equations restores causality 
but may not guarantee stability \cite{Huovinen:2008teS}. 

Hydrodynamic analysis of the spectra and azimuthal anisotropy of 
particles produced in heavy-ion collisions at the Relativistic Heavy 
Ion Collider (RHIC) \cite{Romatschke:2007mqS,Song:2010mgS} and 
recently at the Large Hadron Collider (LHC) \cite 
{Luzum:2010agS,Qiu:2011hfS} suggests that the matter formed in these 
collisions is strongly-coupled quark-gluon plasma (QGP). Although IS 
hydrodynamics has been quite successful in modelling relativistic 
heavy ion collisions, there are several inconsistencies and 
approximations in its formulation which prevent proper understanding 
of the thermodynamic and transport properties of the QGP. The 
standard derivation of IS equations using the second-law of 
thermodynamics contains unknown transport coefficients related to 
relaxation times of the dissipative quantities viz., the bulk 
viscous pressure, the particle diffusion current and the shear 
stress tensor \cite{Israel:1979wpS}. While IS equations derived from 
kinetic theory can provide reliable values for the shear relaxation 
time ($\tau_\pi$), the bulk relaxation time ($\tau_\Pi$) remains 
ambiguous. Moreover, IS derivation of second-order hydrodynamics 
from kinetic theory relies on additional approximations and 
assumptions: Grad's 14-moment approximation for the single particle 
distribution function \cite{Israel:1979wpS,GradS} and use of the 
second moment of the Boltzmann equation (BE) to obtain evolution 
equations for dissipative quantities \cite 
{Israel:1979wpS,Baier:2006umS}. 

Apart from these problems in the formulation, IS theory suffers from 
several other shortcomings. In one-dimensional Bjorken scaling 
expansion \cite{Bjorken:1982qrS}, IS theory leads to negative 
longitudinal pressure \cite {Martinez:2009mfS,Rajagopal:2009ywS} which 
limits its application within a certain temperature range. Further, 
the scaling solutions of IS equations when compared with transport 
results show disagreement for shear viscosity to entropy density 
ratio, $\eta/s>0.5$ indicating the breakdown of the second-order 
theory \cite {Huovinen:2008teS,El:2008yyS}. Moreover, in the study of 
identical particle correlations, the experimentally observed 
$1/\sqrt{m_T}$ scaling of the Hanburry Brown-Twiss (HBT) radii ($m_T$
being the transverse mass of the hadron pair), which is also 
predicted by the ideal hydrodynamics, is broken when viscous 
corrections to the distribution function are included \cite 
{Teaney:2003kpS}. The correct formulation of the relativistic 
dissipative fluid dynamics is thus far from settled and is currently 
under intense investigation \cite 
{Huovinen:2008teS,Baier:2006umS,El:2008yyS,Betz:2008meS,El:2009vjS,Denicol:2010xnS,Denicol:2012cnS,Martinez:2010scS,Jaiswal:2014isaS}.

In this synopsis we report on some major progress we have made in 
the formulation of relativistic dissipative fluid dynamics within 
the framework of kinetic theory. The problem pertaining to $\tau_\Pi$
has been solved by considering entropy four-current defined using 
Boltzmann H-function \cite{Jaiswal:2013fcS}. Using this method, 
hydrodynamic evolution, production of thermal dileptons and 
subsequent hadronization of the strongly interacting matter has been 
studied \cite{Bhalerao:2013ahaS}. An alternate derivation of the 
dissipative equations, which does not make use of the 14-moment 
approximation as well as the second moment of BE, has also been 
outlined \cite{Jaiswal:2013npaS}. The form of viscous corrections to 
the distribution function is derived up to second-order in gradients 
which restores the observed $1/\sqrt{m_T}$ scaling of the HBT radii 
\cite{Bhalerao:2013pzaS}. Finally, with the motivation to improve the 
IS theory beyond its present scope, two rigorous investigations have 
been outlined in this synopsis: (a) Derivation of a novel 
third-order evolution equation for shear stress tensor \cite
{Jaiswal:2013vtaS}, and (b) Derivation of second-order dissipative 
equations from the BE where the collision term is modified to 
include non-local effects \cite{Jaiswal:2012qmS}.
 
This synopsis is organized in the following manner. In Section 2, 
relativistic kinetic theory and dissipative fluid dynamics are 
outlined. Section 3 describes a derivation of the dissipative 
hydrodynamic equations using the second law of thermodynamics 
wherein all the second-order transport coefficients get determined 
uniquely within a single theoretical framework. In Section 4, the 
results obtained using the methodology of Section 3 have been 
applied to study particle spectra. In Section 5, an alternate 
derivation of the dissipative equations which does not make use of 
the two major approximation/assumption namely, Grad's 14-moment 
approximation and second moment of BE, inherent in IS theory, has 
been outlined. In Section 6, the form of second-order viscous 
corrections to the distribution function is derived and the effects 
of these corrections on particle spectra and HBT radii are compared 
with those due to the traditional Grad's 14-moment approximation. 
The derivation of Section 5 has been extended to third-order in 
Section 7. In Section 8, the collision term in the BE is modified to 
include non-local effects and subsequently second-order dissipative 
equations have been derived using this modified BE. Finally, in 
Section 9 a summary is provided. 

%%%%%%%%%%%%%%%%%%%%%%%%%%%%%%%%%%%%%%%%%%%%%%%%%%%%%%%%%%%%%%%%%%%%%%%%

\section*{\Large{2. Relativistic kinetic theory and fluid dynamics}} 

The various formulations of relativistic dissipative hydrodynamics, 
outlined in this synopsis, are obtained within the framework of 
relativistic kinetic theory. We briefly outline here the salient 
features of relativistic kinetic theory and dissipative 
hydrodynamics which have been employed in the subsequent calculations.

Macroscopic properties of a many-body system are governed by the 
interactions among its constituent particles and the external 
constraints on the system. Kinetic theory presents a statistical 
framework in which the macroscopic quantities are expressed in terms 
of single-particle phase-space distribution function. Various 
currents controlling the hydrodynamic evolution of the system, such 
as particle four-current ($N^\mu$), energy-momentum tensor 
($T^{\mu\nu}$) and entropy four-current ($S^\mu$) are written as 
\cite{deGrootS}

\begin{align}
N^\mu &= \int dp \ p^\mu f,  \label{PFCS} \\
T^{\mu\nu} &= \int dp \ p^\mu p^\nu f,  \label{EMTS} \\
S^\mu_{r=0} &= -\int dp ~p^\mu f \left(\ln f - 1\right) , \label{EFCCS} \\  
S^\mu_{r=\pm 1} &= -\int dp ~p^\mu  \left(f \ln f + r\tilde f \ln \tilde f\right). \label{EFCQS}
\end{align}
Here, $dp = g d{\bf p}/[(2 \pi)^3\sqrt{{\bf p}^2+m^2}]$, $g$ and $m$ 
being the degeneracy factor and particle rest mass, $p^{\mu}$ is the 
particle four-momentum, $f\equiv f(x,p)$ is the single particle 
phase-space distribution function. The quantity $\tilde f \equiv 1 - 
rf$, where $r = 1,-1,0$ for Fermi, Bose, and Boltzmann gas, 
respectively.

The conserved particle current and the energy-momentum tensor can 
be expressed as 
\begin{equation}\label{NTDS}
N^\mu = nu^\mu + n^\mu,  \quad
T^{\mu\nu} = \epsilon u^\mu u^\nu-(P+\Pi)\Delta ^{\mu \nu} 
+ \pi^{\mu\nu},
\end{equation}
where $n, \epsilon, P$ are respectively number density, energy 
density, pressure, and $\Delta^{\mu\nu}=g^{\mu\nu}-u^\mu u^\nu$ is 
the projection operator on the three-space orthogonal to the 
hydrodynamic four-velocity $u^\mu$ defined in the Landau frame: 
$T^{\mu\nu} u_\nu=\epsilon u^\mu$. For small departures from 
equilibrium, $f(x,p)$ can be written as $f = f_0 + \delta f$. The 
equilibrium distribution function is defined as $f_0 = [\exp(\beta 
u\cdot p -\alpha) + r]^{-1}$ where the inverse temperature 
$\beta=1/T$ and $\alpha=\beta\mu$ ($\mu$ being the chemical 
potential) are defined by the equilibrium matching conditions 
$n\equiv n_0$ and $\epsilon \equiv \epsilon_0$. The scalar product 
is defined as $u\cdot p\equiv u_\mu p^\mu$. The dissipative 
quantities, viz., the bulk viscous pressure ($\Pi$), the particle 
diffusion current ($n^\mu$) and the shear stress tensor 
($\pi^{\mu\nu}$) are respectively
\begin{eqnarray}
\Pi &=& -\frac{\Delta_{\alpha\beta}}{3}\int dp \ p^
\alpha p^\beta\, \delta f ,  \label{BEES} \\
n^\mu &=&  \Delta^{\mu\nu} \int dp \ p_\nu\, \delta f , \label{CCEES} \\
\pi^{\mu\nu} &=& \Delta^{\mu\nu}_{\alpha\beta} \int dp \ p^
\alpha p^\beta\, \delta f. \label{SEES}
\end{eqnarray}
Here $\Delta^{\mu\nu}_{\alpha\beta} = [\Delta^\mu_\alpha 
\Delta^\nu_\beta + \Delta^\mu_\beta \Delta^\nu_\alpha - 
(2/3)\Delta^{\mu\nu} \Delta_{\alpha\beta}]/2$ is the traceless 
symmetric projection operator. 

Conservation of current, $\partial_\mu N^\mu=0$, and energy-momentum 
tensor, $\partial_\mu T^{\mu\nu} =0$, yield the fundamental 
evolution equations for $n$, $\epsilon$ and $u^\mu$
\begin{eqnarray}
\dot n+n\theta + \partial_\mu n^\mu &=& 0, \label{evolnS} \\
\dot\epsilon + (\epsilon+P+\Pi)\theta - \pi^{\mu\nu}\sigma_{\mu\nu} &=& 0,  \label{evoleS} \\
(\epsilon+P+\Pi)\dot u^\alpha - \nabla^\alpha (P+\Pi) + \Delta^\alpha_\nu \partial_\mu \pi^{\mu\nu}  &=& 0. \label{evoluS}
\end{eqnarray}
Here the notations are $\dot A=u^\mu\partial_\mu A$, 
$\theta=\partial_\mu u^\mu$, $\nabla^\alpha = 
\Delta^{\mu\alpha}\partial_\mu$ and 
$\sigma^{\mu\nu}=\Delta^{\mu\nu}_{\alpha\beta}\nabla^\alpha u^\beta$. 
Even if the equation of state relating $\epsilon$ and $P$ is 
provided, the system of Eqs. (\ref{evolnS})-(\ref {evoluS}) is not 
closed unless the evolution equations for the dissipative quantities, 
namely, $\Pi$, $\pi^{\mu\nu}$, $n^\mu$ are specified.

The evolution equations for the dissipative quantities expressed in 
terms of the non-equilibrium distribution function, as in Eqs. (\ref
{BEES})- (\ref{SEES}), can be obtained provided the evolution of 
distribution function is specified from some microscopic 
considerations. Boltzmann equation governs the evolution of the 
single-particle phase-space distribution function $f$ which provides 
a reliably accurate description of the microscopic dynamics. For 
microscopic interactions restricted to $2 \leftrightarrow 2$ elastic 
collisions, the form of the BE is given by
\begin{equation}\label{BES}
p^\mu \partial_\mu f = C[f] = \frac{1}{2} \int dp' dk \ dk' \  W_{pp' \to kk'}(f_k f_{k'} \tilde f_p \tilde f_{p'} - f_p f_{p'} \tilde f_k \tilde f_{k'}),
\end{equation}
where $C[f]$ is the collision functional and $W_{pp' \to kk'}$ is 
the collisional transition rate. The first and second terms within 
the integral of Eq. (\ref{BES}) refer to the processes $kk' \to pp'$ 
and $pp' \to kk'$, respectively. In the relaxation-time 
approximation (RTA), where it is assumed that the effect of the 
collisions is to restore the distribution function to its local 
equilibrium value exponentially, the collision integral reduces to 
$C[f]=-(u\cdot p)\delta f/\tau_R$ \cite{Anderson_WittingS}. The 
results of these discussions will be used in the following sections.

%%%%%%%%%%%%%%%%%%%%%%%%%%%%%%%%%%%%%%%%%%%%%%%%%%%%%%%%%%%%%%%%%%%%%%%%

\section*{\Large{3. Dissipative fluid dynamics from the entropy principle}}

The standard derivation of IS theory invoking the second-law of 
thermodynamics, $\partial_\mu S^\mu \ge 0$, contains unknown 
second-order transport coefficients in the entropy four current 
$S^\mu$. These coefficients have to be determined from an alternate 
theory and as a consequence, the evolution equations remain 
incomplete. In this section, a formal derivation of the dissipative 
hydrodynamic equations is outlined wherein all the second-order 
transport coefficients get determined uniquely within a single 
theoretical framework \cite{Jaiswal:2013fcS}. This is achieved by 
invoking the second law of thermodynamics for the generalized 
entropy four-current expressed in terms of the phase-space 
distribution function given by Grad's 14-moment approximation. 

The starting point for the derivation of the dissipative evolution 
equations is the entropy four-current expression generalized from 
Boltzmann's H-function given in Eqs. (\ref{EFCCS})-(\ref{EFCQS}). The 
divergence of $S^\mu_{r=0,\pm 1}$ leads to
\begin{equation}\label{EFCDS}
\partial_\mu S^\mu = -\int dp ~p^\mu \left(\partial_\mu f\right) 
\ln (f/\tilde f) .
\end{equation}
To proceed further, Grad's 14-moment approximation \cite{GradS} for 
the single-particle distribution in orthogonal basis \cite
{Denicol:2010xnS} has been used
\begin{equation}\label{G14S}
f = f_0 + f_0 \tilde f_0 \phi, ~~~ \phi =  \lambda_\Pi \Pi + \lambda_n n_\alpha p^\alpha 
+ \lambda_\pi \pi_{\alpha\beta} p^\alpha p^\beta .
\end{equation}
The coefficients ($\lambda_\Pi, \lambda_n, \lambda_\pi$) are 
typically assumed to be independent of four-momentum $p^{\mu}$ and 
are functions of $(\epsilon, \alpha, \beta)$. Expanding the 
logarithm in Eq. (\ref{EFCDS}) in terms of $\phi$ and retaining all 
terms up to third-order in gradients (where $\phi$ is linear in 
dissipative quantities), Eq. (\ref{EFCDS}) reduces to
\begin{equation}\label{EFCD2S}
\partial_\mu S^\mu = -\!\int\!  dp ~p^\mu \Big[ 
\phi\left(\partial_\mu f_0\right) - 
\phi^2 (\tilde f_0 -1/2)(\partial_{\mu}f_0) 
+\phi^2 \partial_\mu (f_0 \tilde f_0) + \phi f_0 \tilde f_0 (\partial_\mu \phi)\Big].
\end{equation}
The various momentum integrals in the above equation can be 
performed by tensor decomposing them using hydrodynamic tensor 
degrees of freedom ($u^\mu$ and $g^{\mu\nu}$) with suitable 
coefficients. 

The second law of thermodynamics, $\partial_{\mu}S^{\mu}\ge 0$, is 
guaranteed to be satisfied if linear relationships between 
thermodynamical fluxes and extended thermodynamic forces are imposed 
in Eq. (\ref{EFCD2S}), leading to the following evolution equations 
for bulk, charge current and shear
\begin{align}
\Pi &= -\zeta\Big[ \theta 
+ \beta_0 \dot \Pi 
+ \beta_{\Pi\Pi} \Pi \theta 
+ \alpha_0 \nabla_\mu n^\mu 
+ \psi\alpha_{n\Pi} n_\mu \dot u^\mu
+ \psi\alpha_{\Pi n} n_\mu \nabla^\mu \alpha  \Big], \label{bulkS} \\ 
n^{\mu} &= \lambda \Big[ T \nabla^\mu \alpha 
- \beta_1\dot n^{\langle\mu\rangle} 
- \beta_{nn} n^\mu \theta
+ \alpha_0 \nabla^\mu \Pi
+ \alpha_1 \Delta^\mu_\rho \nabla_\nu \pi^{\rho\nu}
+ \tilde \psi\alpha_{n\Pi} \Pi \dot u^{\langle\mu\rangle} \nonumber \\
&\quad\quad\; + \tilde \psi\alpha_{\Pi n} \Pi \nabla^\mu \alpha
+ \tilde \chi\alpha_{\pi n} \pi_\nu^\mu \nabla^\nu \alpha
+ \tilde \chi\alpha_{n\pi} \pi_\nu^\mu \dot u^\nu \Big], \label{currentS} \\ 
\pi^{\mu\nu} &= 2\eta\Big[ \sigma^{\mu\nu} 
- \beta_2\dot\pi^{\langle\mu\nu\rangle} 
- \beta_{\pi\pi}\theta\pi^{\mu\nu}
- \alpha_1 \nabla^{\langle\mu}n^{\nu\rangle}
- \chi\alpha_{\pi n} n^{\langle\mu} \nabla^{\nu\rangle} \alpha 
- \chi\alpha_{n\pi } n^{\langle\mu} \dot u^{\nu\rangle} \Big] , \label{shearS}
\end{align}
with the coefficients of charge conductivity, bulk and shear 
viscosity, viz. $\lambda, \zeta,\eta \ge 0$. We define the notation 
$A^{\langle\mu\rangle}\equiv \Delta^\mu_\alpha A^\alpha$ and
$B^{\langle\mu\nu\rangle}\equiv 
\Delta^{\mu\nu}_{\alpha\beta}B^{\alpha\beta}$. The 
general expressions for $\beta_1,\alpha_0,\alpha_1$ and 
$\beta_0,\beta_2$ in the classical limit simplify to \cite
{Jaiswal:2013fcS}
\begin{equation}\label{ABS}
\beta_1 = \frac{\epsilon+P}{n^2}, \quad   \alpha_0 = \alpha_1 = \frac{1}{n}, \quad \beta_0 = \frac{1}{P},\quad  
\beta_2 = \frac{3}{\epsilon+P} + \frac{m^2\beta^2P}{2(\epsilon+P)^2}.
\end{equation}
The other coefficients in Eqs. (\ref{bulkS})-(\ref{shearS}) are 
obtained in terms of $\beta_0,\beta_1,\beta_2,\alpha_0,\alpha_1$ and 
their derivatives. These coefficients are obtained consistently 
within the same theoretical framework. In contrast, in the standard 
derivation from entropy principles \cite{Israel:1979wpS}, these 
transport coefficients have to be estimated from an alternate theory.

The viscous relaxation times  are defined as $\tau_{\Pi} = 
\zeta\,\beta_0$ and $\tau_{\pi} = 2\,\eta\,\beta_2$. It is important 
to note that in the photon limit ($m/T\to 0$), $\beta_0$ in Eq. (\ref
{ABS}) and hence $\tau_\Pi$ in the present calculation remain finite 
unlike all other previous calculations where they diverged. In 
the absence of any reliable prediction for the bulk relaxation time 
$\tau_\Pi$, it has been customary to keep it fixed or set it equal 
to the shear relaxation time $\tau_\pi$ or parametrize it in such a 
way that it captures critical slowing-down of the medium near $T_c$ 
due to growing correlation lengths \cite
{Fries:2008tsS,Denicol:2009amS,Song:2009rhS,Rajagopal:2009ywS}. Since 
$\zeta/s$ has a peak near the phase transition, the $\tau_\Pi$ 
obtained here naturally captures the phenomenon of critical 
slowing-down.

For one-dimensional scaling expansion of the matter formed in 
relativistic heavy-ion collisions \cite{Bjorken:1982qrS}, Eqs. (\ref
{evoleS}), (\ref{bulkS}) and (\ref{shearS}) are solved simultaneously 
in the Milne co-ordinate system ($\tau,x,y,\eta$), where $\tau = 
\sqrt{t^2-z^2}$, $\eta=\tanh^{-1}(z/t)$ and $u^\mu=(1,0,0,0)$. 
Recent lattice QCD results for the equation of state \cite
{Bazavov:2009znS} and $\zeta/s$ \cite {Meyer:2007dyS} have been used. 
The results obtained in the present calculations are compared with 
those obtained by considering $\tau_\Pi=\tau_\pi$ and $\tau_\Pi={\rm 
Const.}$ In both these cases, the longitudinal pressure 
($P_L=P+\Pi-\pi$) becomes negative near the phase-transition 
temperature $T_c$ leading to mechanical instabilities such as 
cavitation. In contrast, $\tau_{\Pi}$ obtained in the present 
calculation does not lead to cavitation and guarantees the 
applicability of hydrodynamics up to temperatures well below $T_c$ 
into the hadronic phase.

%%%%%%%%%%%%%%%%%%%%%%%%%%%%%%%%%%%%%%%%%%%%%%%%%%%%%%%%%%%%%%%%%%%%%%%%

\section*{\Large{4. Viscous hydrodynamics and particle production}}

The method developed in the previous section is employed here to 
derive hydrodynamic equations and study hadron and 
dilepton production corresponding to two different forms of the 
non-equilibrium distribution function \cite 
{Denicol:2010xnS,Dusling:2007giS} :
\begin{equation}\label{NEDFS}
f = f_0\left(1+\phi_{1,2}\right), \quad
\phi_1 = \frac{\Pi}{P} + \frac{p^\mu p^\nu \pi_{\mu\nu}}{2(\epsilon+P)T^2}, \quad
\phi_2 = \frac{p^\mu p^\nu}{2(\epsilon+P)T^2}\left(\pi_{\mu\nu}+\frac{2}{5}\Pi\Delta_{\mu\nu}\right).
\end{equation}
As in the previous section, the evolution equations for bulk 
pressure and shear stress tensor are obtained as
\begin{equation}
\Pi = -\zeta\left[ \theta + \beta_0 \dot \Pi + \frac{4}{3}\beta_0 \theta\Pi \right], \quad 
\pi^{\mu\nu} = 2\eta\left[ \sigma^{\mu\nu} - \beta_2\dot\pi^{\langle\mu\nu\rangle} - \frac{4}{3}\beta_2 \theta\pi^{\mu\nu} \right], \label{viscousS}
\end{equation}
where the transport coefficients corresponding to the two cases in 
Eq. (\ref{NEDFS}) are found to be
\begin{equation}\label{betasS}
\beta_0^{(1)} = \frac{1}{P},\quad  
\beta_0^{(2)} = \frac{18}{5(\epsilon+P)} + \frac{3m^2\beta^2P}{5(\epsilon+P)^2},\quad
\beta_2^{(1)} = \beta_2^{(2)} = \frac{3}{\epsilon+P} + \frac{m^2\beta^2P}{2(\epsilon+P)^2}.
\end{equation}
The evolution equations thus obtained are used to study the 
transverse momentum spectra of hadrons and thermal dileptons 
\cite{Bhalerao:2013ahaS}. 

The Cooper-Frye freeze-out prescription to obtain hadronic spectra 
is given by \cite{Cooper:1974mvS}
\begin{equation}
\frac{dN}{d^2p_Tdy} = \frac{g}{(2\pi)^3} \int p_\mu d\Sigma^\mu f.
\label{hadronsS}
\end{equation}
where, $d\Sigma^\mu$ represents the volume element on the freeze-out 
hypersurface. The rate of thermal dilepton production is \cite{rvogtBookS}
\begin{equation}
\frac{dN}{d^4xd^4p}=\int\frac{d^3\textbf{p}_1}{(2\pi)^3} \frac{d^3\textbf{p}_2}{(2\pi)^3} f(E_1,T) f(E_2,T) 
v_{rel} g^2\sigma(M^2)\delta^4(p-p_1-p_2),
\label{dileptonsS}
\end{equation}
where $p_i=(E_i,\textbf{p}_i)$ are the four momenta of the initial 
particles with masses $m_i$, 
$M^2=(E_1+E_2)^2-(\textbf{p}_1+\textbf{p}_2)^2$, 
$v_{rel}=M\sqrt{(M^2-4m_i^2)}/(2E_1E_2)$ denotes the relative 
velocity and $\sigma(M^2)$ is the thermal dilepton production cross 
section. For consistency, we use the same non-equilibrium 
distribution function in the calculation of the particle spectra as 
in the derivation of the evolution equations.

Within a one-dimensional scaling expansion of the matter formed in 
relativistic heavy-ion collisions, we observed that the transport 
coefficients obtained in Eq. (\ref{betasS}) do not lead to 
cavitation. We also demonstrate that for the two cases described in 
Eq. (\ref{NEDFS}) the transverse momentum spectra exhibit appreciable 
differences for hadron and especially for dileptons \cite 
{Bhalerao:2013ahaS}. Further we find that an inconsistent treatment 
of the distribution function in hydrodynamic evolution and freezeout 
affects the particle spectra significantly.

%%%%%%%%%%%%%%%%%%%%%%%%%%%%%%%%%%%%%%%%%%%%%%%%%%%%%%%%%%%%%%%%%%%%%%%%

\section*{\Large{5. Dissipative fluid dynamics from Boltzmann equation within relaxation-time approximation}}

Israel-Stewart's derivation of second-order dissipative 
hydrodynamics from kinetic theory is based on two strong 
approximation/assumption viz. Grad's 14-moment approximation for the 
distribution function and the use of the second moment of the 
Boltzmann equation (BE) to obtain evolution equations for 
dissipative quantities \cite{Israel:1979wpS}. In this section, an 
alternate derivation of hydrodynamic equations for dissipative 
quantities has been outlined \cite{Jaiswal:2013npaS} which does not 
make use of these assumptions. Instead, the iterative solution of 
BE in relaxation-time approximation (RTA) has been used for the 
distribution function and the evolution equations for the 
dissipative quantities have been derived directly from their 
definitions.

Boltzmann equation with RTA for the collision term can be written as 
\cite{Anderson_WittingS}
\begin{equation}\label{RBES}
p^\mu\partial_\mu f =  -\frac{u\cdot p}{\tau_R}(f-f_0)~,
\end{equation}
In order to solve the above equation, the particle distribution 
function is expanded about its equilibrium value in powers of 
space-time gradients.
\begin{equation}\label{CEES}
f = f_0 + \delta f, \quad \delta f= \delta f^{(1)} + \delta f^{(2)} + \cdots ,
\end{equation}
where $\delta f^{(1)}$ is first-order in gradients, $\delta f^{(2)}$ 
is second-order, etc. The Boltzmann equation, (\ref{RBES}), in the 
form $f=f_0-(\tau_R/u\cdot p)\,p^\mu\partial_\mu f$, can be solved 
iteratively as
\begin{equation}\label{F1F2S}
f_1 = f_0 -\frac{\tau_R}{u\cdot p} \, p^\mu \partial_\mu f_0, \quad f_2 = f_0 -\frac{\tau_R}{u\cdot p} \, p^\mu \partial_\mu f_1, ~~\, \cdots
\end{equation}
where $f_1=f_0+\delta f^{(1)}$ and $f_2=f_0+\delta f^{(1)}+\delta 
f^{(2)}$. To first and second-order in gradients, we obtain
\begin{equation}\label{SOCS}
\delta f^{(1)} = -\frac{\tau_R}{u\cdot p} \, p^\mu \partial_\mu f_0~, \quad
\delta f^{(2)} = \frac{\tau_R}{u\cdot p}p^\mu p^\nu\partial_\mu\Big(\frac{\tau_R}{u\cdot p} \partial_\nu f_0\Big)~. 
\end{equation}

The first-order dissipative equations can be obtained 
from Eqs. (\ref{BEES})-(\ref{SEES}) using $\delta f = \delta f^{(1)}$ 
from Eq. (\ref{SOCS}) and performing the integrals
\begin{align}
\Pi &= -\frac{\Delta_{\alpha\beta}}{3}\!\!\int\!\! dp \, p^\alpha p^\beta \left[-\frac{\tau_R}{u.p} \, p^\mu \partial_\mu f_0\right] 
= -\tau_R\beta_\Pi\theta , \label{FOBES}\\
n^\mu &= \Delta^\mu_\alpha \!\!\int\!\! dp \, p^\alpha \left[-\frac{\tau_R}{u.p} \, p^\mu \partial_\mu f_0\right] 
= \tau_R\beta_n\nabla^\mu\alpha , \label{FOCES}\\
\pi^{\mu\nu} &= \Delta^{\mu\nu}_{\alpha\beta}\int dp \ p^\alpha p^\beta \left[-\frac{\tau_R}{u.p} \, p^\mu \partial_\mu f_0\right] 
= 2\tau_R\beta_\pi\sigma^{\mu\nu} , \label{FOSES}
\end{align}
where,
\begin{align}
\beta_\Pi =&~ \frac{1}{3}\left(1-3c_s^2\right)(\epsilon+P) - \frac{2}{9}(\epsilon-3P) 
- \frac{m^4}{9}\left<(u.p)^{-2}\right>_{0} , \label{BBS}\\
\beta_n =&~ - \frac{n^2}{\beta(\epsilon+P)} + \frac{2\left<1\right>_0}{3\beta} + \frac{m^2}{3\beta}\left<(u.p)^{-2}\right>_0 ,\label{BCS}\\
\beta_\pi =&~ \frac{4P}{5} + \frac{\epsilon-3P}{15} - \frac{m^4}{15}\left<(u.p)^{-2}\right>_0 .\label{BSS}
\end{align}
Here, $\left<\cdots\right>_0=\int dp(\cdots)f_0$, and 
$c_s^2=(\partial P/\partial\epsilon)_{s/n}$ is the speed of sound 
squared ($s$ being the entropy density).

Second-order evolution equations can also be obtained similarly by 
substituting $\delta f=\delta f^{(1)}+\delta f^{(2)}$ from Eq. 
(\ref{SOCS}) in Eqs. (\ref{BEES})-(\ref{SEES}). The second-order 
equations obtained after performing the integrals are
\begin{align}
\frac{\Pi}{\tau_R} =& -\dot{\Pi}
-\beta_{\Pi}\theta 
-\delta_{\Pi\Pi}\Pi\theta
+\lambda_{\Pi\pi}\pi^{\mu\nu}\sigma_{\mu \nu }
-\tau_{\Pi n}n\cdot\dot{u}
-\lambda_{\Pi n}n\cdot\nabla\alpha
-\ell_{\Pi n}\partial\cdot n ~, \label{BULKS}\\
\frac{n^{\mu}}{\tau_R} =& -\dot{n}^{\langle\mu\rangle}
+\beta_{n}\nabla^{\mu}\alpha
-n_{\nu}\omega^{\nu\mu}
-\lambda_{nn}n^{\nu}\sigma_{\nu}^{\mu}
-\delta_{nn}n^{\mu}\theta 
+\lambda_{n\Pi}\Pi\nabla^{\mu}\alpha
-\lambda_{n\pi}\pi^{\mu\nu}\nabla_{\nu}\alpha  \nonumber \\
&-\tau_{n\pi}\pi_{\nu}^{\mu}\dot{u}^{\nu}
+\tau_{n\Pi}\Pi\dot{u}^{\mu}
+\ell_{n\pi}\Delta^{\mu\nu}\partial_{\gamma}\pi_{\nu}^{\gamma}
-\ell_{n\Pi}\nabla^{\mu}\Pi~,  \label{HEATS} \\
\frac{\pi^{\mu\nu}}{\tau_R} =& -\dot{\pi}^{\langle\mu\nu\rangle}
+2\beta_{\pi}\sigma^{\mu\nu}
+2\pi_{\gamma}^{\langle\mu}\omega^{\nu\rangle\gamma}
-\tau_{\pi\pi}\pi_{\gamma}^{\langle\mu}\sigma^{\nu\rangle\gamma}
-\delta_{\pi\pi}\pi^{\mu\nu}\theta 
+\lambda_{\pi\Pi}\Pi\sigma^{\mu\nu}
-\tau_{\pi n}n^{\langle\mu}\dot{u}^{\nu\rangle }  \nonumber \\
&+\lambda_{\pi n}n^{\langle\mu}\nabla ^{\nu\rangle}\alpha
+\ell_{\pi n}\nabla^{\langle\mu}n^{\nu\rangle } ~. \label{SHEARS}
\end{align}
All the coefficients in the above equations have been calculated in 
terms of the thermodynamic variables. In one-dimensional scaling 
expansion of the viscous medium, the evolution of pressure 
anisotropy obtained from solving the second-order equations derived 
here shows reasonably good agreement with those obtained using 
parton cascade BAMPS simulation for relativistic heavy-ion 
collisions \cite{El:2009vjS}. It is also demonstrated that heuristic 
inclusion of higher-order corrections in shear evolution equation 
significantly improves the agreement with transport calculation \cite
{Jaiswal:2013npaS}. This concurrence also suggests that RTA for the 
collision term in BE is reasonably accurate when applied to 
heavy-ion collisions.

%%%%%%%%%%%%%%%%%%%%%%%%%%%%%%%%%%%%%%%%%%%%%%%%%%%%%%%%%%%%%%%%%%%%%%%%

\section*{\Large{6. Effect of viscous corrections on hadronic spectra and Hanbury Brown-Twiss radii}}

In this section, we obtain the form of viscous corrections to the 
distribution function, Eq. (\ref{SOCS}), in terms of the hydrodynamic 
quantities. Further, we study the effect of these corrections on the 
hadronic spectra and Hanbury Brown-Twiss (HBT) radii and compare 
with the results obtained using Grad's 14-moment approximation \cite
{GradS}, Eq. (\ref{G14S}).

For a system of massless particles at vanishing chemical potential, 
Eq. (\ref{SOCS}) can be rewritten up to second order in gradients as 
\cite{Bhalerao:2013pzaS}
\begin{align}
\delta f_1 \!+ \delta f_2 \!=\ & \frac{f_0\beta}{\beta_\pi} \bigg[\ 
\frac{1}{2(u\cdot p)}\left\{ p^\alpha p^\beta \pi_{\alpha\beta}
-2\tau_\pi\, p^\alpha p^\beta \pi^\gamma_\alpha\, \omega_{\beta\gamma}
+\frac{5}{7\beta_\pi}\, p^\alpha p^\beta \pi^\gamma_\alpha\, \pi_{\beta\gamma}
-\frac{2\tau_\pi}{3}\, p^\alpha p^\beta \pi_{\alpha\beta}\theta \right\} \nonumber\\
&-\frac{(u\cdot p)}{70\beta_\pi}\, \pi^{\alpha\beta}\pi_{\alpha\beta}
-\frac{\tau_\pi}{5}\, p^\alpha \left(\nabla^\beta\pi_{\alpha\beta}\right)
+\frac{6\tau_\pi}{5}\, p^\alpha\dot u^\beta\pi_{\alpha\beta}
+\frac{3\tau_\pi}{(u\cdot p)^2}\, p^\alpha p^\beta p^\gamma \pi_{\alpha\beta}\dot u_\gamma \nonumber\\
&+\frac{\beta+(u\cdot p)^{-1}}{4(u\cdot p)^2\beta_\pi}\, p^\alpha p^\beta p^\gamma p^\delta \pi_{\alpha\beta}\pi_{\gamma\delta}
-\frac{\tau_\pi}{2(u\cdot p)^2}\, p^\alpha p^\beta p^\gamma \left(\nabla_\gamma\pi_{\alpha\beta}\right) \bigg].
\label{SOVCS}
\end{align}
The first term on the RHS of the above equation corresponds to the 
first-order correction and the rest are all of second order.

For one-dimensional scaling expansion of the viscous medium, we 
evolve the system using Eqs. (\ref{evoleS}) and (\ref{SHEARS}) up to 
the freeze-out temperature. Subsequently, employing corrections to 
the distribution function from Eq. (\ref{SOVCS}), the particle 
spectra are obtained using Eq. (\ref{hadronsS}) and the HBT radii 
are calculated using the formula
\begin{equation}
R_L^2(K_T) = \frac{\int K_\mu d\Sigma^\mu f(x,K)z^2}{\int K_\mu d\Sigma^\mu f(x,K)}.
\end{equation}

We find that although the effect of the second-order correction is 
small, the effect of viscous corrections on spectra and HBT radii 
using Eq. (\ref{SOVCS}) is considerably different from that using 
Grad's expansion. While Grad's 14-moment approximation results in 
the breakdown of the experimentally observed and ideal hydrodynamic 
prediction of $1/\sqrt{m_T}$ scaling of the HBT radii \cite 
{Teaney:2003kpS}, we show that this scaling can be restored by using 
the form of the non-equilibrium distribution function obtained in 
Eq. (\ref{SOVCS}) \cite{Bhalerao:2013pzaS}. Moreover, while Grad's 
approximation results in imaginary HBT radii for large transverse 
momenta, we find that the form in Eq. (\ref{SOVCS}) is well behaved 
showing convergence at second-order.

%%%%%%%%%%%%%%%%%%%%%%%%%%%%%%%%%%%%%%%%%%%%%%%%%%%%%%%%%%%%%%%%%%%%%%%%

\section*{\Large{7. Third-order dissipative fluid dynamics}}

In Section 5, it was found that a heuristic inclusion of 
higher-order terms in hydrodynamic equations improves the agreement 
with transport calculations. In this section, the treatment of the 
Section 5 is extended to derive a full third-order evolution 
equation of shear stress tensor for the case of massless Boltzmann 
gas, relevant for gluon dominated QGP \cite{Jaiswal:2013vtaS}.

Rewriting the BE in RTA, Eq. (\ref{RBES}) as $\delta\dot f = -\dot 
f_0 - p^\gamma\nabla_\gamma f/(u\cdot p) - \delta f/\tau_R$, the 
evolution of the shear stress tensor can be obtained from Eq. (\ref
{SEES}) as
\begin{equation} 
\dot\pi^{\langle\mu\nu\rangle}  = 
-\Delta^{\mu\nu}_{\alpha\beta}\int dp \, p^\alpha p^\beta\left(\dot f_0 + \frac{1}{u\cdot p}p^\gamma\nabla_\gamma f\right)
- \frac{\pi^{\mu\nu}}{\tau_R}. \label{SOSES}
\end{equation}
For the dissipative equations to be third-order in gradients the 
distribution function in right hand side of Eqs. (\ref{SOSES}) need 
to be computed only till second-order ($f=f_2$), Eq. (\ref{F1F2S}). 
After performing the integrations, the third-order evolution 
equation for shear stress tensor is finally obtained as
\begin{align}
\dot{\pi}^{\langle\mu\nu\rangle} \!=& -\frac{\pi^{\mu\nu}}{\tau_{\pi}}
+2\beta_{\pi}\sigma^{\mu\nu}
+2\pi_{\gamma}^{\langle\mu}\omega^{\nu\rangle\gamma}
-\frac{10}{7}\pi_{\gamma}^{\langle\mu}\sigma^{\nu\rangle\gamma} 
-\frac{4}{3}\pi^{\mu\nu}\theta
+\tau_{\pi}\bigg[\frac{50}{7}\pi^{\rho\langle\mu}\omega^{\nu\rangle\gamma}\sigma_{\rho\gamma}
-\frac{10}{63}\pi^{\mu\nu}\theta^2 \nonumber \\
&-\frac{76}{245}\pi^{\mu\nu}\sigma^{\rho\gamma}\sigma_{\rho\gamma}
-\frac{44}{49}\pi^{\rho\langle\mu}\sigma^{\nu\rangle\gamma}\sigma_{\rho\gamma}
-\frac{2}{7}\pi^{\rho\langle\mu}\omega^{\nu\rangle\gamma}\omega_{\rho\gamma}
-\frac{2}{7}\omega^{\rho\langle\mu}\omega^{\nu\rangle\gamma}\pi_{\rho\gamma}
+\frac{26}{21}\pi_{\gamma}^{\langle\mu}\omega^{\nu\rangle\gamma}\theta \nonumber \\
&-\frac{2}{3}\pi_{\gamma}^{\langle\mu}\sigma^{\nu\rangle\gamma}\theta\bigg]
-\frac{24}{35}\nabla^{\langle\mu}\left(\pi^{\nu\rangle\gamma}\dot u_{\gamma}\tau_\pi\right)
+\frac{6}{7}\nabla_{\gamma}\left(\tau_\pi\dot u^{\gamma}\pi^{\langle\mu\nu\rangle}\right)
+\frac{4}{35}\nabla^{\langle\mu}\left(\tau_\pi\nabla_\gamma\pi^{\nu\rangle\gamma}\right) \nonumber \\
&-\frac{2}{7}\nabla_{\gamma}\left(\tau_\pi\nabla^{\langle\mu}\pi^{\nu\rangle\gamma}\right)
-\frac{1}{7}\nabla_{\gamma}\left(\tau_\pi\nabla^{\gamma}\pi^{\langle\mu\nu\rangle}\right)
+\frac{12}{7}\nabla_{\gamma}\left(\tau_\pi\dot u^{\langle\mu}\pi^{\nu\rangle\gamma}\right) .
\label{TOSHEARS}
\end{align}

In the Bjorken scenario, the results obtained by solving the 
third-order equation derived here show an improved agreement with 
the exact solution of BE compared to second-order results. It is 
also demonstrated that the present derivations shows better 
agreement with the BAMPS \cite{Jaiswal:2013vtaS} compared to an 
alternate third-order derivation from entropy considerations.

%%%%%%%%%%%%%%%%%%%%%%%%%%%%%%%%%%%%%%%%%%%%%%%%%%%%%%%%%%%%%%%%%%%%%%%%

\section*{\Large{8. Nonlocal generalization of the collision term and dissipative fluid dynamics}}

All formulations of second-order dissipative hydrodynamics that 
employ the Boltzmann equation make a strict assumption of local 
collision term in the configuration space. In this section, a formal 
derivation of the dissipative hydrodynamic equations within kinetic 
theory has been presented using a nonlocal collision term in the 
Boltzmann equation \cite{Jaiswal:2012qmS}. New 
second-order terms have been obtained and the coefficients of the 
terms in the widely used traditional IS equations are also altered. 

The starting point of this new derivation is the relativistic 
Boltzmann equation, Eq. (\ref{BES}). Traditionally, the collision 
term $C[f]$ in this equation is assumed to be a purely local 
functional of $f(x,p)$, independent of $\partial_\mu f$. This 
locality assumption is a powerful restriction \cite{Israel:1979wpS} 
which is relaxed by including the gradients of $f(x,p)$ in $C[f]$.
\begin{equation}\label{MBES}
p^\mu \partial_\mu f = C_m[f] 
=  C[f] + \partial_\mu(A^\mu f) + \partial_\mu\partial_\nu(B^{\mu\nu}f) + \cdots,
\end{equation}
The collision term in Eq. (\ref{BES}) assumes that the two processes 
$kk' \to pp'$ and $pp' \to kk'$ occur at the same space-time point 
$x^\mu$. This however is not realistic and a spacetime separation 
$\xi^\mu$ is provided between the two collisions. With this 
viewpoint, the second term in $C[f]$ of Eq. (\ref{BES}) involves 
$f(x-\xi,p)f(x-\xi,p')\tilde f(x-\xi,k)\tilde f(x-\xi,k')$, which on 
Taylor expansion at $x^\mu$ up to second order in $\xi^\mu$, results 
in the modified Boltzmann equation (\ref{MBES}) with
\begin{equation}\label{coeff1S}
A^\mu = \frac{1}{2} \int dp' dk \ dk' \ \xi^\mu W_{pp' \to kk'}
f_{p'} \tilde f_k \tilde f_{k'},\quad 
B^{\mu\nu} = -\frac{1}{4} \int dp' dk \ dk' \ \xi^\mu \xi^\nu W_{pp' \to kk'}
f_{p'} \tilde f_k \tilde f_{k'}.
\end{equation}
The momentum dependence of the coefficients $A^\mu$ and $B^{\mu\nu}$ 
can be made explicit by expressing them in terms of the available 
tensors $p^\mu$ and the metric $g^{\mu\nu} \equiv {\rm 
diag}(1,-1,-1,-1)$ as $A^\mu = ap^\mu$ and $B^{\mu\nu}= 
b_1g^{\mu\nu} + b_2 p^\mu p^\nu$. The coefficients $a$, $b_1$ and 
$b_2$ are functions of $x^\mu$. To constrain $\xi^\mu$, macroscopic 
conservation equations are demanded to hold for $C_m[f]$. 
Conservation of current and energy-momentum implies vanishing zeroth 
and first moments of the collision term $C_m[f]$. Moreover, the 
arbitrariness in $\xi^\mu$ requires that these conditions be 
satisfied at each order in $\xi^\mu$. This leads to three constraint 
equations for the coefficients ($a, b_1, b_2$), namely $\partial_\mu 
a = 0$,
\begin{align}\label{paramS}
\partial^2\left(b_1 \langle1\rangle_0 \right) 
+ \partial_\mu \partial_\nu\left( b_2 \langle p^\mu p^\nu\rangle_0\right) = 0, \quad
u_\alpha \partial_\mu \partial_\nu \left( b_2 \langle p^\mu p^\nu p^\alpha\rangle_0 \right)  
+ u_\alpha \partial^2 \left(b_1 n u^\alpha \right) = 0.
\end{align}

In order to obtain the evolution equations for the dissipative 
quantities, the approach used to derive third-order evolution 
equation in the previous section has been followed. The comoving 
derivative of the dissipative quantities can be written directly 
from their definition, Eqs. (\ref{BEES})-(\ref{SEES}), as
\begin{equation}\label{BE2S}
\dot\Pi = -\frac{\Delta_{\alpha\beta}}{3}\int dp \ 
p^\alpha p^\beta \delta\dot f, \quad
\dot n^{\langle\mu\rangle} =  \Delta^{\mu\nu} \int dp \ p_\nu \delta\dot f, \quad
\dot\pi^{\langle\mu\nu\rangle} = \Delta^{\mu\nu}_{\alpha\beta} \int dp \ 
p^\alpha p^\beta \delta\dot f,
\end{equation}

Writing Eq. (\ref{MBES}) in the form $\delta\dot f = - \dot f_0 - 
(p^\mu\nabla_\mu f - C_m[f])/(u\cdot p)$ and using Grad's 14 moment 
approximation we finally obtain the second-order evolution equations 
for the dissipative quantities as
\begin{align}
\dot\Pi = &~ -\frac{\Pi}{\tau_\Pi'}
- \beta_\Pi' \theta
+ \tau_{\Pi n} n \cdot \dot u - l_{\Pi n} \partial \cdot n
- \delta_{\Pi\Pi} \Pi\theta 
+ \lambda_{\Pi n} n \cdot \nabla \alpha
+ \lambda_{\Pi\pi} \pi_{\mu\nu} \sigma^{\mu\nu} \nonumber\\
& + \Lambda_{\Pi\dot u} \dot u \cdot \dot u
+ \Lambda_{\Pi\omega} \omega_{\mu\nu} \omega^{\nu\mu} + (8 \ {\rm terms}) , \label{bulk1S}\\
\dot n^{\langle\mu\rangle} = &~ -\frac{n^\mu}{\tau_n'}
- \beta_n' \nabla^\mu\alpha
+ \lambda_{nn} n_\nu \omega^{\nu\mu}
- \delta_{nn} n^\mu \theta
+ l_{n \Pi}\nabla^\mu \Pi
- l_{n \pi}\Delta^{\mu\nu} \partial_\gamma \pi^\gamma_\nu
- \tau_{n \Pi} \Pi \dot u^\mu \nonumber \\
&- \tau_{n \pi}\pi^{\mu \nu} \dot u_\nu
+\lambda_{n\pi}n_\nu \pi^{\mu \nu}
+ \lambda_{n \Pi}\Pi n^\mu 
+  \Lambda_{n \dot u} \omega^{\mu \nu} \dot u_\nu
+ \Lambda_{n \omega} \Delta^\mu_\nu \partial_\gamma \omega^{\gamma \nu}
+ (9 \ {\rm terms}), \label{heat1S}\\
\dot\pi^{\langle\mu\nu\rangle} =&~ -\frac{\pi^{\mu\nu}}{\tau_\pi'}
- \beta_\pi' \sigma^{\mu\nu}
+ \tau_{\pi n} n^{\langle\mu}\dot u^{\nu\rangle}
+ l_{\pi n} \nabla^{\langle \mu}n^{\nu\rangle} 
+ \lambda_{\pi\pi} \pi_\rho^{\langle \mu} \omega ^{\nu\rangle \rho}
- \lambda_{\pi n} n^{\langle\mu} \nabla^{\nu\rangle} \alpha
- \tau_{\pi\pi} \pi_\rho^{\langle\mu} \sigma^{\nu\rangle\rho} \nonumber\\
&- \delta_{\pi\pi} \pi^{\mu\nu}\theta
+ \Lambda_{\pi\dot u} \dot u^{\langle \mu} \dot u^{\nu\rangle}
+ \Lambda_{\pi\omega} \omega_\rho^{\langle \mu} \omega^{\nu\rangle\rho}
+ \chi_1 \dot b_2 \pi^{\mu\nu}
+ \chi_2 \dot u^{\langle \mu} \nabla^{\nu\rangle} b_2
+ \chi_3 \nabla^{\langle \mu} \nabla^{\nu\rangle} b_2. \label{shear1S}
\end{align}
The ``8 terms" (``9 terms'') involve second-order, linear scalar 
(vector) combinations of derivatives of $b_1,b_2$. Within 
one-dimensional scaling expansion, the solution of the above equation 
with small initial corrections due to $a,\ b_1,\ b_2$, (nonlocal 
hydrodynamics) exhibits appreciable deviation from the local theory 
\cite{Jaiswal:2012qmS}. This clearly demonstrate the 
importance of the nonlocal effects, which should be incorporated in 
transport calculations as well.

%%%%%%%%%%%%%%%%%%%%%%%%%%%%%%%%%%%%%%%%%%%%%%%%%%%%%%%%%%%%%%%%%%%%%%%%

\section*{\Large{9. Summary}}

This synopsis provides an outline of theoretical formulations of 
relativistic dissipative fluid dynamics from various approaches. 
Several longstanding problems in the formulation as well as in the 
application of relativistic hydrodynamics relevant to heavy-ion 
collisions have been addressed here. The evolution equations for the 
dissipative quantities along with the second-order transport 
coefficients have been derived using the second law of 
thermodynamics within a single theoretical framework. In particular, 
the problem pertaining to the relaxation time for the evolution of 
bulk viscous pressure has been solved here. Subsequently, using the 
same method for two different forms of non-equilibrium 
single-particle distribution functions, viscous evolution equations 
have been derived and applied to study the particle production and 
transverse momentum spectra of hadrons and thermal dileptons. 

An alternate formulation of second-order dissipative hydrodynamics 
has been outlined in which iterative solution of the Boltzmann 
equation for non-equilibrium distribution function is employed 
instead of the 14-moment ansatz most commonly used in the 
literature. The equations for the dissipative quantities have been 
obtained directly from their definitions rather than an arbitrary 
moment of Boltzmann equation in the traditional Israel-Stewart 
formulation. Using the iterative solution of Boltzmann equation, the 
form of second-order viscous correction to the distribution function 
has been derived. The effects of these corrections on particle 
spectra and HBT radii are compared to those due to 14-moment ansatz. 
This method has been further extended to obtain third-order 
evolution equation for shear stress tensor. 

Finally, the collision term in the Boltzmann equation corresponding 
to $2\to2$ elastic collisions has been modified to include the 
gradients of the distribution function. This non-local collision 
term has then been used to derive second-order evolution equations 
for the dissipative quantities. The numerical significance of these 
new formulations has been demonstrated within the framework of 
one-dimensional boost-invariant Bjorken expansion of the matter 
formed in relativistic heavy-ion collisions.

%%%%%%%%%%%%%%%%%%%%%%%%%%%%%%%%%%%%%%%%%%%%%%%%%%%%%%%%%%%%%%%%%%%%%%%%

\begin{singlespace*}

\end{singlespace*}

\newpage

%%%%%%%%%%%%%%%%%%%%%%%%%%%%%%%%%%%%%%%%%%%%%%%%%%%%%%%%%%%%%%%%%%%%%%%%

\begin{center}
\Large{\bf LIST OF PUBLICATIONS}
\end{center}
\vspace{0.1cm}
\section*{\Large{Publications contributing to this thesis}}
\begin{enumerate}
\item
	Rajeev S. Bhalerao, \underline{Amaresh Jaiswal}, Subrata Pal, 
	and V. Sreekanth, {\it ``Relativistic viscous hydrodynamics for 
	heavy-ion collisions: A comparison between Chapman-Enskog and 
	Grad's methods"} {\bf Phys.\ Rev.\ C 89}, 054903 (2014) 
	[arXiv:1312.1864].
	 
\item
	Rajeev S. Bhalerao, \underline{Amaresh Jaiswal}, Subrata Pal, 
	and V. Sreekanth, {\it ``Particle production in relativistic 
	heavy-ion collisions: A consistent hydrodynamic approach"}, {\bf 
	Phys.\ Rev.\ C 88}, 044911 (2013) [arXiv:1305.4146].

\item
	\underline{Amaresh Jaiswal}, {\it ``Relativistic third-order 
	dissipative fluid dynamics from kinetic theory"}, {\bf Phys.\ 
	Rev.\ C 88}, 021903(R) (2013) [arXiv:1305.3480].
	 
\item
	\underline{Amaresh Jaiswal}, {\it ``Relativistic dissipative 
	hydrodynamics from kinetic theory with relaxation-time 
	approximation"}, {\bf Phys.\ Rev.\ C 87}, 051901(R) (2013) 
	[arXiv:1302.6311].
	 
\item
	 \underline{Amaresh Jaiswal}, Rajeev S. Bhalerao, and Subrata 
	 Pal, {\it ``Complete relativistic second-order dissipative 
	 hydrodynamics from the entropy principle"}, {\bf Phys.\ Rev.\ C 
	 87}, 021901(R) (2013) [arXiv:1302.0666].
	 
\item
	\underline{Amaresh Jaiswal}, Rajeev S. Bhalerao, and Subrata 
	Pal, {\it ``New relativistic dissipative fluid dynamics from 
	kinetic theory"}, {\bf Phys.\ Lett.\ B 720}, 347 (2013) 
	[arXiv:1204.3779].
\end{enumerate}
\vspace{-0.1cm}
\section*{\Large{Publications not contributing to this thesis}}
\begin{enumerate}
\item
	\underline{Amaresh Jaiswal}, Radoslaw Ryblewski, and Michael 
	Strickland, {\it ``Transport coefficients for bulk viscous 
	evolution in the relaxation time approximation"}, Submitted to 
	{\bf Phys.\ Rev.\ C} (2014) [arXiv:1407.7231].
\end{enumerate}
\vspace{-0.1cm}
\section*{\Large{Conference proceedings}}
\begin{enumerate}
\item
	\underline{Amaresh Jaiswal}, {\it ``Relaxation-time 
	approximation and relativistic viscous hydrodynamics from 
	kinetic theory"}, To appear in {\bf Nucl.\ Phys.\ A}, 
	Proceedings of the XXIV International Conference on 
	Ultrarelativistic Nucleus-Nucleus Collisions, Quark-Matter 2014, 
	[arXiv:1407.0837].

\item
	\underline{Amaresh Jaiswal}, {\it ``Relativistic third-order 
	viscous hydrodynamics"}, To appear in {\bf Proceedings of the 
	International Conference on Matter at Extreme Conditions : Then 
	\& Now} (2014).
	
\item
	\underline{Amaresh Jaiswal}, Rajeev S. Bhalerao, and Subrata 
	Pal, {\it ``Boltzmann H-theorem and relativistic second-order 
	dissipative hydrodynamics"}, {\bf Proceedings of the DAE Symp. 
	on Nucl. Phys. 58} (2013) pp. 684-685.
	
\item
	\underline{Amaresh Jaiswal}, Rajeev S. Bhalerao and Subrata Pal, 
	{\it ``Boltzmann equation with a nonlocal collision term and the 
	resultant dissipative fluid dynamics"}, {\bf J.\ Phys.\ Conf.\ 
	Ser.\  422}, 012003 (2013) [arXiv:1210.8427].

\item
	\underline{Amaresh Jaiswal}, Rajeev S. Bhalerao, and Subrata 
	Pal, {\it ``New derivation of relativistic dissipative fluid 
	dynamics"}, {\bf Proceedings of the DAE Symp. on Nucl. Phys. 57} 
	(2012) pp. 760-761.
	 
\item
	\underline{Amaresh Jaiswal}, Rajeev S. Bhalerao, and Subrata 
	Pal, {\it ``Relativistic hydrodynamics from Boltzmann equation 
	with modified collision term"}, {\bf Proceedings of the  QGP 
	Meet 2012}, Narosa Publication, New Delhi, India 
	[arXiv:1303.1892].
\end{enumerate}
\end{synopsis}

%% file: Chapter1.tex
%#######################################################################
\chapter{Introduction}
%#######################################################################

%The study of the extremes of temperature and density has always been 
%the primary focus of modern physics. 
Nuclear physics is the branch of modern physics that deals with the 
study of the constituents and interactions of atomic nuclei. Much of 
current research in high energy nuclear physics relates to the study 
of nuclei under extreme conditions of temperature and density. 
Investigation of the thermodynamic and transport properties of the 
nuclear matter at extremely high temperatures (trillions of Kelvin, 
million times hotter than the core of the sun) and high densities 
(quadrillion times that of water) has gained widespread interest and 
is a topic of extensive research in recent times, see \cite 
{Rischke:2003mt} and references therein.

The nucleus of an atom is made up of nucleons, i.e., neutrons and 
protons, which belong to a larger group of particles collectively 
known as hadrons. Hadrons interact among themselves through strong 
force and constitute the building blocks of all known nuclear 
matter. In the early 1930's, the only hadrons measured 
experimentally were the neutrons and protons. They were considered 
to be elementary particles which interacted by exchange of force 
carriers called pions \cite{Yukawa}. Experimentally, pions were 
detected later in 1947 by Lattes {\it et al.} \cite{Lattes}. However, 
during the next decade, multitude of new hadrons were discovered 
which led to the conclusion that they could not be all elementary 
particles, but instead, should have an inner substructure. In 1968, 
deep inelastic scattering experiments performed at the Stanford 
Linear Accelerator Center (SLAC) located at California, USA, 
conclusively proved that the proton was not an elementary particle 
\cite{Bloom,Breidenbach}, but appeared to be made up of point-like 
particles, originally called partons by Feynman \cite{Feynman}.

It is now well established that Quantum Chromodynamics (QCD) is the 
fundamental theory of strong interactions. According to QCD, hadrons 
can be described in terms of elementary particles called quarks and 
gluons \cite{Gell-Mann}. Gluons are the mediators of the strong 
force in a way similar to photons that are the mediators of the 
electromagnetic force. Analogous to the electric charge carried by 
electromagnetically interacting particles, strongly interacting 
objects also carry a so-called color charge, or simply color. 
However, in contrast to electromagnetic quantum field theory where 
there is only one charge namely the electric charge, there are three 
colors in QCD \cite{Greenberg,Han,Swartz:1995hc}. This property of 
QCD allows gluons to interact with each other, unlike photons. 
Additionally, QCD enjoys two other very interesting properties \cite 
{Gross:1973id,Politzer:1973fx,Peskin}: 
\begin{enumerate}
\item {\bf Confinement}: It is a property of QCD that does not allow 
particles with a color charge to exist as an asymptotic state. In 
other words, it is the phenomenon that color charged particles (such 
as quarks) cannot be isolated, and therefore cannot be directly 
observed. Therefore in the vacuum, quarks must always combine to 
form colorless bound states, i.e., hadrons. Within the framework of 
QCD, all the hadrons observed experimentally so far can be described 
as bound states formed by quarks.
\item {\bf Asymptotic freedom}: This property of QCD causes 
interactions between quarks and gluons to become asymptotically 
weaker as energy increases and distance decreases. This implies that 
at very high energies, quarks and gluons should behave as almost 
free particles. Therefore at very high energies, the QCD matter can 
be treated as a weakly coupled system and approximative schemes like 
perturbation theory become applicable.
\end{enumerate}

The existence of both confinement and asymptotic freedom has led to 
many speculations about the thermodynamic and transport properties 
of QCD. Due to confinement, the nuclear matter must be made of 
hadrons at low energies, hence it is expected to behave as a weakly 
interacting gas of hadrons. On the other hand, at very high energies 
asymptotic freedom implies that quarks and gluons interact only 
weakly and the nuclear matter is expected to behave as a weakly 
coupled gas of quarks and gluons. In between these two 
configurations there must be a phase transition where the hadronic 
degrees of freedom disappear and a new state of matter, in which the 
quark and gluon degrees of freedom manifest directly over a certain 
volume, is formed. This new phase of matter, referred to as 
Quark-Gluon Plasma (QGP), is expected to be created when 
sufficiently high temperatures or densities are reached \cite 
{Lee,Collins:1974ky}.

The QGP is believed to have existed in the very early universe (a 
few microseconds after the Big Bang), or some variant of which 
possibly still exists in the inner core of a neutron star where it 
is estimated that the density can reach values ten times higher than 
those of ordinary nuclei. It was conjectured theoretically that such 
extreme conditions can also be realized on earth, in a controlled 
experimental environment, by colliding two heavy nuclei with 
ultra-relativistic energies \cite{Baumgardt:1975qv}. This may 
transform a fraction of the kinetic energies of the two colliding 
nuclei into heating the QCD vacuum within an extremely small volume 
where temperatures million times hotter than the core of the sun may 
be achieved. 

%%%%%%%%%%%%%%%%%%%%%%%%%%%%%%%%%%%%%%%%%%%%%%%%%%
\begin{figure}[t]
\begin{center}
\includegraphics[scale=0.7]{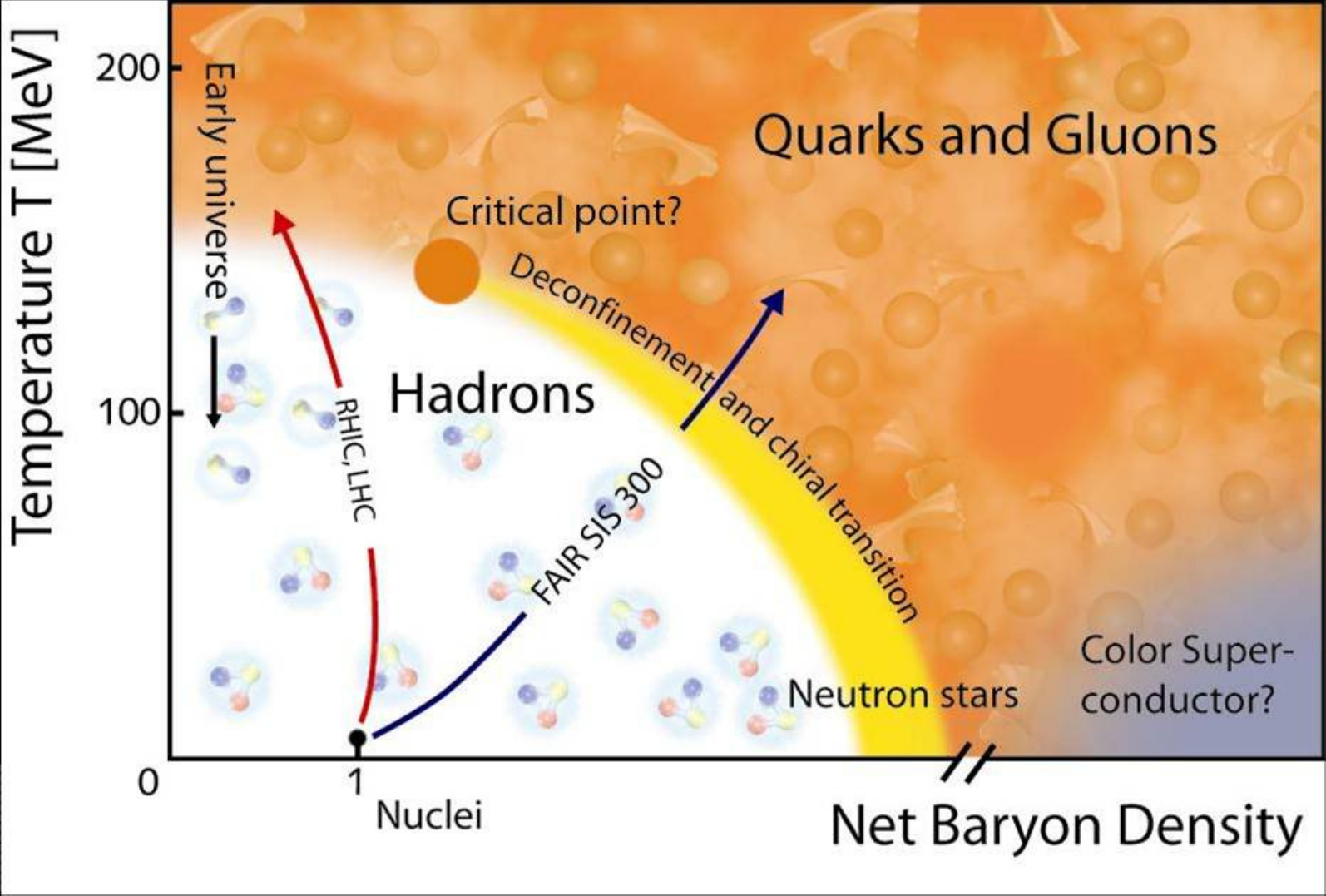}
\end{center}
\vspace{-0.4cm} 
\caption[Illustration of the QCD phase diagram]{Schematic phase 
diagram of the QCD matter. The net baryon density on x-axis is 
normalized to that of the normal nuclear matter \cite{gsi.de}.}
\label{Phases}
\end{figure}
%%%%%%%%%%%%%%%%%%%%%%%%%%%%%%%%%%%%%%%%%%%%%%%%%%

With the advent of modern accelerator facilities, ultra-relativistic 
heavy-ion collisions have provided an opportunity to systematically 
create and study different phases of the bulk nuclear matter. It is 
widely believed that the QGP phase is formed in heavy-ion collision 
experiments at Relativistic Heavy-Ion Collider (RHIC) located at 
Brookhaven National Laboratory, USA and Large Hadron Collider (LHC) 
at European Organization for Nuclear Research (CERN), Geneva. A 
number of indirect evidences found at the Super Proton Synchrotron 
(SPS) at CERN, strongly suggested the formation of a {\it ``new 
state of matter"} \cite{CERN}, but quantitative and clear results 
were only obtained at RHIC energies \cite{Tannenbaum:2006ch,Kolb,Gyulassy, 
Tomasik,Muller,Arsene:2004fa,Adcox:2004mh,Back:2004je,Adams:2005dq}, 
and recently at LHC energies \cite{Aamodt:2010pa,Aamodt:2010pb, 
Aamodt:2010cz,ALICE:2011ab}. The regime with relatively large baryon 
chemical potential will be probed by the upcoming experimental 
facilities like Facility for Anti-proton and Ion Research (FAIR) at 
GSI, Darmstadt. An illustration of the QCD phase diagram and the 
regions probed by these experimental facilities is shown in Fig. 
\ref{Phases} \cite{gsi.de}.

\begin{figure}[t]
\begin{center}
\includegraphics[scale=0.7]{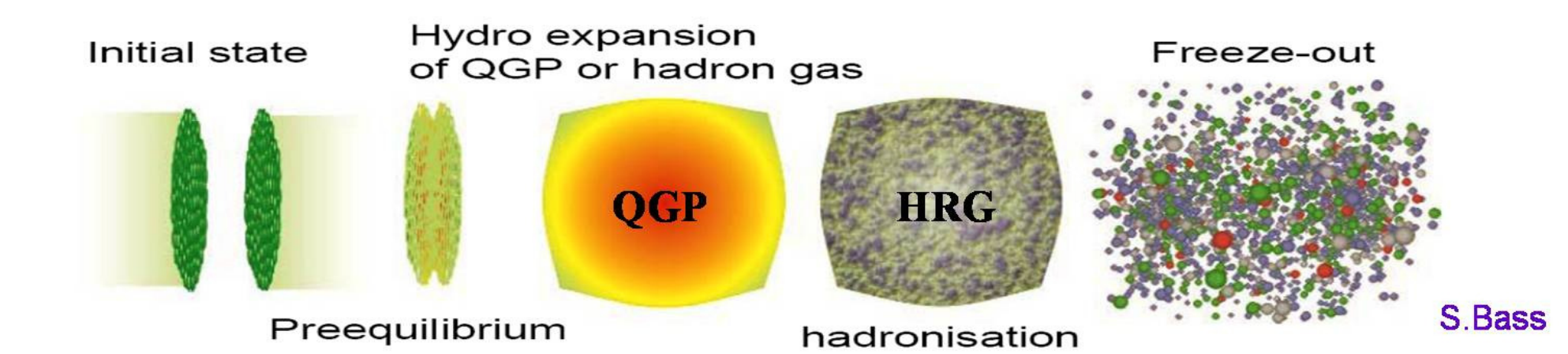}
\end{center}
\vspace{-0.4cm} 
\caption[Various stages of relativistic heavy ion collisions]
{Various stages of ultra-relativistic heavy ion collisions \cite
{duke.edu}.}
\label{Stages}
\end{figure}

It is possible to create hot and dense nuclear matter with very high 
energy densities in relatively large volumes by colliding 
ultra-relativistic heavy ions. In these conditions, the nuclear 
matter created may be close to (local) thermodynamic equilibrium, 
providing the opportunity to investigate the various phases and the 
thermodynamic and transport properties of QCD. It is important to 
note that, even though it appears that a deconfined state of matter 
is formed in these colliders, investigating and extracting the 
transport properties of QGP from heavy-ion collisions is not an easy 
task since it cannot be observed directly. Experimentally, it is 
only feasible to measure energy and momenta of the particles 
produced in the final stages of the collision, when nuclear matter 
is already relatively cold and non-interacting. Hence, in order to 
study the thermodynamic and transport properties of the QGP, the 
whole heavy ion collision process from the very beginning till the 
end has to be modelled: starting from the stage where two highly 
Lorentz contracted heavy nuclei collide with each other, the 
formation and thermalization of the QGP or de-confined phase in the 
initial stages of the collision, its subsequent space-time 
evolution, the phase transition to the hadronic or confined phase of 
matter, and eventually, the dynamics of the cold hadronic matter 
formed in the final stages of the collision. The different stages of 
ultra-relativistic heavy ion collisions are schematically 
illustrated in Fig. \ref{Stages} \cite{duke.edu}.

%%%%%%%%%%%%%%%%%%%%%%%%%%%%%%%%%%%%%%%%%%%%%%%%%%
%%%%%%%%%%%%%%%%%%%%%%%%%%%%%%%%%%%%%%%%%%%%%%%%%%

Assuming that thermalization is achieved in the early stages of 
heavy-ion collisions and that the interaction between the quarks is 
strong enough to maintain local thermodynamic equilibrium during the 
subsequent expansion, the time evolution of the QGP and hadronic 
matter can be described by the laws of fluid dynamics \cite 
{Stoecker:1986ci,Rischke:1995ir,Rischke:1995mt,Shuryak:2003xe}. 
Fluid dynamics, also loosely referred to as hydrodynamics, is an 
effective approach through which a system can be described by 
macroscopic variables, such as local energy density, pressure, 
temperature and flow velocity. Application of viscous hydrodynamics 
to high-energy heavy-ion collisions has evoked widespread interest 
ever since a surprisingly small value for the shear viscosity to 
entropy density ratio $\eta/s$ was estimated from the analysis of 
the elliptic flow data \cite{Romatschke:2007mq}. Indeed the 
estimated $\eta/s$ was close to the conjectured lower bound 
$\eta/s|_{\rm KSS} = 1/4\pi$ from ADS/CFT\cite 
{Policastro:2001yc,Kovtun:2004de}. This led to the claim that the 
QGP formed at RHIC was the most perfect fluid ever observed. A 
precise estimate of $\eta/s$ is vital to the understanding of the 
properties of the QCD matter and is presently a topic of intense 
investigation, see \cite{Chaudhuri} and references therein.

%%%%%%%%%%%%%%%%%%%%%%%%%%%%%%%%%%%%%%%%%%%%%%%%%%%%%%%%%%%%%%%%%%%%%%%%

\section{Relativistic fluid dynamics}

The physical characterization of a system consisting of many degrees 
of freedom is in general a non-trivial task. For instance, the 
mathematical formulation of a theory describing the microscopic 
dynamics of a system containing a large number of interacting 
particles is one of the most challenging problems of theoretical 
physics. However, it is possible to provide an effective macroscopic 
description, over large distance and time scales, by taking into 
account only the degrees of freedom that are relevant at these 
scales. This is a consequence of the fact that on macroscopic 
distance and time scales the actual degrees of freedom of the 
microscopic theory are imperceptible. Most of the microscopic 
variables fluctuate rapidly in space and time, hence only average 
quantities resulting from the interactions at the microscopic level 
can be observed on macroscopic scales. These rapid fluctuations lead 
to very small changes of the average values, and hence are not 
expected to contribute to the macroscopic dynamics. On the other 
hand, variables that do vary slowly, such as the conserved 
quantities, are expected to play an important role in the effective 
description of the system. Fluid dynamics is one of the most common 
examples of such a situation. It is an effective theory describing 
the long-wavelength, low frequency limit of the underlying 
microscopic dynamics of a system. 

A fluid is defined as a continuous system in which every 
infinitesimal volume element is assumed to be close to thermodynamic 
equilibrium and to remain near equilibrium throughout its evolution. 
Hence, in other words, in the  neighbourhood of each point in space, 
an infinitesimal volume called fluid element is defined in which the 
matter is assumed to be homogeneous, i.e., any spatial gradients can 
be ignored, and is described by a finite number of thermodynamic 
variables. This implies that each fluid element must be large 
enough, compared to the microscopic distance scales, to guarantee 
the proximity to thermodynamic equilibrium, and, at the same time, 
must be small enough, relative to the macroscopic distance scales, 
to ensure the continuum limit. The co-existence of both continuous 
(zero volume) and thermodynamic (infinite volume) limits within a 
fluid volume might seem paradoxical at first glance. However, if the 
microscopic and the macroscopic length scales of the system are 
sufficiently far apart, it is always possible to establish the 
existence of a volume that is small enough compared to the 
macroscopic scales, and at the same time, large enough compared to 
the microscopic ones. In this thesis, we will assume the existence 
of a clear separation between microscopic and macroscopic scales to 
guarantee the proximity to local thermodynamic equilibrium. 

Relativistic fluid dynamics has been quite successful in explaining 
the various collective phenomena observed in astrophysics, cosmology 
and the physics of high-energy heavy-ion collisions. In cosmology 
and certain areas of astrophysics, one needs a fluid dynamics 
formulation consistent with the General Theory of Relativity \cite 
{Ibanez}. On the other hand, a formulation based on the Special 
Theory of Relativity is quite adequate to treat the evolution of the 
strongly interacting matter formed in high-energy heavy-ion 
collisions when it is close to a local thermodynamic equilibrium. In 
fluid dynamical approach, although no detailed knowledge of the 
microscopic dynamics is needed, however, knowledge of the equation 
of state relating pressure, energy density and baryon density is 
required. The collective behaviour of the hot and dense quark-gluon 
plasma created in ultra-relativistic heavy-ion collisions has been 
studied quite extensively within the framework of relativistic fluid 
dynamics. In application of fluid dynamics, it is natural to first 
employ the simplest version which is ideal hydrodynamics [26, 27] 
which neglects the viscous effects and assumes that local 
equilibrium is always perfectly maintained during the fireball 
expansion. Microscopically, this requires that the microscopic 
scattering time be much shorter than the macroscopic expansion 
(evolution) time. In other words, ideal hydrodynamics assumes that 
the mean free path of the constituent particles is much smaller than 
the system size. However, as all fluids are dissipative in nature 
due to the quantum mechanical uncertainty principle \cite 
{Danielewicz:1984ww}, the ideal fluid results serve only as a 
benchmark when dissipative effects become important.

When discussing the application of relativistic dissipative fluid 
dynamics to heavy-ion collision, one is faced with yet another 
predicament: the theory of relativistic dissipative fluid dynamics 
is not yet conclusively established. In fact, introducing 
dissipation in relativistic fluids is not at all a trivial task and 
still remains one of the important topics of research in high-energy 
physics. Therefore, in order to quantify the transport properties of 
the QGP from experiment and confirm the claim that it is indeed the 
most perfect fluid ever created, the theoretical foundations of 
relativistic dissipative fluid dynamics must be first addressed and 
clearly understood.

%%%%%%%%%%%%%%%%%%%%%%%%%%%%%%%%%%%%%%%%%%%%%%%%%%%%%%%%%%%%%%%%%%%%%%%%

\section{Problems in relativistic dissipative fluid dynamics}

Ideal hydrodynamics assumes that local thermodynamic equilibrium is 
perfectly maintained and each fluid element is homogeneous, i.e., 
spatial gradients are absent (zeroth order in gradient expansion). 
If this is not satisfied, dissipative effects come into play. The 
earliest theoretical formulations of relativistic dissipative 
hydrodynamics also known as first-order theories, are due to Eckart 
\cite{Eckart:1940zz} and Landau-Lifshitz \cite{Landau}. However, 
these formulations, collectively called relativistic Navier-Stokes 
(NS) theory, suffer from acausality and numerical instability. The 
reason for the acausality is that in the gradient expansion the 
dissipative currents are linearly proportional to gradients of 
temperature, chemical potential, and velocity, resulting in 
parabolic equations. Thus, in Navier-Stokes theory the gradients 
have an instantaneous influence on the dissipative currents. Such 
instantaneous effects tend to violate causality and cannot be 
allowed in a covariant setup, leading to the instabilities 
investigated in Refs. \cite{Hiscock:1983zz,Hiscock:1985zz,Hiscock:1987zz}.

The second-order Israel-Stewart (IS) theory \cite{Israel:1979wp}, 
restores causality but may not guarantee stability \cite 
{Huovinen:2008te}. The acausality problems were solved by 
introducing a time delay in the creation of the dissipative currents 
from gradients of the fluid-dynamical variables. In this case, the 
dissipative quantities become independent dynamical variables 
obeying equations of motion that describe their relaxation towards 
their respective Navier-Stokes values. The resulting equations are 
hyperbolic in nature which preserves causality. Israel-Stewart 
theory has been widely applied to ultra-relativistic heavy-ion 
collisions in order to describe the time evolution of the QGP and 
the subsequent freeze-out process of the hadron resonance gas.

Hydrodynamic analysis of the spectra and azimuthal anisotropy of 
particles produced in heavy-ion collisions at RHIC \cite 
{Romatschke:2007mq,Song:2010mg} and recently at LHC \cite 
{Luzum:2010ag,Qiu:2011hf} suggests that the matter formed in these 
collisions is strongly-coupled quark-gluon plasma (sQGP). Although 
IS hydrodynamics has been quite successful in modelling relativistic 
heavy ion collisions, there are several inconsistencies and 
approximations in its formulation which prevent proper understanding 
of the thermodynamic and transport properties of the QGP. The 
standard derivation of IS equations using the second-law of 
thermodynamics contains unknown transport coefficients related to 
relaxation times of the dissipative quantities viz., the bulk 
viscous pressure, the particle diffusion current and the shear 
stress tensor \cite{Israel:1979wp}. While IS equations derived from 
kinetic theory can provide reliable values for the shear relaxation 
time ($\tau_\pi$), the bulk relaxation time ($\tau_\Pi$) still 
remains ambiguous. 

Israel and Stewart's derivation of second-order hydrodynamics from 
kinetic theory relies on two additional approximations and 
assumptions: 
\begin{enumerate}
\item {\bf Grad's 14-moment approximation}: For small departures 
from equilibrium, the single-particle distribution function is 
obtained by using a truncated expansion in a Taylor-like series in 
powers of particle four-momenta $p^\mu$  \cite{Israel:1979wp,Grad}. 
This approximation contains fourteen dynamic variables hence the 
name 14-moment approximation. Here it is implicitly assumed that the 
power series in momenta is convergent and is truncated at quadratic 
order.
\item {\bf Choice of second moment of the Boltzmann equation}: In a 
theory with conserved charges the integral over momenta (or zeroth 
moment) of the Boltzmann equation (BE) leads to conservation of 
charge current. The first moment of the BE, i.e., momentum integral 
of the BE weighted with $p^\mu$, gives the conservation of the 
energy-momentum tensor. The derivation of second-order fluid 
dynamics from kinetic theory by Israel and Stewart is based on the 
assumption that the second moment of BE must contain information 
about the non-equilibrium (or dissipative) dynamics of the system 
\cite{Israel:1979wp,Baier:2006um}. This choice is arbitrary in the 
sense that higher moments of BE combined with the 14-moment 
approximation lead to different evolution equations for the 
dissipative quantities.
\end{enumerate}

Apart from these problems in the formulation, IS theory suffers from 
several other shortcomings. In one-dimensional Bjorken scaling 
expansion \cite{Bjorken:1982qr}, IS theory leads to negative 
longitudinal pressure \cite{Martinez:2009mf,Rajagopal:2009yw} which 
limits its application within a certain temperature range. Further, 
the scaling solutions of IS equations when compared with transport 
results show disagreement for shear viscosity to entropy density 
ratio, $\eta/s>0.5$ indicating the breakdown of the second-order 
theory \cite{Huovinen:2008te,El:2008yy}. Moreover, in the study of 
identical particle pair-correlations, the experimentally observed 
$1/\sqrt{m_T}$ scaling of the Hanburry Brown-Twiss (HBT) radii ($m_T$
being the transverse mass of the hadron pair), which is also 
predicted by the ideal hydrodynamics, is broken when viscous 
corrections to the distribution function are included \cite 
{Teaney:2003kp}. The correct formulation of the relativistic 
dissipative fluid dynamics is thus far from settled and is currently 
under intense investigation \cite {Huovinen:2008te,Baier:2006um, 
El:2008yy,Betz:2008me,El:2009vj,Martinez:2010sc,Denicol:2010xn, 
Denicol:2012cn,Jaiswal:2014isa}.

In this thesis, we report on some major progress we have made in the 
formulation of relativistic dissipative fluid dynamics within the 
framework of kinetic theory. The problem pertaining to the bulk 
pressure relaxation time, $\tau_\Pi$, has been solved by considering 
entropy four-current defined using Boltzmann H-function \cite 
{Jaiswal:2013fc}. Using this method, hydrodynamic evolution, 
production of thermal dileptons and subsequent hadronization of the 
strongly interacting matter have been studied \cite 
{Bhalerao:2013aha}. An alternate derivation of the dissipative 
equations, which does not make use of the 14-moment approximation as 
well as the second moment of BE, has also been presented \cite 
{Jaiswal:2013npa}. The form of viscous corrections to the 
distribution function is derived up to second-order in gradients 
which restores the observed $1/\sqrt{m_T}$ scaling of the HBT radii 
\cite{Bhalerao:2013pza}. Finally, with the motivation to improve the 
IS theory beyond its present scope, two rigorous investigations have 
been presented in this thesis: (a) Derivation of a novel third-order 
evolution equation for shear stress tensor \cite{Jaiswal:2013vta, 
Jaiswal:2014raa}, and (b) Derivation of second-order dissipative 
equations from the BE where the collision term is modified to 
include non-local effects \cite {Jaiswal:2012qm,Jaiswal:2012dd, 
Jaiswal:2013jja}.

%%%%%%%%%%%%%%%%%%%%%%%%%%%%%%%%%%%%%%%%%%%%%%%%%%%%%%%%%%%%%%%%%%%%%%%%

\section{Organization of the thesis} 

The derivation of a relativistic fluid-dynamical theory consistent 
with causality, which is applicable to the physics of 
ultra-relativistic heavy-ion collisions, is the main purpose of this 
thesis. This thesis is organized in the following manner: 

In Chapter 2, we review relativistic fluid dynamics from a 
phenomenological perspective. We start by deriving the equations of 
motion of an ideal relativistic fluid and introduce dissipation in a 
phenomenological manner. Next, the equations of relativistic 
Navier-Stokes theory are derived via the second law of 
thermodynamics, and then subsequently extended to Israel-Stewart 
theory. Then we briefly discuss relativistic kinetic theory and 
express various hydrodynamic quantities in terms of single-particle, 
phase-space distribution function. Finally, this chapter concludes 
with a discussion about the evolution of the phase-space 
distribution function via Boltzmann equation.

In Chapter 3, we present a derivation of relativistic dissipative 
hydrodynamic equations, which invokes the second law of 
thermodynamics for the entropy four-current expressed in terms of 
the single-particle phase-space distribution function obtained from 
Grad's 14-moment approximation. In this derivation all the 
second-order transport coefficients are uniquely determined within a 
single theoretical framework. In particular, this removes the 
long-standing ambiguity in the relaxation time for bulk viscous 
pressure. We find that in the one-dimensional scaling expansion, 
these transport coefficients prevent the occurrence of cavitation 
(negative pressure) even for rather large values of the bulk 
viscosity estimated in lattice QCD. 

In Chapter 4, using the derivation methodology of Chapter 3, we 
derive relativistic viscous hydrodynamic equations for two different 
forms of the non-equilibrium single-particle distribution function. 
These equations are used to study thermal dilepton and hadron 
spectra within longitudinal scaling expansion of the matter formed 
in relativistic heavy-ion collisions. We observe that an 
inconsistent treatment of the nonequilibrium effects influences the 
particle production significantly.

In Chapter 5, starting from the Boltzmann equation with the 
relaxation-time approximation for the collision term and using 
Chapman-Enskog like expansion for distribution function close to 
equilibrium, we derive hydrodynamic evolution equations for the 
dissipative quantities directly from their definitions. This 
derivation does not make use of the two major 
approximations/assumptions namely, Grad's 14-moment approximation 
and second moment of BE, inherent in IS theory. In the case of 
one-dimensional scaling expansion, we demonstrate that our results 
are in better agreement with numerical solution of Boltzmann 
equation as compared to Israel-Stewart results and also show that 
including approximate higher-order corrections in viscous evolution 
significantly improves this agreement. 

In Chapter 6, we derive the form of viscous corrections to the 
distribution function up to second-order in gradients by employing 
iterative solution of Boltzmann equation in relaxation time 
approximation. Within one dimensional scaling expansion, we 
demonstrate that while Grad's 14-moment approximation leads to the 
violation of the observed $1/\sqrt{m_T}$ scaling of HBT radii, the 
viscous corrections obtained here does not exhibit such unphysical 
behaviour.

In Chapter 7, we present the derivation of a novel third-order 
hydrodynamic evolution equation for shear stress tensor from kinetic 
theory. We quantify the significance of the new derivation within 
one-dimensional scaling expansion and demonstrate that the results 
obtained using third-order viscous equations derived here provide a 
very good approximation to the exact solution of Boltzmann equation 
in relaxation time approximation. We also show that our results are 
in better agreement with transport results when compared with an 
existing third-order calculation based on the second-law of 
thermodynamics. 

In Chapter 8, starting with the relativistic Boltzmann equation 
where the collision term is generalized to include nonlocal effects 
via gradients of the phase-space distribution function, and using 
Grad's 14-moment approximation for the distribution function, we 
derive equations for the relativistic dissipative fluid dynamics. 
This method generates all the second-order terms that are allowed by 
symmetry, some of which have been missed by the traditional 
approaches based on the 14-moment approximation. We find that 
nonlocality of the collision term has a rather strong influence on 
the evolution of the viscous medium via hydrodynamic equations.

Finally, in Chapter 9 we summarize our results and also discuss the 
future perspectives for further studies.

%%%%%%%%%%%%%%%%%%%%%%%%%%%%%%%%%%%%%%%%%%%%%%%%%%%%%%%%%%%%%%%%%%%%%%%%

\section{Conventions and notations used} 

In this thesis, unless stated otherwise, all physical quantities are 
expressed in terms of natural units, where, $\hbar = c = k_{\rm B} = 
1$, with $\hbar=h/2\pi$ where $h$ is the Planck constant, $c$ the 
velocity of light, and $k_{\rm B}$ the Boltzmann constant. Unless 
stated otherwise, the spacetime is always taken to be Minkowskian 
where the metric tensor is given by 
$g_{\mu\nu}=\mbox{diag}(+1,-1,-1,-1)$. Apart from Minkowskian 
coordinates $x^\mu=(t,x,y,z)$, we will also regularly employ Milne 
coordinate system $x^\mu=(\tau,x,y,\eta_s)$ or 
$x^\mu=(\tau,r,\varphi,\eta_s)$, with proper time 
$\tau=\sqrt{t^2-z^2}$, the radial coordinate $r=\sqrt{x^2+y^2}$, the 
azimuthal angle $\varphi=\tan^{-1}(y/x)$, and spacetime rapidity 
$\eta_s=\tanh^{-1}(z/t)$. Hence, $t=\tau\cosh\eta_s$, 
$x=r\cos\varphi$, $y=r\sin\varphi$, and $z=\tau\sinh\eta_s$. For the 
coordinate system $x^\mu=(\tau,x,y,\eta_s)$, the metric becomes 
$g_{\mu\nu}=\mbox{diag}(1,-1,-1,-\tau^2)$, whereas for 
$x^\mu=(\tau,r,\varphi,\eta_s)$, the metric is 
$g_{\mu\nu}=\mbox{diag}(1,-1,-r^2,-\tau^2)$.

Roman letters are used to indicate indices that vary from 1-3 and 
Greek letters for indices that vary from 0-3. Covariant and 
contravariant four-vectors are denoted as $p_\mu$ and $p^\mu$, 
respectively. The notation $p\cdot q\equiv p_\mu q^\mu$ represents 
scalar product of a covariant and a contravariant four-vector. 
Tensors without indices shall always correspond to Lorentz scalars. 
We follow Einstein summation convention, which states that repeated 
indices in a single term are implicitly summed over all the values 
of that index.

We denote the fluid four-velocity by $u^\mu$ and the Lorentz 
contraction factor by $\gamma$. The projector onto the space 
orthogonal to $u^\mu$ is defined as: $\Delta^{\mu\nu}\equiv 
g^{\mu\nu}-u^\mu u^\nu$. Hence, $\Delta^{\mu\nu}$ satisfies the 
conditions $\Delta^{\mu\nu} u_\mu=\Delta^{\mu\nu} u_\nu=0$ with 
trace $\Delta^\mu_\mu=3$. The partial derivative $\partial^{\mu}$ 
can then be decomposed as:
\begin{equation*}
\partial^{\mu} = \nabla^\mu + u^{\mu} D, \quad \mathrm{where~}\quad \nabla^\mu\equiv\Delta^{\mu \nu} \partial_{\nu} \quad\mathrm{and}\quad D\equiv u^{\mu}\partial_{\mu}.
\end{equation*}
In the fluid rest frame, $D$ reduces to the time derivative and 
$\nabla^\mu$ reduces to the spacial gradient. Hence, the notation 
$\dot{f}\equiv Df$ is also commonly used. We also frequently use the 
symmetric, anti-symmetric and angular brackets notations defined as
\begin{eqnarray*}
A_{(\mu} B_{\nu)} &\equiv&\frac{1}{2}\left(A_\mu B_\nu+A_\nu
B_\mu\right),
\\
A_{[\mu} B_{\nu]} &\equiv&\frac{1}{2}\left(A_\mu B_\nu-A_\nu
B_\mu\right),
\\
A_{\langle \mu} B_{\nu\rangle} &\equiv&
\Delta^{\alpha\beta}_{\mu\nu} A_\alpha B_\beta.
\end{eqnarray*}
where, 
\begin{equation*}
\Delta^{\alpha\beta}_{\mu\nu}\equiv\frac{1}{2}\left(\Delta^\alpha_\mu \Delta^\beta_\nu +
\Delta^\alpha_\nu \Delta^\beta_\mu-\frac{2}{3} \Delta^{\alpha \beta}
\Delta_{\mu \nu}\right)
\end{equation*}
is the traceless symmetric projection operator orthogonal to $u^\mu$ 
satisfying the conditions 
$\Delta^{\alpha\beta}_{\mu\nu}\Delta_{\alpha\beta}= 
\Delta^{\alpha\beta}_{\mu\nu}\Delta^{\mu\nu}=0$.

Using the above notations, the commonly used local fluid rest frame 
variables in dissipative viscous hydrodynamics are expressed in 
terms of the energy momentum tensor $T^{\mu\nu}$, charge 
four-current $N^{\mu}$ and entropy four-current $S^{\mu}$ as follows:
\begin{eqnarray*}
n & \equiv & \makebox[1.5in][l]{$u_\mu N^\mu$} \mbox{net charge density};\hspace{3.0cm}\\
n^\mu &\equiv& \makebox[1.5in][l]{$\btu^\mu_\nu N^\nu$} \mbox{net flow of charge};\\
\eps &\equiv& \makebox[1.5in][l]{$u_\mu T^{\mu\nu} u_\nu$} \mbox{energy density};\\
P + \Pi &\equiv& \makebox[1.5in][l]{$\displaystyle -1/3
\btu_{\mu\nu} T^{\mu\nu}$ } \mbox{P: thermal pressure, $\Pi$: bulk
pressure;
}\\
h &\equiv& \makebox[1.5in][l]{$\displaystyle (\epsilon+P)/n$ } \mbox{enthalpy;
}\\
\pi^{\mu\nu} &\equiv& \makebox[1.5in][l]{$T^{\langle\mu\nu\rangle}$
}\mbox{shear stress
tensor};\\
h^\mu &\equiv&  \makebox[1.5in][l]{$u_\nu T^{\nu\lambda}
\btu^\mu_\lambda $}
\mbox{energy flow};\\
q^\mu &\equiv& \makebox[1.5in][l]{$h^\mu - h\, n^\mu$}\mbox{heat flow};\\
s &\equiv& \makebox[1.5in][l]{$u_\mu\, S^\mu $}\mbox{entropy density};\\
\Phi^\mu &\equiv& \makebox[1.5in][l]{$\btu^\mu_\nu S^\nu $}
\mbox{entropy flux};\\
c_s^2 &\equiv& \makebox[1.5in][l]{$(dP/d\epsilon)_{s/n}$}\mbox{adiabatic speed of sound squared}.
\end{eqnarray*}

We also define the following scalar and tensors constructed from the 
gradients of the fluid four-velocity $u^\mu$:
\begin{eqnarray*}
\theta &\equiv& \makebox[3.in][l]{$\partial \cdot u$}
\mbox{expansion rate},\\
\sigma^{\mu\nu}  &\equiv& \makebox[3.in][l]{$\nabla^{\La \mu}u^{\nu \Ra}=\frac{1}{2}(\nabla^\mu u^\nu+\nabla^\nu
u^\mu)-\frac{1}{3}\nabla^{\mu\nu}\partial_\alpha u^\alpha$}
\mbox{velocity stress tensor}, \\
\omega^{\mu\nu}  &\equiv& \makebox[3.in][l]{$\nabla^{[\mu}
u^{\nu]}$} \mbox{vorticity tensor}.
\end{eqnarray*}

%% file: Chapter2.tex
%#######################################################################
\chapter{Thermodynamics, relativistic fluid dynamics and kinetic theory} 
%#######################################################################

The most appealing feature of relativistic fluid dynamics is the 
fact that it is simple and general. It is simple in the sense that 
all the information of the system is contained in its thermodynamic 
and transport properties, i.e., its equation of state and transport 
coefficients. Fluid dynamics is also general because it relies on 
only one assumption: the system remains close to local thermodynamic 
equilibrium throughout its evolution. Although the hypothesis of 
proximity to local equilibrium is quite strong, it saves us from 
making any further assumption regarding the description of the 
particles and fields, their interactions, the classical or quantum 
nature of the phenomena involved etc. In this chapter, we review the 
basic aspects of thermodynamics and discuss relativistic fluid 
dynamics from a phenomenological perspective. The salient features 
of kinetic theory in the context of fluid dynamics will also be 
discussed. The concepts introduced in this Chapter will be required 
in the following Chapters to derive dissipative hydrodynamic 
equations for applications in high-energy heavy-ion physics.

This chapter is organized as follows: In Sec. 2.1, we introduce the 
basic laws of thermodynamics and derive the thermodynamic relations 
that will be used later in this thesis. Section 2.2 contains a brief 
review of relativistic ideal fluid dynamics. We derive the general 
form of the conserved currents of an ideal fluid and their equations 
of motion. In Sec. 2.3, we postulate the thermodynamic relations in 
a covariant notation using the definition of hydrodynamic 
four-velocity from the previous section. In Sec. 2.4 we introduce 
dissipation in fluid dynamics, explain the basic aspects of 
dissipative fluid dynamics and derive a covariant version of 
Navier-Stokes theory using the second law of thermodynamics. We 
discuss the problems of Navier-Stokes theory in the relativistic 
regime, i.e., the acausality and instability of this theory. We also 
review Israel-Stewart theory and show how to derive causal fluid 
dynamical equations from the second law of thermodynamics. Finally, 
Sec. 2.5 contains a discussion about the relativistic kinetic 
theory, where we express fluid dynamical currents in terms of 
single-particle phase-space distribution function. We also outline 
the basic aspects of relativistic Boltzmann equation and its 
implications on the evolution of the distribution function.

%%%%%%%%%%%%%%%%%%%%%%%%%%%%%%%%%%%%%%%%%%%%%%%%%%%%%%%%%%%%%%%%%%%%%%%%

\section{Thermodynamics}

Thermodynamics is an empirical description of the macroscopic or 
large-scale properties of matter and it makes no hypotheses about 
the small-scale or microscopic structure. It is concerned only with 
the average behaviour of a very large number of microscopic 
constituents, and its laws can be derived from statistical 
mechanics. A thermodynamic system can be described in terms of a 
small set of extensive variables, such as volume ($V$), the total 
energy ($E$), entropy ($S$), and number of particles ($N$), of the 
system. Thermodynamics is based on four phenomenological laws that 
explain how these quantities are related and how they change with 
time \cite{Fermi,Reif,Reichl}.
\begin{itemize}
\item {\bf Zeroth Law}: If two systems are both in thermal 
equilibrium with a third system then they are in thermal equilibrium 
with each other. This law helps define the notion of temperature.
\item {\bf First Law}: All the energy transfers must be accounted 
for to ensure the conservation of the total energy of a 
thermodynamic system and its surroundings. This law is the principle 
of conservation of energy.
\item {\bf Second Law}: An isolated physical system spontaneously 
evolves towards its own internal state of thermodynamic equilibrium. 
Employing the notion of entropy, this law states that the change in 
entropy of a closed thermodynamic system is always positive or zero. 
\item {\bf Third Law}: Also known an Nernst's heat theorem, states 
that the difference in entropy between systems connected by a 
reversible process is zero in the limit of vanishing temperature. In 
other words, it is impossible to reduce the temperature of a system 
to absolute zero in a finite number of operations. 
\end{itemize}

The first law of thermodynamics postulates that the changes in the 
total energy of a thermodynamic system must result from: {\bf (1)} 
heat exchange, {\bf (2)} the mechanical work done by an external 
force, and {\bf (3)} from particle exchange with an external medium. 
Hence the conservation law relating the small changes in state 
variables, $E$, $V$, and $N$ is
\begin{equation}\label{FLTI}
\delta E = \delta Q - P\delta V + \mu\,\delta N,
\end{equation}
where $P$ and $\mu$ are the pressure and chemical potential, 
respectively, and $\delta Q$ is the amount of heat exchange.

The heat exchange takes into account the energy variations due to 
changes of internal degrees of freedom that are not described by the 
state variables. The heat itself is not a state variable since it 
can depend on the past evolution of the system and may take several 
values for the same thermodynamic state. However, when dealing with 
reversible processes (in time), it becomes possible to assign a 
state variable related to heat. This variable is the entropy, $S$ , 
and is defined in terms of the heat exchange as $\delta Q = T\delta 
S$, with the temperature $T$ being the proportionality constant. 
Then, when considering variations between equilibrium states that 
are infinitesimally close to each other, it is possible to write the 
first law of thermodynamics in terms of differentials of the state 
variables,
\begin{equation}\label{FLT}
dE = TdS - PdV + \mu\, dN.
\end{equation}
Hence, using Eq. (\ref{FLT}), the intensive quantities, $T$, $\mu$ 
and $P$, can be obtained in terms of partial derivatives of the 
entropy as
\begin{equation}\label{DSDEDNDV}
\left.{\frac{\partial S}{\partial E}}\right\vert_{N,V} = \frac{1}{T}, \qquad
\left.{\frac{\partial S}{\partial V}}\right\vert_{N,E} = \frac{P}{T}, \qquad
\left.{\frac{\partial S}{\partial N}}\right\vert_{E,V} = -\frac{\mu}{T}.
\end{equation}

The entropy is mathematically defined as an extensive and additive 
function of the state variables, which means that
\begin{equation}\label{EEFSV}
S(\lambda E, \lambda V, \lambda N) = \lambda S(E, V, N).
\end{equation}
Differentiating both sides with respect to $\lambda$, we obtain
\begin{equation}\label{DSDL}
S = E\left.{\frac{\partial S}{\partial\lambda E}}\right\vert_{\lambda N,\lambda V} 
+ V\left.{\frac{\partial S}{\partial\lambda V}}\right\vert_{\lambda N,\lambda E} 
+ N\left.{\frac{\partial S}{\partial\lambda N}}\right\vert_{\lambda E,\lambda V},
\end{equation}
which holds for any arbitrary value of $\lambda$. Setting $\lambda=1$
and using Eq. (\ref{DSDEDNDV}), we obtain the so-called Euler's 
relation
\begin{equation}\label{TDER}
E = - PV + TS + \mu\, N.
\end{equation}
Using Euler's relation, Eq. (\ref{TDER}), along with the first law 
of thermodynamics, Eq. (\ref{FLT}), we arrive at the Gibbs-Duhem 
relation 
\begin{equation}\label{GDR}
VdP = SdT + Nd\mu.
\end{equation}

In terms of energy, entropy and number densities defined as 
$\epsilon\equiv E/V$, $s\equiv S/V$, and $n\equiv N/V$ respectively, 
the Euler's relation, Eq. (\ref{TDER}) and Gibbs-Duhem relation, Eq. 
(\ref{GDR}), reduce to 
\begin{align}
\epsilon &= - P + Ts + \mu\, n  \label{ERD} \\
dP &= s\,dT + n\,d\mu. \label{GDRD}
\end{align}
Differentiating Eq.(\ref{ERD}) and using Eq. (\ref{GDRD}), we obtain
the relation analogous to first law of thermodynamics
\begin{equation}\label{ERGDR}
d\epsilon = Tds + \mu\, dn \quad\Rightarrow\quad ds = \frac{1}{T}\, d\epsilon - \frac{\mu}{T}\, dn.
\end{equation}
It is important to note that all the densities defined above 
$(\epsilon,~s,~n)$ are intensive quantities.

The equilibrium state of a system is defined as a stationary state 
where the extensive and intensive variables of the system do not 
change. We know from the second law of thermodynamics that the 
entropy of an isolated thermodynamic system must either increase or 
remain constant. Hence, if a thermodynamic system is in equilibrium, 
the entropy of the system being an extensive variable, must remain 
constant. On the other hand, for a system that is out of 
equilibrium, the entropy must always increase. This is an extremely 
powerful concept that will be extensively used in this chapter to 
constrain and derive the equations of motion of a dissipative fluid. 
This concludes a brief outline of the basics of thermodynamics; for 
a more detailed review, see Ref. \cite{Reichl}. In the next section, 
we introduce and derive the equations of relativistic ideal fluid 
dynamics. 

%%%%%%%%%%%%%%%%%%%%%%%%%%%%%%%%%%%%%%%%%%%%%%%%%%%%%%%%%%%%%%%%%%%%%%%%

\section{Relativistic ideal fluid dynamics}

An ideal fluid is defined by the assumption of local thermal 
equilibrium, i.e., all fluid elements must be exactly in 
thermodynamic equilibrium \cite{Landau,Weinberg}. This means that at 
each space-time coordinate of the fluid $x\equiv x^\mu$, there can 
be assigned a temperature $T(x)$, a chemical potential $\mu(x)$, 
and a collective four-velocity field, 
\begin{equation}\label{umu}
u^\mu(x)\equiv \frac{dx^\mu}{d\tau}. 
\end{equation}
The proper time increment $d\tau$ is given by the line element
\begin{equation}\label{dtau2}
(d\tau)^2 = g_{\mu\nu}dx^\mu dx^\nu = (dt)^2 - (d\vec x)^2 
= (dt)^2\left[ 1 - (\vec v)^2\right],
\end{equation}
where $\vec v \equiv d\vec x/dt$. This implies that
\begin{equation}\label{umuf}
u^\mu(x) = \frac{dt}{d\tau}\frac{dx^\mu}{dt} = \gamma(\vec v)\binom{1}{\vec v}
\end{equation}
where $\gamma(\vec v) = 1/\sqrt{1-\vec v^2}$. In the 
non-relativistic limit, we obtain $u^\mu(x)=(1,\vec v)$. It is 
important to note that the four-vector $u^\mu(x)$ only contains 
three independent components since it obeys the relation
\begin{equation}\label{umuumu}
u^2 \equiv u^\mu(x) g_{\mu\nu} u^\nu(x) = \gamma^2(\vec v)\left( 1 - \vec v^2 \right) = 1.
\end{equation}
The quantities $T$, $\mu$ and $u^\mu$ are often referred to as the 
primary fluid-dynamical variables.

The state of a fluid can be completely specified by the densities 
and currents associated with conserved quantities, i.e., energy, 
momentum, and (net) particle number. For a relativistic fluid, the 
state variables are the energy- momentum tensor, $T^{\mu\nu}$, and 
the (net) particle four-current, $N^\mu$. To obtain the general form 
of these currents for an ideal fluid, we first define the local rest 
frame (LRF) of the fluid. In this frame, $\vec v=0$, and the 
energy-momentum tensor, $T^{\mu\nu}_{LRF}$, the (net) particle 
four-current, $N^\mu_{LRF}$, and the entropy four-current, 
$S^\mu_{LRF}$, should have the characteristic form of a system in 
static equilibrium. In other words, in local rest frame, there is no 
flow of energy ($T^{i0}_{LRF}=0$), the force per unit surface 
element is isotropic ($T^{ij}_{LRF}=\delta^{ij} P$) and there is no 
particle and entropy flow ($\vec N=0$ and $\vec S=0$). Consequently, 
the energy-momentum tensor, particle and entropy four-currents in 
this frame take the following simple forms
\begin{equation}\label{TNSLRF}
T^{\mu\nu}_{LRF} = \left( \begin{array}{cccc}
\epsilon & 0 & 0 & 0 \\
0 & P & 0 & 0 \\
0 & 0 & P & 0 \\
0 & 0 & 0 & P \end{array} 
\right) \!,\quad
N^\mu_{LRF} = \left( \begin{array}{c}
n  \\
0  \\
0  \\
0  \end{array} 
\right) \!,\quad
S^\mu_{LRF} = \left( \begin{array}{c}
s  \\
0  \\
0  \\
0  \end{array} 
\right) \!.
\end{equation}

For an ideal relativistic fluid, the general form of the 
energy-momentum tensor, $T^{\mu\nu}_{(0)}$, (net) particle 
four-current, $N^\mu_{(0)}$, and the entropy four-current, 
$S^\mu_{(0)}$, has to be built out of the hydrodynamic tensor 
degrees of freedom, namely the vector, $u^\mu$, and the metric 
tensor, $g_{\mu\nu}$. Since $T^{\mu\nu}_{(0)}$ should be symmetric 
and transform as a tensor, and, $N^\mu_{(0)}$ and $S^\mu_{(0)}$ 
should transform as a vector, under Lorentz transformations, the 
most general form allowed is therefore
\begin{equation}\label{TNSTD}
T^{\mu\nu}_{(0)} = c_1 u^\mu u^\nu + c_2 g^{\mu\nu} ,\quad
N^\mu_{(0)} = c_3 u^\mu,\quad
S^\mu_{(0)} = c_4 u^\mu.
\end{equation}
In the local rest frame, $\vec v=0 \Rightarrow u^\mu=(1,\vec 0)$. 
Hence in this frame, Eq. (\ref{TNSTD}) takes the form
\begin{equation}\label{TNSLRFID}
T^{\mu\nu}_{(0)LRF} = \left( \begin{array}{cccc}
c_1+c_2 & 0 & 0 & 0 \\
0 & -c_2 & 0 & 0 \\
0 & 0 & -c_2 & 0 \\
0 & 0 & 0 & -c_2 \end{array} 
\right) \!\!,\quad
N^\mu_{(0)LRF} = \left( \begin{array}{c}
c_3  \\
0  \\
0  \\
0  \end{array} 
\right) \!\!,\quad
S^\mu_{(0)LRF} = \left( \begin{array}{c}
c_4  \\
0  \\
0  \\
0  \end{array} 
\right) \!\!.
\end{equation}
By comparing the above equation with the corresponding general 
expressions in the local rest frame, Eq. (\ref {TNSLRF}), one 
obtains the following expressions for the coefficients
\begin{equation}\label{c1c2c3c4}
c_1 = \epsilon + P, \quad c_2 = -P, \quad c_3 = n, \quad c_4 = s.
\end{equation}  
The conserved currents of an ideal fluid can then be expressed as
\begin{equation}\label{CCIF}
T^{\mu\nu}_{(0)} = \epsilon u^\mu u^\nu - P \Delta^{\mu\nu} ,\quad
N^\mu_{(0)} = n u^\mu,\quad
S^\mu_{(0)} = s u^\mu,
\end{equation}
where $\Delta^{\mu\nu}=g^{\mu\nu}-u^\mu u^\nu$ is the projection 
operator onto the three-space orthogonal to $u^\mu$, and satisfies 
the following properties of an orthogonal projector,
\begin{equation}\label{DMUNU}
u_\mu\Delta^{\mu\nu} = \Delta^{\mu\nu}u_\nu = 0, \quad
\Delta^\mu_\rho \Delta^{\rho\nu} = \Delta^{\mu\nu}, \quad
\Delta^\mu_\mu = 3.
\end{equation} 

The dynamical description of an ideal fluid is obtained using the 
conservation laws of energy, momentum and (net) particle number. 
These conservation laws can be mathematically expressed using the 
four-divergences of energy-momentum tensor and particle four-current 
which leads to the following equations,
\begin{equation}\label{IFDCE}
\partial_\mu T^{\mu\nu}_{(0)} = 0, \quad 
\partial_\mu N^\mu_{(0)} = 0,
\end{equation}
where the partial derivative $\partial_\mu \equiv \partial/\partial 
x^\mu$ transforms as a covariant vector under Lorentz 
transformations. Using the four-velocity, $u^\mu$, and the 
projection operator, $\Delta^{\mu\nu}$, the derivative, 
$\partial_\mu$, can be projected along and orthogonal to $u^\mu$
\begin{equation}\label{PFDAOU}
D \equiv u^\mu \partial_\mu, \quad \nabla_\mu \equiv \Delta_\mu^\rho \partial_\rho,
\quad\Rightarrow\quad  \partial_\mu = u_\mu D + \nabla_\mu.
\end{equation}
Projection of energy-momentum conservation equation along and 
orthogonal to $u^\mu$ together with the conservation law for 
particle number, leads to the equations of motion of ideal fluid 
dynamics,
\begin{align}
u_\mu\partial_\nu T^{\mu\nu}_{(0)} = 0 &\quad\Rightarrow\quad 
D\epsilon + (\epsilon+P)\theta = 0,   \label{IFDE1}\\ 
\Delta^\alpha_\mu\partial_\nu T^{\mu\nu}_{(0)} = 0 &\quad\Rightarrow\quad
(\epsilon+P)Du^\alpha -\nabla^\alpha P = 0,   \label{IFDE2}\\
\partial_\mu N^\mu_{(0)} = 0 &\quad\Rightarrow\quad
Dn + n\theta = 0,   \label{IFDE3}
\end{align}
where $\theta\equiv\partial_\mu u^\mu$. It is important to note that 
an ideal fluid is described by four fields, $\epsilon$, $P$, $n$, 
and $u_\mu$, corresponding to six independent degrees of freedom. The 
conservation laws, on the other hand, provide only five equations of 
motion. The equation of state of the fluid, $P=P(n,\epsilon)$, 
relating the pressure to other thermodynamic variables has to be 
specified to close this system of equations. The existence of 
equation of state is guaranteed by the assumption of local thermal 
equilibrium and hence the equations of ideal fluid dynamics are 
always closed.

%%%%%%%%%%%%%%%%%%%%%%%%%%%%%%%%%%%%%%%%%%%%%%%%%%%%%%%%%%%%%%%%%%%%%%%%

\section{Covariant thermodynamics}

In the following, we re-write the equilibrium thermodynamic 
relations derived in Sec. 2.1, Eqs. (\ref{ERD}), (\ref {GDRD}), and 
(\ref{ERGDR}), in a covariant form \cite
{Israel:1979wp,Israel:1976tn}. For this purpose, it is convenient to 
introduce the following notations
\begin{equation}\label{CTND}
\beta \equiv \frac{1}{T}, \quad \alpha \equiv \frac{\mu}{T}, \quad \beta^\mu \equiv \frac{u^\mu}{T}.
\end{equation}
In these notations, the covariant version of the Euler's relation, 
Eq. (\ref{ERD}), and the Gibbs-Duhem relation, Eq. (\ref{GDRD}), can 
be postulated as,
\begin{align}
S^\mu_{(0)} &= P\beta^\mu + \beta_\nu T^{\mu\nu}_{(0)} - \alpha N^\mu_{(0)}, \label{CER}\\
d\left(P\beta^\mu\right) &= N^\mu_{(0)} d\alpha - T^{\mu\nu}_{(0)} d\beta_\nu, \label{CGDR}
\end{align}
respectively. The above equations can then be used to derive a 
covariant form of the first law of thermodynamics, Eq. (\ref{ERGDR}),
\begin{equation}\label{CFLT}
dS^\mu_{(0)} = \beta_\nu dT^{\mu\nu}_{(0)} - \alpha dN^\mu_{(0)}.
\end{equation}

The covariant thermodynamic relations were constructed in a such a 
way that when Eqs. (\ref{CER}), (\ref{CGDR}) and (\ref{CFLT}) are 
contracted with $u_\mu$, 
\begin{align}
u_\mu \left[ S^\mu_{(0)} - P\beta^\mu - \beta_\nu T^{\mu\nu}_{(0)} + \alpha N^\mu_{(0)} \right] &=0  
\quad\Rightarrow\quad s + \alpha n -\beta (\epsilon+P) = 0, \\
u_\mu \left[ d\left(P\beta^\mu\right) - N^\mu_{(0)} d\alpha + T^{\mu\nu}_{(0)} d\beta_\nu \right] &=0 
\quad\Rightarrow\quad d(\beta P) - nd\alpha + \epsilon d\beta = 0, \\
u_\mu \left[ dS^\mu_{(0)} - \beta_\nu dT^{\mu\nu}_{(0)} + \alpha dN^\mu_{(0)} \right] &=0
\quad\Rightarrow\quad ds - \beta d\epsilon + \alpha dn = 0,
\end{align}
we obtain the usual thermodynamic relations, Eqs. (\ref{ERD}), (\ref 
{GDRD}), and (\ref{ERGDR}). Here we have used the property of the 
fluid four-velocity, $u_\mu u^\mu=1\Rightarrow u_\mu du^\mu=0$. The 
projection of Eqs. (\ref{CER}), (\ref{CGDR}) and (\ref{CFLT}) onto 
the three-space orthogonal to $u^\mu$ just leads to trivial identities,
\begin{align}
\Delta^\alpha_\mu \left[ S^\mu_{(0)} - P\beta^\mu - \beta_\nu T^{\mu\nu}_{(0)} + \alpha N^\mu_{(0)} \right] &=0  
\quad\Rightarrow\quad 0 = 0, \\
\Delta^\alpha_\mu \left[ d\left(P\beta^\mu\right) - N^\mu_{(0)} d\alpha + T^{\mu\nu}_{(0)} d\beta_\nu \right] &=0 
\quad\Rightarrow\quad 0 = 0, \\
\Delta^\alpha_\mu \left[ dS^\mu_{(0)} - \beta_\nu dT^{\mu\nu}_{(0)} + \alpha dN^\mu_{(0)} \right] &=0
\quad\Rightarrow\quad 0 = 0.
\end{align}
From the above equations we conclude that the covariant 
thermodynamic relations do not contain more information than the 
usual thermodynamic relations.

The first law of thermodynamics, Eq. (\ref{CFLT}), leads to the 
following expression for the entropy four-current divergence,
\begin{equation}\label{CTEFCD}
\partial_\mu S^\mu_{(0)} = \beta_\mu\partial_\nu T^{\mu\nu}_{(0)} - \alpha\partial_\mu N^\mu_{(0)}.
\end{equation}
After employing the conservation of energy-momentum and net particle 
number, Eq. (\ref{IFDCE}), the above equation leads to the 
conservation of entropy, $\partial_\mu S^\mu_{(0)} = 0$. It is 
important to note that within equilibrium thermodynamics, the 
entropy conservation is a natural consequence of energy-momentum and 
particle number conservation, and the first law of thermodynamics. 
The equation of motion for the entropy density is then obtained as
\begin{equation}\label{CTECE}
\partial_\mu S^\mu_{(0)} = 0 \quad\Rightarrow\quad Ds + s\theta = 0.
\end{equation}
We observe that the rate equation of the entropy density in the 
above equation is identical to that of the net particle number, Eq. 
(\ref{IFDE3}). Therefore, we conclude that for ideal hydrodynamics, 
the ratio of entropy density to number density ($s/n$) is a constant 
of motion.

%%%%%%%%%%%%%%%%%%%%%%%%%%%%%%%%%%%%%%%%%%%%%%%%%%%%%%%%%%%%%%%%%%%%%%%%

\section{Relativistic dissipative fluid dynamics}

The derivation of relativistic ideal fluid dynamics proceeds by 
employing the properties of the Lorentz transformation, the 
conservation laws, and most importantly, by imposing local 
thermodynamic equilibrium. It is important to note that while the 
properties of Lorentz transformation and the conservation laws are 
robust, the assumption of local thermodynamic equilibrium is a 
strong restriction. The deviation from local thermodynamic 
equilibrium results in dissipative effects, and, as all fluids are 
dissipative in nature due to the uncertainty principle \cite 
{Danielewicz:1984ww}, the assumption of local thermodynamic 
equilibrium is never strictly realized in practice. In the 
following, we consider a more general theory of fluid dynamics that 
attempts to take into account the dissipative processes that must 
happen, because a fluid can never maintain exact local thermodynamic 
equilibrium throughout its dynamical evolution. 

Dissipative effects in a fluid originate from irreversible 
thermodynamic processes that occur during the motion of the fluid. 
In general, each fluid element may not be in equilibrium with the 
whole fluid, and, in order to approach equilibrium, it 
exchanges heat with its surroundings. Moreover, the fluid elements 
are in relative motion and can also dissipate energy by friction. 
All these processes must be included in order to obtain a reasonable 
description of a relativistic fluid.

The earliest covariant formulation of dissipative fluid dynamics 
were due to Eckart \cite{Eckart:1940zz}, in 1940, and, later, by 
Landau and Lifshitz \cite{Landau}, in 1959. The formulation of these 
theories, collectively known as first-order theories (order of 
gradients), was based on a covariant generalization of the 
Navier-Stokes theory. The Navier-Stokes theory, at that time, had 
already become a successful theory of dissipative fluid dynamics. It 
was employed efficiently to describe a wide variety of 
non-relativistic fluids, from weakly coupled gases such as air, to 
strongly coupled fluids such as water. Hence, a relativistic 
generalisation of Navier-Stokes theory was considered to be the most 
effective and promising way to describe relativistic dissipative 
fluids. 

The formulation of relativistic dissipative hydrodynamics turned out 
to be more subtle since the relativistic generalisation of 
Navier-Stokes theory is intrinsically unstable \cite 
{Hiscock:1983zz,Hiscock:1985zz,Hiscock:1987zz}. The source of such 
instability is attributed to the inherent acausal behaviour of this 
theory \cite{Denicol:2008ha,Pu:2009fj}. A straightforward 
relativistic generalisation of Navier-Stokes theory allows signals 
to propagate with infinite speed in a medium. While in 
non-relativistic theories, this does not give rise to an intrinsic 
problem and can be ignored, in relativistic systems where causality 
is a physical property that is naturally preserved, this feature 
leads to intrinsically unstable equations of motion. Nevertheless, 
it is instructive to review the first-order theories as they are an 
important initial step to illustrate the basic features of 
relativistic dissipative fluid-dynamics.

As in the case of ideal fluids, the basic equations governing the 
motion of dissipative fluids are also obtained from the conservation 
laws of energy-momentum and (net) particle number,
\begin{equation}\label{DFDCE}
\partial_\mu T^{\mu\nu} = 0, \quad 
\partial_\mu N^\mu = 0.
\end{equation}

However, for dissipative fluids, the energy-momentum tensor is no 
longer diagonal and isotropic in the local rest frame. Moreover, due 
to diffusion, the particle flow is expected to appear in the local 
rest frame of the fluid element. To account for these effects, 
dissipative currents $\tau^{\mu\nu}$ and $n^\mu$ are added to the 
previously derived ideal currents, $T^{\mu\nu}_{(0)}$ and 
$N^\mu_{(0)}$,
\begin{equation}\label{CCDF}
T^{\mu\nu} = T^{\mu\nu}_{(0)} + \tau^{\mu\nu} = \epsilon u^\mu u^\nu - P \Delta^{\mu\nu} + \tau^{\mu\nu},\quad
N^\mu = N^\mu_{(0)} + n^\mu = n u^\mu + n^\mu,
\end{equation}
where, $\tau^{\mu\nu}$ is required to be symmetric 
($\tau^{\mu\nu}=\tau^{\nu\mu}$) in order to satisfy angular momentum 
conservation. The main objective then becomes to find the dynamical 
or constitutive equations satisfied by these dissipative currents.

%-----------------------------------------------------------------------

\subsection{Matching conditions}

The introduction of the dissipative currents causes the equilibrium 
variables to be ill-defined, since the fluid can no longer be 
considered to be in local thermodynamic equilibrium. Hence, in a 
dissipative fluid, the thermodynamic variables can only be defined 
in terms of an artificial equilibrium state, constructed such that 
the thermodynamic relations are valid as if the fluid were in local 
thermodynamic equilibrium. The first step to construct such an 
equilibrium state is to define $\epsilon$ and $n$ as the total 
energy and particle density in the local rest frame of the fluid, 
respectively. This is guaranteed by imposing the so-called matching 
or fitting conditions \cite{Israel:1979wp},
\begin{equation}\label{MFC}
\epsilon \equiv u_\mu u_\nu T^{\mu\nu} ,\quad
n \equiv u_\mu N^\mu .
\end{equation}
These matching conditions enforces the following constraints on the 
dissipative currents
\begin{equation}\label{RMFC}
u_\mu u_\nu \tau^{\mu\nu} = 0 ,\quad
u_\mu n^\mu = 0 .
\end{equation}
Subsequently, using $n$ and $\epsilon$, an artificial equilibrium 
state can be constructed with the help of the equation of state. It 
is however important to note that while the energy and particle 
densities are physically defined, all the other thermodynamic 
quantities ($s,~P,~T,~\mu,\cdots $) are defined only in terms of 
an artificial equilibrium state and do not necessarily retain their 
usual physical meaning.

%-----------------------------------------------------------------------

\subsection{Tensor decompositions of dissipative quantities}

To proceed further, it is convenient to decompose $\tau^{\mu\nu}$ in 
terms of its irreducible components, i.e., a scalar, a four-vector, 
and a traceless and symmetric second-rank tensor. Moreover, this 
tensor decomposition must be consistent with the matching or 
orthogonality condition, Eq. (\ref {RMFC}), satisfied by 
$\tau^{\mu\nu}$. To this end, we introduce another projection 
operator, the double symmetric, traceless projector orthogonal to 
$u^\mu$,
\begin{equation}\label{DSTPO}
\Delta^{\mu\nu}_{\alpha\beta} \equiv \frac{1}{2}\left(\Delta^{\mu}_{\alpha}\Delta^{\nu}_{\beta} 
+ \Delta^{\mu}_{\beta}\Delta^{\nu}_{\alpha} - \frac{2}{3}\Delta^{\mu\nu}\Delta_{\alpha\beta}\right),
\end{equation}
with the following properties,
\begin{equation}\label{PDSTPO}
\Delta^{\mu\nu}_{~~~\alpha\beta} = \Delta^{~~~\mu\nu}_{\alpha\beta}, \quad
\Delta^{\mu\nu}_{\rho\sigma}\Delta^{\rho\sigma}_{\alpha\beta} = \Delta^{\mu\nu}_{\alpha\beta}, \quad
u_\mu\Delta^{\mu\nu}_{\alpha\beta} = g_{\mu\nu}\Delta^{\mu\nu}_{\alpha\beta} = 0, \quad
\Delta^{\mu\nu}_{\mu\nu} = 5.
\end{equation}
The parentheses in the above equation denote symmetrization of the 
Lorentz indices, i.e., $A^{(\mu\nu)}\equiv(A^{\mu\nu}+A^{\nu\mu})/2$. 
The dissipative current $\tau^{\mu\nu}$ then can be tensor 
decomposed in its irreducible form by using $u^\mu$, 
$\Delta^{\mu\nu}$ and $\Delta^{\mu\nu}_{\alpha\beta}$ as
\begin{equation}\label{TDT}
\tau^{\mu\nu} \equiv -\Pi\Delta^{\mu\nu} + 2u^{(\mu}h^{\nu)} + \pi^{\mu\nu},
\end{equation}
where we have defined
\begin{equation}\label{DDQ}
\Pi \equiv -\frac{1}{3}\Delta_{\alpha\beta}\tau^{\alpha\beta}, \quad
h^\mu \equiv \Delta^\mu_\alpha u_\beta\tau^{\alpha\beta}, \quad
\pi^{\mu\nu} \equiv \Delta^{\mu\nu}_{\alpha\beta}\tau^{\alpha\beta}.
\end{equation}
The scalar $\Pi$ is the bulk viscous pressure, the vector $h^\mu$ is 
the energy-diffusion four-current, and the second-rank tensor 
$\pi^{\mu\nu}$ is the shear-stress tensor. The properties of the 
projection operators $\Delta^\mu_\alpha$ and 
$\Delta^{\mu\nu}_{\alpha\beta}$ imply that both $h^\mu$ and 
$\pi^{\mu\nu}$ are orthogonal to $u^\mu$ and, additionally, 
$\pi^{\mu\nu}$ is traceless. Armed with these definitions, all the 
irreducible hydrodynamic fields are expressed in terms of $N^\mu$ 
and $T^{\mu\nu}$ as
\begin{align}\label{IHFD}
\epsilon &= u_\alpha u_\beta T^{\alpha\beta}, \quad
n = u_\alpha N^\alpha, \quad
\Pi = -P -\frac{1}{3}\Delta_{\alpha\beta}T^{\alpha\beta}, \nonumber\\
h^\mu &= u_\alpha T^{\langle\mu\rangle\alpha}, \quad
n^\mu = N^{\langle\mu\rangle}, \quad
\pi^{\mu\nu} = T^{\langle\mu\nu\rangle},
\end{align}
where the angular bracket notations are defined as, 
$A^{\langle\mu\rangle}\equiv\Delta^\mu_\alpha A^\alpha$ and 
$B^{\langle\mu\nu\rangle}\equiv\Delta^{\mu\nu}_{\alpha\beta}B^{\alpha\beta}$.

We observe that $T^{\mu\nu}$ is a symmetric second-rank tensor with 
ten independent components and $N^\mu$ is a four-vector; overall 
they have fourteen independent components. Next we count the number 
of independent components in the tensor decompositions of 
$T^{\mu\nu}$ and $N^\mu$. Since $n^\mu$ and $h^\mu$ are orthogonal 
to $u^\mu$, they can have only three independent components each. 
The shear-stress tensor $\pi^{\mu\nu}$ is symmetric, traceless and 
orthogonal to $u^\mu$, and hence, can have only five independent 
components. Together with $u^\mu$, $\epsilon$, $n$ and $\Pi$, which 
have in total six independent components ($P$ is related to 
$\epsilon$ via equation of state), we count a total of seventeen 
independent components, three more than expected. The reason being 
that so far, the velocity field $u^\mu$ was introduced as a general 
normalized four-vector and was not specified. Hence $u^\mu$ has to 
be defined to reduce the number of independent components to the 
correct value.

%-----------------------------------------------------------------------

\subsection{Definition of the velocity field}

In the process of formulating the theory of dissipative fluid 
dynamics, the next important step is to fix $u^\mu$. In the case of 
ideal fluids, the local rest frame was defined as the frame in which 
there is, simultaneously, no net energy and particle flow. While the 
definition of local rest frame was unambiguous for ideal fluids, 
this definition is no longer possible in the case of dissipative 
fluids due to the presence of both energy and particle diffusion. 
From a mathematical perspective, the fluid velocity can be defined 
in numerous ways. However, from the physical perspective, there are 
two natural choices. The \emph{Eckart definition} \cite
{Eckart:1940zz}, in which the velocity is defined by the flow of 
particles
\begin{equation}\label{EDF}
N^\mu = nu^\mu \quad \Rightarrow \quad n^\mu = 0,
\end{equation}
and the \emph{Landau definition} \cite{Landau}, in which the 
velocity is specified by the flow of the total energy
\begin{equation}\label{LDF}
u_\nu T^{\mu\nu} = \epsilon u^\mu \quad \Rightarrow \quad h^\mu = 0.
\end{equation}

We note that the above two definitions of $u^\mu$ impose different 
constraints on the dissipative currents. In the Eckart definition 
the particle diffusion is always set to zero, while in the Landau 
definition, the energy diffusion is zero. In other words, the Eckart 
definition of the velocity field eliminates any diffusion of 
particles whereas the Landau definition eliminates any diffusion of 
energy. In this thesis, we shall always use the Landau definition, 
Eq. (\ref{LDF}). The conserved currents in this frame take the 
following form
\begin{equation}\label{CCDLF}
T^{\mu\nu} =  \epsilon u^\mu u^\nu - (P+\Pi) \Delta^{\mu\nu} + \pi^{\mu\nu},\quad
N^\mu = n u^\mu + n^\mu.
\end{equation}

As done for ideal fluids, the energy-momentum conservation equation 
in Eq. (\ref{DFDCE}) is decomposed parallel and orthogonal to $u^\mu$. 
Using Eq. (\ref{CCDLF}) together with the conservation law for 
particle number in Eq. (\ref{DFDCE}), leads to the equations of 
motion for dissipative fluids,
\begin{align}
u_\mu\partial_\nu T^{\mu\nu} = 0 &\quad\Rightarrow\quad 
\dot\epsilon + (\epsilon+P+\Pi)\theta - \pi^{\mu\nu}\sigma_{\mu\nu}= 0,   \label{DFDE1}\\ 
\Delta^\alpha_\mu\partial_\nu T^{\mu\nu} = 0 &\quad\Rightarrow\quad
(\epsilon+P+\Pi)\dot u^\alpha -\nabla^\alpha (P+\Pi) + \Delta^\alpha_\mu\partial_\nu\pi^{\mu\nu} = 0,   \label{DFDE2}\\
\partial_\mu N^\mu = 0 &\quad\Rightarrow\quad
\dot n + n\theta + \partial_\mu n^\mu = 0,   \label{DFDE3}
\end{align}
where $\dot A\equiv DA=u^\mu\partial_\mu A$, and the shear tensor 
$\sigma^{\mu\nu}\equiv\nabla^{\langle\mu}u^{\nu\rangle} 
=\Delta^{\mu\nu}_{\alpha\beta}\nabla^\alpha u^\beta$.

We observe that while there are fourteen total independent 
components of $T^{\mu\nu}$ and $N^\mu$, Eqs. (\ref {DFDE1})-(\ref
{DFDE3}) constitute only five equations. Therefore, in order to 
derive the complete set of equations for dissipative fluid dynamics, 
one still has to obtain the additional nine equations of motion that 
will close Eqs. (\ref{DFDE1})-(\ref {DFDE3}). Eventually, this 
corresponds to finding the closed dynamical or constitutive 
relations satisfied by the dissipative tensors $\Pi$, $n^\mu$ and 
$\pi^{\mu\nu}$.

%-----------------------------------------------------------------------

\subsection{Relativistic Navier-Stokes theory}

In the presence of dissipative currents, the entropy is no longer a 
conserved quantity, i.e., $\partial_\mu S^\mu \neq 0$. Since the 
form of the entropy four-current for a dissipative fluid is not 
known \emph{a priori}, it is not trivial to obtain its equation. We 
proceed by recalling the form of the entropy four-current for ideal 
fluids, Eq. (\ref{CER}), and extending it for dissipative fluids,
\begin{equation}\label{EFCDF}
S^\mu = P\beta^\mu + \beta_\nu T^{\mu\nu} - \alpha N^\mu.
\end{equation}
The above extension remains valid because an artificial equilibrium 
state was constructed using the matching conditions to satisfy the 
thermodynamic relations as if in equilibrium. This was the key step 
proposed by Eckart, Landau and Lifshitz in order to derive the 
relativistic Navier-Stokes theory \cite{Eckart:1940zz,Landau}. The 
next step is to calculate the entropy generation, $\partial_\mu 
S^\mu$, in dissipative fluids. To this end, we substitute the form 
of $T^{\mu\nu}$ and $N^\mu$ for dissipative fluids from Eq. (\ref
{CCDLF}) in Eq. (\ref{EFCDF}). Taking the divergence and using Eqs. 
(\ref{DFDE1})-(\ref{DFDE3}), we obtain
\begin{equation}\label{CTEFCDR1}
\partial_\mu S^\mu = -\beta\Pi\theta - n^\mu\nabla_\mu\alpha + \beta\pi^{\mu\nu}\sigma_{\mu\nu}.
\end{equation}

The relativistic Navier-Stokes theory can then be obtained by 
applying the second law of thermodynamics to each fluid element, 
i.e., by requiring that the entropy production $\partial_\mu S^\mu$ 
must always be positive,
\begin{equation}\label{DEFCDE}
-\beta\Pi\theta - n^\mu\nabla_\mu\alpha + \beta\pi^{\mu\nu}\sigma_{\mu\nu} \geq 0.
\end{equation}
The above inequality can be satisfied for all possible fluid 
configurations if one assumes that the bulk viscous pressure $\Pi$, 
the particle-diffusion four-current $n^\mu$, and the shear-stress 
tensor $\pi^{\mu\nu}$ are linearly proportional to $\theta$, 
$\nabla^\mu\alpha$, and $\sigma^{\mu\nu}$, respectively. This leads to
\begin{equation}\label{RNSE}
\Pi = -\zeta\theta, \quad
n^\mu = \kappa\nabla^\mu\alpha, \quad
\pi^{\mu\nu} = 2\eta\sigma^{\mu\nu},
\end{equation}
where the proportionality coefficients $\zeta$, $\kappa$ and $\eta$ 
refer to the bulk viscosity, the particle diffusion, and the shear 
viscosity, respectively. Substituting the above equation in Eq. (\ref
{CTEFCDR1}), we observe that the source term for entropy production 
becomes a quadratic function of the dissipative currents
\begin{equation}\label{STEP}
\partial_\mu S^\mu = \frac{\beta}{\zeta}\, \Pi^2 - \frac{1}{\kappa}\, n_\mu n^\mu + \frac{\beta}{2\eta}\, \pi_{\mu\nu}\pi^{\mu\nu}.
\end{equation}
In the above equation, since $n^\mu$ is orthogonal to the timelike 
four-vector $u^\mu$, it is spacelike and hence $n_\mu n^\mu<0$. 
Moreover, $\pi^{\mu\nu}$ is symmetric in its Lorentz indices, and in 
the local rest frame $\pi^{0\mu}=\pi^{\mu0}=0$. Since the trace of 
the square of a symmetric matrix is always positive, therefore 
$\pi_{\mu\nu}\pi^{\mu\nu}>0$. Hence, as long as 
$\zeta,\kappa,\eta\geq 0$, the entropy production is always 
positive. Constitutive relations for the dissipative quantities, Eq. 
(\ref{RNSE}), along with Eqs. (\ref{DFDE1})-(\ref {DFDE3}) are known 
as the relativistic Navier-Stokes equations.

The relativistic Navier-Stokes theory in this form was obtained 
originally by Landau and Lifshitz \cite{Landau}. A similar theory 
was derived independently by Eckart, using a different definition of 
the fluid four-velocity \cite{Eckart:1940zz}. However, as already 
mentioned, the Navier-Stokes theory is acausal and, consequently, 
unstable. The source of the acausality can be understood from the 
constitutive relations satisfied by the dissipative currents, Eq. 
(\ref{RNSE}). The linear relations between dissipative currents and 
gradients of the primary fluid-dynamical variables imply that any 
inhomogeneity of $\alpha$ and $u^\mu$, immediately results in 
dissipative currents. This instantaneous effect is not allowed in a 
relativistic theory which eventually causes the theory to be 
unstable. Several theories have been developed to incorporate 
dissipative effects in fluid dynamics without violating causality: 
Grad-Israel-Stewart theory \cite{Israel:1979wp,Grad,Israel:1976tn}, 
the divergence-type theory \cite{Muller:1967zza,Muller:1999in}, 
extended irreversible thermodynamics \cite{Jou}, Carter's theory 
\cite{Carter}, \"Ottinger-Grmela theory \cite{Grmela:1997zz}, among 
others. Israel and Stewart's formulation of causal relativistic 
dissipative fluid dynamics is the most popular and widely used; in 
the following we briefly review their approach. 

%-----------------------------------------------------------------------

\subsection{Causal fluid dynamics: Israel-Stewart theory}

The main idea behind the Israel-Stewart formulation was to apply the 
second law of thermodynamics to a more general expression of the 
non-equilibrium entropy four-current \cite
{Israel:1979wp,Grad,Israel:1976tn}. In equilibrium, the entropy 
four-current was expressed exactly in terms of the primary 
fluid-dynamical variables, Eq. (\ref{CER}). Strictly speaking, the 
nonequilibrium entropy four-current should depend on a larger 
number of independent dynamical variables, and, a direct extension 
of Eq. (\ref{CER}) to Eq. (\ref{EFCDF}) is, in fact, incomplete. A 
more realistic description of the entropy four-current can be 
obtained by considering it to be a function not only of the primary 
fluid-dynamical variables, but also of the dissipative currents. The 
most general off-equilibrium entropy four-current is then given by 
\begin{equation}\label{EFCMGF}
S^\mu = P\beta^\mu + \beta_\nu T^{\mu\nu} - \alpha N^\mu - Q^\mu\left(\delta N^\mu, \delta T^{\mu\nu}\right).
\end{equation}
where $Q^\mu$ is a function of deviations from local equilibrium, 
$\delta N^\mu\equiv N^\mu-N^\mu_{(0)}$, $\delta T^{\mu\nu}\equiv 
T^{\mu\nu}-T^{\mu\nu}_{(0)}$. Using Eq. (\ref{CCDLF}) and 
Taylor-expanding $Q^\mu$ to second order in dissipative fluxes, we 
obtain
\begin{equation}\label{AEFCC2}
S^\mu = su^\mu - \alpha n^\mu - \left(\beta_0\Pi^2 - \beta_1 n_\nu n^\nu 
+ \beta_2\pi_{\rho\sigma} \pi^{\rho\sigma}\right) \frac{u^\mu}{2T} 
- \left(\alpha_0\Pi\Delta^{\mu\nu} + \alpha_1\pi^{\mu\nu}\right)\frac{n_\nu}{T}
+ {\cal O}(\delta^3),
\end{equation}
where ${\cal O}(\delta^3)$ denotes third order terms in the 
dissipative currents and 
$\beta_0,~\beta_1,~\beta_2,~\alpha_0,~\alpha_1$ are the 
thermodynamic coefficients of the Taylor expansion and are 
complicated functions of the temperature and chemical potential.

We observe that the existence of second-order contributions to the 
entropy four-current in Eq. (\ref{AEFCC2}) should lead to 
constitutive relations for the dissipative quantities which are 
different from relativistic Navier-Stokes theory obtained previously 
by employing the second law of thermodynamics. The relativistic 
Navier-Stokes theory can then be understood to be valid only up to 
first order in the dissipative currents (hence also called 
first-order theory). Next, we re-calculate the entropy production, 
$\partial_\mu S^\mu$, using the more general entropy four-current 
given in Eq. (\ref{AEFCC2}),
\begin{align}\label{EFCD3C2}
\partial_\mu S^\mu = 
& - \beta\Pi\left[ \theta + \beta_0\dot\Pi 
+ \beta_{\Pi\Pi} \Pi\theta 
+ \alpha_0 \nabla_\mu n^{\mu} 
+ \psi\alpha_{n\Pi } n_\mu \dot u^\mu 
+ \psi\alpha_{\Pi n} n_\mu \nabla^{\mu}\alpha  \right] \nonumber \\
&- \beta n^\mu \Big[ T\nabla_\mu \alpha - \beta_1\dot n_\mu 
- \beta_{nn} n_\mu \theta  
+ \alpha_0\nabla_{\mu}\Pi
+ \alpha_1\nabla_\nu \pi^\nu_\mu 
+ \tilde \psi\alpha_{n\Pi } \Pi\dot u_\mu \nonumber \\
&\qquad\quad~~+ \tilde \psi\alpha_{\Pi n}\Pi\nabla_{\mu}\alpha 
+ \tilde \chi\alpha_{\pi n} \pi^\nu_\mu \nabla_\nu \alpha  
+ \tilde \chi\alpha_{n\pi } \pi^\nu_\mu \dot u_\nu \Big] \nonumber \\
& + \beta\pi^{\mu\nu}\left[ \sigma_{\mu\nu} 
- \beta_2\dot\pi_{\mu\nu} 
- \beta_{\pi\pi}\theta\pi_{\mu\nu}
- \alpha_1 \nabla_{\langle\mu}n_{\nu\rangle} 
- \chi\alpha_{\pi n}n_{\langle\mu}\nabla_{\nu\rangle}\alpha 
- \chi\alpha_{n\pi }n_{\langle\mu}\dot u_{\nu\rangle} \right],
\end{align}
As argued before, the only way to explicitly satisfy the second law 
of thermodynamics is to ensure that the entropy production is a 
positive definite quadratic function of the dissipative currents. 

The second law of thermodynamics, $\partial_{\mu}S^{\mu}\ge 0$, is 
guaranteed to be satisfied if we impose linear relationships between 
thermodynamical fluxes and extended thermodynamic forces, leading to 
the following evolution equations for bulk pressure, 
particle-diffusion four-current and shear stress tensor,
\begin{align}
\Pi =\,& -\zeta\left[ \theta 
+ \beta_0 \dot \Pi 
+ \beta_{\Pi\Pi} \Pi \theta 
+ \alpha_0 \nabla_\mu n^\mu  
+ \psi\alpha_{n\Pi} n_\mu \dot u^\mu
+ \psi\alpha_{\Pi n} n_\mu \nabla^\mu \alpha  \right], \label{bulkC2} \\ 
n^{\mu} =\ & \lambda \Big[ T \nabla^\mu \alpha 
- \beta_1\dot n^{\langle\mu\rangle} 
- \beta_{nn} n^\mu \theta
+ \alpha_0 \nabla^\mu \Pi 
+ \alpha_1 \Delta^\mu_\rho \nabla_\nu \pi^{\rho\nu}
+ \tilde \psi\alpha_{n\Pi} \Pi \dot u^{\langle\mu\rangle} \nonumber \\
&\quad~ + \tilde \psi\alpha_{\Pi n} \Pi \nabla^\mu \alpha 
+ \tilde \chi\alpha_{\pi n} \pi_\nu^\mu \nabla^\nu \alpha
+ \tilde \chi\alpha_{n\pi} \pi_\nu^\mu \dot u^\nu \Big], \label{currentC2} \\ 
\pi^{\mu\nu} =\ & 2\eta\Big[ \sigma^{\mu\nu} 
- \beta_2\dot\pi^{\langle\mu\nu\rangle} 
- \beta_{\pi\pi}\theta\pi^{\mu\nu}
- \alpha_1 \nabla^{\langle\mu}n^{\nu\rangle}
- \chi\alpha_{\pi n} n^{\langle\mu} \nabla^{\nu\rangle} \alpha 
- \chi\alpha_{n\pi } n^{\langle\mu} \dot u^{\nu\rangle} \Big] , \label{shearC2}
\end{align}
where $\lambda\equiv\kappa/T$. This implies that the dissipative 
currents must satisfy the dynamical equations,
\begin{align}
\dot\Pi + \frac{\Pi}{\tau_\Pi} =\,& -\frac{1}{\beta_0}\left[ \theta  
+ \beta_{\Pi\Pi} \Pi \theta 
+ \alpha_0 \nabla_\mu n^\mu  
+ \psi\alpha_{n\Pi} n_\mu \dot u^\mu
+ \psi\alpha_{\Pi n} n_\mu \nabla^\mu \alpha  \right], \label{bulkC2F} \\ 
\dot n^{\langle\mu\rangle} + \frac{n^\mu}{\tau_n} =\ & \frac{1}{\beta_1}\Big[ T \nabla^\mu \alpha  
- \beta_{nn} n^\mu \theta
+ \alpha_0 \nabla^\mu \Pi 
+ \alpha_1 \Delta^\mu_\rho \nabla_\nu \pi^{\rho\nu}
+ \tilde \psi\alpha_{n\Pi} \Pi \dot u^{\langle\mu\rangle} \nonumber \\
&\quad~ + \tilde \psi\alpha_{\Pi n} \Pi \nabla^\mu \alpha 
+ \tilde \chi\alpha_{\pi n} \pi_\nu^\mu \nabla^\nu \alpha
+ \tilde \chi\alpha_{n\pi} \pi_\nu^\mu \dot u^\nu \Big], \label{currentC2F} \\ 
\dot\pi^{\langle\mu\nu\rangle} + \frac{\pi^{\mu\nu}}{\tau_\pi} =\ & \frac{1}{\beta_2}\Big[ \sigma^{\mu\nu}  
- \beta_{\pi\pi}\theta\pi^{\mu\nu}
- \alpha_1 \nabla^{\langle\mu}n^{\nu\rangle}
- \chi\alpha_{\pi n} n^{\langle\mu} \nabla^{\nu\rangle} \alpha 
- \chi\alpha_{n\pi } n^{\langle\mu} \dot u^{\nu\rangle} \Big] . \label{shearC2F}
\end{align}
The above equations for the dissipative quantities are 
relaxation-type equations with the relaxation times defined as
\begin{equation}\label{RTC2}
\tau_{\Pi} \equiv \zeta\,\beta_0, \quad
\tau_{n} \equiv \lambda\,\beta_1 = \kappa\,\beta_1/T, \quad
\tau_{\pi} \equiv 2\,\eta\,\beta_2 ,
\end{equation}
Since the relaxation times must be positive, the Taylor expansion 
coefficients $\beta_0$, $\beta_1$ and $\beta_2$ must all be larger 
than zero. 

The most important feature of the Israel-Stewart theory is the 
presence of relaxation times corresponding to the dissipative 
currents. These relaxation times indicate the time scales within 
which the dissipative currents react to hydrodynamic gradients, in 
contrast to the relativistic Navier-Stokes theory where this process 
occurs instantaneously. The introduction of such relaxation 
processes restores causality and transforms the dissipative currents 
into independent dynamical variables that satisfy partial 
differential equations instead of constitutive relations. However, 
it is important to note that this welcome feature comes with a 
price: five new parameters, $\beta_0$, $\beta_1$, $\beta_2$, 
$\alpha_0$ and $\alpha_1$, are introduced in the theory. These 
coefficients cannot be determined within the present framework, 
i.e., within the framework of thermodynamics alone, and as a 
consequence the evolution equations remain incomplete. Microscopic 
theories, such as kinetic theory, have to be invoked in order to 
determine these coefficients. In the next section, we review the 
basics of relativistic kinetic theory and Boltzmann transport 
equation.

%%%%%%%%%%%%%%%%%%%%%%%%%%%%%%%%%%%%%%%%%%%%%%%%%%%%%%%%%%%%%%%%%%%%%%%%

\section{Relativistic kinetic theory}

Macroscopic properties of a many-body system are governed by the 
interactions among its constituent particles and the external 
constraints on the system. Kinetic theory presents a statistical 
framework in which the macroscopic quantities are expressed in terms 
of single-particle phase-space distribution function. The various 
formulations of relativistic dissipative hydrodynamics, presented in 
this thesis, are obtained within the framework of relativistic 
kinetic theory. In the following, we briefly outline the salient 
features of relativistic kinetic theory and dissipative 
hydrodynamics which have been employed in the subsequent 
calculations \cite{deGroot}.

Let us consider a system of relativistic particles, each having rest 
mass $m$, momentum $\vec p$ and energy $p^0$. Therefore from 
relativity, we have, $p^0=\sqrt{(\vec p)^2+m^2}$. For a large number 
of particles, we introduce a single-particle distribution function 
$f(x,p)$ which gives the distribution of the four-momentum 
$p=p^\mu=(p^0,\vec p)$ at each space-time point such that 
$f(x,p)\Delta^3x\Delta^3p$ gives the average number of particles at 
a given time $t$ in the volume element $\Delta^3x$ at point $\vec x$ 
with momenta in the range $(\vec p,\vec p+\Delta\vec p)$. However, 
this definition of the single-particle phase-space distribution 
function $f(x,p)$ assumes that, while on one hand, the number of 
particles contained in $\Delta^3x$ is large, on the other hand, 
$\Delta^3x$ is small compared to macroscopic point of view. 

The particle density $n(x)$ is introduced to describe, in general, a 
non-uniform system, such that $n(x)\Delta^3x$ is the average number 
of particles in volume $\Delta^3x$ at $(\vec x,~t)$. Similarly, 
particle flow $\vec j(x)$ is defined as the particle current. With 
the help of the distribution function, the particle density and 
particle flow are given by
\begin{equation}\label{PDPFC2}
n(x) = \int d^3p~f(x,p), \quad 
\vec j(x) = \int d^3p~\vec v\,f(x,p),
\end{equation}
where $\vec v=\vec p/p^0$ is the particle velocity. These two local 
quantities, particle density and particle flow constitute a 
four-vector field $N^\mu=(n,\vec j)$, called particle four-flow, and 
can be written in a unified way as
\begin{equation}\label{PFFC2}
N^\mu(x) = \int \frac{d^3p}{p^0}\,p^\mu\,f(x,p).
\end{equation}
Note that since $d^3p/p^0$ is a Lorentz invariant quantity, $f(x,p)$ 
should be a scalar in order that $N^\mu$ transforms as a four-vector.

Since the energy per particle is $p^0$, the average energy density 
and the energy flow can be written in terms of the distribution 
function as
\begin{equation}\label{AEEFC2}
T^{00}(x) = \int d^3p~p^0\,f(x,p), \quad T^{0i}(x) = \int d^3p~p^0\,v^i\,f(x,p).
\end{equation}
The momentum density is defined as the average value of particle 
momenta $p^i$, and, the momentum flow or pressure tensor is defined 
as the flow in direction $j$ of momentum in direction $i$. For these 
two quantities, we have
\begin{equation}\label{MDMFC2}
T^{i0}(x) = \int d^3p~p^i\,f(x,p), \quad T^{ij}(x) = \int d^3p~p^i\,v^j\,f(x,p).
\end{equation}
Combining all these in a compact covariant form using $v^i=p^i/p^0$, 
we obtain the energy-momentum tensor of a macroscopic system
\begin{equation}\label{EMTC2}
T^{\mu\nu}(x) = \int \frac{d^3p}{p^0}\,p^\mu\,p^\nu\,f(x,p).
\end{equation}
Observe that the above definition of the energy momentum tensor 
corresponds to second moment of the distribution function, and 
hence, it is a symmetric quantity.

The H-function introduced by Boltzmann implies that the 
nonequilibrium local entropy density of a system can be written as
\begin{equation}\label{EDC2}
s(x) = -\int d^3p~f(x,p)\left[ \ln f(x,p) - 1\right].
\end{equation}
The entropy flow corresponding to the above entropy density is
\begin{equation}\label{EFC2}
\vec S(x) = -\int d^3p~\vec v\,f(x,p)\left[ \ln f(x,p) - 1 \right].
\end{equation}
These two local quantities, entropy density and entropy flow 
constitute a four-vector field $S^\mu=(s,\vec S)$, called entropy 
four-flow, and can be written in a unified way as
\begin{equation}\label{EFCC2}
S^\mu(x) = -\int \frac{d^3p}{p^0}\, p^\mu\,f(x,p)\left[ \ln f(x,p) - 1 \right].
\end{equation}
The above definition of entropy four-current is valid for a system 
comprised of Maxwell-Boltzmann gas. This expression can also be 
extended to a system consisting of particles obeying Fermi-Dirac 
statistics ($r=1$), or Bose-Einstein statistics ($r=-1$) as
\begin{equation}\label{EFCQC2}
S^\mu(x) = -\int \frac{d^3p}{p^0}\, p^\mu \left[ f(x,p) \ln f(x,p) + r\tilde f(x,p) \ln \tilde f(x,p) \right], 
\end{equation}
where $\tilde f \equiv 1 - rf$. The expressions for the entropy 
four-current given in Eqs. (\ref {EFCC2}) and (\ref{EFCQC2}) can be 
used to formulate the generalized second law of thermodynamics 
(entropy law), and, define thermodynamic equilibrium.

For small departures from equilibrium, $f(x,p)$ can be written as 
$f=f_0+\delta f$. The equilibrium distribution function $f_0$ is 
defined as
\begin{equation}\label{EDFC2}
f_0(x,p) = \frac{1}{\exp(\beta u\cdot p -\alpha) + r}, 
\end{equation}
where the scalar product is defined as $u\cdot p\equiv u_\mu p^\mu$ 
and $r=0$ for Maxwell-Boltzmann statistics. Note that in 
equilibrium, i.e., for $f(x,p)=f_0(x,p)$, the particle four-flow 
and energy momentum tensor given in Eqs. (\ref{PFFC2}) and (\ref 
{EMTC2}) reduce to that of ideal hydrodynamics $N^\mu_{(0)}$ and 
$T^{\mu\nu}_{(0)}$. Therefore using Eq. (\ref{CCDLF}), the 
dissipative quantities, viz., the bulk viscous pressure $\Pi$, the 
particle diffusion current $n^\mu$, and the shear stress tensor 
$\pi^{\mu\nu}$ can be written as
\begin{equation}\label{BPSC2}
\Pi = -\frac{1}{3}\,\Delta_{\alpha\beta}\int \frac{d^3p}{p^0}\, p^\alpha p^\beta\, \delta f, \quad
n^\mu =  \Delta^{\mu\nu} \int \frac{d^3p}{p^0}\, p_\nu\, \delta f, \quad
\pi^{\mu\nu} = \Delta^{\mu\nu}_{\alpha\beta} \int \frac{d^3p}{p^0}\, p^\alpha p^\beta \delta f.
\end{equation}

The evolution equations for the dissipative quantities expressed in 
terms of the non-equilibrium distribution function, Eq. (\ref 
{BPSC2}), can be obtained provided the evolution of distribution 
function is specified from some microscopic considerations. 
Boltzmann equation governs the evolution of the phase-space 
distribution function which provides a reliably accurate description 
of the microscopic dynamics. For microscopic interactions restricted 
to $2 \leftrightarrow 2$ elastic collisions, the form of the 
Boltzmann equation is given by
\begin{equation}\label{BEC2}
p^\mu \partial_\mu f = C[f] = \frac{1}{2} \int dp' dk \ dk' \  W_{pp' \to kk'}(f_k f_{k'} \tilde f_p \tilde f_{p'} - f_p f_{p'} \tilde f_k \tilde f_{k'}),
\end{equation}
where $dp\equiv d^3p/p^0$, $C[f]$ is the collision functional and 
$W_{pp' \to kk'}$ is the collisional transition rate. The first and 
second terms within the integral of Eq. (\ref{BEC2}) refer to the 
processes $kk'\to pp'$ and $pp'\to kk'$, respectively. In the 
relaxation-time approximation, where it is assumed that the effect 
of the collisions is to restore the distribution function to its 
local equilibrium value exponentially, the collision integral 
reduces to $C[f]=-(u\cdot p)\delta f/\tau_R$ \cite 
{Anderson_Witting}. The results of these discussions will be used in 
the following chapters.

%% file: Chapter3.tex
%#######################################################################
\chapter{Boltzmann H-theorem and relativistic dissipative fluid dynamics}
%#######################################################################

%%%%%%%%%%%%%%%%%%%%%%%%%%%%%%%%%%%%%%%%%%%%%%%%%%%%%%%%%%%%%%%%%%%%%%%%

\section{Introduction}

Implementation of viscous hydrodynamics to study ultra-relativistic 
heavy-ion collisions has evoked widespread interest ever since a 
surprisingly small value of the shear viscosity to entropy density 
ratio, $\eta/s$, was estimated from the analysis of the elliptic 
flow data \cite{Romatschke:2007mq}. A precise estimate of $\eta/s$ 
is vital to the understanding of the properties of the QCD matter. 
However, the extraction of $\eta/s$ from hydrodynamic modelling of 
high-energy heavy-ion collisions is fraught with many uncertainties. 
Apart from the uncertainties prevailing in setting up the boundary 
conditions, there are ambiguities arising from the formulation of 
dissipative fluid dynamics equations itself.

In this chapter, we provide a solution to one of the major 
uncertainties that hinders an accurate extraction of the viscous 
corrections to the ideal fluid behaviour, namely the inadequate 
knowledge of the second-order transport coefficients. In the 
standard derivation of second-order evolution equations for 
dissipative quantities from the requirement of positive divergence 
of the entropy four-current, the most general algebraic form of the 
entropy current is parametrized in terms of unknown thermodynamic 
coefficients \cite{Israel:1979wp}. These coefficients which are 
related to relaxation times and coupling lengths of the shear and 
bulk pressures and heat current, however, remain undetermined within 
the framework of thermodynamics alone \cite{Muronga:2003ta}. While 
kinetic theory for massless particles \cite{Baier:2006um} and 
strongly coupled ${\cal N}=4$ supersymmetric Yang-Mills theory \cite
{Baier:2007ix} predict different shear relaxation times $\tau_\pi = 
3/2\pi T$ and $(2-\ln 2)/2\pi T$, respectively, for $\eta/s = 1/4\pi$, 
the bulk relaxation time $\tau_\Pi$ remains completely ambiguous. 
Hence {\it ad hoc} choices have been made for the value of $\tau_\Pi$ 
in hydrodynamic studies \cite
{Fries:2008ts,Denicol:2009am,Song:2009rh,Roy:2011pk,Rajagopal:2009yw}.

Lattice QCD studies for gluonic plasma in fact predict large values 
of bulk viscosity to entropy density ratio, $\zeta/s$, of about 
(6-25) $\eta/s|_{\rm KSS}$ near the QCD phase-transition temperature 
$T_c$ \cite{Meyer:2007dy}. This would translate into large values of 
the bulk pressure and bulk relaxation time, and may affect the 
evolution of the system significantly \cite
{Denicol:2009am,Song:2009rh}. Further, the large bulk pressure could 
result in a negative longitudinal pressure leading to mechanical 
instabilities (cavitation) whereby the fluid breaks up into droplets 
\cite{Torrieri:2008ip,Rajagopal:2009yw,Bhatt:2010cy}. Thus the 
theoretical uncertainties arising from the absence of reliable 
estimates for the second-order transport coefficients should be 
eliminated for a proper understanding of the system evolution.

We present here a formal derivation of the dissipative hydrodynamic 
equations where all the second-order transport coefficients get 
determined uniquely within a single theoretical framework. This is 
achieved by invoking the second law of thermodynamics for the 
generalized entropy four-current obtained using Boltzmann's 
H-function in terms of the phase-space distribution function, where 
the nonequilibrium distribution function is given by Grad's 
14-moment approximation. Significance of these coefficients is 
demonstrated in one-dimensional scaling expansion of the viscous 
medium.

Hydrodynamic evolution of a medium is governed by the conservation 
equations for the energy-momentum tensor and the particle four-flow 
\cite{deGroot}. We recall the expressions of the energy-momentum 
tensor and the particle four-flow from the previous chapter
\begin{align}\label{NTDC3}
T^{\mu\nu} &= \int dp \ p^\mu p^\nu f = \epsilon u^\mu u^\nu-(P+\Pi)\Delta ^{\mu \nu} 
+ \pi^{\mu\nu},  \nonumber\\
N^\mu &= \int dp \ p^\mu f = nu^\mu + n^\mu,
\end{align}
where $dp = g d{\bf p}/[(2 \pi)^3\sqrt{{\bf p}^2+m^2}]$, $g$ and $m$ 
being the degeneracy factor and particle rest mass, $p^{\mu}$ is the 
particle four-momentum, $f\equiv f(x,p)$ is the single-particle 
phase-space distribution function. The above integral expressions 
assume the system to be dilute so that the effects of interaction 
are small \cite{deGroot}. In the above tensor decompositions, 
$\epsilon, P, n$ are respectively energy density, pressure, net 
number density, and the dissipative quantities are the bulk viscous 
pressure $(\Pi)$, shear stress tensor $(\pi^{\mu\nu})$ and particle 
diffusion current $(n^\mu)$. Here $\Delta^{\mu\nu}=g^{\mu\nu}-u^\mu 
u^\nu$ is the projection operator on the three-space orthogonal to 
the hydrodynamic four-velocity $u^\mu$ defined in the Landau frame: 
$T^{\mu\nu} u_\nu=\epsilon u^\mu$. 

Energy-momentum conservation, $\partial_\mu T^{\mu\nu} =0$ and current
conservation, $\partial_\mu N^{\mu}=0$ yield the fundamental evolution
equations for $\epsilon$, $u^\mu$ and $n$.
\begin{align}\label{evolC3}
D\epsilon + (\epsilon+P+\Pi)\partial_\mu u^\mu - \pi^{\mu\nu}\nabla_{(\mu} u_{\nu)} &= 0,  \nonumber\\
(\epsilon+P+\Pi)D u^\alpha - \nabla^\alpha (P+\Pi) + \Delta^\alpha_\nu \partial_\mu \pi^{\mu\nu}  &= 0,  \nonumber\\
Dn + n\partial_\mu u^\mu + \partial_\mu n^{\mu} &=0.
\end{align}
We use the standard notation $A^{(\alpha}B^{\beta )} = (A^\alpha
B^\beta + A^\beta B^\alpha)/2$, $D=u^\mu\partial_\mu$, and
$\nabla^\alpha = \Delta^{\mu\alpha}\partial_\mu$. Even if the
equation of state is given, the system of Eqs. (\ref{evolC3}) is not
closed unless the evolution equations for the dissipative quantities
$\Pi$, $\pi^{\mu\nu}$, $n^\mu$ are specified.

Traditionally the dissipative equations have been obtained by 
invoking the second law of thermodynamics, viz., $\partial_\mu S^\mu 
\geq 0$, from the algebraic form of the entropy four-current $S^\mu$ 
\cite{Israel:1979wp,Baier:2006um,Muronga:2003ta}. We recall that 
$S^\mu$ can be expressed in terms of hydrodynamic variables as 
obtained in Eqs. (\ref {EFCMGF}) and (\ref{AEFCC2})
\begin{align}\label{AEFCC3}
S^\mu =& \ P\beta u^\mu - \alpha N^\mu + \beta u_\nu T^{\mu \nu}-Q^\mu(\delta N^\mu, \delta T^{\mu \nu}) \nonumber \\
=& \ s u^\mu - \frac{\mu n^\mu}{T} - \left(\beta_0\Pi^2 - \beta_1 n_\nu n^\nu 
+ \beta_2\pi_{\rho\sigma} \pi^{\rho\sigma}\right) \frac{u^\mu}{2T} 
- \left(\alpha_0\Pi\Delta^{\mu\nu} + \alpha_1\pi^{\mu\nu}\right)\frac{n_\nu}{T}.
\end{align}
Here $\beta=1/T$ is the inverse temperature, $\mu$ is the chemical
potential, $\alpha=\beta \mu$, and $Q^\mu$ is a function of deviations
from local equilibrium. The second equality is obtained by
using the definition of the equilibrium entropy density
$s=\beta (\epsilon+P-\mu n)$ and 
Taylor-expanding $Q^\mu$ to second order in dissipative fluxes.
In this expansion,
$\beta_i(\epsilon,n) \geq 0$ and $\alpha_i(\epsilon,n) \geq 0$ are the
thermodynamic coefficients corresponding to pure and mixed terms. 
These coefficients can be obtained within the kinetic theory approach
such as the IS theory \cite{Israel:1979wp}. However, it
is important to note that they cannot be determined
solely from thermodynamics using Eq. (\ref{AEFCC3}) and as a consequence 
the evolution equations remain incomplete.

%%%%%%%%%%%%%%%%%%%%%%%%%%%%%%%%%%%%%%%%%%%%%%%%%%%%%%%%%%%%%%%%%%%%%%%%

\section{Boltzmann's H-function and dissipative equations}

In contrast to the above approach, our starting point for the 
derivation of the dissipative evolution equations is the entropy 
four-current expression generalized from Boltzmann's H-function, 
Eqs. (\ref{EFCC2}) and (\ref{EFCQC2}):
\begin{eqnarray}\label{EFCC3}
S^\mu_{r=0} &=& -\int dp ~p^\mu \left( f \ln f - 1 \right) , \nonumber \\  
S^\mu_{r=\pm 1} &=& -\int dp ~p^\mu \left( f \ln f + r\tilde f \ln \tilde f\right),\quad
\end{eqnarray}
where $\tilde f \equiv 1 - rf$ and $r = 1,-1,0$ for Fermi, Bose, and
Boltzmann gas, respectively. The divergence of $S^\mu_{r=0,\pm 1}$ leads to
\begin{equation}\label{EFCDC3}
\partial_\mu S^\mu = -\int dp ~p^\mu \left(\partial_\mu f\right) \ln\! \left(\frac{f}{\tilde f}\right) .
\end{equation}

For small departures from equilibrium, $f$ can be written as $f = 
f_0 + \delta f$. The equilibrium distribution functions are defined 
as $f_0 = [\exp(\beta u\cdot p -\alpha) + r]^{-1}$, where $\beta=1/T$
and $\alpha=\mu/T$ are obtained from the equilibrium matching 
conditions $n\equiv n_0$ and $\epsilon \equiv \epsilon_0$.

To proceed further, we take recourse to Grad's 14-moment 
approximation for $\delta f$ which can be obtained from a 
Taylor-like expansion in the powers of momenta \cite
{Grad,Israel:1979wp}
\begin{equation}\label{Grad14C3}
\delta f = f_0\tilde f_0\left[\varepsilon(x) + \varepsilon_\alpha(x)p^\alpha
+ \varepsilon_{\alpha\beta}(x)p^\alpha p^\beta \right],
\end{equation}
where $\varepsilon$'s are the momentum-independent coefficients in 
the expansion, which, however, may depend on thermodynamic and 
dissipative quantities. The above expression for $\delta f$ can be 
written in an orthogonal basis \cite{Denicol:2010xn,Jaiswal:2012qm}
\begin{equation}\label{Grad14OC3}
\delta f = f_0\tilde f_0\left[\varepsilon(x) + \varepsilon_\alpha(x) p^{\langle\alpha\rangle}
+ \varepsilon_{\alpha\beta}(x)p^{\langle\alpha}p^{\beta\rangle} \right],
\end{equation}
where the notations, $A^{\langle\mu\rangle} = 
\Delta^{\mu}_{\nu}A^{\nu}$ and $B^{\langle\mu\nu\rangle} = 
\Delta^{\mu\nu}_{\alpha\beta}B^{\alpha\beta}$ represent space-like 
and traceless symmetric projections respectively, both orthogonal to 
$u^{\mu}$, where $\Delta^{\mu\nu}_{\alpha\beta} = 
[\Delta^{\mu}_{\alpha}\Delta^{\nu}_{\beta} + 
\Delta^{\mu}_{\beta}\Delta^{\nu}_{\alpha} - 
(2/3)\Delta^{\mu\nu}\Delta_{\alpha\beta}]/2$. 
In this orthogonal expansion, we obtain
\begin{equation}\label{G14C3}
\delta f \equiv f_0 \tilde f_0 \phi, \quad \phi =  \lambda_\Pi\Pi + \lambda_n n_\alpha\, p^\alpha 
+ \lambda_\pi\pi_{\alpha\beta}\, p^\alpha p^\beta.
\end{equation}
The coefficients ($\lambda_\Pi, \lambda_n, \lambda_\pi$) are 
assumed to be independent of four-momentum $p^{\mu}$ and are 
functions of $(\epsilon, \alpha, \beta)$. 

From Eqs. (\ref{EFCDC3}) and (\ref{G14C3}), we obtain
\begin{equation}\label{EFCD1C3}
\partial_\mu S^\mu = -\int dp ~ p^\mu  \left(\partial_\mu f\right)\!
\left[ \ln \! \left( \frac{f_0}{\tilde f_0} \right) + \ln \!
\left( 1 + \frac{\phi}{1-rf_0\phi}\right) \right].
\end{equation}
The $\phi$-independent terms on the right vanish due to 
energy-momentum and current conservation equations. To obtain 
second-order evolution equations for dissipative quantities, one 
should consider $S^\mu$ up to the same order. Hence $\partial_\mu 
S^\mu$ necessarily becomes third-order. Expanding the $\phi$
-dependent terms in Eq. (\ref{EFCD1C3}) and retaining all terms up to 
third order in gradients (where $\phi$ is linear in dissipative 
quantities), we get
\begin{equation}\label{EFCD2C3}
\partial_\mu S^\mu = -\int  dp ~p^\mu \left[ 
\phi(\partial_\mu f_0) - \phi^2(\tilde f_0 -1/2)(\partial_{\mu}f_0) 
+\phi^2 \partial_\mu(f_0 \tilde f_0) + \phi f_0 \tilde f_0 (\partial_\mu \phi) \right].
\end{equation}
The various integrals in the above equation can be decomposed into 
hydrodynamic tensor degrees of freedom via the definitions:
\begin{equation}\label{TDIC3}
I^{\mu_1\mu_2\cdots\mu_n} \equiv \int dp \ p^{\mu_1}  \cdots p^{\mu_n} f_0 
= I_{n0} u^{\mu_1}  \cdots u^{\mu_n} + I_{n1} (\Delta^{\mu_1\mu_2} u^{\mu_3} \cdots u^{\mu_n} + \mathrm{perms.}) + \cdots,
\end{equation}
where `perms' denotes all non-trivial permutations of the Lorentz 
indices. We similarly define the integrals 
$J^{\mu_1\mu_2\cdots\mu_n}$ and $K^{\mu_1\mu_2\cdots\mu_n}$ such that
\begin{align}\label{TDJKC3}
J^{\mu_1\mu_2\cdots\mu_n} &\equiv \int dp \ p^{\mu_1}  \cdots p^{\mu_n} f_0 \tilde f_0
= J_{n0} u^{\mu_1}  \cdots u^{\mu_n} + J_{n1} (\Delta^{\mu_1\mu_2} u^{\mu_3} \cdots u^{\mu_n} + \mathrm{perms.}) + \cdots, \nonumber \\
K^{\mu_1\mu_2\cdots\mu_n} &\equiv \int dp \ p^{\mu_1}  \cdots p^{\mu_n} f_0 \tilde f_0^2
= K_{n0} u^{\mu_1}  \cdots u^{\mu_n} + K_{n1} (\Delta^{\mu_1\mu_2} u^{\mu_3} \cdots u^{\mu_n} + \mathrm{perms.}) + \cdots.
\end{align} 

The coefficients $I_{nq}$, $J_{nq}$ and $K_{nq}$ can be obtained by 
suitable contractions of the integrals $I^{\mu_1\mu_2\cdots\mu_n}$, 
$J^{\mu_1\mu_2\cdots\mu_n}$ and $K^{\mu_1\mu_2\cdots\mu_n}$, 
repectively, and are related to each other by
\begin{align}\label{ICRC3}
2K_{nq} &= J_{nq} + \frac{1}{\beta}\big[- J_{n-1,q-1} + (n-2q)J_{n-1,q}\big], \nonumber \\
J_{nq} &= \frac{1}{\beta}\big[-I_{n-1,q-1} + (n-2q)I_{n-1,q} \big],
\end{align}
and also satisfy the differential relations
\begin{align}\label{ICDRC3}
2K_{nq} &= J_{nq} - \frac{d}{d\beta}J_{n-1,q} = J_{nq} + \frac{d}{d\alpha}J_{nq}, \nonumber \\
J_{nq} &= -\frac{d}{d\beta}I_{n-1,q} = \frac{d}{d\alpha}I_{nq}. 
\end{align}
With the help of these relations and Grad's 14-moment approximation,
Eq. (\ref{EFCD2C3}) reduces to
\begin{align}\label{EFCD3C3}
\partial_\mu S^\mu = 
& - \beta\Pi\left[ \theta + \beta_0\dot\Pi 
+ \beta_{\Pi\Pi} \Pi\theta 
+ \alpha_0 \nabla_\mu n^{\mu} 
+ \psi\alpha_{n\Pi } n_\mu \dot u^\mu 
+ \psi\alpha_{\Pi n} n_\mu \nabla^{\mu}\alpha  \right] \nonumber \\
&- \beta n^\mu \Big[ T\nabla_\mu \alpha - \beta_1\dot n_\mu 
- \beta_{nn} n_\mu \theta  
+ \alpha_0\nabla_{\mu}\Pi
+ \alpha_1\nabla_\nu \pi^\nu_\mu 
+ \tilde \psi\alpha_{n\Pi } \Pi\dot u_\mu \nonumber \\
&\qquad\quad~~+ \tilde \psi\alpha_{\Pi n}\Pi\nabla_{\mu}\alpha 
+ \tilde \chi\alpha_{\pi n} \pi^\nu_\mu \nabla_\nu \alpha  
+ \tilde \chi\alpha_{n\pi } \pi^\nu_\mu \dot u_\nu \Big] \nonumber \\
& + \beta\pi^{\mu\nu}\left[ \sigma_{\mu\nu} 
- \beta_2\dot\pi_{\mu\nu} 
- \beta_{\pi\pi}\theta\pi_{\mu\nu}
- \alpha_1 \nabla_{\langle\mu}n_{\nu\rangle} 
- \chi\alpha_{\pi n}n_{\langle\mu}\nabla_{\nu\rangle}\alpha 
- \chi\alpha_{n\pi }n_{\langle\mu}\dot u_{\nu\rangle} \right],
\end{align}
where $\alpha_i,~\beta_i,~\alpha_{XY},~\beta_{XX}$ are known functions
of $\beta,~\alpha$ and the integral coefficients $I_{nq},~J_{nq}$ and
$K_{nq}$. Two new parameters $\psi$ and $\chi$ with $\tilde \psi =
1-\psi$ and $\tilde \chi = 1-\chi$ are introduced to `share' the
contributions stemming from the cross terms of $\Pi$ and
$\pi^{\mu\nu}$ with $n^{\mu}$.

The second law of thermodynamics, $\partial_{\mu}S^{\mu}\ge 0$, is 
guaranteed to be satisfied if we impose linear relationships between 
thermodynamical fluxes and extended thermodynamic forces, leading to 
the following evolution equations for bulk pressure, charge current 
and shear stress tensor
\begin{align}
\Pi =\,& -\zeta\left[ \theta 
+ \beta_0 \dot \Pi 
+ \beta_{\Pi\Pi} \Pi \theta 
+ \alpha_0 \nabla_\mu n^\mu  
+ \psi\alpha_{n\Pi} n_\mu \dot u^\mu
+ \psi\alpha_{\Pi n} n_\mu \nabla^\mu \alpha  \right], \label{bulkC3} \\ 
n^{\mu} =\ & \lambda \Big[ T \nabla^\mu \alpha 
- \beta_1\dot n^{\langle\mu\rangle} 
- \beta_{nn} n^\mu \theta
+ \alpha_0 \nabla^\mu \Pi 
+ \alpha_1 \Delta^\mu_\rho \nabla_\nu \pi^{\rho\nu}
+ \tilde \psi\alpha_{n\Pi} \Pi \dot u^{\langle\mu\rangle} \nonumber \\
&\quad~ + \tilde \psi\alpha_{\Pi n} \Pi \nabla^\mu \alpha 
+ \tilde \chi\alpha_{\pi n} \pi_\nu^\mu \nabla^\nu \alpha
+ \tilde \chi\alpha_{n\pi} \pi_\nu^\mu \dot u^\nu \Big], \label{currentC3} \\ 
\pi^{\mu\nu} =\ & 2\eta\Big[ \sigma^{\mu\nu} 
- \beta_2\dot\pi^{\langle\mu\nu\rangle} 
- \beta_{\pi\pi}\theta\pi^{\mu\nu}
- \alpha_1 \nabla^{\langle\mu}n^{\nu\rangle}
- \chi\alpha_{\pi n} n^{\langle\mu} \nabla^{\nu\rangle} \alpha 
- \chi\alpha_{n\pi } n^{\langle\mu} \dot u^{\nu\rangle} \Big] , \label{shearC3}
\end{align}
with the coefficients of charge conductivity, bulk and shear 
viscosity, viz. $\lambda, \zeta,\eta \ge 0$. The coefficients of 
particle diffusion $\kappa$ can be written in terms of the 
coefficient of charge conductivity $\lambda$ as $\kappa=\lambda T$. 
It may be noted that although the forms of the Eqs. 
(\ref{bulkC3})-(\ref {shearC3}) are the same as in the standard 
Israel-Stewart theory \cite {Israel:1979wp,Muronga:2003ta}, Eqs. 
(\ref{bulkC2})-(\ref{shearC2}), all the transport coefficients are 
explicitly determined in the present derivation:
\begin{align}
\beta_0 &= \lambda^2_\Pi J_{10}/\beta,\quad 
\beta_1 = -\lambda^2_n J_{31}/\beta,\quad 
\beta_2 = 2\lambda^2_\pi J_{52}/\beta, \nonumber \\
\alpha_0 &= \lambda_\Pi \lambda_n J_{21}/\beta,\quad 
\alpha_1 = -2\lambda_\pi \lambda_n J_{42}/\beta. \label{alphasC3}
\end{align}
As a consequence, the relaxation times defined as,
\begin{equation}\label{RTC3}
\tau_{\Pi} = \zeta\,\beta_0, \quad
\tau_{n} = \lambda\,\beta_1, \quad
\tau_{\pi} = 2\,\eta\,\beta_2 ,
\end{equation}
can be obtained directly.
With $\lambda_\Pi = -1/J_{21}$, $\lambda_n=1/J_{21}$, 
$\lambda_{\pi}=1/(2J_{42})$, $n=I_{10}$, $\epsilon=I_{20}$, and
$P=-I_{21}$, the expressions for $\beta_1,\alpha_0,\alpha_1$ simplify to
\begin{equation}\label{B1A0A1C3}
\beta_1 = (\epsilon+P)/n^2, \quad   \alpha_0 = \alpha_1 = 1/n. 
\end{equation}
For a classical Boltzmann gas ($\tilde f_0=1$), 
the coefficients $\beta_0$ and $\beta_2$ take the simple forms
\begin{equation}\label{B0B2C3}
\beta_0 = 1/P,\quad  
\beta_2 = 3/(\epsilon+P) + m^2\beta^2P/[2(\epsilon+P)^2].
\end{equation}
Equations (\ref{bulkC3})-(\ref{shearC3}) in conjunction with the 
second-order transport coefficients (\ref{B1A0A1C3}) and (\ref
{B0B2C3}) constitute one of the main results in this derivation. 
These coefficients are obtained consistently within the same 
theoretical framework. In contrast, in the standard derivation from 
entropy principles \cite{Israel:1979wp}, the transport coefficients 
have to be estimated from an alternate theory. For instance, in the 
Israel-Stewart derivation based on kinetic theory, these involve 
complicated expressions which in the photon limit ($m \beta \to 0$) 
reduce to \cite{Israel:1976tn}
\begin{equation}\label{ISTC3}
\beta_0^{IS} = 216/(m^4 \beta^4 P), \quad \beta_2^{IS} = 3/4P.
\end{equation}
An alternate derivation from kinetic theory (KT) using directly the 
definition of dissipative currents yields 
\cite{Denicol:2010xn}
\begin{align}\label{DKRC3}
\beta_0^{KT} =& \left[ \left(\frac{1}{3}-c_s^2 \right)(\epsilon+P)-\frac{2}{9}(\epsilon-3P) 
 - \frac{m^4}{9}\mean{(u.p)^{-2}} \right]^{-1},  \nonumber \\
\beta_2^{KT} =& \, \frac{1}{2} \left[ \frac{4P}{5}
+\frac{1}{15}(\epsilon-3P)
- \frac{m^4}{15}\mean{(u.p)^{-2}}\right]^{-1},
\end{align}
where $c_s$ is the speed of sound and $\mean{\cdots}\equiv \int dp(\cdots)f_0$.
A field-theoretical (FT) approach gives \cite{Huang:2011ez}
\begin{align}\label{HKC3}
\beta_0^{FT} =& \left[ \left(\frac{1}{3}-c_s^2 \right)(\epsilon+P)-\frac{a}{9}(\epsilon-3P)\right]^{-1},
\nonumber \\
\beta_2^{FT} =& \, 1/[2(3-a)P],
\end{align}
where $a=2$ for charged scalar bosons and $a=3$ for fermions.  We 
find that our expression for $\beta_2$ (Eq. (\ref{B0B2C3})) in the 
massless limit, agrees with the IS result (Eq. (\ref{ISTC3})) and 
also with those obtained in Refs. \cite{Baier:2006um,El:2009vj}. 
Thus the shear relaxation times $\tau_\pi$ (Eq. (\ref{RTC3})) 
obtained here and in these studies are also identical. As $\beta_0$ 
in Eqs. (\ref{ISTC3})-(\ref{HKC3}) diverge in the massless limit, so 
does the bulk relaxation time $\tau_\Pi$ (Eq. (\ref{RTC3})), thereby 
stopping the evolution of the bulk pressure. It is important to note 
that $\beta_0$ in Eq. (\ref{B0B2C3}) and hence $\tau_\Pi$ in the 
present calculation remain finite in this limit. For a more detailed 
comparison of IS, KT and FT results, the reader is referred to \cite
{Denicol:2010br}. The two parameters $\psi$ and $\chi$ occurring in 
Eq. (\ref{EFCD3C3}) remain undetermined as in \cite{Israel:1979wp}; 
however, these do not contribute to the scaling expansion.

%%%%%%%%%%%%%%%%%%%%%%%%%%%%%%%%%%%%%%%%%%%%%%%%%%%%%%%%%%%%%%%%%%%%%%%%

\section{Numerical results and discussions}

To demonstrate the numerical significance of the new coefficients
derived here, we consider the evolution equations in the
boost-invariant Bjorken hydrodynamics at vanishing net baryon number
density \cite{Bjorken:1982qr}. In terms of the coordinates
($\tau,x,y,\eta_s$) where $\tau = \sqrt{t^2-z^2}$ and
$\eta_s=\tanh^{-1}(z/t)$, the initial four-velocity becomes
$u^\mu=(1,0,0,0)$. For this scenario $n^\mu=0$ and the evolution
equations for $\epsilon$, $\Phi \equiv -\tau^2\pi^{\eta_s\eta_s}$ 
and $\Pi$ reduce to (see Appendix A for details)
\begin{align}
\frac{d\epsilon}{d\tau} &= -\frac{1}{\tau}\left(\epsilon + P + \Pi -\Phi\right), \label{BEDC3} \\
\tau_{\pi}\frac{d\Phi}{d\tau} &= \frac{4\eta}{3\tau} - \Phi - \frac{4\tau_{\pi}}{3\tau}\Phi, \label{BshearC3} \\
\tau_{\Pi}\frac{d\Pi}{d\tau} &= -\frac{\zeta}{\tau} - \Pi - \frac{4\tau_{\Pi}}{3\tau}\Pi. \label{BbulkC3}
\end{align}
Noting that $\beta_0=1/P$, $\beta_2=3/(\epsilon+P)$ and $s=(\epsilon+P)/T$,
the relaxation times defined in Eq. (\ref{RTC3}) reduce to
\begin{equation}\label{MtausC3} 
\tau_{\Pi} = \frac{\epsilon+P}{PT}\left(\frac{\zeta}{s}\right), \quad
\tau_{\pi} = \frac{6}{T}\left(\frac{\eta}{s}\right).
\end{equation}

\begin{figure}[t] 
\begin{center} 
\includegraphics[scale=0.5]{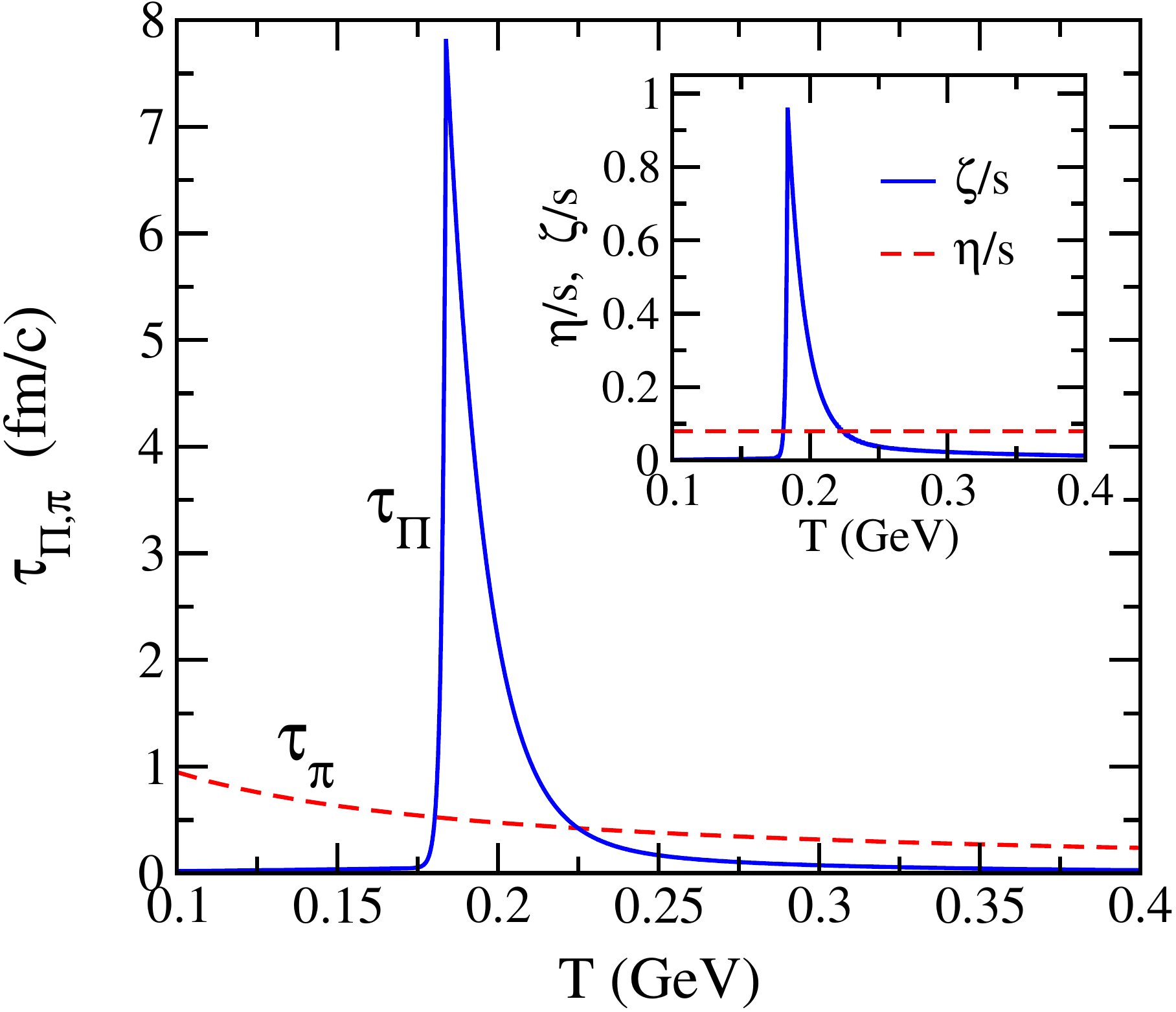} 
\end{center} 
\vspace{-0.4cm} 
\caption[Temperature dependence of bulk and shear relaxation 
  times]{Temperature dependence of bulk and shear relaxation times. 
  Inset shows $\zeta/s$ (see text) and $\eta/s = 1/4\pi$.}
\label{tauTC3}
\end{figure}

We have used the state-of-the-art equation of state
\cite{Huovinen:2009yb}, which is based on a recent lattice QCD result
\cite{Bazavov:2009zn}. For $\zeta/s$ at $T \geq T_c \approx 184$ MeV,
we use the parametrized form \cite{Rajagopal:2009yw} of the lattice
QCD results of Meyer \cite{Meyer:2007dy} which suggest a peak near
$T_c$. At $T<T_c$, the sharp drop in $\zeta/s$ reflects its extremely
small value found in the hadron resonance gas model \cite{Prakash:1993bt};
see inset of Fig. \ref{tauTC3}. For the $\eta/s$ ratio, we use the
minimal KSS bound \cite {Kovtun:2004de} value of $1/4\pi$.

In the absence of any reliable prediction for the bulk relaxation 
time $\tau_\Pi$, it has been customary to keep it fixed \cite 
{Song:2009rh,Roy:2011pk} or set it equal to the shear relaxation 
time $\tau_\pi$ \cite{Fries:2008ts,Rajagopal:2009yw} or parametrize 
it in such a way that it captures critical slowing-down of the 
medium near $T_c$ due to growing correlation lengths \cite 
{Denicol:2009am,Song:2009rh}. Since $\zeta/s$ has a peak near the 
phase transition, the $\tau_\Pi$ obtained here (Eq. (\ref{MtausC3})) 
and shown in Fig. \ref{tauTC3}, {\it naturally} captures the 
phenomenon of critical slowing-down.

The evolution equations (\ref{BEDC3})-(\ref{BbulkC3}) are solved 
simultaneously with an initial temperature $T_0 = 310$ MeV \cite
{Rajagopal:2009yw} and initial time $\tau_0 = 0.5$ fm/c typical for 
the RHIC beam energy scan. We take initial values for bulk stress 
and shear stress, $\Pi= \Phi = 0$ GeV/fm$^3$ which corresponds to an 
isotropic initial pressure configuration.

\begin{figure}[t] 
\begin{center} 
\includegraphics[scale=0.5]{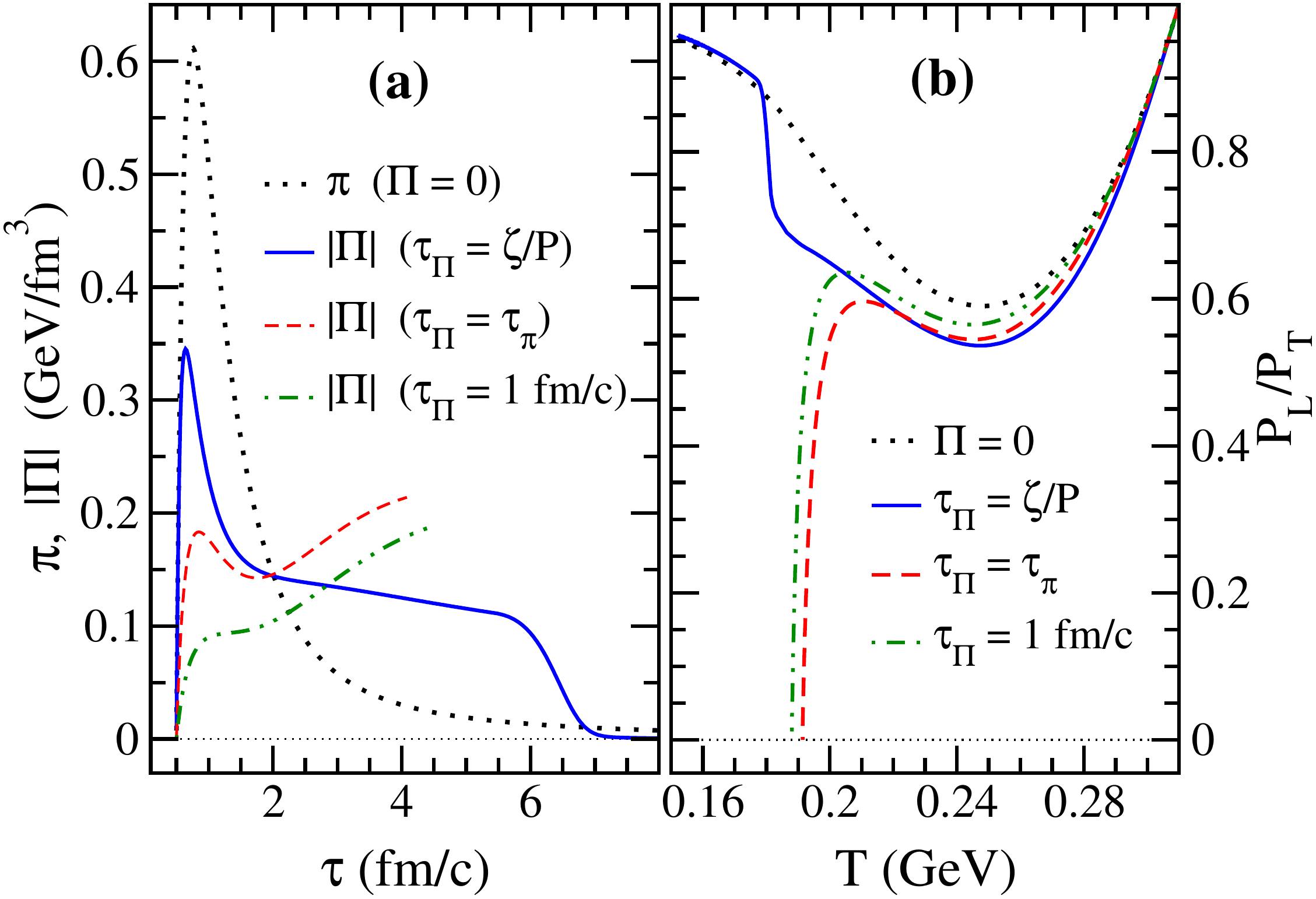} 
\end{center} 
\vspace{-0.4cm} 
\caption[Evolution of viscous pressures and pressure 
  anisotropy]{(a) Time evolution of shear stress in the absence of 
  bulk ($\Pi=0$) and magnitude of bulk stress for $\tau_\Pi=\zeta/P$ 
  and $\tau_\Pi = \tau_\pi$. The arrow indicates the time when $T_c$ 
  is reached. (b) Temperature dependence of pressure anisotropy, 
  $P_L/P_T$, for these three cases. The results are for initial 
  $T=310$ MeV, $\tau_0=0.5$ fm/c and $\eta/s = 1/4\pi$. The 
  evolution is stopped when $P_L$ vanishes.}
\label{PiphiC3}
\end{figure}

Figure \ref{PiphiC3}(a) shows time evolution of the shear pressure
$\Phi$ and the magnitude of the bulk pressure $\Pi$. At early times
$\tau \lesssim 2$ fm/c or equivalently at $T \gtrsim 1.2 T_c$, shear
dominates bulk. This implies that eccentricity-driven elliptic flow
which develops early in the system would be controlled more by the
shear pressure \cite{Song:2009rh}. At later times (when $T \sim T_c$),
the large value of $\zeta/s$ makes the bulk pressure dominant. This
leads to sizeable entropy generation (Eq. (\ref{EFCD3C3})) and
consequently enhanced particle production.

Figure \ref{PiphiC3}(a) also compares the $\Pi$ evolution for bulk
relaxation time, $\tau_{\Pi}$, calculated from Eq. (\ref{MtausC3})
(solid line) and $\tau_{\Pi} = \tau_{\pi}$ (dashed line) and 
$\tau_{\Pi} = 1$ fm/c (dashed-dotted line). At early
times, the larger value of $\tau_{\Pi}$ in the latter cases (see
Fig. \ref{tauTC3}) results in a relatively smaller growth of $|\Pi|$ as
evident from Eq. (\ref{BbulkC3}). Near $T_c$, the rapid increase in
$\zeta/s$ causes $|\Pi|$ to increase. Subsequently the longitudinal
pressure $P_L= (P+\Pi-\Phi)$ vanishes leading to cavitation
\cite{Fries:2008ts,Rajagopal:2009yw,Torrieri:2008ip,Bhatt:2010cy}. In
contrast, with our $\tau_{\Pi}$, this rise in $\zeta/s$ is
overcompensated by a faster increase in $\tau_{\Pi}$ thereby slowing
down the evolution of $\Pi$. This behaviour prevents the onset of
cavitation and guarantees the applicability of hydrodynamics with bulk
and shear up to temperatures well below $T_c$ into the hadronic
phase. Furthermore, this slowing down of the medium followed by its
rapid expansion, has the right trend to explain the identical-pion
correlation measurements (Hanbury Brown-Twiss puzzle)
\cite{Paech:2006st,Pratt:2008qv}.

The absence of cavitation in the present calculation is clearly 
evident in Fig. \ref{PiphiC3}(b) which shows the variation of 
pressure anisotropy, $P_L/P_T = (P+\Pi-\Phi)/(P+\Pi+\Phi/2)$, with 
temperature. Near $T_c$, the longitudinal pressure $P_L$ vanishes if 
one assumes $\tau_{\Pi} = \tau_{\pi}$ (dashed line) or a constant 
value $\tau_{\Pi} = 1$ fm/c (dashed-dotted line) leading to 
cavitation, whereas it is found to be positive for all temperatures 
with $\tau_\Pi$ derived here (solid line). In fact, we have found 
that in the latter case, cavitation is completely avoided for the 
entire range of $\zeta/s$ values ($0.5 < \zeta/s < 2.0$ near $T_c$) 
estimated in lattice QCD \cite{Meyer:2007dy}. The sizeable 
difference between the $\Pi = 0$ case (dot-dashed line) and the 
$\tau_\Pi=\zeta/P$ case (solid line) clearly underscores the 
importance of bulk pressure near $T_c$, which can have significant 
implications for the elliptic flow $v_2$ \cite{Denicol:2009am} thus 
affecting the extraction of $\eta/s$. Further, the large bulk 
pressure when incorporated in the freezeout prescription could also 
affect the final particle abundances and spectra.

We have also found that the evolution of $\Pi$ is insensitive to the
choice of initial conditions such as $\Pi(\tau_0)=0$ and the
Navier-Stokes value $-\zeta(T_0)/\tau_0$. This is due to very small
$\tau_\Pi$ at early times (or higher temperatures) which causes $\Pi$
to quickly lose the memory of its initial condition and to relax to
the same value at $\tau \gtrsim 1$ fm/c.

%%%%%%%%%%%%%%%%%%%%%%%%%%%%%%%%%%%%%%%%%%%%%%%%%%%%%%%%%%%%%%%%%%%%%%%%

\section{Summary and conclusions}

To summarize, we have presented a new derivation of the relativistic 
dissipative hydrodynamic equations from entropy considerations. We 
arrive at the same form of dissipative evolution equations as in the 
standard derivation but with all second-order transport coefficients 
such as the relaxation times and the entropy flux coefficients 
determined consistently within the same framework. We find that in 
the Bjorken scenario, although the bulk pressure can be large, the 
relaxation time derived here prevents the onset of cavitation due to 
the critical slowing down of bulk evolution near $T_c$. 

In the next chapter, we employ the method developed here to derive 
relativistic viscous hydrodynamic equations for two different forms 
of the non-equilibrium single-particle distribution function. These 
equations are used to study thermal dilepton and hadron spectra 
within longitudinal scaling expansion of the matter formed in 
relativistic heavy-ion collisions. For consistency, the same 
non-equilibrium distribution function will be used in the particle 
production prescription as in the derivation of the viscous 
evolution equations.

%% file: Chapter4.tex
%########################################################################
\chapter{A consistent hydrodynamic approach to particle production}
%########################################################################

%%%%%%%%%%%%%%%%%%%%%%%%%%%%%%%%%%%%%%%%%%%%%%%%%%%%%%%%%%%%%%%%%%%%%%%%

\section{Introduction}

Evolution of the strongly-interacting matter produced in high-energy
heavy-ion collisions, when the system is close to local thermodynamic
equilibrium, has been studied extensively within the framework of the
relativistic dissipative hydrodynamics; for a recent review see
Ref. \cite{Heinz:2013th}. As the system expands and becomes dilute
enough the hydrodynamic description breaks down, leading to a
freezeout or a transition from the hydrodynamic description to a
particle description \cite{Cooper:1974mv}. The dissipative effects are
important not only during the hydrodynamic evolution, but also in the
particle production \cite{Teaney:2003kp}, and both have to be treated
in a consistent manner. Moreover, the transport coefficients and
relaxation times which constitute an external input to the
hydrodynamic equations need to be in conformity with the theoretical
framework used to derive the hydrodynamic equations
\cite{Jaiswal:2013fc}. Ad hoc choices or inconsistent treatments could
significantly affect the final-state particle yields, spectra and
other observables derived from them.

As already mentioned in the previous chapters, hydrodynamics is 
formulated as an order-by-order expansion in gradients of the 
hydrodynamic four-velocity $u^\mu$ where the ideal hydrodynamics is 
zeroth order and relativistic Navier-Stokes equation is first order 
in gradients; the latter violates causality. Derivation of the 
(causal) second-order dissipative hydrodynamic equations proceeds in 
a variety of ways \cite{Romatschke:2009im}. For instance, in the 
derivations based on kinetic theory the non-equilibrium phase-space 
distribution function, $f(x,p)$, needs to be specified. This is 
commonly achieved by taking recourse to Grad's 14-moment 
approximation \cite{Grad}. The hydrodynamic equations are then 
derived by suitable coarse-graining of the microscopic dynamics. For 
consistency, the same $f(x,p)$ ought to be used in the 
particle-production prescription \cite 
{Cooper:1974mv,McLerran:1984ay} as well. This important 
consideration has been overlooked in several hydrodynamic studies of 
heavy-ion collisions.

An alternate derivation of hydrodynamic equations starts from a
generalized entropy four-current, $S^\mu$, expressed in terms of a few
unknown coefficients and then invokes the second law of thermodynamics
($\partial_\mu S^\mu \geq 0$) \cite{Romatschke:2009im}. These
coefficients which are related to relaxation times for shear and bulk
pressures remain undetermined, and have to be obtained from kinetic
theory \cite{Israel:1979wp,Muronga:2003ta}. Even then the bulk
relaxation time remains ambiguous. Ideally, a single theoretical
framework should give rise to dissipative evolution equations as well
as determine these unknown coefficients \cite{Jaiswal:2013fc}. The
bulk relaxation time obtained in Ref. \cite{Jaiswal:2013fc} exhibits
critical slowing down near the QCD phase transition and does not lead
to cavitation.

In this chapter, we employ the method of the previous chapter based 
on the entropy four-current to derive second-order viscous 
hydrodynamics corresponding to two different forms of the 
non-equilibrium distribution function. These distribution functions 
are formally different and one of them is used here for the first 
time to study the particle production in heavy-ion collisions. For 
consistency, we use the same non-equilibrium distribution function 
in the calculation of the particle spectra as in the derivation of 
the evolution equations. We perform a comparative numerical study of 
these two formalisms in the Bjorken scaling expansion. As an 
application, we study the production of thermal dileptons and 
hadrons in various scenarios.

%%%%%%%%%%%%%%%%%%%%%%%%%%%%%%%%%%%%%%%%%%%%%%%%%%%%%%%%%%%%%%%%%%%%%%%%

\section{Viscous hydrodynamics}

From Eq. (\ref{EFCC2}), the entropy four-current for particles 
obeying the Boltzmann statistics is given by \cite{deGroot}
\begin{equation}\label{EFCC4}
S^\mu(x) = -\int dp ~p^\mu f \left(\ln f - 1\right),
\end{equation}
where $dp = g d{\bf p}/[(2 \pi)^3\sqrt{{\bf p}^2+m^2}]$, $g$ and $m$
being the degeneracy factor and the particle rest mass, $p^{\mu}$ is
the particle four-momentum, and $f\equiv f(x,p)$ is the
single-particle phase-space distribution function. For a system close
to equilibrium, $f$ can be written as $f = f_0 + \delta f \equiv
f_0(1+\phi)$, where the equilibrium distribution function is defined
as $f_0 = \exp(-\beta u\cdot p)$. Here $\beta \equiv 1/T$ is the
inverse temperature, $u^\mu$ is defined in the Landau frame
\cite{deGroot}, and we have assumed the baryo-chemical potential to be
zero.

The divergence of $S^\mu$ reads
\begin{align}\label{EFCDC4}
\partial_\mu S^\mu &= -\int dp ~p^\mu \left(\partial_\mu f\right) 
\ln f \nonumber \\
&= -\int dp~p^\mu \left[\phi(1+\phi/2)(\partial_\mu f_0)+\phi(\partial_\mu \phi)f_0\right],
\end{align}
where in the second equality terms up to third order in gradients have
been retained.

To proceed further, the non-equilibrium part of the distribution
function $\delta f \equiv f_0\phi$ needs to be specified. In the
present chapter, we consider two different forms of $\phi$. The first
form is obtained using Grad's 14-moment approximation \cite{Grad} for
the single-particle distribution function in orthogonal basis \cite
{Denicol:2012cn}. We propose
\begin{equation}\label{phi1C4}
\phi_1 = \frac{\Pi}{P} + \frac{p^\mu p^\nu \pi_{\mu\nu}}{2(\epsilon+P)T^2}, 
\end{equation}
where corrections up to second order in momenta are present. Equation
(\ref{phi1C4}) has not been used before to study particle production in
heavy-ion collisions. The second form is obtained by considering the
corrections which are only quadratic in momenta \cite{Dusling:2007gi}:
\begin{equation}\label{phi2C4}
\phi_2 = \frac{p^\mu p^\nu}{2(\epsilon+P)T^2}\left(\pi_{\mu\nu}
+\frac{2}{5}\Pi\Delta_{\mu\nu}\right).
\end{equation}
In Eqs. (\ref{phi1C4}) and (\ref{phi2C4}), $\epsilon$ and $P$ are the
thermodynamic energy density and pressure, $\Pi$ the bulk viscous
pressure, $\pi^{\mu\nu}$ the shear stress tensor, and
$\Delta^{\mu\nu}=g^{\mu\nu}-u^\mu u^\nu$. The energy-momentum tensor
can be expressed in terms of these quantities as $T^{\mu\nu} =
\epsilon u^\mu u^\nu-(P+\Pi)\Delta^{\mu \nu} + \pi^{\mu\nu}$. Note
that although the contributions due to shear in Eqs. (\ref{phi1C4}) and
(\ref{phi2C4}) are identical, those due to bulk viscosity are distinct.
In the following, we shall refer to analyses performed using
Eqs. (\ref{phi1C4}) and (\ref{phi2C4}) as `Case 1' and `Case 2',
respectively.

Performing the integrals in Eq. (\ref{EFCDC4}) as outlined in
Ref. \cite{Jaiswal:2013fc}, we obtain
\begin{equation}\label{EFCD3C4}
\partial_\mu S^\mu = 
- \beta\Pi\left[ \theta +\! \beta_0\dot\Pi 
+ \frac{4}{3}\beta_0 \theta\Pi \right] 
+ \beta\pi^{\mu\nu}\left[ \sigma_{\mu\nu} 
- \beta_2\dot\pi_{\mu\nu} 
- \frac{4}{3}\beta_2 \theta\pi_{\mu\nu} \right],
\end{equation}
where $\beta_0$ and $\beta_2$ are functions of thermodynamic 
quantities $\epsilon$ and $T$, $\dot X \equiv u^\mu \partial_\mu X$, 
$\theta=\partial_\mu u^\mu$, and $\sigma^{\mu\nu}=\nabla^{\langle \mu
\nu \rangle}$. The notation $A^{\langle\mu\nu\rangle} = 
\Delta^{\mu\nu}_{\alpha\beta}A^{\alpha\beta}$, where 
$\Delta^{\mu\nu}_{\alpha\beta} = 
[\Delta^{\mu}_{\alpha}\Delta^{\nu}_{\beta} + 
\Delta^{\mu}_{\beta}\Delta^{\nu}_{\alpha} - 
(2/3)\Delta^{\mu\nu}\Delta_{\alpha\beta}]/2$, represents the 
traceless symmetric projection orthogonal to $u^{\mu}$.

The second law of thermodynamics, $\partial_{\mu}S^{\mu}\ge 0$, is
guaranteed to be satisfied if linear relationships between
thermodynamical fluxes and extended thermodynamic forces are imposed.
This leads to the following evolution equations for bulk and shear
\begin{align}
\Pi &= -\zeta\left[ \theta 
+ \beta_0 \dot \Pi 
+ \frac{4}{3}\beta_0 \theta\Pi \right], \label{bulkC4} \\ 
\pi^{\mu\nu} &= 2\eta\left[ \sigma^{\mu\nu} 
- \beta_2\dot\pi^{\langle\mu\nu\rangle} 
- \frac{4}{3}\beta_2 \theta\pi^{\mu\nu} \right] , \label{shearC4}
\end{align}
where the coefficients of bulk and shear viscosity satisfy $\zeta,\eta
\ge 0$. The bulk and shear relaxation times defined as $\tau_{\Pi} =
\zeta \beta_0$ and $\tau_{\pi} = 2 \eta \beta_2$, can be obtained
directly from the transport coefficients $\beta_0$ and $\beta_2$ which
are determined explicitly in the above derivations.

For Case 1, the coefficients $\beta_0$ and $\beta_2$ become
\begin{equation}\label{betas1C4}
\beta_0^{(1)} = 1/P,\quad  
\beta_2^{(1)} = 3/(\epsilon+P) + m^2\beta^2P/[2(\epsilon+P)^2],
\end{equation}
whereas for Case 2, they reduce to
\begin{equation}\label{betas2C4}
\beta_{0}^{(2)} = \frac{18}{5(\epsilon+P)} + \frac{3m^2\beta^2P}{5(\epsilon+P)^2},\quad
\beta_{2}^{(2)} = \beta_2^{(1)}.
\end{equation}
We note that although the relaxation time corresponding to shear
($\beta_2$) is the same for both the cases, that corresponding to bulk
($\beta_0$) is different. We stress that these coefficients have been
obtained consistently within a single theoretical framework. This is
in contrast to the standard derivation \cite {Israel:1979wp}, where
the transport coefficients have to be estimated from an alternate
theory.

%%%%%%%%%%%%%%%%%%%%%%%%%%%%%%%%%%%%%%%%%%%%%%%%%%%%%%%%%%%%%%%%%%%%%%%%

\section{Thermal dilepton and hadron production}
 
Particle production is influenced by viscosity in two ways: first
through the viscous hydrodynamic evolution of the system and second
through corrections to the particle production rate via the
non-equilibrium distribution function
\cite{Teaney:2003kp}. Hydrodynamic evolution was considered in the
previous section; here we will concentrate on the thermal dilepton and
hadron production rates in heavy-ion collisions. While the hadrons are
emitted mostly in the final stages of the evolution, the dileptons are
emitted at all stages and thus probe the entire temperature history of
the system.

In the quark-gluon plasma (QGP) phase, the dominant production
mechanism for dileptons is $q\bar q\rightarrow \gamma^* \rightarrow
l^+ l^-$, whereas in the hadronic phase the main contribution arises
from $\pi^+ \pi^-\rightarrow \rho^0 \rightarrow l^+ l^-$. The dilepton
production rate for these processes is given by \cite{rvogt}
\begin{equation}
\frac{dN}{d^4x d^4p}=g^2 \int\frac{d^3\textbf{p}_1}{(2\pi)^3} \frac{d^3\textbf{p}_2}{(2\pi)^3} f(E_1,T) f(E_2,T) 
v_{rel} \sigma(M^2)\delta^4(p-p_1-p_2),
\label{dprC4}
\end{equation}
where $p_i=(E_i,\textbf{p}_i)$ are the four momenta of the incoming
particles having equal masses $m_i$ and relative velocity
$v_{rel}=M(M^2-4m_i^2)^{1/2}/2E_1E_2$. Further, $M$ and $\sigma(M^2)$
are the dilepton invariant mass and production cross section,
respectively. Substituting for $f=f_0+\delta f$ and retaining only the
terms linear in $\delta f$, the dilepton production rate can be
expressed as a sum of contributions due to ideal, shear and bulk:
\begin{eqnarray}
 \frac{dN}{d^4xd^4p}
= \frac{dN^{(0)}}{d^4xd^4p} + \frac{dN^{(\pi)}}{d^4xd^4p} + \frac{dN^{(\Pi)}}{d^4xd^4p}. 
\label{dilrateC4}
\end{eqnarray}
For the case $M\gg T\gg m_i$, the equilibrium distribution functions
can be approximated by the Maxwell-Boltzmann form $f(E,T)= \exp(-E/T)$
and $v_{rel}\simeq M^2/2E_1E_2$. In the QGP phase (for $q\bar q$
annihilation) we have $M^2g^2\sigma(M^2)=(80\pi/9)\alpha^2$ (with
$N_f$=2 and $N_c=3$) and in the hadronic phase (for $\pi^+\pi^-$
annihilation) we have $M^2g^2\sigma(M^2)=(4\pi/3)\alpha^2
|F_\pi(M^2)|^2$ \cite{rvogt}. The electromagnetic pion form factor is
$|F_\pi(M^2)|^2=m_\rho^4/[(m_\rho^2-M^2)^2+m_\rho^2 \Gamma_\rho^2]$,
where $m_\rho=775$ MeV and $\Gamma_\rho=149$ MeV are the mass and
decay width of the $\rho(770)$ meson \cite{Song:2010fk}.

With the above approximations, the integrals in Eq. (\ref{dilrateC4})
can be performed. The ideal part is given by \cite{rvogt}
\begin{equation}
  \frac{dN^{(0)}}{d^4xd^4p}=\frac{1}{2}~\frac{M^2g^2\sigma(M^2)}{(2\pi)^5}~e^{-p_0/T}.\label{dilrate0C4}
\end{equation}
The shear viscosity contribution is the same for $\phi_1$ and
$\phi_2$, Eqs. (\ref{phi1C4}) and (\ref{phi2C4}), and is given by
\cite{Dusling:2008xj}
\begin{equation}
\frac{dN^{(\pi)}}{d^4xd^4p}=
\frac{2}{3}\left(\frac{p^\mu p^\nu}{2sT^3}\pi_{\mu\nu} \right)\frac{dN^{(0)}}{d^4xd^4p},
\label{rateshC4}
\end{equation}
where $s=(\epsilon+P)/T$ is the equilibrium entropy density. The bulk
viscosity contributions for $\phi_1$ is
\begin{equation}
\frac{dN^{(\Pi)}_1}{d^4xd^4p}
=\frac{\Pi}{P}\,\frac{dN^{(0)}}{d^4xd^4p},\label{rate1C4}
\end{equation}
and that for $\phi_2$ can be expressed as \cite{Bhatt:2011kx}
\begin{equation}
\frac{dN^{(\Pi)}_2}{d^4xd^4p}=
\frac{2}{5 sT^3} \left(\frac{M^2}{12}g^{\alpha\beta}-\frac{1}{3} p^\alpha p^\beta \right) \Delta_{\alpha\beta}\Pi \, \frac{dN^{(0)}}{d^4xd^4p}.
\label{rate2C4}
\end{equation}

The hadron spectra are obtained using the Cooper-Frye freezeout
prescription \cite{Cooper:1974mv}
\begin{equation}
\frac{dN}{d^2p_Tdy} = \frac{g}{(2\pi)^3} \int p_\mu d\Sigma^\mu f(x,p),
\label{CFC4}
\end{equation}
where, $d\Sigma^\mu$ represents the element of the three-dimensional
freezeout hypersurface and $f(x,p)$ represents the phase-space
distribution function at freezeout.

For the two cases discussed above we shall study the evolution of the
hydrodynamic variables and their influence on the dilepton and
hadron production rates.

%%%%%%%%%%%%%%%%%%%%%%%%%%%%%%%%%%%%%%%%%%%%%%%%%%%%%%%%%%%%%%%%%%%%%%%%

\section{Bjorken scenario}

We consider the evolution of the system in longitudinal scaling
expansion at vanishing net baryon number density
\cite{Bjorken:1982qr}. In terms of the Milne coordinates
($\tau,r,\varphi,\eta_s$), where $\tau = \sqrt{t^2-z^2}$ and
$\eta_s=\tanh^{-1}(z/t)$, and with $u^\mu=(1,0,0,0)$, evolution
equations for $\epsilon$, $\Phi \equiv -\tau^2 \pi^{\eta_s \eta_s}$
and $\Pi$ become (see Appendix A for details)
\begin{align}
\frac{d\epsilon}{d\tau} &= -\frac{1}{\tau}\left(\epsilon + P + \Pi -\Phi\right), \label{BEDC4} \\
\tau_{\pi}\frac{d\Phi}{d\tau} &= \frac{4\eta}{3\tau} - \Phi - \frac{4\tau_{\pi}}{3\tau}\Phi, \label{BshearC4} \\
\tau_{\Pi}\frac{d\Pi}{d\tau} &= -\frac{\zeta}{\tau} - \Pi - \frac{4\tau_{\Pi}}{3\tau}\Pi \label{BbulkC4}.
\end{align}
The bulk and shear relaxation times $\tau_{\Pi} = \zeta \beta_0$ and
$\tau_{\pi} = 2 \eta \beta_2$, reduce to
\begin{equation}\label{MtausC4}
\tau_{\Pi}^{(1)} = \frac{\epsilon+P}{PT} \left(\frac{\zeta}{s} \right), \quad
\tau_{\Pi}^{(2)} = \frac{18}{5T}\left(\frac{\zeta}{s}\right), \quad
\tau_{\pi} = \frac{6}{T}\left(\frac{\eta}{s}\right),
\end{equation}
for the two different forms of $\phi$, Eqs. (\ref{phi1C4}) and
(\ref{phi2C4}).

Once the temperature evolution is known from the hydrodynamical model,
the total dilepton spectrum is obtained by integrating the total rate
over the space-time evolution of the system
\begin{align}\label{totalC4}
\frac{dN_{1,2}}{d^2p_TdM^2dy}= A_\perp
\int_{\tau_0}^{\tau_{fo}}
d\tau ~\tau \int_{-\infty}^{\infty}d\eta_s
\left(\frac{1}{2}\frac{dN_{1,2}}{d^4xd^4p}\right),
\end{align}
where $A_\perp$ is the transverse area of the overlap zone of the 
colliding nuclei and, $\tau_0$ and $\tau_{fo}$ are the initial and 
freezeout times for the hydrodynamic evolution. We note that for the 
Bjorken expansion, $d^4x=A_\perp d\eta_s\tau d\tau$. For central 
($b=0$) collisions, $A_\perp=\pi R_A^2$ where $R_A=1.2 A^{1/3}$ is the 
nuclear radius, $A$ being the mass number of the colliding nuclei.

In $(\tau,r,\varphi,\eta_s)$ coordinates, particle four momentum is
$p^{\mu} = (m_T \cosh(y-\eta_s),~p_T \cos(\varphi_p- \varphi),~p_T
\sin (\varphi_p-\varphi)/r,~m_T \sinh(y-\eta_s)/\tau)$, where $m_T^2$
= $p_T^2+m^2$. The other factors appearing in the rate expressions,
Eqs. (\ref{rateshC4})-(\ref{CFC4}), are then given by
\begin{eqnarray}\label{visc-fcts1C4}
 p^{\alpha}p^{\beta} \pi_{\alpha\beta} 
&=& \frac{\Phi}{2} p_T^2-\Phi ~m_T^2~\sinh^2(y-\eta_s),\\
 p^\alpha p^\beta \Delta_{\alpha\beta} 
&=& -p_T^2-m_T^2~\sinh^2(y-\eta_s). \label{visc-fcts2C4}
\end{eqnarray}

Similar to the dilepton spectra, the hadronic spectra can also be
split up into three parts. Writing the momentum flux through the
hypersurface element as $p_\mu d\Sigma^\mu = m_T \cosh(y-\eta_s) \tau
d\eta_s r dr d\varphi$, and performing the $\eta_s$ integration, we
get for the ideal case,
\begin{equation}
\frac{dN^{(0)}}{d^2p_Tdy} = \frac{g}{4\pi^3}m_T\tau_{fo} A_\perp K_1(z_m),
\end{equation}
where $K_n$ are the modified Bessel functions of the second kind and
$z_m\equiv m_T/T$. The contribution due to the shear viscosity to the
hadron production reduces to
\begin{equation}
\frac{dN^{(\pi)}}{d^2p_Tdy} = \frac{\Phi}{4(\epsilon+P)} \left[z_p^2 - 2z_m \frac{K_2(z_m)}{K_1(z_m)}\right]\frac{dN^{(0)}}{d^2p_Tdy},
\end{equation}
where $z_p\equiv p_T/T$. The bulk viscosity contribution in Case 1,
Eq. (\ref{phi1C4}), is calculated to be
\begin{equation}
\label{hrate1C4}
\frac{dN^{(\Pi)}_1}{d^2p_Tdy} = \frac{\Pi}{P}\frac{dN^{(0)}}{d^2p_Tdy} ,
\end{equation}
whereas in Case 2, Eq. (\ref{phi2C4}), it reduces to
\begin{equation}
\label{hrate2C4}
\frac{dN^{(\Pi)}_2}{d^2p_Tdy} =  \frac{-\Pi}{5(\epsilon+P)}\left[z_p^2 + z_m \frac{K_2(z_m)}{K_1(z_m)}\right]\frac{dN^{(0)}}{d^2p_Tdy}.
\end{equation}
Here we have used the recurrence relation
$K_{n+1}(z)=2nK_n(z)/z+K_{n-1}(z)$. It is important to note that the
bulk viscosity contribution in Case 1 is negative, whereas that in
Case 2 is positive ($\Pi<0$).

%%%%%%%%%%%%%%%%%%%%%%%%%%%%%%%%%%%%%%%%%%%%%%%%%%%%%%%%%%%%%%%%%%%%%%%%

\section{Numerical results and discussion}

We now present numerical results for the Bjorken expansion of the 
medium for the initial temperature $T_0=310$ MeV and time 
$\tau_0=0.5$ fm/c, typical for the Relativistic Heavy-Ion Collider.  
The freezeout temperature was taken to be $T_{fo}=160$ MeV.  Initial 
pressure configuration was assumed to be isotropic: $\Phi=0=\Pi$. We 
employ the equation of state of Refs. \cite
{Huovinen:2009yb,Bazavov:2009zn} based on a recent lattice QCD 
simulation. The shear viscosity to entropy density ratio $\eta/s$ 
was taken to be $1/4\pi$ corresponding to the conjectured lower 
bound obtained in Ref. \cite{Kovtun:2004de}. For the bulk viscosity 
to entropy density ratio $\zeta/s$ at $T \geq T_c \approx 184$ MeV 
we adopted a parametrized form of the lattice QCD result; see Refs. 
\cite{Meyer:2007dy,Rajagopal:2009yw}.  For $T<T_c$, we parametrized 
$\zeta/s$ given in Ref. \cite{Prakash:1993bt}.

\begin{figure}[t]
\begin{center}
\includegraphics[scale=0.5]{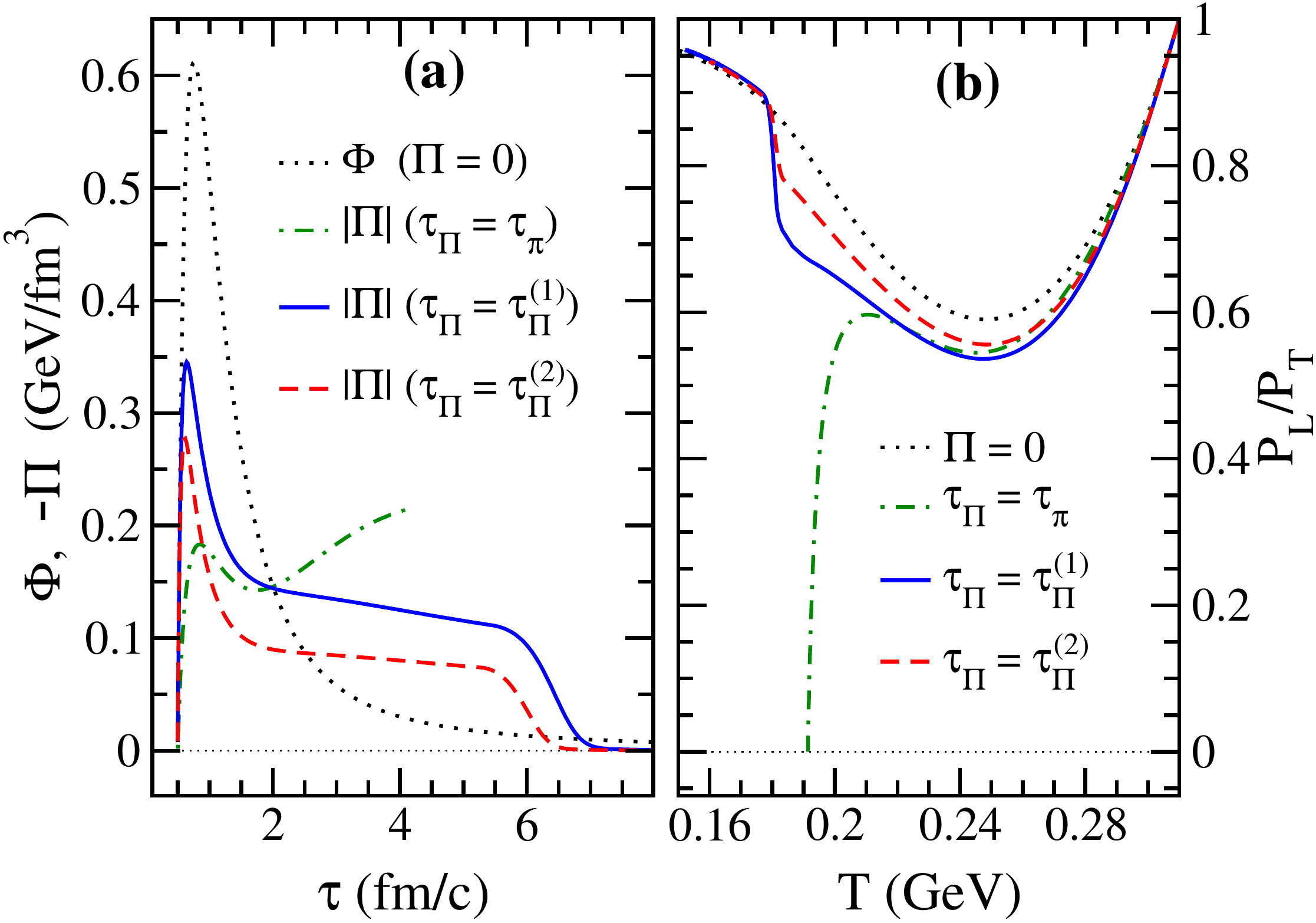}
\end{center}
\vspace{-0.4cm}
\caption[Evolution of shear and bulk viscous pressures, and pressure 
  anisotropy]{(a) Time evolution of shear, $\Phi$ and bulk, $\Pi$ 
  viscous pressures, and (b) temperature dependence of the ratio of 
  the longitudinal to transverse pressures $P_L/P_T$, for the 
  various bulk relaxation times $\tau_\Pi$ defined in Eq. (\ref
  {MtausC4}). Note that for $\tau_\Pi=\tau_\pi$, cavitation ($P_L<0$) 
  sets in.}
\label{plptC4}\end{figure}

Figure \ref{plptC4}(a) presents the time evolution of shear ($\Phi$) and
bulk ($\Pi$) viscous pressures for the various bulk relaxation times
$\tau_\Pi$ defined in Eq. (\ref{MtausC4}). At times $\tau \gtrsim 3$
fm/c, corresponding to temperatures $T \lesssim 1.2 ~T_c$, the bulk
dominates the shear pressure which can influence the particle
production appreciably. The widely used choice $\tau_\Pi = \tau_\pi$
(dot-dashed line) leads to vanishing longitudinal pressure $P_L$ and
cavitation \cite{Rajagopal:2009yw} as is evident in
Fig. \ref{plptC4}(b). On the other hand, $\tau_\Pi=\tau_\Pi^{(1,2)}$
does not lead to cavitation as discussed in \cite{Jaiswal:2013fc}. As
$\tau_\Pi^{(1)} > \tau_\Pi^{(2)}$ at all times, the magnitude of $\Pi$
is found to be larger in Case 1 (solid line). This leads to enhanced
pressure anisotropy, i.e., a larger departure of
$P_L/P_T=(P+\Pi-\Phi)/(P+\Pi+\Phi/2)$ from unity.

\begin{figure}[t]
\begin{center}
\includegraphics[scale=0.5]{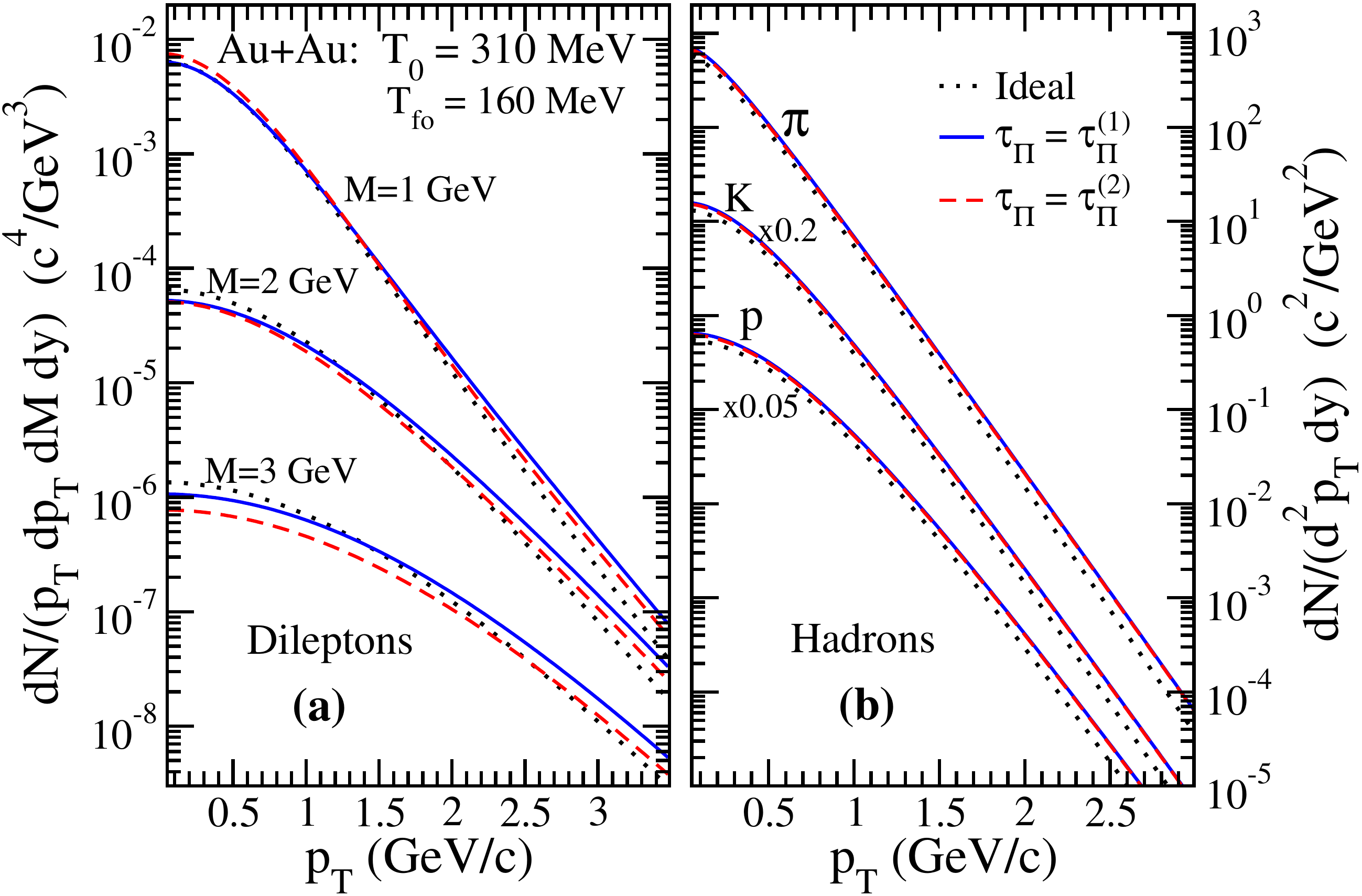}
\end{center}
\vspace{-0.4cm}
\caption[Particle spectra as a function of the transverse momentum]
  {Particle spectra as a function of the transverse momentum $p_T$, 
  for ideal and viscous hydrodynamics with bulk relaxation times 
  $\tau_\Pi$ defined in Eq. (\ref{MtausC4}) for (a) dileptons of 
  invariant mass $M=1,~2,~3$ GeV/$c^2$, and (b) hadrons.}
\label{dnptC4}
\end{figure}

\begin{figure}[t]
\begin{center}
\includegraphics[scale=0.5]{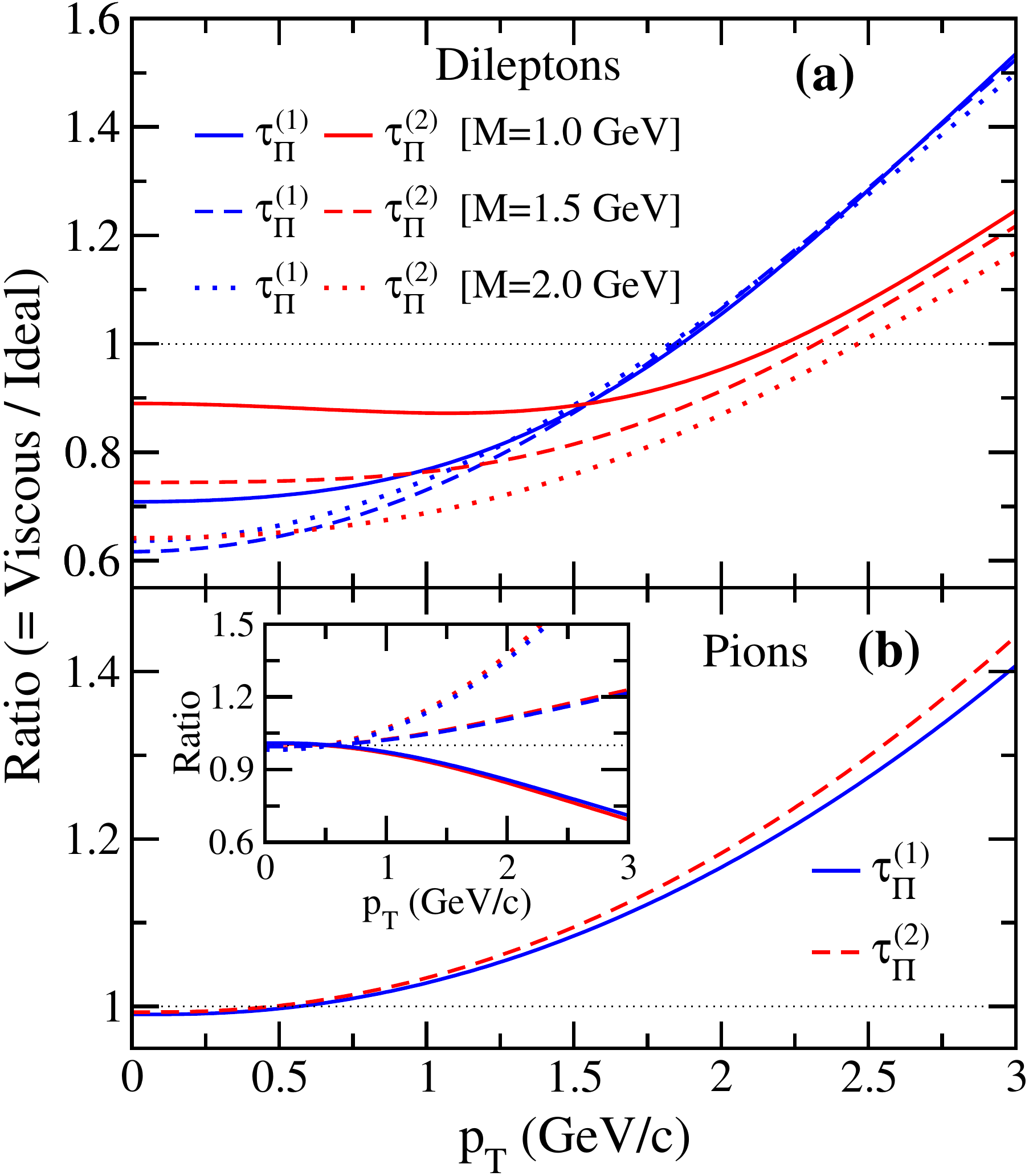}
\end{center}
\vspace{-0.4cm}
\caption[Ratios of particle yields]{Ratios of particle yields for 
  viscous and ideal hydrodynamics as a function of $p_T$, for the 
  two bulk relaxation times $\tau_\Pi$ defined in Eq. (\ref 
  {MtausC4}) for (a) dileptons of invariant mass $M=1,~1.5,~2$ 
  GeV/$c^2$, and (b) pions. Inset: Pion yields in various evolution 
  and production scenarios scaled by the consistent second-order 
  calculation for Case 1 (blue) and Case 2 (red). Solid lines: 
  second-order evolution with ideal production rate; Dashed lines: 
  second-order evolution with first-order correction to the 
  production rate; Dotted lines: ideal evolution with first-order 
  correction to the production rate.}
\label{ratioC4}
\end{figure}

Figure \ref{dnptC4} displays dilepton and hadron transverse momentum
spectra for the two choices of $\tau_\Pi$, in comparison with the
ideal hydrodynamic calculation, and Fig. \ref{ratioC4} shows the same
spectra normalized by the ideal case. Note the enhancement of the
dilepton spectra at high $p_T$, and suppression at low $p_T$ compared
to the ideal case. The high-$p_T$ dileptons emerge predominantly at
early times when the temperature and density are high. Viscosity slows
down the cooling of the system \cite{Muronga:2003ta} producing
relatively larger number of hard dileptons. We observe that at high
$p_T$, the viscous correction to the dilepton production rate due to
shear is positive and dominates that due to bulk. The low-$p_T$
dileptons are produced mainly at later stages of the evolution when
the negative correction due to the bulk viscosity dominates
(Fig. \ref{plptC4}) leading to the suppression of the spectra compared
to the ideal case. Further for Case 2 (red lines), the $p_T^2$
dependence of the viscous correction, Eqs. (\ref{rate2C4}) and
(\ref{visc-fcts2C4}), implies a smaller enhancement (suppression) at
high (low) $p_T$, compared to Case 1 (blue lines). The $M$-dependent
splitting is consistent with Eqs. (\ref{rate1C4})-(\ref{rate2C4}).

Figure \ref{ratioC4}(b) shows the pion spectra scaled by the ideal case
for the two choices of $\tau_\Pi$. The negative contribution from the
bulk viscous correction for Case 1, Eq. (\ref{hrate1C4}), causes
suppression of the ratio relative to Case 2, Eq. (\ref{hrate2C4}), where
the correction is positive. More massive hadrons display qualitatively
similar behaviour. Interestingly, at high $p_T$, dileptons and hadrons
display opposite trends for $\tau_\Pi^{(1)}$ and $\tau_\Pi^{(2)}$
(Fig. \ref{ratioC4}).

\begin{figure}[t]
\begin{center}
\includegraphics[scale=0.5]{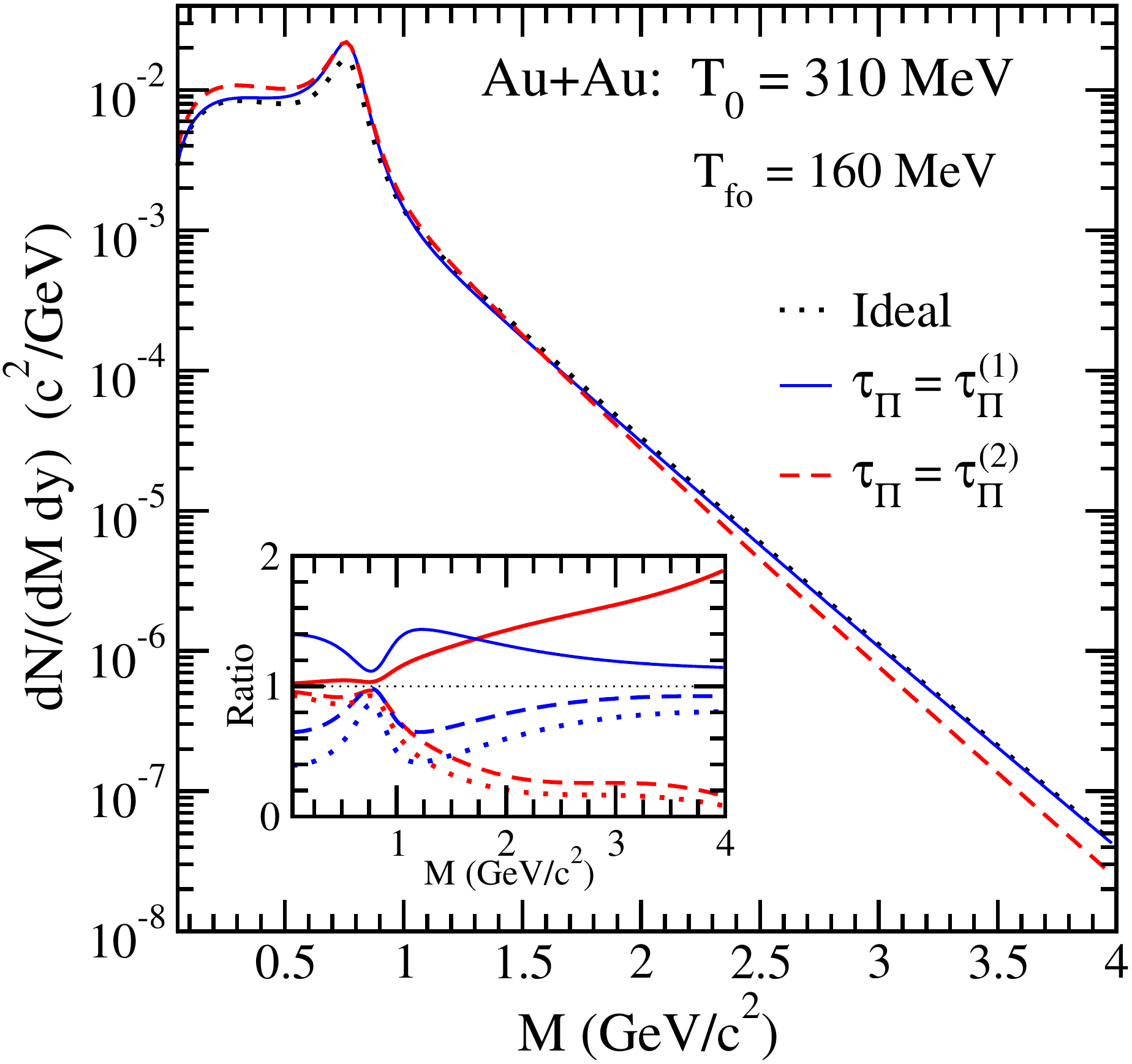}
\end{center}
\vspace{-0.4cm}
\caption[Dilepton yields as a function of the invariant mass]{Dilepton 
  yields as a function of the invariant mass $M$, in ideal and 
  viscous hydrodynamics with the two bulk relaxation times $\tau_\Pi$
  defined in Eq. (\ref{MtausC4}). Inset: Same as Fig. \ref{ratioC4} 
  inset but for dileptons.}
\label{dndmC4}
\end{figure}

Finally, Fig. \ref{dndmC4} shows the dilepton invariant mass spectra for
the two cases (Eqs. (\ref{phi1C4})-(\ref{phi2C4})) in comparison with the
ideal case. Results based on Case 1 (blue solid) are almost the same
as those obtained in the ideal case at all invariant masses. This is
essentially due to the fact that the invariant mass spectrum is
dictated by the yields at small $p_T$ where the two are nearly
identical (Fig. \ref{dnptC4}(a)). For Case 2 (red dashed) the spectrum
exhibits enhanced low-mass and suppressed high-mass dilepton
yields. This again can be traced back to the trend seen in
Fig. \ref{dnptC4}(a). Note that the peak at $M=0.77$ GeV corresponds to
the dilepton production from the $\rho(770)$ decay.

In contrast to the consistent approach adopted here, in
Refs. \cite{Monnai:2009ad,Dusling:2008xj}, ideal hydrodynamical
evolution was followed by particle production with non-ideal $f(x,p)$
up to first order in gradients. On the other hand, in
Refs. \cite{Bhatt:2011kx,Denicol:2009am,Dusling:2011fd}, although the
evolution was according to the second-order viscous hydrodynamics, the
freezeout procedure involved ideal \cite{Denicol:2009am} or
Navier-Stokes \cite{Bhatt:2011kx,Dusling:2011fd} corrections to the
$f(x,p)$. To illustrate the differences arising due to inconsistent
approaches, we show in the insets of Figs. \ref{ratioC4} and \ref{dndmC4},
pion and dilepton production rates in various evolution and production
scenarios scaled by the rate obtained in a consistent second-order
calculation. We find that the results deviate from unity significantly
which may have important implications for the on-going efforts to
extract transport properties of QGP within hydrodynamic framework.

%%%%%%%%%%%%%%%%%%%%%%%%%%%%%%%%%%%%%%%%%%%%%%%%%%%%%%%%%%%%%%%%%%%%%%%%

\section{Summary and conclusions}

We have derived viscous hydrodynamic equations for two different 
forms of the non-equilibrium distribution function, and have 
consistently used the same distribution function in the particle 
production prescription. In the Bjorken scaling expansion, we found 
appreciable differences between these two cases, for both dilepton 
and hadron production rates. Further, we showed that the dilepton 
and pion yields are significantly affected if the viscous effects in 
hydrodynamic evolution and particle production are not mutually 
consistent.

The derivation of second-order dissipative hydrodynamics from the 
entropy principles, as discussed in the present and previous 
chapter, is not complete in the sense that it misses several terms 
in the dynamical equations for dissipative quantities, compared to 
the derivations based on Boltzmann equation \cite
{Romatschke:2009im}. Moreover, as the non-equilibrium distribution 
function can be obtained by solving the Boltzmann equation, Grad's 
14-moment approximation is unnecessary in the formulation of 
dissipative hydrodynamics based on Boltzmann equation. In the next 
chapter, we derive second-order hydrodynamic evolution equations for 
the dissipative quantities, directly from their definitions, by 
solving the Boltzmann kinetic equation iteratively to obtain the 
non-equilibrium distribution function. 

%% file: Chapter5.tex
%########################################################################
\chapter{Chapman-Enskog expansion and relativistic dissipative hydrodynamics}
%########################################################################

%%%%%%%%%%%%%%%%%%%%%%%%%%%%%%%%%%%%%%%%%%%%%%%%%%%%%%%%%%%%%%%%%%%%%%%%

\section{Introduction}

The earliest theoretical formulation of relativistic dissipative 
hydrodynamics also known as first-order theories (order of 
gradients), are due to Eckart \cite{Eckart:1940zz} and 
Landau-Lifshitz \cite{Landau}. The Chapman-Enskog (CE) expansion has 
been the most common method to obtain first-order hydrodynamics from 
Boltzmann Equation (BE) \cite{Chapman}. However, as discussed in 
Chapter 2, these theories involve parabolic differential equations 
and suffer from acausality and numerical instability. The derivation 
of second-order fluid-dynamics by Israel and Stewart (IS) from 
kinetic theory uses extended Grad's method \cite{Israel:1979wp}. The 
approach by Israel and Stewart may not guarantee stability but 
solves the acausality problem \cite{Huovinen:2008te} at the cost of 
introducing two additional approximations: \textbf{(a)} 14-moment 
approximation for the distribution function and, \textbf{(b)} use 
of second moment of BE to obtain evolution equations for dissipative 
quantities.

Grad's method, originally proposed for non-relativistic systems, was 
modified by Israel and Stewart so that it could be applicable to the 
relativistic case. In this extension, known as 14-moment 
approximation, the distribution function is Taylor expanded in 
powers of four-momenta around its local equilibrium value, see 
Chapter 3. Truncating the Taylor expansion at second-order in 
momenta results in 14 unknowns that have to be determined to 
describe the distribution function. This expansion implicitly 
assumes a converging series in powers of momenta. In addition, it is 
assumed that the order of expansion in 14-moment approximation 
(expanded as a series in momenta) coincides with that of gradient 
expansion of hydrodynamics. This is evident because Grad's 
approximation truncated at second-order in momenta is not consistent 
with second-order hydrodynamics.

Another assumption inherent in IS derivation is the choice of second 
moment of the BE to extract the equation of motion for the 
dissipative quantities. This choice is arbitrary in the sense that 
once the distribution function is specified, any moment of the BE 
will lead to a closed set of equations for the dissipative currents 
but with different transport coefficients. In fact, it has been 
pointed out in Ref. \cite{Denicol:2010xn} that instead of this 
ambiguous choice of the second-moment of BE by IS, the dissipative 
quantities can be obtained directly from their definition. 
Consistent and accurate formulation of relativistic dissipative 
hydrodynamics is still unresolved and is currently an active 
research area \cite{Denicol:2010xn,Jaiswal:2013fc,Jaiswal:2012qm,El:2009vj,Denicol:2012cn}.

In this chapter, we present an alternative derivation of 
hydrodynamic equations for dissipative quantities which do not make 
use of both these assumptions. We revisit the CE expansion of the 
distribution function using BE in Relaxation Time Approximation 
(RTA). The RTA for the collision term in BE is based on the 
assumption that the effect of the collisions is to exponentially 
restore the distribution function to its local equilibrium value. 
Although the information about the microscopic interactions of the 
constituent particles is not retained here, it is a reasonably good 
approximation to describe a system which is close to local 
equilibrium. Using this expansion, we derive the first and 
second-order equations of motion for the dissipative quantities from 
their definition. In one-dimensional boost-invariant Bjorken 
scenario, we demonstrate that our second-order results are in better 
agreement with transport results as compared to those obtained by 
using IS equations. We also illustrate that heuristic incorporation 
of higher-order corrections in viscous evolution equation 
significantly improves this agreement.

\section{Chapman-Enskog expansion}

Fluid dynamics is best described as a long-wavelength, low-frequency 
limit of an underlying microscopic theory. Further, BE governs the 
temporal evolution of single particle phase-space distribution function 
$f\equiv f(x,p)$ which provides a reliably accurate description of 
the microscopic dynamics in the dilute limit. With this motivation, 
our starting point for the derivation of hydrodynamic equations is 
relativistic BE with RTA for the collision 
term \cite{Anderson_Witting}
\begin{equation}\label{RBEC5}
p^\mu\partial_\mu f =  -\frac{u\!\cdot\!p}{\tau_R}(f-f_0)~,
\end{equation}
where, $p^{\mu}$ is the particle four-momentum, $u_\mu$ is the fluid 
four-velocity and $\tau_R$ is the relaxation time. We define the 
scalar product $u\cdot p\equiv u_\mu p^\mu$. The equilibrium 
distribution functions for Fermi, Bose, and Boltzmann particles 
($r=1,-1,0$) are
\begin{equation}\label{EDFC5}
f_0=\frac{1}{\exp(\beta\,u\!\cdot\!p-\alpha)+r}~.
\end{equation}
Here, $\beta=1/T$ is the inverse temperature and $\alpha=\mu/T$ is 
the ratio of chemical potential to temperature.

In the CE expansion, the particle distribution function is expanded 
about its equilibrium value in powers of space-time gradients.
\begin{equation}\label{CEEC5}
f = f_0 + \delta f, \quad \delta f= \delta f^{(1)} + \delta f^{(2)} + \cdots ,
\end{equation}
where $\delta f^{(1)}$ is first-order in gradients, $\delta f^{(2)}$ 
is second-order and so on. The Boltzmann equation, (\ref{RBEC5}), in the 
form $f=f_0-(\tau_R/u\cdot p)\,p^\mu\partial_\mu f$, can be solved 
iteratively as \cite{Jaiswal:2013npa,Romatschke:2011qp}
\begin{equation}\label{F1F2C5}
f_1 = f_0 -\frac{\tau_R}{u\!\cdot\!p} \, p^\mu \partial_\mu f_0, \quad f_2 = f_0 -\frac{\tau_R}{u\!\cdot\!p} \, p^\mu \partial_\mu f_1, \quad \cdots
\end{equation}
where $f_1=f_0+\delta f^{(1)}$ and $f_2=f_0+\delta f^{(1)}+\delta 
f^{(2)}$. To first and second-order in gradients, we obtain
\begin{align}
\delta f^{(1)} &= -\frac{\tau_R}{u\!\cdot\!p} \, p^\mu \partial_\mu f_0, \label{FOCC5} \\
\delta f^{(2)} &= \frac{\tau_R}{u\!\cdot\!p}p^\mu p^\nu\partial_\mu\Big(\frac{\tau_R}{u\!\cdot\!p} \partial_\nu f_0\Big). \label{SOCC5}
\end{align}

For the sake of comparison, we also write down the Grad's 14-moment 
expansion \cite{Grad} in orders of momenta as suggested by IS 
\cite{Israel:1979wp} in orthogonal basis \cite{Denicol:2012cn},
\begin{equation}\label{FMEC5}
\delta f= f_0 \tilde f_0\left( \lambda_\Pi \Pi + \lambda_n n_\alpha p^\alpha 
+ \lambda_\pi \pi_{\alpha\beta} p^\alpha p^\beta \right) + \mathcal{O}(p^3),
\end{equation}
where, $\tilde f_0=1-rf_0$ and $\lambda_\Pi$, $\lambda_n$, 
$\lambda_\pi$ are determined from the definition of the dissipative 
quantities, Eqs. (\ref{FBEC5})-(\ref{FSEC5}). Since hydrodynamics 
involves expansion in orders of gradients, hence for consistency, CE 
should be preferred over 14-moment approximation in derivation of 
hydrodynamic equations. 

\section{Relativistic hydrodynamics}

The conserved energy-momentum tensor and particle current can be 
expressed in terms of distribution function, as described in Chapter 
2,
\begin{align}\label{NTDC5}
T^{\mu\nu} &= \int dp \ p^\mu p^\nu f = \epsilon u^\mu u^\nu-(P+\Pi)\Delta ^{\mu \nu} 
+ \pi^{\mu\nu},  \nonumber\\
N^\mu &= \int dp \ p^\mu f = nu^\mu + n^\mu,
\end{align}
where $dp = g d{\bf p}/[(2 \pi)^3\sqrt{{\bf p}^2+m^2}]$, $g$ and $m$ 
being the degeneracy factor and particle mass. In the tensor 
decompositions, $\epsilon, P, n$ are respectively energy density, 
pressure, net number density, and $\Delta^{\mu\nu}=g^{\mu\nu}-u^\mu 
u^\nu$ is the projection operator on the three-space orthogonal to 
the hydrodynamic four-velocity $u^\mu$ defined in the Landau frame: 
$T^{\mu\nu} u_\nu=\epsilon u^\mu$. The metric tensor is 
$g^{\mu\nu}\equiv\mathrm{diag}(+,-,-,-)$. The bulk viscous pressure 
$(\Pi)$, shear stress tensor $(\pi^{\mu\nu})$ and particle diffusion 
current $(n^\mu)$ are the dissipative quantities. 

Energy-momentum conservation, $\partial_\mu T^{\mu\nu} =0$ and 
current conservation, $\partial_\mu N^{\mu}=0$,  yields the 
fundamental evolution equations for $n$, $\epsilon$ and $u^\mu$
\begin{align}\label{evolC5}
\dot\epsilon + (\epsilon+P+\Pi)\theta - \pi^{\mu\nu}\nabla_{(\mu} u_{\nu)} &= 0,  \nonumber\\
(\epsilon+P+\Pi)\dot u^\alpha - \nabla^\alpha (P+\Pi) + \Delta^\alpha_\nu \partial_\mu \pi^{\mu\nu}  &= 0, \nonumber\\
\dot n + n\theta + \partial_\mu n^{\mu} &=0~. 
\end{align}
We use the standard notation $\dot A=u^\mu\partial_\mu A$ for 
co-moving derivative, $\nabla^\alpha=\Delta^{\mu\alpha}\partial_\mu$ 
for space-like derivative, $\theta=\partial_\mu u^\mu$ for expansion 
scalar and $A^{(\alpha}B^{\beta )}=(A^\alpha B^\beta + A^\beta 
B^\alpha)/2$ for symmetrization.

Even if the equation of state relating $\epsilon$ and $P$ is 
provided, the system of Eqs. (\ref{evolC5}) is not closed unless the 
dissipative quantities $\Pi$, $n^\mu$ and $\pi^{\mu\nu}$ are 
specified. To obtain the expressions for these dissipative 
quantities, we write them using Eq. (\ref{NTDC5}) in terms of away 
from equilibrium part of the distribution functions, $\delta f$, as
\begin{align}
\Pi &= -\frac{\Delta_{\alpha\beta}}{3} \int dp \, p^\alpha p^\beta \delta f , \label{FBEC5}\\
n^\mu &= \Delta^\mu_\alpha \int dp \, p^\alpha \delta f , \label{FCEC5}\\
\pi^{\mu\nu} &= \Delta^{\mu\nu}_{\alpha\beta} \int dp \, p^\alpha p^\beta \delta f ,\label{FSEC5}
\end{align}
where
$\Delta^{\mu\nu}_{\alpha\beta} = [\Delta^{\mu}_{\alpha}\Delta^{\nu}_{\beta} + \Delta^{\mu}_{\beta}\Delta^{\nu}_{\alpha} - (2/3)\Delta^{\mu\nu}\Delta_{\alpha\beta}]/2$.

The first-order dissipative equations can be obtained from Eqs. 
(\ref{FBEC5})-(\ref{FSEC5}) using $\delta f = \delta f^{(1)}$ from Eq. 
(\ref{FOCC5})
\begin{align}
\Pi &= -\frac{\Delta_{\alpha\beta}}{3}\!\!\int\!\! dp \, p^\alpha p^\beta \left(-\frac{\tau_R}{u\!\cdot\!p} \, p^\gamma \partial_\gamma f_0\right), \label{FOBEC5}\\
n^\mu &= \Delta^\mu_\alpha \!\!\int\!\! dp \, p^\alpha \left(-\frac{\tau_R}{u\!\cdot\!p} \, p^\gamma \partial_\gamma f_0\right), \label{FOCEC5}\\
\pi^{\mu\nu} &= \Delta^{\mu\nu}_{\alpha\beta}\int dp \ p^\alpha p^\beta \left(-\frac{\tau_R}{u\!\cdot\!p} \, p^\gamma \partial_\gamma f_0\right). \label{FOSEC5}
\end{align}
Assuming the relaxation time $\tau_R$ to be independent of 
four-momenta, the integrals in Eqs. (\ref{FOBEC5})-(\ref{FOSEC5}) 
reduce to
\begin{equation}\label{FOEC5}
\Pi = -\tau_R\beta_\Pi\theta, ~~  n^\mu = \tau_R\beta_n\nabla^\mu\alpha, ~~ \pi^{\mu\nu} = 2\tau_R\beta_\pi\sigma^{\mu\nu},
\end{equation}
where 
$\sigma^{\mu\nu}=\Delta^{\mu\nu}_{\alpha\beta}\nabla^{\alpha}u^\beta$. 
The coefficients $\beta_\Pi,~\beta_n$ and $\beta_\pi$ are 
found to be
\begin{align}
\beta_\Pi =&~ \frac{1}{3}\left(1-3c_s^2\right)(\epsilon+P) - \frac{2}{9}(\epsilon-3P) - \frac{m^4}{9}\left<(u\!\cdot\!p)^{-2}\right>_0, \label{BBC5}\\
\beta_n =&~ - \frac{n^2}{\beta(\epsilon+P)} + \frac{2\left<1\right>_{0^-}}{3\beta} + \frac{m^2}{3\beta}\left<(u\!\cdot\!p)^{-2}\right>_0, \label{BCC5}\\
\beta_\pi =&~ \frac{4P}{5} + \frac{\epsilon-3P}{15} - \frac{m^4}{15}\left<(u\!\cdot\!p)^{-2}\right>_0 ,\label{BSC5}
\end{align}
where $\left<\cdots\right>_0=\int dp(\cdots)f_0$, and 
$c_s^2=(dP/d\epsilon)_{s/n}$ is the adiabatic speed of sound squared 
($s$ being the entropy density). It is interesting to note that 
these coefficients are in perfect agreement with those obtained in 
the Ref. \cite {Denicol:2010xn} in which the evolution equations are 
derived directly from their definition. This is due to the fact that 
in Ref. \cite{Denicol:2010xn}, the coefficients $\beta_\Pi,~\beta_n$ 
and $\beta_\pi$, are associated with first-order terms and do not 
involve 14-moment approximation. In the massless limit, 
$\beta_\pi=4P/5$ is also in agreement with that obtained in Ref. 
\cite {Romatschke:2011qp} employing CE expansion in BE with 
medium-dependent masses.

In the process to obtain second-order equations, we discover that CE 
expansion for the distribution function does not support derivation 
of hydrodynamic evolution equations from arbitrary moment choice of 
BE. Using the definition of dissipative quantities to obtain their 
evolution equations comes naturally when employing CE expansion as 
demonstrated while deriving first-order equations, Eq. (\ref{FOEC5}). 
Second-order evolution equations can also be obtained similarly by 
substituting $\delta f=\delta f^{(1)}+\delta f^{(2)}$ from Eqs. (\ref
{FOCC5}) and (\ref{SOCC5}) in Eqs. (\ref {FBEC5})-(\ref{FSEC5}).
\begin{align} 
\frac{\Pi}{\tau_R} =& \
\frac{\Delta_{\alpha\beta}}{3}\int dp \, p^\alpha p^\beta \left[\frac{p^\gamma}{u\!\cdot\!p}\partial_\gamma f_0 
- \frac{p^\gamma p^\rho}{u\!\cdot\!p}\partial_\gamma\Big(\frac{\tau_R}{u\!\cdot\!p}\partial_\rho f_0\Big)\right], \label{SOBEC5} \\
\frac{n^\mu}{\tau_R} =& 
-\Delta^\mu_\alpha \int dp \, p^\alpha \left[\frac{p^\gamma}{u\!\cdot\!p}\partial_\gamma f_0 
- \frac{p^\gamma p^\rho}{u\!\cdot\!p}\partial_\gamma\Big(\frac{\tau_R}{u\!\cdot\!p}\partial_\rho f_0\Big)\right], \label{SOCEC5} \\
\frac{\pi^{\mu\nu}}{\tau_R} =& 
-\Delta^{\mu\nu}_{\alpha\beta}\int dp \, p^\alpha p^\beta \left[\frac{p^\gamma}{u\!\cdot\!p}\partial_\gamma f_0 
- \frac{p^\gamma p^\rho}{u\!\cdot\!p}\partial_\gamma\Big(\frac{\tau_R}{u\!\cdot\!p}\partial_\rho f_0\Big)\right]. \label{SOSEC5}
\end{align}

The derivatives of equilibrium distribution function ($\partial_\mu 
f_0,~\partial_\mu \partial_\nu f_0$) appearing in above equations 
can be obtained by successively differentiating Eq. (\ref{EDFC5}). The 
momentum integrations can be decomposed into hydrodynamic tensor 
degrees of freedom via the definitions:
\begin{equation}\label{TDIC5}
I^{\mu_1\cdots\mu_n}_{(m)} \equiv \int \frac{dp}{(u\!\cdot\! p)^m} p^{\mu_1}\cdots p^{\mu_n} f_0  
= I_{n0}^{(m)} u^{\mu_1}\cdots u^{\mu_n}  
+ I_{n1}^{(m)} (\Delta^{\mu_1\mu_2} u^{\mu_3} \cdots u^{\mu_n} + \mathrm{perms}) + \cdots,
\end{equation}
where `perms' denotes all non-trivial permutations of the Lorentz 
indices. We similarly define $J^{\mu_1\mu_2\cdots\mu_n}_{(m)}$ where 
the momentum integrals are weighted with $f_0 \tilde f_0$, and are 
tensor decomposed with coefficients $J_{nq}^{(m)}$. 

After performing the integration, the relaxation time appearing on 
the right hand side of Eqs. (\ref{SOBEC5})-(\ref{SOSEC5}) are absorbed 
using the first-order equations for the dissipative quantities, Eq. 
(\ref{FOEC5}). Using the identity $\nabla^\mu\beta=-\beta\dot 
u^\mu+[n/(\epsilon+P)]\nabla^\mu\alpha+\mathcal{O}(\delta^2)$, the 
terms containing derivatives of the relaxation time cancel each 
other up to second-order in gradients and hence the right hand side 
of Eqs. (\ref{SOBEC5})-(\ref{SOSEC5}) can be made independent of 
$\tau_R$, see Appendix B. The second-order evolution equations of 
the dissipative quantities are finally obtained as
\begin{align}
\frac{\Pi}{\tau_R} =& -\dot{\Pi}
-\beta_{\Pi}\theta 
-\delta_{\Pi\Pi}\Pi\theta
+\lambda_{\Pi\pi}\pi^{\mu\nu}\sigma_{\mu \nu }  
-\tau_{\Pi n}n\cdot\dot{u}
-\lambda_{\Pi n}n\cdot\nabla\alpha
-\ell_{\Pi n}\partial\cdot n ~, \label{BULKC5}\\
\frac{n^{\mu}}{\tau_R} =& -\dot{n}^{\langle\mu\rangle}
+\beta_{n}\nabla^{\mu}\alpha
-n_{\nu}\omega^{\nu\mu}
-\lambda_{nn}n^{\nu}\sigma_{\nu}^{\mu}
-\delta_{nn}n^{\mu}\theta  
+\lambda_{n\Pi}\Pi\nabla^{\mu}\alpha
-\lambda_{n\pi}\pi^{\mu\nu}\nabla_{\nu}\alpha \nonumber \\
&-\tau_{n\pi}\pi_{\nu}^{\mu}\dot{u}^{\nu}  
+\tau_{n\Pi}\Pi\dot{u}^{\mu}
+\ell_{n\pi}\Delta^{\mu\nu}\partial_{\gamma}\pi_{\nu}^{\gamma}
-\ell_{n\Pi}\nabla^{\mu}\Pi~,  \label{HEATC5} \\
\frac{\pi^{\mu\nu}}{\tau_R} =& -\dot{\pi}^{\langle\mu\nu\rangle}
+2\beta_{\pi}\sigma^{\mu\nu}
+2\pi_{\gamma}^{\langle\mu}\omega^{\nu\rangle\gamma}
-\tau_{\pi\pi}\pi_{\gamma}^{\langle\mu}\sigma^{\nu\rangle\gamma} 
-\delta_{\pi\pi}\pi^{\mu\nu}\theta 
+\lambda_{\pi\Pi}\Pi\sigma^{\mu\nu}
-\tau_{\pi n}n^{\langle\mu}\dot{u}^{\nu\rangle }  \nonumber \\
&+\lambda_{\pi n}n^{\langle\mu}\nabla ^{\nu\rangle}\alpha
+\ell_{\pi n}\nabla^{\langle\mu}n^{\nu\rangle } ~, \label{SHEARC5}
\end{align}
where $\omega^{\mu\nu}=(\nabla^\mu u^\nu-\nabla^\nu u^\mu)/2$ is the 
definition of the vorticity tensor. All the coefficients in the 
above equations have been obtained in terms of $\beta$ and the 
integral coefficients $I_{nq}^{(m)}$ and $J_{nq}^{(m)}$, see 
Appendix B. It is clear that in Eqs. (\ref{BULKC5})-(\ref{SHEARC5}), 
the Boltzmann relaxation time $\tau_R$ can be replaced by those of 
the individual dissipative quantities $\tau_\Pi,~\tau_n,~\tau_\pi$. 
At this stage, it seems as though the three relaxation times 
$\tau_\Pi,~\tau_n,~\tau_\pi$ are all equal to $\tau_R$. This is 
because the collision term in the BE, Eq. (\ref{RBEC5}) is written in 
RTA which does not entirely capture the microscopic interactions. 
This apparent equality vanishes if the first-order equation, Eq. 
(\ref{FOEC5}) is compared with the relativistic Navier-Stokes 
equations for dissipative quantities ($\Pi=-\zeta\theta$, 
$n^\mu=\kappa\nabla^\mu \alpha$ and 
$\pi^{\mu\nu}=2\eta\sigma^{\mu\nu}$). The dissipative relaxation 
times are then obtained in terms of first-order transport 
coefficients $\zeta,~\kappa$ and $\eta$ which can be calculated 
independently taking into account the full microscopic behaviour of 
the system \cite {Kovtun:2004de,Meyer:2007dy}. 

We remark that although the form of the evolution equations for 
dissipative quantities obtained here, Eqs. (\ref{BULKC5})-(\ref 
{SHEARC5}), are the same as those obtained in the previous 
calculations using both 14-moment approximation and second moment of 
BE \cite{Betz:2008me}, the coefficients obtained are different. In 
the following discussion, we refer to the results in Ref. \cite
{Betz:2008me} as the IS results although the power counting scheme 
differs from the one employed originally by Israel and Stewart. 

For the special case of a system consisting of single species of 
massless Boltzmann gas, we find that
\begin{equation}\label{TDCEC5}
\beta_\pi = \frac{4P}{5}, \quad \tau_{\pi\pi} = \frac{10}{7}, \quad \delta_{\pi\pi} = \frac{4}{3};
\end{equation}
while these coefficients obtained via IS approach are 
\cite{Betz:2008me}
\begin{equation}\label{TDISC5}
\beta_\pi^{\rm{IS}} = \frac{2P}{3}, \quad \tau_{\pi\pi}^{\rm{IS}} = 2, \quad \delta_{\pi\pi}^{\rm{IS}} = \frac{4}{3}.
\end{equation}
In this limit, although the coefficients of $\pi^{\mu\nu}\theta$ are 
same for both the cases ($\delta_{\pi\pi}=\delta_{\pi\pi}^{\rm{IS}}$), 
the coefficient of $\sigma^{\mu\nu}$ and 
$\pi_{\gamma}^{\langle\mu}\sigma^{\nu\rangle\gamma}$ are different 
($\beta_\pi\ne\beta_\pi^{\rm{IS}},~\tau_{\pi\pi}\ne\tau_{\pi\pi}^{\rm{I
S}}$).

We note that CE expansion, as opposed to 14-moment approximation, 
can be done iteratively to arbitrarily higher orders. Hence using CE 
expansion, dissipative hydrodynamic equations of any order can in 
principle be derived. To obtain $n$th-order evolution equations for 
dissipative quantities, $\delta f=\delta f^{(1)}+\delta 
f^{(2)}+\cdots+\delta f^{(n)}$ should be used in Eqs. (\ref{FBEC5})- 
(\ref{FSEC5}). For instance, substitution of $\delta f=\delta 
f^{(1)}+\delta f^{(2)}+\delta f^{(3)}$ in Eqs. (\ref{FBEC5})-(\ref 
{FSEC5}) will eventually lead to third-order evolution equations. 
Derivation of third-order hydrodynamics, as outlined above, is done 
in Chapter 7.

%%%%%%%%%%%%%%%%%%%%%%%%%%%%%%%%%%%%%%%%%%%%%%%%%%%%%%%%%%%%%%%%%%%%%%%%

\section{Numerical results and discussions}

To demonstrate the numerical significance of the new coefficients 
derived here, we consider evolution in the boost invariant Bjorken 
case of a massless Boltzmann gas ($\epsilon=3P$) at vanishing net 
baryon number density \cite {Bjorken:1982qr}. In terms of the Milne 
co-ordinates $(\tau,x,y,\eta_s)$, where $\tau = \sqrt{t^2-z^2}$ and 
$\eta_s=\tanh^{-1}(z/t)$, the initial four-velocity becomes 
$u^\mu=(1,0,0,0)$. For this scenario, $\Pi=n^\mu=0$, and the 
evolution equations for $\epsilon$, 
$\Phi\equiv-\tau^2\pi^{\eta_s\eta_s}$ reduces to (see Appendix A for 
details)
\begin{align}
\frac{d\epsilon}{d\tau} &= -\frac{1}{\tau}\left(\epsilon + P  -\Phi\right), \label{BEDC5} \\
\frac{d\Phi}{d\tau} &= - \frac{\Phi}{\tau_R} + \beta_\pi\frac{4}{3\tau} - \lambda\frac{\phi}{\tau}. \label{BshearC5}
\end{align}
The second-order transport coefficients simplify to
\begin{equation}\label{BTCC5}
\lambda \equiv \frac{1}{3}\tau_{\pi\pi}+\delta_{\pi\pi} = \frac{38}{21}, \quad \lambda^{\rm{IS}}=2.
\end{equation}

\begin{figure}[t]
\begin{center}
\includegraphics[scale=0.5]{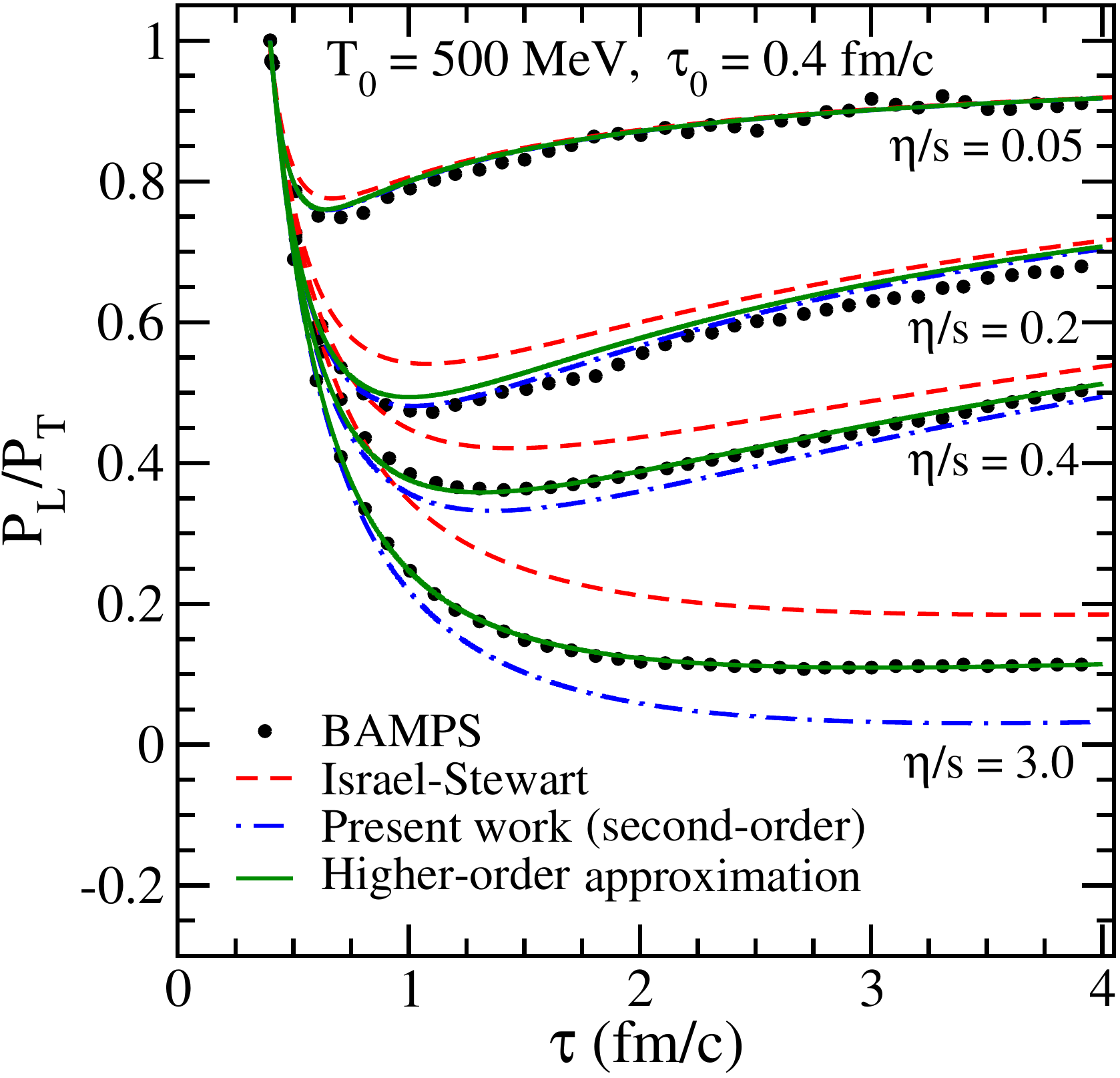}
\end{center}
\vspace{-0.4cm}
\caption[Time evolution of pressure anisotropy in various cases]{Time 
  evolution of $P_L/P_T$ in BAMPS (dots), IS (dashed lines), present 
  work (dashed-dotted line), and a heuristic higher-order 
  approximation (solid line) for isotropic initial pressure 
  configuration $(\Phi_0=0)$.}
\label{PLPTC5}
\end{figure}

Initial temperature $T_0=500$ MeV at proper time $\tau_0=0.4$ fm/c 
are chosen to solve the coupled differential Eqs. (\ref{BEDC5}) and 
(\ref{BshearC5}). These values correspond to LHC initial conditions 
\cite{El:2007vg}. We assume isotropic initial pressure configuration 
i.e. $\Phi_0=0$. Fig. \ref{PLPTC5}, shows the proper time dependence of 
pressure anisotropy defined as $P_L/P_T= (P-\Phi)/(P+\Phi/2)$. The 
dashed and dashed-dotted lines represent the results from IS theory 
and our second-order results, respectively. The dots correspond to 
the results of a transport model, the Boltzmann Approach of 
MultiParton Scatterings (BAMPS), which is based on parton cascade 
simulations \cite {El:2009vj,Xu:2004mz}. The calculations in BAMPS 
are performed with variable values for the cross section such that 
the shear viscosity to entropy density ratio is a constant.

We note that the results from IS theory always overestimate the 
pressure anisotropy as compared to the transport results even for 
viscosities as small as $\eta/s=0.05$. It is evident from the figure 
that our results are in better agreement with BAMPS as compared to 
the results of IS. For very high viscosity, i.e., for $\eta/s=3.0$, 
although at early times we have a better agreement with BAMPS as 
compared to IS, at later times there is a large deviation. This 
disagreement may be attributed to the fact that the present 
hydrodynamic calculation is terminated at second-order in gradients. 
Inclusion of higher-order corrections may improve the agreement of 
dissipative hydrodynamic calculation results with those obtained 
using BAMPS as illustrated in the following.

In Ref. \cite{El:2009vj}, while performing a third-order calculation 
it was demonstrated that within one-dimensional scaling expansion, 
the higher-order gradient terms can acquire the form 
$(\frac{\Phi}{\epsilon})^n\frac{\epsilon}{\tau}$, where, $n=r-1$ for 
$r$th-order corrections. The other forms of higher-order corrections 
is reducible to this structure through lower-order evolution 
equations. Here we assume a similar heuristic expression for 
higher-order corrections
\begin{equation}
\frac{d\Phi}{d\tau} = - \frac{\Phi}{\tau_R} + \beta_\pi\frac{4}{3\tau} 
- \lambda\frac{\Phi}{\tau} - \chi\frac{\Phi^2}{\beta_\pi\tau}, \label{HOAC5}
\end{equation}
where the coefficient $\chi$ contains corrections to shear stress 
evolution due to higher-order gradients. This coefficient can be 
obtained by demanding that the above equation be valid for a free 
streaming of particles in the limit of infinite shear viscosity 
($\eta\to\infty$). In this limit, $\tau_R\to\infty$, and within 
one-dimensional scaling expansion the energy density evolves as 
$\dot\epsilon=-\epsilon/\tau$ which implies that $\dot P=-P/\tau$. 
For this case, using Eq. (\ref{BEDC5}), we arrive at $\Phi=\epsilon/3=P$
which indicates disappearance of the longitudinal pressure. 
Substituting all these in Eq. (\ref{HOAC5}), we obtain $\chi=36/175$.

Fig. \ref{PLPTC5}, also shows $P_L/P_T$ evolution for the results 
obtained after including higher-order corrections (solid lines). We 
observe that the incorporation of higher-order corrections 
significantly improves the agreement with BAMPS. It is important to 
note that the BAMPS calculations are performed with the form of the 
collision term that captures the realistic microscopic interactions 
whereas the derivation of dissipative hydrodynamic equations in this 
chapter uses RTA for the collision term. Within CE formalism, more 
sophisticated ways exist for solving the BE, for eg., by using 
variational methods \cite{Chapman} or by considering momentum 
dependent relaxation time \cite {Prakash:1993bt,Teaney:2009qa}. It 
is, in principle, possible to derive second-order dissipative 
hydrodynamic evolution equations using momentum dependent relaxation 
time provided the dependence is specified explicitly. While this is 
left for future work, we observe that the near perfect agreement of 
the BAMPS results with those obtained using higher-order corrections 
clearly suggest that the momentum independent relaxation time for 
the BE used in the present derivation is sufficiently reliable for 
the range of $\eta/s$ considered here. However, the results obtained 
by using a momentum dependent relaxation time may show a better 
agreement with BAMPS data already at second-order.

RTA for the collision term assumes that the effect of the collisions 
is to restore the distribution function to its local equilibrium 
value exponentially. This is a very good approximation as long as 
the deviations from local equilibrium are small. As discussed above, 
we find that for the range of $\eta/s$ considered here, the 
deviation from equilibrium is not so large because the RTA is still 
valid. It is also important to note that large values of $\eta/s$ 
($> 0.4$) are not relevant to the physics of strongly coupled systems 
like Quark Gluon Plasma (QGP). The QGP formed at RHIC and LHC 
behaves as a near perfect fluid with a small estimated 
$\eta/s\approx 0.08-0.2$ \cite {Romatschke:2007mq,Song:2010mg}. 
Using second-order evolution equations derived here, we get 
reasonably good agreement with BAMPS results for $\eta/s\le 0.4$ 
(Fig. \ref{PLPTC5}). This suggests that BE with RTA for the collision 
term can be successfully applied in understanding the hydrodynamic 
behaviour of QGP formed in relativistic heavy-ion collisions. 

%%%%%%%%%%%%%%%%%%%%%%%%%%%%%%%%%%%%%%%%%%%%%%%%%%%%%%%%%%%%%%%%%%%%%%%%

\section{Summary and conclusions}

To summarize, we have presented a new derivation of relativistic 
second-order hydrodynamics from BE. We use Chapman-Enskog expansion 
for out of equilibrium distribution function instead of 14-moment 
approximation and derive evolution equations for dissipative 
quantities directly from their definitions rather than employing 
second moment of Boltzmann equation. In this new approach, we get 
rid of two powerful assumptions of Israel-Stewart kind of derivation 
which is 14-moment approximation and choice of second moment of 
Boltzmann equation. Although the form of the evolution equation 
remains the same, the coefficients are found to be different. For 
small $\eta/s$, our second-order results show reasonably good 
agreement with the parton cascade BAMPS for the $P_L/P_T$ evolution. 
We find that heuristic inclusion of higher-order corrections in 
shear evolution equation significantly improves the agreement with 
transport calculation for large $\eta/s$ as well. This concurrence 
also suggests that relaxation time approximation for the collision 
term in Boltzmann equation is reasonably accurate when applied to 
heavy-ion collisions.

A very important consequence of Chapman-Enskog like expansion is 
that, unlike Grad's approximation which is linear in dissipative 
quantities, higher-order nonlinear corrections to the equilibrium 
distribution function can also be obtained. This has interesting 
implications for the formulation of dissipative hydrodynamics as 
well as its implementation to the physics of high-energy heavy-ion 
collisions. For example, a different form of nonequilibrium 
distribution function in the particle production prescription \cite 
{Cooper:1974mv}, may affect the observables significantly. In the 
next chapter, we derive an explicit expression for the viscous 
correction to the equilibrium distribution function up to 
second-order in gradients by employing Chapman-Enskog like 
expansion. We compare the hadronic spectra and longitudinal Hanbury 
Brown-Twiss (HBT) radii obtained using this alternate form of the 
viscous correction and Grad's 14-moment approximation.

%% file: Chapter6.tex
%########################################################################
\chapter{Chapman-Enskog vs Grad's methods: Effects on spectra and HBT radii}
%########################################################################

%%%%%%%%%%%%%%%%%%%%%%%%%%%%%%%%%%%%%%%%%%%%%%%%%%%%%%%%%%%%%%%%%%%%%%%%

\section{Introduction}

The Standard Model of relativistic heavy-ion collisions relies on
relativistic hydrodynamics to simulate the intermediate-stage
evolution of the high-energy-density fireball formed in these
collisions \cite{Heinz:2013th}. Recent simulations generally make use
of some version of the M\"uller-Israel-Stewart second-order theory of
causal dissipative hydrodynamics \cite{Israel:1979wp,Muronga:2003ta}.
Hydrodynamics has achieved remarkable success in explaining, for
example, the observed mass ordering of the elliptic flow
\cite{Adams:2005dq,Adcox:2004mh,Song:2013qma}, higher harmonics of the
azimuthal anisotropic flow \cite{Adare:2011tg,Chatrchyan:2013kba}, and
the ridge and shoulder structure in long-range rapidity correlations
\cite{ATLAS:2012at}. The recently measured correlators between event
planes of different harmonics \cite{Jia:2012sa} too can be understood
qualitatively within event-by-event hydrodynamics \cite{Qiu:2012uy}.
Notwithstanding these successes, the basic formulation of the
dissipative hydrodynamic equations continues to be an area of
considerable activity, largely because of the ambiguities arising due
to the variety of ways in which these equations can be derived
\cite{El:2009vj,Romatschke:2009im,Denicol:2012cn,Jaiswal:2012qm,Jaiswal:2013fc,Jaiswal:2013npa,Bazow:2013ifa}.
                                    
For a system that is out of equilibrium, the existence of
thermodynamic gradients results in thermodynamic forces, which give
rise to various transport phenomena. To quantify these nonequilibrium
effects, it is convenient to first specify the nonequilibrium
phase-space distribution function $f(x,p)$ and then calculate the
various transport coefficients. In the context of hydrodynamics, two
most commonly used methods to determine the form of the distribution
function close to local thermodynamic equilibrium are: (1) Grad's
14-moment approximation \cite{Grad} and (2) the Chapman-Enskog method
\cite{Chapman}. Although both the methods involve expanding $f(x,p)$
around the equilibrium distribution function $f_0(x,p)$, there are
important differences.

In the relativistic version of Grad's 14-moment approximation, the
small deviation from equilibrium is usually approximated by means of a
Taylor-like series expansion in momenta truncated at quadratic order
\cite{Israel:1979wp,Romatschke:2009im}. Further, the 14 coefficients
in this expansion are assumed to be linear in dissipative
fluxes. However, it is not apparent why a power series in momenta
should be convergent and whether one is justified in making such
an ansatz, without a small expansion parameter.

The Chapman-Enskog method, on the other hand, aims at obtaining a
perturbative solution of the Boltzmann transport equation using the
Knudsen number (ratio of mean free path to a typical macroscopic
length) as a small expansion parameter. This is equivalent to making a
gradient expansion about the local equilibrium distribution function
\cite{deGroot}. This method of obtaining the form of the
nonequilibrium distribution function is consistent
\cite{Jaiswal:2013npa} with dissipative hydrodynamics which is also
formulated as a gradient expansion.

The above two methods have been compared and shortcomings of Grad's 
approximation have been pointed out in the literature \cite
{velasco,Calzetta:2013vma,Tsumura:2013uma}. In spite of these 
shortcomings, the derivations of relativistic second-order 
dissipative hydrodynamic equations, as well as particle-production 
prescriptions, rely almost exclusively on Grad's approximation. The 
Chapman-Enskog method, on the other hand, has seldom been employed 
in the hydrodynamic modelling of the relativistic heavy-ion 
collisions. The focus of this chapter is to explore the 
applicability of the latter method.

In this chapter, the Boltzmann equation in the relaxation-time 
approximation is solved iteratively, which results in a 
Chapman-Enskog-like expansion of the nonequilibrium distribution 
function. Truncating the expansion at the second order, we derive an 
explicit expression for the viscous correction to the equilibrium 
distribution function. We compare the hadronic spectra and 
longitudinal Hanbury Brown-Twiss (HBT) radii obtained using the form 
of the viscous correction derived here and Grad's 14-moment 
approximation, within one dimensional scaling expansion. We find 
that at large transverse momenta, the present method yields smaller 
hadron multiplicities. We also show analytically that while Grad's 
approximation leads to the violation of the experimentally observed 
$1/\sqrt{m_T}$ scaling of HBT radii \cite
{Bearden:2001sy,Adcox:2002uc}, the viscous correction obtained here 
does not exhibit such unphysical behaviour. Finally, we demonstrate 
the rapid convergence of the Chapman-Enskog-like expansion up to 
second order.

%%%%%%%%%%%%%%%%%%%%%%%%%%%%%%%%%%%%%%%%%%%%%%%%%%%%%%%%%%%%%%%%%%%%%%%%

\section{Relativistic viscous hydrodynamics}

Within the framework of relativistic hydrodynamics, the variables that
characterize the macroscopic state of a system are the energy-momentum
tensor, $T^{\mu\nu}$, particle four-current, $N^\mu$, and entropy
four-current, $S^\mu$. The local conservation of net charge
($\partial_\mu N^\mu = 0$) and energy-momentum ($\partial_\mu
T^{\mu\nu}=0$) lead to the equations of motion of a relativistic
fluid, whereas, the second law of thermodynamics requires
$\partial_\mu S^\mu \ge 0$. For a system with no net conserved
charges, hydrodynamic evolution is governed only by the conservation
equations for energy and momentum.

The energy-momentum tensor of a macroscopic system can be expressed in
terms of a single-particle phase-space distribution function, and can
be tensor decomposed into hydrodynamic degrees of freedom \cite
{deGroot}. Here we restrict ourselves to a system of massless
particles (ultrarelativistic limit) for which the bulk viscosity
vanishes, leading to
\begin{align}\label{NTDC6}
T^{\mu\nu} &= \!\int\! dp \ p^\mu p^\nu\, f(x,p) = \epsilon u^\mu u^\nu 
- P\Delta ^{\mu \nu} + \pi^{\mu\nu},
\end{align}
where $dp\equiv g d{\bf p}/[(2 \pi)^3|\bf p|]$, $g$ being the
degeneracy factor, $p^\mu$ is the particle four-momentum, and $f(x,p)$
is the phase-space distribution function. In the tensor
decomposition, $\epsilon$, $P$, and $\pi^{\mu\nu}$ are energy density,
thermodynamic pressure, and shear stress tensor, respectively.
The projection operator $\Delta^{\mu\nu}\equiv g^{\mu\nu}-u^\mu u^\nu$
is orthogonal to the hydrodynamic four-velocity $u^\mu$ defined in the
Landau frame: $T^{\mu\nu} u_\nu=\epsilon u^\mu$. The metric tensor is
Minkowskian, $g^{\mu\nu}\equiv\mathrm{diag}(+,-,-,-)$.

The evolution equations for $\epsilon$ and $u^\mu$,
\begin{align}\label{evolC6}
\dot\epsilon + (\epsilon+P)\theta - \pi^{\mu\nu}\nabla_{(\mu} u_{\nu)} &= 0,  \nonumber\\
(\epsilon+P)\dot u^\alpha - \nabla^\alpha P + \Delta^\alpha_\nu \partial_\mu \pi^{\mu\nu}  &= 0,
\end{align}
are obtained from the conservation of the energy-momentum tensor.  We
use the standard notation $\dot A\equiv u^\mu\partial_\mu A$ for
comoving derivative, $\theta\equiv \partial_\mu u^\mu$ for
expansion scalar, $A^{(\alpha}B^{\beta )}\equiv (A^\alpha B^\beta +
A^\beta B^\alpha)/2$ for symmetrization, and $\nabla^\alpha\equiv
\Delta^{\mu\alpha}\partial_\mu$ for space-like derivatives. In the
ultrarelativistic limit, the equation of state relating energy density
and pressure is $\epsilon=3P\propto\beta^{-4}$. The inverse
temperature, $\beta\equiv1/T$, is determined by the Landau matching
condition $\epsilon=\epsilon_0$ where $\epsilon_0$ is the equilibrium
energy density. In this limit, the derivatives of $\beta$,
\begin{align}
\dot\beta &= \frac{\beta}{3}\theta - \frac{\beta}{12P}\pi^{\rho\gamma}\sigma_{\rho\gamma}, \label{evol1C6} \\
\nabla^\alpha\beta &= -\beta\dot u^\alpha - \frac{\beta}{4P} \Delta^\alpha_\rho \partial_\gamma \pi^{\rho\gamma}, \label{evol2C6}
\end{align}
can be obtained from Eq. (\ref{evolC6}), where
$\sigma^{\rho\gamma}\equiv\nabla^{(\rho}u^{\gamma)}-(\theta/3)\Delta^{\rho
\gamma}$ is the velocity stress tensor \cite{Jaiswal:2013vta}. The
above identities will be used later in the derivations of viscous
corrections to the distribution function and shear evolution equation.

For a system close to local thermodynamic equilibrium, the phase-space
distribution function can be written as $f=f_0+\delta f$, where the
deviation from equilibrium is assumed to be small $(\delta f\ll
f)$. Here $f_0$ represents the equilibrium distribution function of
massless Boltzmann particles at vanishing chemical potential,
$f_0=\exp(-\beta\,u\cdot p)$. From Eq. (\ref{NTDC6}), the shear stress
tensor, $\pi^{\mu\nu}$, can be expressed in terms of the
nonequilibrium part of the distribution function, $\delta f$, as
\cite{Romatschke:2009im}
\begin{align}\label{FSEC6}
\pi^{\mu\nu} &= \Delta^{\mu\nu}_{\alpha\beta} \int dp \, p^\alpha p^\beta\, \delta f,
\end{align}
where $\Delta^{\mu\nu}_{\alpha\beta}\equiv
\Delta^{\mu}_{(\alpha}\Delta^{\nu}_{\beta)} -
(1/3)\Delta^{\mu\nu}\Delta_{\alpha\beta}$ is a traceless symmetric
projection operator orthogonal to $u^\mu$. To make further progress,
the form of $\delta f$ has to be determined. In the following, we
adopt a Chapman-Enskog-like expansion for the distribution function,
to obtain $\delta f$ order-by-order in gradients, by solving the
Boltzmann equation iteratively in the relaxation-time approximation.

%%%%%%%%%%%%%%%%%%%%%%%%%%%%%%%%%%%%%%%%%%%%%%%%%%%%%%%%%%%%%%%%%%%%%%%%

\section{Chapman-Enskog expansion}

Determination of the nonequilibrium phase-space distribution 
function is one of the central problems in statistical mechanics. 
This can be achieved by solving a kinetic equation such as the 
Boltzmann equation. Our starting point is the relativistic Boltzmann 
equation with the relaxation-time approximation for the collision 
term, as given in Eq. (\ref{RBEC5}),
\begin{equation}\label{RBEC6}
p^\mu\partial_\mu f = C[f] = 
-\left(u\!\cdot\! p\right) \frac{\delta f}{\tau_R},
\end{equation}
where $\tau_R$ is the relaxation time. We recall that the zeroth and
first moments of the collision term, $C[f]$, should vanish to ensure
the conservation of particle current and energy-momentum tensor \cite
{deGroot}. This requires that $\tau_R$ is independent of momenta, and
$u^\mu$ is defined in the Landau frame \cite
{Anderson_Witting}. Therefore, within the relaxation-time
approximation, Landau frame is mandatory and not a choice.

Exact solutions of the Boltzmann equation are possible only in rare 
circumstances. The most common technique of generating an 
approximate solution to the Boltzmann equation is the Chapman-Enskog 
expansion where the distribution function is expanded about its 
equilibrium value in powers of space-time gradients, as done in Eq. 
(\ref{CEEC5}),
\begin{equation}\label{CEEC6}
f = f_0 + \delta f, \quad \delta f= \delta f^{(1)} + \delta f^{(2)} + \cdots,
\end{equation}
where $\delta f^{(n)}$ is n$th$-order in derivatives. As 
done in the previous chapter, the Boltzmann equation can be 
solved iteratively by rewriting Eq. (\ref{RBEC6}) in the form 
$f=f_0-(\tau_R/u\cdot p)\,p^\mu\partial_\mu f$ \cite
{Romatschke:2011qp,Jaiswal:2013npa,Teaney:2013gca}. We obtain
\begin{equation}\label{F1F2C6}
f_1 = f_0 -\frac{\tau_R}{u\cdot p} \, p^\mu \partial_\mu f_0, \quad 
f_2 = f_0 -\frac{\tau_R}{u\cdot p} \, p^\mu \partial_\mu f_1, ~~\, \cdots
\end{equation}
where $f_n=f_0+\delta f^{(1)}+\delta f^{(2)}+\cdots+\delta f^{(n)}$.
To first- and second-orders in derivatives, we have
\begin{equation}\label{FSOCC6}
\delta f^{(1)} = -\frac{\tau_R}{u\cdot p} \, p^\mu \partial_\mu f_0, \quad
\delta f^{(2)} = \frac{\tau_R}{u\cdot p}p^\mu p^\nu\partial_\mu\Big(\frac{\tau_R}{u\cdot p} \partial_\nu f_0\Big).
\end{equation}
In the next section, the above expressions for $\delta f$ along with
Eq. (\ref{FSEC6}) will be used in the derivation of the evolution
equation for the shear stress tensor.

%%%%%%%%%%%%%%%%%%%%%%%%%%%%%%%%%%%%%%%%%%%%%%%%%%%%%%%%%%%%%%%%%%%%%%%%

\section{Viscous evolution equation}

In order to complete the set of hydrodynamic equations,
Eq. (\ref{evolC6}), we need to derive an expression for the shear stress
tensor, $\pi^{\mu\nu}$. The first-order expression for $\pi^{\mu\nu}$
can be obtained from Eq. (\ref{FSEC6}) using $\delta f = \delta f^{(1)}$
from Eq. (\ref{FSOCC6}),
\begin{align}
\pi^{\mu\nu} &= \Delta^{\mu\nu}_{\alpha\beta}\int dp \ p^\alpha p^\beta \left(-\frac{\tau_R}{u.p} \, p^\gamma \partial_\gamma\, f_0\right) . \label{FOSEC6}
\end{align}
Using Eqs. (\ref{evol1C6}) and (\ref{evol2C6}) and keeping only those
terms which are first-order in gradients, the integral in the above
equation reduces to
\begin{equation}\label{FOEC6}
\pi^{\mu\nu} = 2\tau_R\beta_\pi\sigma^{\mu\nu},
\end{equation}
where $\beta_\pi = 4P/5$ \cite{Jaiswal:2013npa}.

The second-order evolution equation for shear stress tensor can also
be obtained in a similar way by using $\delta f = \delta
f^{(1)}+\delta f^{(2)}$ from Eq. (\ref{FSOCC6}) in
Eq. (\ref{FSEC6}). Performing the integrations, we get
\cite{Jaiswal:2013vta,Jaiswal:2014raa}
\begin{equation}\label{SOSHEARC6}
\dot{\pi}^{\langle\mu\nu\rangle} + \frac{\pi^{\mu\nu}}{\tau_R} = 2\beta_{\pi}\sigma^{\mu\nu}
+2\pi_\gamma^{\langle\mu}\omega^{\nu\rangle\gamma}
-\frac{10}{7}\pi_\gamma^{\langle\mu}\sigma^{\nu\rangle\gamma} 
-\frac{4}{3}\pi^{\mu\nu}\theta,
\end{equation}
where $\omega^{\mu\nu}\equiv(\nabla^\mu u^\nu-\nabla^\nu u^\mu)/2$ is
the vorticity tensor, and we have used Eq. (\ref{FOEC6}). It is clear
from the form of the above equation that the relaxation time $\tau_R$
can be identified with the shear relaxation time $\tau_\pi$. By
comparing the first-order evolution Eq. (\ref{FOEC6}) with the
relativistic Navier-Stokes equation
$\pi^{\mu\nu}=2\eta\sigma^{\mu\nu}$, we obtain
$\tau_\pi=\eta/\beta_\pi$, where $\eta$ is the coefficient of shear
viscosity.

%%%%%%%%%%%%%%%%%%%%%%%%%%%%%%%%%%%%%%%%%%%%%%%%%%%%%%%%%%%%%%%%%%%%%%%%

\section{Corrections to the distribution function}

In this section, we derive the expression for the nonequilibrium 
part of the distribution function, $\delta f$, up to second order in 
gradients of $u^\mu$. For this purpose, we employ Eq. (\ref{FSOCC6}) 
which was obtained using a Chapman-Enskog-like expansion. We then 
recall the derivation of the standard Grad's 14-moment approximation 
for $\delta f$, and compare these two expressions.

Using Eqs. (\ref{evol1C6}) and (\ref{evol2C6}) for the derivatives 
of $\beta$, and Eq. (\ref{SOSHEARC6}) for $\sigma^{\mu\nu}$, in Eq. 
(\ref{FSOCC6}), we arrive at the form of the second-order viscous 
correction to the distribution function:
\begin{align}
\delta f \!=\ &  \frac{f_0\beta}{2\beta_\pi(u\!\cdot\!p)}\, p^\alpha p^\beta \pi_{\alpha\beta}
-\frac{f_0\beta}{\beta_\pi} \bigg[\frac{\tau_\pi}{u\!\cdot\!p}\, p^\alpha p^\beta \pi^\gamma_\alpha\, \omega_{\beta\gamma} 
-\frac{5}{14\beta_\pi (u\!\cdot\!p)}\, p^\alpha p^\beta \pi^\gamma_\alpha\, \pi_{\beta\gamma}
+\frac{\tau_\pi}{3(u\!\cdot\!p)}\, p^\alpha p^\beta \pi_{\alpha\beta}\theta  \nonumber\\
&-\frac{6\tau_\pi}{5}\, p^\alpha\dot u^\beta\pi_{\alpha\beta}
+\frac{(u\!\cdot\!p)}{70\beta_\pi}\, \pi^{\alpha\beta}\pi_{\alpha\beta}
+\frac{\tau_\pi}{5}\, p^\alpha \left(\nabla^\beta\pi_{\alpha\beta}\right)
-\frac{3\tau_\pi}{(u\!\cdot\!p)^2}\, p^\alpha p^\beta p^\gamma \pi_{\alpha\beta}\dot u_\gamma \nonumber\\
&+\frac{\tau_\pi}{2(u\!\cdot\!p)^2}\, p^\alpha p^\beta p^\gamma \left(\nabla_\gamma\pi_{\alpha\beta}\right)
-\frac{\beta+(u\!\cdot\!p)^{-1}}{4(u\!\cdot\!p)^2\beta_\pi}\, \left(p^\alpha p^\beta \pi_{\alpha\beta}\right)^2\bigg]
+{\cal O}(\delta^3), \label{SOVCC6} \\ 
\equiv~& \delta f_1 + \delta f_2 +{\cal O}(\delta^3). \label{SOVCEC6}
\end{align}
The first term on the right-hand side of Eq. (\ref{SOVCC6}) corresponds
to the first-order correction, $\delta f_1$, whereas the terms within
square brackets are of second order, $\delta f_2$ (see Appendix C).
Note that $\delta f_1 \ne \delta f^{(1)}$ and $\delta f_2 \ne \delta
f^{(2)}$, due to the nonlinear nature of Eqs. (\ref{evol1C6}), (\ref
{evol2C6}), and (\ref{SOSHEARC6}). It is straightforward to show that the
form of $\delta f$ in Eq. (\ref{SOVCC6}) satisfies the matching
condition $\epsilon =\epsilon_0$ and the Landau frame definition
$u_\nu T^{\mu \nu} = \epsilon u^\mu$ \cite{deGroot}, i.e.,
\begin{equation}\label{checksC6}
\int dp\, (u\cdot p)^2\, \delta f = 0, \quad \int dp\, \Delta_{\mu\alpha}u_\beta\, p^\alpha p^\beta\, \delta f = 0,
\end{equation}
order-by-order in gradients, see Appendix C.

On the other hand, Grad's 14-moment approximation for $\delta f$ can
be obtained from a Taylor-like expansion in the powers of momenta
\cite{Israel:1979wp,Romatschke:2009im}
\begin{equation}
 \delta f_G = f_0 \left[\varepsilon(x)+\varepsilon_\alpha(x)p^\alpha+\varepsilon_{\alpha\beta}(x)p^\alpha p^\beta \right],
\end{equation}
where $\varepsilon$'s are the momentum-independent coefficients in the
expansion, which, however, may depend on thermodynamic and dissipative
quantities. For a system of massless particles with no net conserved
charges, i.e., in the absence of bulk viscosity and charge diffusion
current, the above equation reduces to
\begin{equation}
 \delta f_G = \frac{f_0 \beta^2}{10 \beta_\pi}\, p^\alpha p^\beta \pi_{\alpha\beta}, \label{CGC6}
\end{equation}
where the coefficient is obtained using Eq. (\ref{FSEC6}). We observe
that unlike Eq. (\ref{SOVCC6}) for the Chapman-Enskog case,
Eq. (\ref{CGC6}) for Grad's is linear in shear stress tensor. However,
it is important to note that both the forms of $\delta f$, i.e.,
$\delta f_1$ and $\delta f_G$, lead to identical evolution equations
for the shear stress tensor, Eq. (\ref{SOSHEARC6}), with the same
coefficients \cite{Denicol:2012cn,Jaiswal:2013vta}.

%%%%%%%%%%%%%%%%%%%%%%%%%%%%%%%%%%%%%%%%%%%%%%%%%%%%%%%%%%%%%%%%%%%%%%%%

\section{Bjorken scenario}
In order to model the hydrodynamical evolution of the matter formed in
the heavy-ion collision experiments, we use the Bjorken prescription
\cite{Bjorken:1982qr} for one-dimensional expansion. We consider
the evolution of a system of massless particles ($\epsilon=3P$) at
vanishing net baryon number density. In terms of the Milne coordinates
($\tau,r,\varphi,\eta_s$), where $\tau = \sqrt{t^2-z^2}$,
$r=\sqrt{x^2+y^2}$, $\varphi=\tan^{-1}(y/x)$, and
$\eta_s=\tanh^{-1}(z/t)$, and with $u^\mu=(1,0,0,0)$, evolution
equations for $\epsilon$ and $\Phi \equiv -\tau^2 \pi^{\eta_s \eta_s}$
become (see Appendix A for details)
\begin{align}
\frac{d\epsilon}{d\tau} &= -\frac{1}{\tau}\left(\epsilon + P -\Phi\right), \label{BEDC6} \\
\frac{d\Phi}{d\tau} &= - \frac{\Phi}{\tau_\pi} + \beta_\pi\frac{4}{3\tau} - \lambda\frac{\Phi}{\tau}. \label{BshearC6}
\end{align}
The transport coefficients appearing in the above equation reduce to \cite{Jaiswal:2013npa}
\begin{equation}\label{BTCC6}
\tau_\pi = \frac{\eta}{\beta_\pi}, \quad \beta_\pi = \frac{4P}{5}, \quad \lambda = \frac{38}{21}.
\end{equation}

In $(\tau,r,\varphi,\eta_s)$ coordinates, the components of particle
four momenta are given by
\begin{align}
p^\tau &= m_T \cosh(y-\eta_s), \quad p^r = p_T \cos(\varphi_p- \varphi), \\
p^\varphi &= p_T\sin(\varphi_p-\varphi)/r, \quad p^{\eta_s} = m_T \sinh(y-\eta_s)/\tau, \nonumber
\end{align}
where $m_T^2=p_T^2+m^2$, $p_T$ is the transverse momentum, $y$ the
particle rapidity, and $\varphi_p$ the azimuthal angle in the momentum
space. We note that for the Bjorken expansion, $\theta=1/\tau$, $\dot
u^{\mu}=0$, $\omega^{\mu\nu}=0$ and $p_\mu d\Sigma^\mu = m_T
\cosh(y-\eta_s) \tau d\eta_s r dr d\varphi$. In this scenario, the
non-vanishing factors appearing in Eq. (\ref{SOVCC6}) reduce to $u\cdot
p = m_T \cosh(y-\eta_s)$, $\pi_{\alpha\beta}\pi^{\alpha\beta}=
3\Phi^2/2$, and
\begin{align}
p^{\alpha}p^{\beta} \pi_{\alpha\beta} &= \frac{\Phi}{2}\, p_T^2-\Phi\, m_T^2\, \sinh^2(y-\eta_s), \nonumber\\
p^{\alpha}p^{\beta} \pi^\gamma_\alpha \pi_{\gamma\beta} &= -\frac{\Phi^2}{4}\, p_T^2-\Phi^2\, m_T^2\, \sinh^2(y-\eta_s), \nonumber\\
p^\alpha p^\beta p^\gamma \nabla_\alpha\pi_{\beta\gamma} &= 2\, \frac{\Phi}{\tau}\, m_T^3\, \sinh^2(y-\eta_s)\cosh(y-\eta_s), \nonumber\\
p^\alpha\nabla^\beta\pi_{\alpha\beta} &= -\frac{\Phi}{\tau}\, m_T\, \cosh(y-\eta_s). \label{visc-fctsC6}
\end{align}

Within the framework of the relativistic hydrodynamics, observables
pertaining to heavy-ion collisions are influenced by viscosity in two
ways: first through the viscous hydrodynamic evolution of the system
and second through corrections to the particle production rate via the
nonequilibrium distribution function \cite{Teaney:2003kp}.
Hydrodynamic evolution and the nonequilibrium corrections to the
distribution function were considered in the previous sections; in the
following sections, we focus on two observables, namely
transverse-momentum spectra and HBT radii of hadrons.

%%%%%%%%%%%%%%%%%%%%%%%%%%%%%%%%%%%%%%%%%%%%%%%%%%%%%%%%%%%%%%%%%%%%%%%%

\section{Hadronic spectra}

The hadron spectra can be obtained using the Cooper-Frye freezeout
prescription \cite{Cooper:1974mv}
\begin{equation}\label{CFC6}
\frac{dN}{d^2p_Tdy} = \frac{g}{(2\pi)^3} \int p_\mu d\Sigma^\mu f(x,p),
\end{equation}
where $p^\mu$ is the particle four momentum, $d\Sigma^\mu$ represents
the element of the three-dimensional freezeout hypersurface and
$f(x,p)$ represents the phase-space distribution function at
freezeout.

For the ideal freezeout case ($f=f_0$), we get
\begin{equation}\label{ICFC6}
\frac{dN^{(0)}}{d^2p_Tdy} = \frac{g}{4 \pi^3}\, m_T\, \tau\, A_\perp\, K_1,
\end{equation}
where $A_\perp$ denotes the transverse area of the overlap zone of
colliding nuclei and $K_n \equiv K_n(z_m)$ are the modified Bessel
functions of the second kind with argument $z_m\equiv
m_T/T$. In Eq. (\ref{ICFC6}) and hereafter, the hydrodynamical quantities such as
$T,~\tau,~\Phi,~P$, etc. correspond to their values at freezeout. The
expression for hadron production up to first order ($f=f_0+\delta
f_1$) is obtained as
\begin{equation}\label{FCFC6}
\frac{d N^{(1)}}{d^2p_Tdy} = \left[1+ \frac{\Phi}{4\beta_\pi z_m} \left\{z_p^2\, \frac{K_0}{K_1} - 2z_m \right\}\right]\frac{dN^{(0)}}{d^2p_Tdy},
\end{equation}
where $z_p\equiv p_T/T$. Here we have used the recurrence relation
$K_{n+1}(z)=2nK_n(z)/z+K_{n-1}(z)$. The derivation of the hadron
spectra up to second order, $dN^{(2)} / d^2p_Tdy$, (by setting
$f=f_0+\delta f_1+\delta f_2$) is presented in the Appendix C.

For comparison, we also present the result for hadron production
obtained using Grad's 14-moment approximation ($f=f_0 + \delta f_G$)
\cite{Teaney:2003kp,Bhalerao:2013aha}
\begin{equation}\label{GCFC6}
\frac{d N^{(G)}}{d^2p_Tdy} = \left[1+\frac{\Phi}{20\beta_\pi} \left\{z_p^2 - 2z_m \frac{K_2}{K_1}  \right\}\right]\frac{dN^{(0)}}{d^2p_Tdy}.
\end{equation}

 \begin{figure}[t]
 \begin{center}
 \includegraphics[scale=0.5]{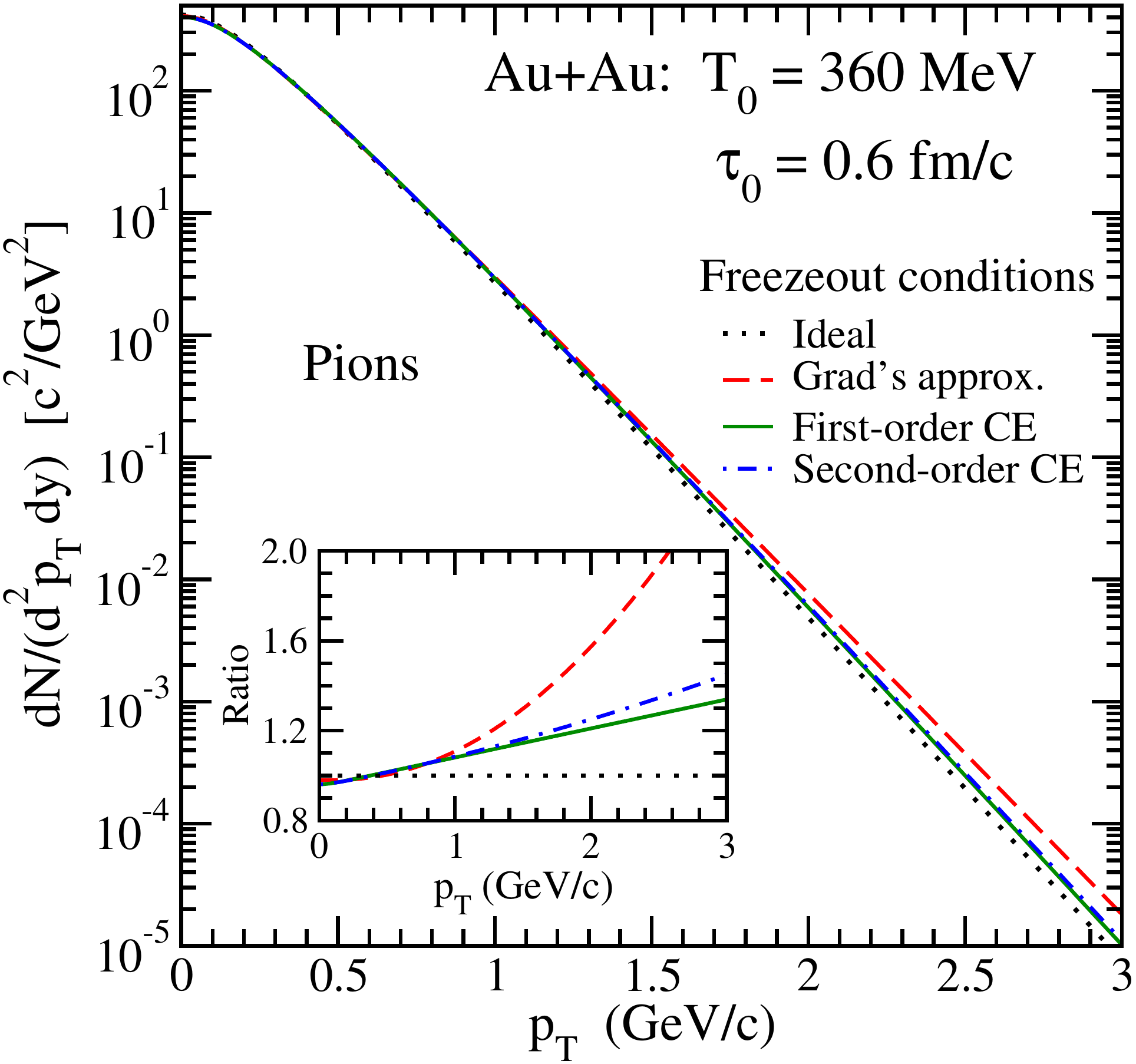}
 \end{center}
 \vspace{-0.4cm}
 \caption[Pion spectra as a function of the transverse momentum]{Pion 
   spectra as a function of the transverse momentum $p_T$, obtained 
   with the second-order hydrodynamic evolution, followed by 
   freezeout in various scenarios: ideal, Grad's 14-moment 
   approximation, first- and second-order Chapman-Enskog. Inset: 
   Pion yields in the above four cases scaled by the corresponding 
   values in the ideal case.}
 \label{spectraC6}
 \end{figure}

We solve the evolution equations (\ref{BEDC6})-(\ref{BshearC6}) with
initial temperature $T_0=360$ MeV, time $\tau_0=0.6$ fm/$c$, and
isotropic pressure configuration $\Phi_0=0$, corresponding to central
($b=0$) Au-Au collisions at the Relativistic Heavy-Ion Collider. The
system is evolved with shear viscosity to entropy density ratio
$\eta/s=1/4\pi$ corresponding to the KSS lower bound
\cite{Kovtun:2004de}, until the freezeout temperature $T=150$ MeV is
reached. In order to study the effects of the various forms of $\delta
f$ via the freezeout prescription, Eq. (\ref{CFC6}), we evolve the
system using the second-order viscous hydrodynamic equations
(\ref{BEDC6}) and (\ref{BshearC6}) in all the cases.

In Fig. \ref{spectraC6}, we present the pion transverse-momentum spectra
for the four freezeout conditions discussed above, namely ideal,
first- and second-order Chapman-Enskog, and Grad's 14-moment
approximation. We observe that nonideal freezeout conditions tend to
increase the high-$p_T$ particle production. While the Chapman-Enskog
corrections are small, Grad's 14-moment approximation results in
rather large corrections to the ideal case. This is clearly evident in
the inset where we show the pion yields in the four cases scaled by
the values in the ideal case. These features can be easily understood
from Eqs. (\ref{FCFC6}) and (\ref{GCFC6}): The first-order Chapman-Enskog
correction is essentially linear in $p_T$ whereas that due to Grad is
quadratic. The second-order Chapman-Enskog correction is small,
indicating rapid convergence of the expansion up to second order.

%%%%%%%%%%%%%%%%%%%%%%%%%%%%%%%%%%%%%%%%%%%%%%%%%%%%%%%%%%%%%%%%%%%%%%%%

\section{HBT radii}

HBT interferometry provides a powerful tool to unravel the space-time
structure of the particle emitting sources in heavy-ion collisions,
because of its ability to measure source sizes, lifetimes and
particle emission durations \cite{Lisa:2005dd}. The source function,
$S(x,K)$ for on-shell particle emission is defined such that it
satisfies
\begin{equation}\label{DSFC6}
\frac{dN}{d^2K_Tdy} \equiv \int d^4x\, S(x,K).
\end{equation}
Comparing the above equation with Eq. (\ref{CFC6}), we see that the
source function is restricted to the freezeout hypersurface and is
given by
\begin{equation}\label{SFC6}
S(x,K) = \frac{g}{(2\pi)^3} \int p_\mu d\Sigma^\mu(x') f(x',p) \delta^4(x-x').
\end{equation}
At relatively small momenta, certain space-time variances of the
source function can be obtained, to a good approximation, from the
correlation between particle pairs \cite{Wiedemann:1999qn}.
Space-time averages with respect to the source function are defined as
\begin{equation}\label{ASFC6}
\mean{\alpha}_K \equiv \frac{\int d^4x\, S(x,K)\alpha}{\int d^4x\, S(x,K)}
=\frac{\int K_\mu d\Sigma^\mu f(x,K)\alpha}{\int K_\mu d\Sigma^\mu f(x,K)},
\end{equation}
where $K_\mu$ is the pair four-momentum.

The longitudinal HBT radius, $R_L$, is calculated in terms of the
transverse momentum, $K_T$, of the identical-particle pair
\cite{Wiedemann:1999qn}:
\begin{equation}\label{HBTC6}
R_L^2(K_T) = \frac{\int K_\mu d\Sigma^\mu f(x,K)z^2}{\int K_\mu d\Sigma^\mu f(x,K)}.
\end{equation}
In the central-rapidity region, the pair four momentum is given by
$K^\mu=(K^\tau, K^r, K^\varphi, K^{\eta_s})=(m_T,K_T,0,0)$. The
integration measure is given by $K_\mu d\Sigma^\mu = m_T \cosh(\eta_s)
\tau d\eta_s r dr d\varphi$ with $m_T=\sqrt{K_T^2+m_p^2}$, $m_p$
being the particle mass. Using the relation $z=\tau\,\sinh(\eta_s)$, we
get
\begin{align}\label{HBTBC6}
 R_L^2(K_T) &= \tau^2\left[\frac{\int K_\mu d\Sigma^\mu f(x,K){\cosh^2(\eta_s)}}
{\int K_\mu d\Sigma^\mu f(x,K)}-1\right], \nonumber \\
&\equiv \tau^2\left[\frac{N[f]}
{D[f]}-1\right].
\end{align}
Note that the integral, $D[f]$, in the denominator in the above
equation is the same as that occurring in the Cooper-Frye prescription
for particle production, Eq. (\ref{CFC6}), and was already calculated in
the previous section. We next calculate the integral, $N[f]$, in the
numerator.

In the ideal case, $f=f_0$, we have
\begin{equation}\label{Nf0C6}
N[f_0] = \frac{2 A_\perp \tau z_m}{4\beta}\left(K_3+3K_1\right).
\end{equation}
This leads to the well known result of Hermann and Bertsch 
\cite{Herrmann:1994rr}
\begin{equation}\label{IHBTC6}
(R_L^2)^{(0)} = \frac{\tau^2}{z_m }\, \frac{K_2}{K_1},
\end{equation}
which for large values of $z_m$ results in the Makhlin-Sinyukov
formula $(R_L^2)^{(0)} = \tau^2 T / m_T$
\cite{Makhlin:1987gm,Csorgo:1995bi}. Thus in the ideal case,
$(R_L)^{(0)}$ exhibits the so-called $1/\sqrt{m_T}$ scaling.

The first-order calculation requires $N[\delta f_1]$ which is given by
\begin{equation}\label{Ndf1C6}
N[\delta f_1] = \frac{2 A_\perp \tau \Phi}{16\beta\beta_\pi}\Big[\left(2z_p^2+z_m^2\right)K_0 + 2z_p^2K_2 - z_m^2K_4\Big].
\end{equation}
The second-order calculation requires $N[\delta f_2]$ which is given
in the Appendix C. For comparison we also calculate $R_L$ in Grad's
14-moment approximation. This requires $N[\delta f_G]$, which we obtain
as
\begin{equation}\label{NdfGC6}
N[\delta f_G] = \frac{2 A_\perp \tau\Phi z_m}{160\beta\beta_\pi}\Big[\left(2z_p^2-6z_m^2\right)K_1 
+ \left(2z_p^2-z_m^2\right)K_3 - z_m^2K_5\Big].
\end{equation}
 
In the following, we show that the viscous correction to $R_L$ due to
Grad's 14-moment approximation violates the experimentally observed
$1/\sqrt{m_T}$ scaling \cite{Bearden:2001sy,Adcox:2002uc}, whereas it
is preserved in the Chapman-Enskog case. To this end, we calculate the
first-order viscous correction to $R_L$ in both the cases. Expanding the
$R_L$ in Eq. (\ref{HBTC6}) to first order in $\delta f$ and using the
relation $z=\tau \sinh(\eta_s)$ we obtain the ideal contribution
\begin{align}\label{RL0C6}
(R_L^2)^{(0)} =&~  \frac{\int K^\mu d\Sigma_\mu\, f_0 \,\tau^2 \sinh^2(\eta_s)}
{\int K^\mu d\Sigma_{\mu}\, f_0}, 
\end{align}
and the first viscous correction in the two cases
\begin{equation}\label{DRL1GC6}
\left(\delta R_L^2\right)^{(1,G)} = -(R_L^2)^{(0)} \left(  
\frac{dN^{(1,G)}}{d^2K_T}-\frac{dN^{(0)}}{d^2K_T} \right)\!\!\Big/
\frac{dN^{(0)}}{d^2K_T} 
+ \frac{ \int K^\mu d\Sigma_\mu\,\tau^2\sinh^2(\eta_s)\,\delta f_{1,G} } 
{ \int K^\mu d\Sigma_\mu \, f_0 }.
\end{equation}
The ideal radius $(R_L^2)^{(0)}$ was obtained in Eq. (\ref{IHBTC6}).
Viscous corrections due to the Chapman-Enskog method and Grad's
14-moment approximation can be obtained similarly.  Substituting the
viscous correction, $\delta f_1$, from Eq. (\ref{SOVCC6}) into
Eq. (\ref{DRL1GC6}), using the results for the particle spectra,
Eqs. (\ref{ICFC6}), (\ref{FCFC6}) and the ideal radius, Eq. (\ref{IHBTC6}), and
performing the $\eta_s$ integrals, we obtain
\begin{equation}\label{DRL1C6}
\frac{\left(\delta R_L^2\right)^{(1)}}{\left(R_L^2\right)^{(0)}} = -\frac{\Phi}{16\beta_\pi}\left[16 
+ \frac{4z_p^2}{z_m}\left(\frac{K_0}{K_1}-\frac{K_1}{K_2}\right)\right]. 
\end{equation}
Similarly, for Grad's approximation, Eq. (\ref{CGC6}), we obtain
\begin{equation}\label{DRLGC6}
\frac{\left(\delta R_L^2\right)^{(G)}}{\left(R_L^2\right)^{(0)}} = -\frac{\Phi}{20\beta_\pi}
\left[ 20 - 2z_m\left(\frac{K_0}{K_1}-\frac{K_1}{K_2}\right) + 4z_m\frac{K_1}{K_2} \right]. 
\end{equation}

Using the asymptotic expansion of modified Bessel functions of the
second kind \cite{abramowitz},
\begin{equation}\label{ASYMPC6}
K_n(z_m) = \left(\frac{\pi}{2z_m}\right)^{\frac{1}{2}}e^{-z_m}\left[1+\frac{4n^2-1}{8z_m}+\cdots\right],
\end{equation}
for large $z_m$, we have
\begin{equation}\label{IDENTYC6}
\frac{K_0}{K_1}-\frac{K_1}{K_2} = \frac{1}{z_m} + {\mathcal O}\left(\frac{1}{z_m^2}\right).
\end{equation}
Hence, for large values of $z_m$, we find
\begin{align}
\left(\delta R_L^2\right)^{(1)} &= -\frac{5\tau^2T\Phi}{4\beta_\pi m_T},
\label{ASYMPRL1C6}\\
\left(\delta R_L^2\right)^{(G)} &= -\frac{\tau^2T\Phi}{5\beta_\pi m_T}\left( 3+\frac{m_T}{T} \right).
\label{ASYMPRLGC6}
\end{align}
It is clear from the above two equations that the viscous correction
to $R_L$ in the Chapman-Enskog case preserves the $1/\sqrt{m_T}$
scaling, whereas in Grad's 14-moment approximation it grows as
$m_T/T$, and thus violates the scaling \cite{Teaney:2003kp}.

 \begin{figure}[t]
 \begin{center}
 \includegraphics[scale=0.5]{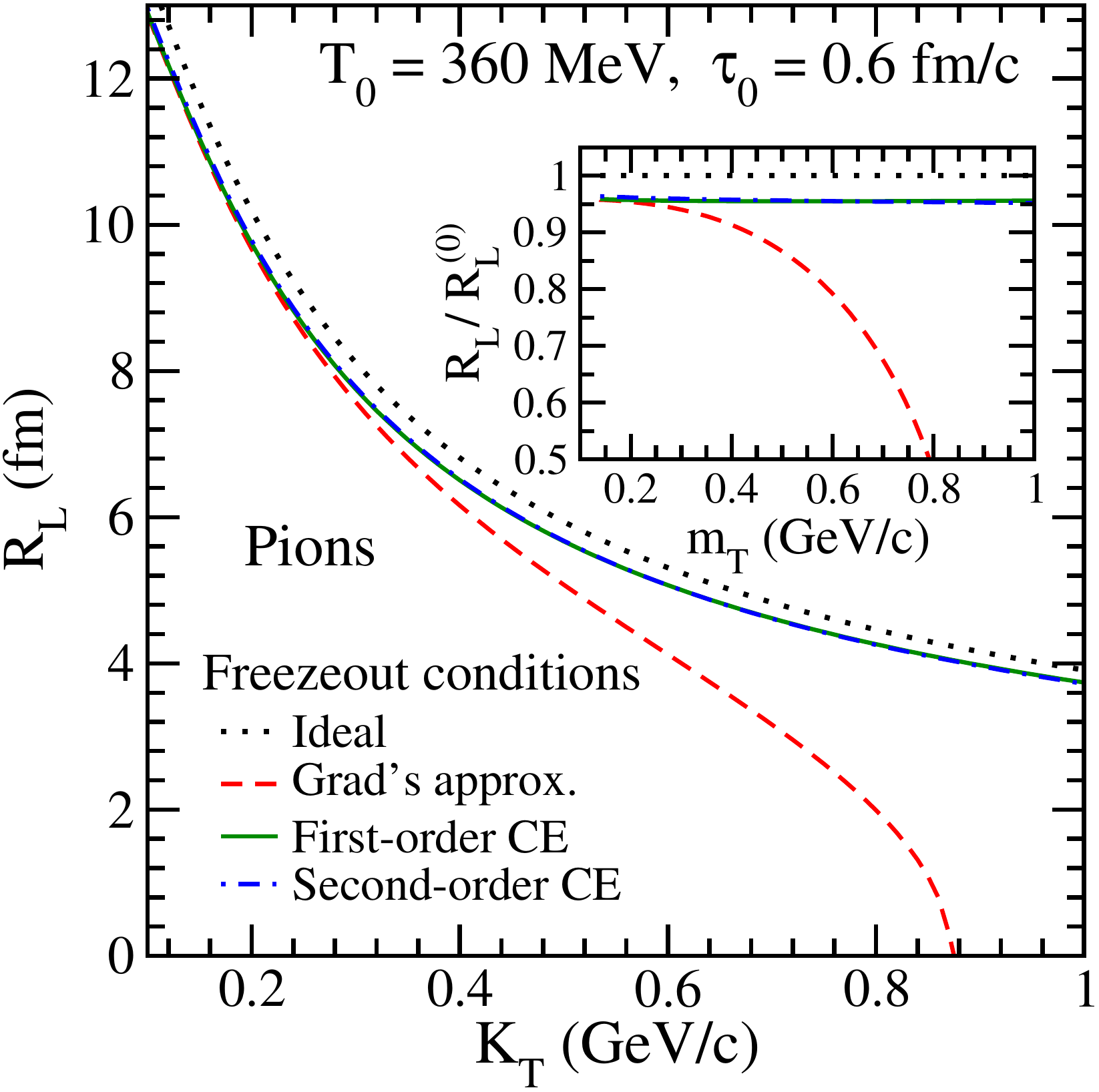}
 \end{center}
 \vspace{-0.4cm}
 \caption[Longitudinal HBT radius as a function of the transverse 
   momentum]{ Longitudinal HBT radius as a function of the 
   transverse momentum $K_T$ of the pion pair, obtained with the 
   second-order hydrodynamic evolution, followed by freezeout in 
   various scenarios: ideal, Grad's 14-moment approximation, first- 
   and second-order Chapman-Enskog. Inset: HBT radius in the above 
   cases scaled by the corresponding values in the ideal case.}
 \label{fhbtC6}
 \end{figure}

Results for the longitudinal HBT radius, $R_L$, for identical-pion
pairs in central Au-Au collisions, for the four cases discussed above,
are displayed in Fig. \ref{fhbtC6}. We note that while there is no
noticeable difference between first- and second-order Chapman-Enskog
results compared to the ideal case, they predict a slightly smaller
value for $R_L$. On the other hand, $R_L$ corresponding to Grad's
approximation exhibits a qualitatively different behaviour and even
becomes imaginary for $K_T \gtrsim 0.9$ GeV/$c$, which is clearly
unphysical. More importantly, the ratio $R_L/R_L^{(0)}$ shown in the
inset of Fig. \ref{fhbtC6}, illustrates that the $1/\sqrt{m_T}$ scaling
which is violated in Grad's approximation, survives in the
Chapman-Enskog case.

%%%%%%%%%%%%%%%%%%%%%%%%%%%%%%%%%%%%%%%%%%%%%%%%%%%%%%%%%%%%%%%%%%%%%%%%

\section{Summary and Conclusions}

We derived the form of the viscous correction to the equilibrium
distribution function, up to second order in gradients, by employing a
Chapman-Enskog-like iterative solution of the Boltzmann equation in
the relaxation time approximation. This approach is in accordance with
the formulation of hydrodynamics which is also a gradient expansion.
We used this form of the viscous correction to calculate the hadronic
transverse-momentum spectra and longitudinal Hanbury Brown-Twiss
radii, and compared them with those obtained in Grad's 14-moment
approximation within the one-dimensional scaling expansion. These
results demonstrate the rapid convergence of the Chapman-Enskog
expansion up to second order, and thus it is sufficient to retain only
the first-order correction in the freezeout prescription. We found
that the Chapman-Enskog method results in softer hadron spectra
compared with Grad's approximation. We further showed that the
experimentally observed $1/\sqrt{m_T}$ scaling of HBT radii which is
also seen in the ideal freezeout calculation, is maintained in the
Chapman-Enskog method. In contrast, the Grad's 14-moment approximation
leads to the violation of this scaling as well as an imaginary value
for $R_L$ at large momenta. For initial conditions typical of
heavy-ion collisions at the Large Hadron Collider ($T_0 = 500$ MeV and
$\tau_0=0.4$ fm/$c$), we have found that the above conclusions remain
unchanged.

Unlike Grad's approximation which is linear in dissipative 
quantities, the form of viscous correction to the distribution 
function obtained here using Chapman-Enskog like expansion contains 
higher-order nonlinear corrections to the equilibrium distribution 
function. This has important implications for the formulation of 
dissipative hydrodynamics as an order-by-order expansion in 
gradients. For example, by calculating the nonequilibrium 
distribution function to any given order, the dissipative evolution 
equation up to that order can be derived. In the next chapter, we 
present the derivation of a novel third-order hydrodynamic evolution 
equation for shear stress tensor from kinetic theory and quantify 
the significance of this new derivation within one-dimensional 
scaling expansion.

We conclude by recalling the well-known form of the viscous correction
due to Grad's 14-moment approximation:
\begin{equation}
 \delta f_G = \frac{f_0 \tilde f_0}{2(\epsilon+P)T^2}\, p^\alpha p^\beta \pi_{\alpha\beta},
\end{equation}
and the alternate form due to Chapman-Enskog method proposed here:
\begin{equation}
 \delta f_{CE}= \frac{5 f_0 \tilde f_0}{8 P T(u \!\cdot\! p)}\, p^\alpha p^\beta \pi_{\alpha\beta},
\end{equation}
where $\tilde f_0 \equiv 1-r f_0$, with $r=1,-1,0$ for Fermi, Bose, and
Boltzmann gases, respectively. In view of the arguments presented in
this chapter, we advocate that the form of $\delta f_{CE}$ proposed here
should be a better alternative for hydrodynamic modelling of
relativistic heavy-ion collisions.
 

%% file: Chapter7.tex
%########################################################################
\chapter{Relativistic third-order viscous fluid dynamics from kinetic theory}
%########################################################################

%%%%%%%%%%%%%%%%%%%%%%%%%%%%%%%%%%%%%%%%%%%%%%%%%%%%%%%%%%%%%%%%%%%%%%%%

\section{Introduction}

Despite the success of IS theory in explaining a wide range of 
collective phenomena observed in heavy-ion collisions, its 
formulation is based on strong assumptions and approximations. The 
original IS theory derived from Boltzmann equation (BE) uses two 
powerful assumptions in the derivation of dissipative equations: use 
of second moment of BE and the 14-moment approximation \cite 
{Israel:1979wp, Grad}. In Ref. \cite {Denicol:2010xn}, although the 
dissipative equations were derived directly from their definitions 
without resorting to second-moment of BE, however the 14-moment 
approximation was still employed. In Chapter 5, it was shown that 
both these assumptions are unnecessary, and instead of 14-moment 
approximation, iterative solution of BE was used to obtain the 
dissipative evolution equations from their definitions.

Apart from these problems in the formulation, IS theory suffers from 
several other shortcomings. In one-dimensional Bjorken scaling 
expansion \cite {Bjorken:1982qr}, for large viscosities or small 
initial time, IS theory has resulted in unphysical effects such as 
reheating of the expanding medium \cite {Muronga:2003ta} and 
negative longitudinal pressure \cite {Martinez:2009mf}. Further, the 
scaling solutions of IS equations when compared with transport 
results show disagreement for $\eta/s>0.5$ indicating the breakdown 
of second-order theory \cite {Huovinen:2008te,El:2008yy}. With this 
motivation, in Ref. \cite {Jaiswal:2012qm}, second-order dissipative 
equations were derived from BE where the collision term was 
generalized to include nonlocal effects via gradients of the 
distribution function. Moreover, in Refs. \cite 
{Jaiswal:2013npa,El:2009vj} it was demonstrated that a heuristic 
inclusion of higher-order corrections led to an improved agreement 
with transport results. In fact, the derivation of higher-order 
constitutive equations from kinetic theory for non-relativistic 
systems has been known for a long time \cite{Cha}. Thus it is of 
interest to improvise the relativistic second-order theory by 
incorporating higher-order corrections. 

In this chapter, we derive a new relativistic third-order evolution 
equation for shear stress tensor from kinetic theory. Without 
resorting to the widely used Grad's 14-moment approximation \cite
{Grad}, we iteratively solve the BE in relaxation time approximation 
(RTA) to obtain nonequilibrium phase-space distribution function. We 
subsequently derive equation of motion for shear stress tensor up-to 
third-order, directly from its definition. Within one-dimensional 
scaling expansion, the results obtained using third-order evolution 
equations derived here shows improved agreement with exact solution 
of BE as compared to second-order equations. We also demonstrate 
that the evolution of pressure anisotropy obtained using our 
equations shows better agreement with the transport results as 
compared to those obtained by using an existing third-order equation 
derived from entropy considerations.

%%%%%%%%%%%%%%%%%%%%%%%%%%%%%%%%%%%%%%%%%%%%%%%%%%%%%%%%%%%%%%%%%%%%%%%%

\section{Relativistic hydrodynamics}

The hydrodynamic evolution of a system is governed by the 
conservation equations for energy and momentum. The conserved 
energy-momentum tensor can be expressed in terms of single-particle, 
phase-space distribution function and tensor decomposed into 
hydrodynamic variables \cite{deGroot}. For a system of massless 
particles, bulk viscosity vanishes leading to
\begin{align}\label{NTDC7}
T^{\mu\nu} &= \!\int\! dp \ p^\mu p^\nu\, f(x,p) = \epsilon u^\mu u^\nu 
- P\Delta ^{\mu \nu} + \pi^{\mu\nu},
\end{align}
where $dp\equiv g d{\bf p}/[(2 \pi)^3|\bf p|]$, $g$ being the 
degeneracy factor, $p^\mu$ is the particle four-momentum and $f(x,p)$
is the phase-space distribution function. In the tensor 
decompositions, $\epsilon$, $P$ and $\pi^{\mu\nu}$ are respectively 
energy density, pressure and the shear stress tensor. The projection 
operator $\Delta^{\mu\nu}\equiv g^{\mu\nu}-u^\mu u^\nu$ is 
orthogonal to the hydrodynamic four-velocity $u^\mu$ defined in the 
Landau frame: $T^{\mu\nu} u_\nu=\epsilon u^\mu$. The metric tensor 
is Minkowskian, $g^{\mu\nu}\equiv\mathrm{diag}(+,-,-,-)$.

Energy-momentum conservation, $\partial_\mu T^{\mu\nu} =0$ yields 
the fundamental evolution equations for $\epsilon$ and $u^\mu$
\begin{align}\label{evolC7}
\dot\epsilon + (\epsilon+P)\theta - \pi^{\mu\nu}\nabla_{(\mu} u_{\nu)} &= 0,  \nonumber\\
(\epsilon+P)\dot u^\alpha - \nabla^\alpha P + \Delta^\alpha_\nu \partial_\mu \pi^{\mu\nu}  &= 0. 
\end{align}
As in the previous chapters, we use the notation $\dot A\equiv 
u^\mu\partial_\mu A$ for comoving derivative, $\theta\equiv 
\partial_\mu u^\mu$ for the expansion scalar, $A^{(\alpha}B^{\beta)}
\equiv (A^\alpha B^\beta + A^\beta B^\alpha)/2$ for symmetrization 
and $\nabla^\alpha\equiv \Delta^{\mu\alpha}\partial_\mu$ for 
space-like derivative. In the massless limit, the energy density and 
pressure are related as $\epsilon=3P\propto\beta^{-4}$. The inverse 
temperature, $\beta\equiv1/T$, is defined by the Landau matching 
condition $\epsilon=\epsilon_0$ where $\epsilon_0$ is the 
equilibrium energy density. In this limit, Eqs. (\ref{evolC7}) can 
be used to obtain the derivatives of $\beta$ as
\begin{equation}\label{evol1C7}
\dot\beta = \frac{\beta}{3}\theta - \frac{\beta}{12P}\pi^{\rho\gamma}\sigma_{\rho\gamma}, \quad
\nabla^\alpha\beta = \!-\beta\dot u^\alpha - \frac{\beta}{4P} \Delta^\alpha_\rho \partial_\gamma \pi^{\rho\gamma} , 
\end{equation}
where 
$\sigma^{\rho\gamma}\equiv\nabla^{(\rho}u^{\gamma)}-(\theta/3)\Delta^{\rho
\gamma}$ is the velocity stress tensor. The above identities will 
be helpful in the derivation of shear evolution equation.

The expression for shear stress tensor ($\pi^{\mu\nu}$) can be 
obtained in terms of the out-of-equilibrium part of the distribution 
function. To this end, we write the nonequilibrium distribution 
function as $f=f_0+\delta f$, where the deviation from equilibrium 
is assumed to be small $(\delta f\ll f)$. The equilibrium 
distribution function represents Boltzmann statistics of massless 
particles at vanishing chemical potential, $f_0=\exp(-\beta\,u\cdot 
p)$, where $u\cdot p \equiv u_\mu p^\mu$. From Eq. (\ref{NTDC7}), 
$\pi^{\mu\nu}$ can be expressed in terms of $\delta f$ as
\begin{align}\label{FSEC7}
\pi^{\mu\nu} &= \Delta^{\mu\nu}_{\alpha\beta} \int dp \, p^\alpha p^\beta\, \delta f,
\end{align}
where $\Delta^{\mu\nu}_{\alpha\beta}\equiv 
\Delta^{\mu}_{(\alpha}\Delta^{\nu}_{\beta)} - 
(1/3)\Delta^{\mu\nu}\Delta_{\alpha\beta}$ is a traceless symmetric 
projection operator orthogonal to $u^\mu$. To proceed further, the 
form of $\delta f$ has to be specified. In the following, Boltzmann 
equation in RTA will be solved iteratively to obtain $\delta f$ 
order-by-order in gradients.

%%%%%%%%%%%%%%%%%%%%%%%%%%%%%%%%%%%%%%%%%%%%%%%%%%%%%%%%%%%%%%%%%%%%%%%%

\section{Chapman-Enskog expansion}

As demonstrated in the previous chapters, nonequilibrium phase-space 
distribution function can be obtained by solving the one-body 
kinetic equation such as the Boltzmann equation. The most common 
technique of generating solutions to such equations is the 
Chapman-Enskog expansion where the particle distribution function is 
expanded about its equilibrium value in powers of space-time 
gradients \cite{Chapman}, which we repeat for convenience.
\begin{equation}\label{CEEC7}
f = f_0 + \delta f, \quad \delta f= \delta f^{(1)} + \delta f^{(2)} + \cdots,
\end{equation}
where $\delta f^{(1)}$ is first-order in derivatives, $\delta f^{(2)}$ 
is second-order and so on. Subsequently, the relativistic Boltzmann 
equation with relaxation time approximation for the collision term 
\cite{Anderson_Witting},
\begin{equation}\label{RBEC7}
p^\mu\partial_\mu f =  -u\!\cdot\! p\frac{\delta f}{\tau_R} \;\Rightarrow\; 
f=f_0-(\tau_R/u\!\cdot\! p)\,p^\mu\partial_\mu f,
\end{equation}
can be solved iteratively as \cite{Jaiswal:2013npa,Romatschke:2011qp}
\begin{equation}\label{F1F2C7}
f_1 = f_0 -\frac{\tau_R}{u\!\cdot\! p} \, p^\mu \partial_\mu f_0, \quad 
f_2 = f_0 -\frac{\tau_R}{u\!\cdot\! p} \, p^\mu \partial_\mu f_1, \quad \cdots
\end{equation}
where $f_n=f_0+\delta f^{(1)}+\delta f^{(2)}+\cdots+\delta f^{(n)}$. 
To first and second-order in derivatives, we obtain
\begin{equation}\label{FSOCC7}
\delta f^{(1)} = -\frac{\tau_R}{u\!\cdot\! p} \, p^\mu \partial_\mu f_0, \quad
\delta f^{(2)} = \frac{\tau_R}{u\!\cdot\! p}p^\mu p^\nu\partial_\mu\Big(\frac{\tau_R}{u\!\cdot\! p} \partial_\nu f_0\Big).
\end{equation}
The above expressions for nonequilibrium part of the distribution 
function along with Eq. (\ref{FSEC7}) will be used in the derivation 
of shear evolution equations.

As a side remark, note that the RTA for the collision term, 
$C[f]=-(u\cdot p)\delta f/\tau_R$ in Eq. (\ref{RBEC7}), should 
satisfy current and energy-momentum conservation, i.e., the zeroth 
and first moment of the collision term should vanish \cite 
{deGroot}. Assuming the relaxation time $\tau_R$ to be independent 
of momenta, these conservation equations are satisfied only if the 
fluid four-velocity is defined in the Landau frame \cite 
{Anderson_Witting}. Hence, within RTA, the Landau frame is imposed 
and is not a choice.

%%%%%%%%%%%%%%%%%%%%%%%%%%%%%%%%%%%%%%%%%%%%%%%%%%%%%%%%%%%%%%%%%%%%%%%%

\section{Evolution equations for shear stress tensor}

The first-order expression for shear stress tensor can be obtained 
from Eq. (\ref{FSEC7}) using $\delta f = \delta f^{(1)}$ from Eq. (\ref
{FSOCC7}),
\begin{align}
\pi^{\mu\nu} &= \Delta^{\mu\nu}_{\alpha\beta}\int dp \ p^\alpha p^\beta \left(-\frac{\tau_R}{u\!\cdot\! p} \, p^\mu \partial_\mu\, f_0\right) . \label{FOSEC7}
\end{align}
Using Eqs. (\ref{evol1C7}) and keeping only those terms which are 
first-order in gradients, the integrals in the above equation reduce 
to
\begin{equation}\label{FOEC7}
\pi^{\mu\nu} = 2\tau_R\beta_\pi\sigma^{\mu\nu}, \quad \beta_\pi = \frac{4}{5}P .
\end{equation}

To obtain the second-order evolution equation, we follow the 
methodology discussed in Ref. \cite {Denicol:2010xn}. The evolution 
of the shear stress tensor can be obtained by considering the 
comoving derivative of Eq. (\ref{FSEC7}),
\begin{equation}
\dot\pi^{\langle\mu\nu\rangle} = \Delta^{\mu\nu}_{\alpha\beta} \int dp\, p^\alpha p^\beta\, \delta\dot f, \label{SSEC7}
\end{equation}
where the notation $A^{\langle\mu\nu\rangle}\equiv 
\Delta^{\mu\nu}_{\alpha\beta}A^{\alpha\beta}$ represents traceless 
symmetric projection orthogonal to $u^{\mu}$.

The comoving derivative of the nonequilibrium part of the 
distribution function ($\delta\dot f$) can be obtained by rewriting 
Eq. (\ref{RBEC7}) in the form
\begin{equation}\label{DFDC7}
\delta\dot f = -\dot f_0 - \frac{1}{u\!\cdot\! p}p^\gamma\nabla_\gamma f - \frac{\delta f}{\tau_R},
\end{equation}
Using this expression for $\delta\dot f$ in Eq. (\ref{SSEC7}), we obtain
\begin{equation} 
\dot\pi^{\langle\mu\nu\rangle} + \frac{\pi^{\mu\nu}}{\tau_R} = 
- \Delta^{\mu\nu}_{\alpha\beta} \!\int\! dp \, p^\alpha p^\beta \!\left(\dot f_0 + \frac{1}{u\!\cdot\! p}p^\gamma\nabla_\gamma f\right)\!. \label{SOSEC7}
\end{equation}
It is clear that in the above equation, the Boltzmann relaxation 
time $\tau_R$ can be replaced by the shear relaxation time 
$\tau_\pi$. By comparing the first-order evolution Eq. (\ref{FOEC7}) 
with the relativistic Navier-Stokes equation 
$\pi^{\mu\nu}=2\eta\sigma^{\mu\nu}$, the shear relaxation time is 
obtained in terms of the first-order transport coefficient, 
$\tau_\pi=\eta/\beta_\pi$.

Note that for the shear evolution equations to be second-order in 
gradients, the distribution function on the right hand side of Eq. 
(\ref{SOSEC7}) need to be computed only till first-order, i.e., 
$f=f_1=f_0+\delta f^{(1)}$. Using Eq. (\ref{FSOCC7}) for $\delta 
f^{(1)}$ and Eqs. (\ref{evol1C7}) for derivatives of $\beta$, and 
keeping terms up to quadratic order in gradients, the second-order 
shear evolution equation is obtained as \cite {Jaiswal:2013npa}
\begin{equation}\label{SOSHEARC7}
\dot{\pi}^{\langle\mu\nu\rangle} \!+ \frac{\pi^{\mu\nu}}{\tau_\pi}\!= 
2\beta_{\pi}\sigma^{\mu\nu}
\!+2\pi_\gamma^{\langle\mu}\omega^{\nu\rangle\gamma}
\!-\frac{10}{7}\pi_\gamma^{\langle\mu}\sigma^{\nu\rangle\gamma} 
\!-\frac{4}{3}\pi^{\mu\nu}\theta,
\end{equation}
where $\omega^{\mu\nu}\equiv(\nabla^\mu u^\nu-\nabla^\nu u^\mu)/2$ 
is the vorticity tensor. We have used the first-order expression for 
shear stress tensor, Eq. (\ref{FOEC7}), to replace 
$\sigma^{\mu\nu}\to\pi^{\mu\nu}$ such that the relaxation times 
appearing on the right hand side of Eq. (\ref{SOSEC7}) are absorbed.

To derive a third-order evolution equation for shear stress tensor, 
the distribution function on the right hand side of Eq. (\ref 
{SOSEC7}) needs to be computed till second-order ($\delta f=\delta 
f^{(1)}+\delta f^{(2)}$). In order to account for all the 
higher-order terms, Eq. (\ref{SOSHEARC7}) was used to substitute for 
$\sigma^{\mu\nu}$. Employing Eqs. (\ref{evol1C7}) for derivatives of 
$\beta$ and keeping terms up to cubic order in derivatives, we 
finally obtain a unique third-order evolution equation for shear 
stress tensor after a straightforward but tedious algebra
\begin{align}\label{TOSHEARC7}
\dot{\pi}^{\langle\mu\nu\rangle} =& -\frac{\pi^{\mu\nu}}{\tau_\pi}
+2\beta_\pi\sigma^{\mu\nu}
+2\pi_{\gamma}^{\langle\mu}\omega^{\nu\rangle\gamma}
-\frac{10}{7}\pi_\gamma^{\langle\mu}\sigma^{\nu\rangle\gamma} 
-\frac{4}{3}\pi^{\mu\nu}\theta
+\frac{25}{7\beta_\pi}\pi^{\rho\langle\mu}\omega^{\nu\rangle\gamma}\pi_{\rho\gamma}
-\frac{1}{3\beta_\pi}\pi_\gamma^{\langle\mu}\pi^{\nu\rangle\gamma}\theta \nonumber \\
&-\frac{38}{245\beta_\pi}\pi^{\mu\nu}\pi^{\rho\gamma}\sigma_{\rho\gamma}
-\frac{22}{49\beta_\pi}\pi^{\rho\langle\mu}\pi^{\nu\rangle\gamma}\sigma_{\rho\gamma}
-\frac{24}{35}\nabla^{\langle\mu}\left(\pi^{\nu\rangle\gamma}\dot u_\gamma\tau_\pi\right)
+\frac{4}{35}\nabla^{\langle\mu}\left(\tau_\pi\nabla_\gamma\pi^{\nu\rangle\gamma}\right) \nonumber \\
&-\frac{2}{7}\nabla_{\gamma}\left(\tau_\pi\nabla^{\langle\mu}\pi^{\nu\rangle\gamma}\right)
+\frac{12}{7}\nabla_{\gamma}\left(\tau_\pi\dot u^{\langle\mu}\pi^{\nu\rangle\gamma}\right)
-\frac{1}{7}\nabla_{\gamma}\left(\tau_\pi\nabla^{\gamma}\pi^{\langle\mu\nu\rangle}\right)
+\frac{6}{7}\nabla_{\gamma}\left(\tau_\pi\dot u^{\gamma}\pi^{\langle\mu\nu\rangle}\right) \nonumber \\
&-\frac{2}{7}\tau_\pi\omega^{\rho\langle\mu}\omega^{\nu\rangle\gamma}\pi_{\rho\gamma}
-\frac{2}{7}\tau_\pi\pi^{\rho\langle\mu}\omega^{\nu\rangle\gamma}\omega_{\rho\gamma} 
-\frac{10}{63}\tau_\pi\pi^{\mu\nu}\theta^2
+\frac{26}{21}\tau_\pi\pi_\gamma^{\langle\mu}\omega^{\nu\rangle\gamma}\theta.
\end{align}
The above equation constitutes the main result of this chapter. We 
note that Eq. (\ref{TOSHEARC7}) represents only a subset of all 
possible third order terms because bulk viscosity and heat current 
has been neglected.

We compare the third-order shear evolution equation derived here 
with that obtained by El {\it et al.}, in Ref. \cite{El:2009vj}. In 
the latter work, the shear evolution equation was derived by 
invoking second law of thermodynamics from kinetic definition of 
entropy four-current, expanded till third-order in $\pi^{\mu\nu}$. 
For ease of comparison, we write the evolution equation obtained in 
Ref. \cite {El:2009vj} in the form
\begin{equation}\label{TOEFC7}
\dot\pi^{\langle\mu\nu\rangle} = -\frac{\pi^{\mu\nu}}{\tau_\pi'} + 2\beta_\pi'\sigma^{\mu\nu}
-\frac{4}{3}\pi^{\mu\nu}\theta 
+ \frac{5}{36\beta_\pi'}\pi^{\mu\nu}\pi^{\rho\gamma}\sigma_{\rho\gamma}
-\frac{16}{9\beta_\pi'}\pi^{\langle\mu}_{\gamma}\pi^{\nu\rangle\gamma}\theta,
\end{equation}
where $\beta_\pi'=2P/3$ and $\tau_\pi'=\eta/\beta_\pi'$. We observe 
that the right-hand-side of Eq. (\ref{TOEFC7}) contains one 
second-order and two third-order terms compared to three 
second-order and fourteen third-order terms obtained here, i.e., Eq. 
(\ref{TOSHEARC7}). It is well known that the approach based on 
entropy method fails to capture all the terms in the dissipative 
evolution equations even at second-order. Moreover, the discrepancy 
at third-order confirms the fact that the evolution equation 
obtained by invoking second law of thermodynamics is incomplete.

%%%%%%%%%%%%%%%%%%%%%%%%%%%%%%%%%%%%%%%%%%%%%%%%%%%%%%%%%%%%%%%%%%%%%%%%

\section{Numerical results and discussion}

To demonstrate the numerical significance of the third-order shear 
evolution equation derived here, we consider boost-invariant Bjorken 
expansion of a system consisting of massless Boltzmann gas \cite 
{Bjorken:1982qr}. Working in Milne coordinates $(\tau,x,y,\eta_s)$, 
where $\tau = \sqrt{t^2-z^2}$, $\eta_s=\tanh^{-1}(z/t)$, and with 
$u^\mu=(1,0,0,0)$, we observe that only the $\eta_s\eta_s$ component 
of Eq. (\ref{TOSHEARC7}) survives. In this scenario, 
$\omega^{\mu\nu}=\dot u^\mu=\nabla^\mu\tau_\pi=0$, $\theta = 1/\tau$ 
and $\sigma^{\eta_s\eta_s} = -2/(3\tau^3)$. Defining 
$\Phi\equiv-\tau^2\pi^{\eta_s\eta_s}$, we find that
$\pi^{\rho\gamma}\sigma_{\rho\gamma} = \Phi/\tau$, and
\begin{align}\label{identityC7}
&\dot\pi^{\langle\eta_s\eta_s\rangle} = -\frac{1}{\tau^2}\frac{d\Phi}{d\tau}, \quad
\pi^{\langle\eta_s}_{\gamma}\sigma^{\eta_s\rangle\gamma} = -\frac{\Phi}{3\tau^3}, \quad
\pi^{\langle\eta_s}_{\gamma}\pi^{\eta_s\rangle\gamma} = -\frac{\Phi^2}{2\tau^2}, \quad
\pi^{\rho\langle\eta_s}\pi^{\eta_s\rangle\gamma}\sigma_{\rho\gamma} = -\frac{\Phi^2}{2\tau^3}, \nonumber\\
&\nabla^{\langle\eta_s}\nabla_{\gamma}\pi^{\eta_s\rangle\gamma} = \frac{2\Phi}{3\tau^4}, \quad 
\nabla_{\gamma}\nabla^{\langle\eta_s}\pi^{\eta_s\rangle\gamma} = \frac{4\Phi}{3\tau^4}, \quad
\nabla^2\pi^{\langle\eta_s\eta_s\rangle} = \frac{4\Phi}{3\tau^4},
\end{align} 
see Appendix A for details.
Using the above results, evolution of $\epsilon$ and $\Phi$ from Eqs. 
(\ref{evolC7}) and (\ref{TOSHEARC7}) reduces to
\begin{align}
\frac{d\epsilon}{d\tau} &= -\frac{1}{\tau}\left(\epsilon + P - \Phi\right), \label{BEDC7} \\
\frac{d\Phi}{d\tau} &= - \frac{\Phi}{\tau_\pi} + \beta_\pi\frac{4}{3\tau} - \lambda\frac{\Phi}{\tau} - \chi\frac{\Phi^2}{\beta_\pi\tau}. \label{BshearC7}
\end{align}
The term with coefficient $\chi$ in the above equation contains 
correction only due to third-order. The first-order shear 
expression, $\Phi=4\beta_\pi\tau_\pi/3\tau$, has been used to 
rewrite some of the third-order contributions in the form 
$\Phi^2/(\beta_\pi\tau)$. The transport coefficients in our 
calculation simplify to
\begin{equation}\label{BTCC7}
\tau_\pi = \frac{\eta}{\beta_\pi}, \quad \beta_\pi = \frac{4P}{5}, \quad \lambda = \frac{38}{21}, \quad \chi = \frac{72}{245}.
\end{equation}
We compare these transport coefficients with those obtained from 
Eq. (\ref{TOEFC7}), where they reduce to
\begin{equation}\label{BTCEC7}
\tau_\pi' = \frac{\eta}{\beta_\pi'}, \quad \beta_\pi' = \frac{2P}{3}, \quad \lambda' = \frac{4}{3}, \quad \chi' = \frac{3}{4}.
\end{equation}

For comparison, we also state the exact solution of Eq. (\ref{RBEC7}) 
in one-dimensional scaling expansion \cite{Baym:1984np,Florkowski:2013lza}:
\begin{equation}\label{ESBEC7}
f(\tau) = D(\tau,\tau_0)f_{\rm in} + \int_{\tau_0}^{\tau}\frac{d\tau'}{\tau_R(\tau')}D(\tau,\tau')f_0(\tau'),
\end{equation}
where, $f_{\rm in}$ and $\tau_0$ are the initial distribution 
function and proper time respectively, and
\begin{equation}\label{DT2T1C7}
D(\tau_2,\tau_1) = \exp\left[-\int_{\tau_1}^{\tau_2}\frac{d\tau''}{\tau_R(\tau'')}\right].
\end{equation}
The damping function $D(\tau_2,\tau_1)$ has the following properties: 
\begin{equation}\label{DRELC7}
D(\tau,\tau)=1, \quad D(\tau_3,\tau_2)D(\tau_2,\tau_1)=D(\tau_3,\tau_1), \quad \frac{\partial D(\tau_2,\tau_1)}{\partial\tau_2} = -\frac{D(\tau_2,\tau_1)}{\tau_R(\tau_2)}.
\end{equation}
To obtain the exact solution, the Boltzmann relaxation time is taken 
to be the same as the shear relaxation time $(\tau_R=\tau_\pi)$. The 
hydrodynamic quantities can then be calculated by using Eq. (\ref 
{ESBEC7}) for the distribution function in Eq. (\ref{NTDC7}) and 
performing the integrations numerically.

\begin{figure}[t]
\begin{center}
\includegraphics[scale=0.4]{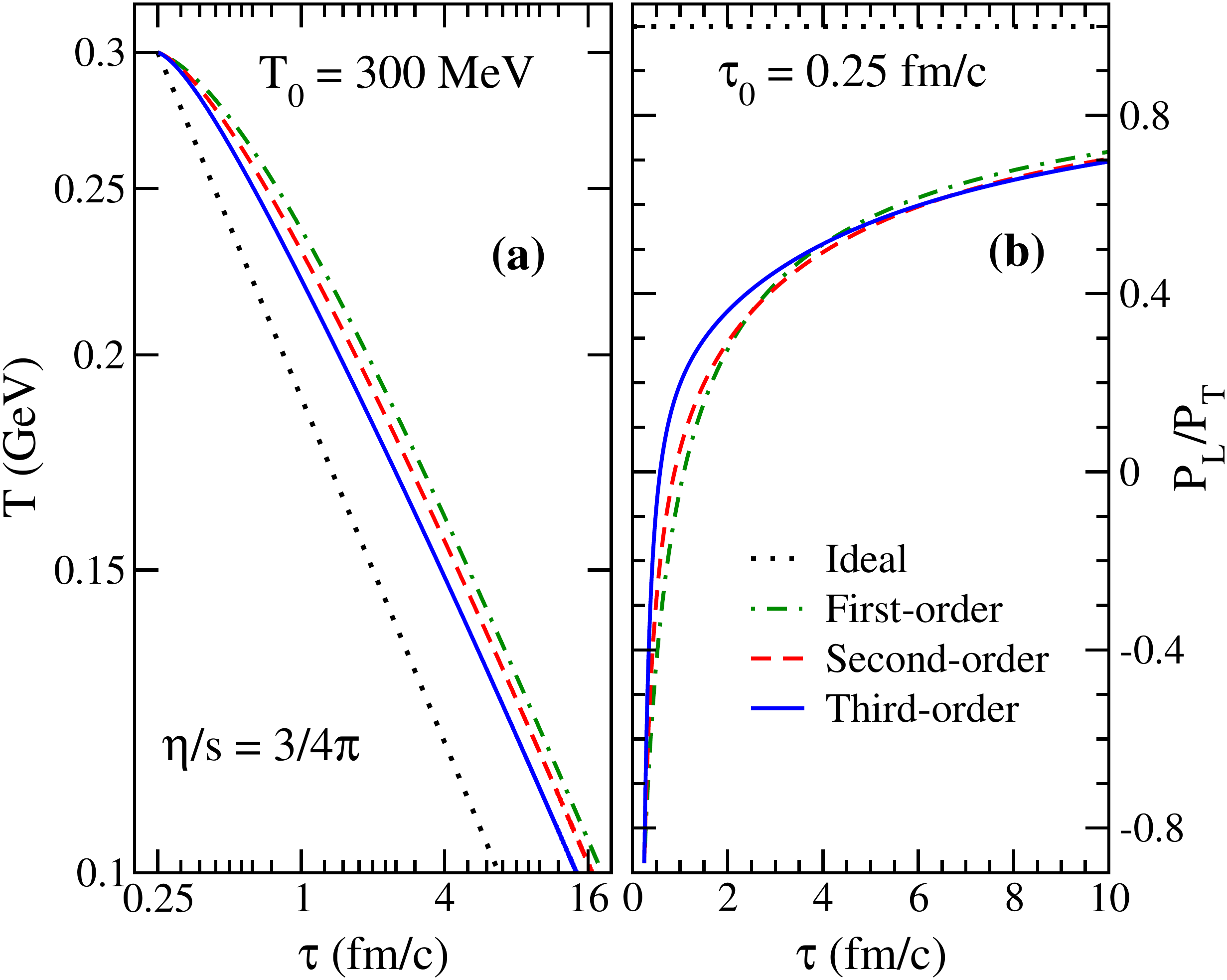}
\end{center}
\vspace{-0.4cm}
\caption[Time evolution of temperature and pressure anisotropy]
  {Time evolution of (a) temperature and (b) pressure anisotropy 
  ($P_L/P_T$), in ideal (dotted line), first-order (dashed-dotted 
  lines), second-order (dashed line) and third-order (solid lines) 
  hydrodynamics, for Navier-Stokes initial condition, 
  $(\Phi_0=4\eta/3\tau_0)$.}   
\label{TPLPTC7}
\end{figure}

To quantify the differences between ideal, first-order, 
second-order, and third-order theories, we solve the evolution 
equations with initial temperature $T_0=300$ MeV at initial time 
$\tau_0=0.25$ fm/c. These values correspond to the Relativistic 
Heavy-Ion Collider initial conditions \cite{El:2007vg}. Figure \ref 
{TPLPTC7} shows proper time evolution of temperature and pressure 
anisotropy $P_L/P_T\equiv(P-\Phi)/(P+\Phi/2)$ in ideal (dotted line), 
first-order (dashed-dotted lines), second-order (dashed line) and 
third-order (solid lines) hydrodynamics. Here we have assumed 
Navier-Stokes initial condition for shear pressure 
$(\Phi_0=4\eta/3\tau_0)$ and solved the evolution equations for a 
representative shear viscosity to entropy density ratio, 
$\eta/s=3/4\pi$.

In Fig. \ref{TPLPTC7} (a), we observe that while ideal hydrodynamics 
predicts a rapid cooling of the system, evolution based on 
third-order equation also shows faster temperature drop compared to 
first-order and second-order evolutions. This implies that the 
thermal photon and dilepton spectra, which are sensitive to 
temperature evolution, may be suppressed by including third-order 
corrections. Moreover, with third-order evolution, the freeze-out 
temperature is attained at an earlier time which may affect the 
hadronic spectra as well. In Fig. \ref{TPLPTC7} (b), note that at 
early times the third-order evolution results in faster 
isotropization of pressure anisotropy compared to first-order and 
second-order. However at later time, the pressure anisotropy 
obtained using second and third-order equations merge indicating the 
convergence of gradient expansion in fluid dynamics.

\begin{figure}[t]
\begin{center}
\includegraphics[scale=0.5]{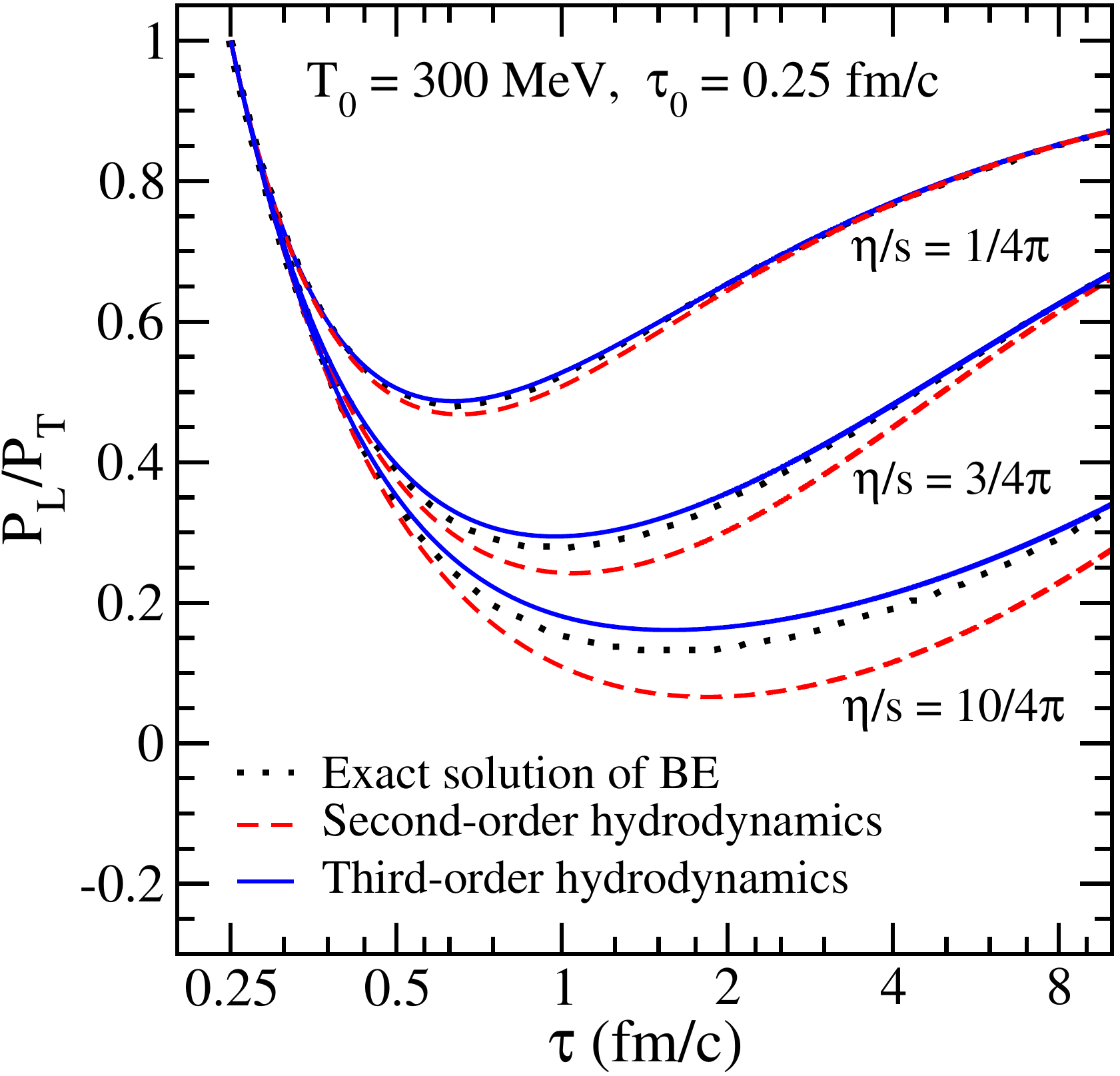}
\end{center}
\vspace{-0.4cm}
\caption[Pressure anisotropy comparison with exact solution of 
  Boltzmann equation]{Time evolution of $P_L/P_T$ obtained using 
  exact solution of Boltzmann equation (dotted line), second-order 
  equations (dashed lines), and third-order equations (solid lines), 
  for isotropic initial pressure configuration $(\Phi_0=0)$ and 
  various $\eta/s$.} 
\label{PLPTRC7}
\end{figure}

Figure \ref{PLPTRC7}, shows the proper time dependence of pressure 
anisotropy for various $\eta/s$ values with isotropic initial 
pressure configuration, i.e., $\Phi_0=0$. The improved agreement of 
third-order results (solid lines) with the exact solution of BE 
(dotted line) as compared to second-order results (dashed line) also 
suggests the convergence of the derivative expansion in hydrodynamics.

\begin{figure}[t]
\begin{center}
\includegraphics[scale=0.5]{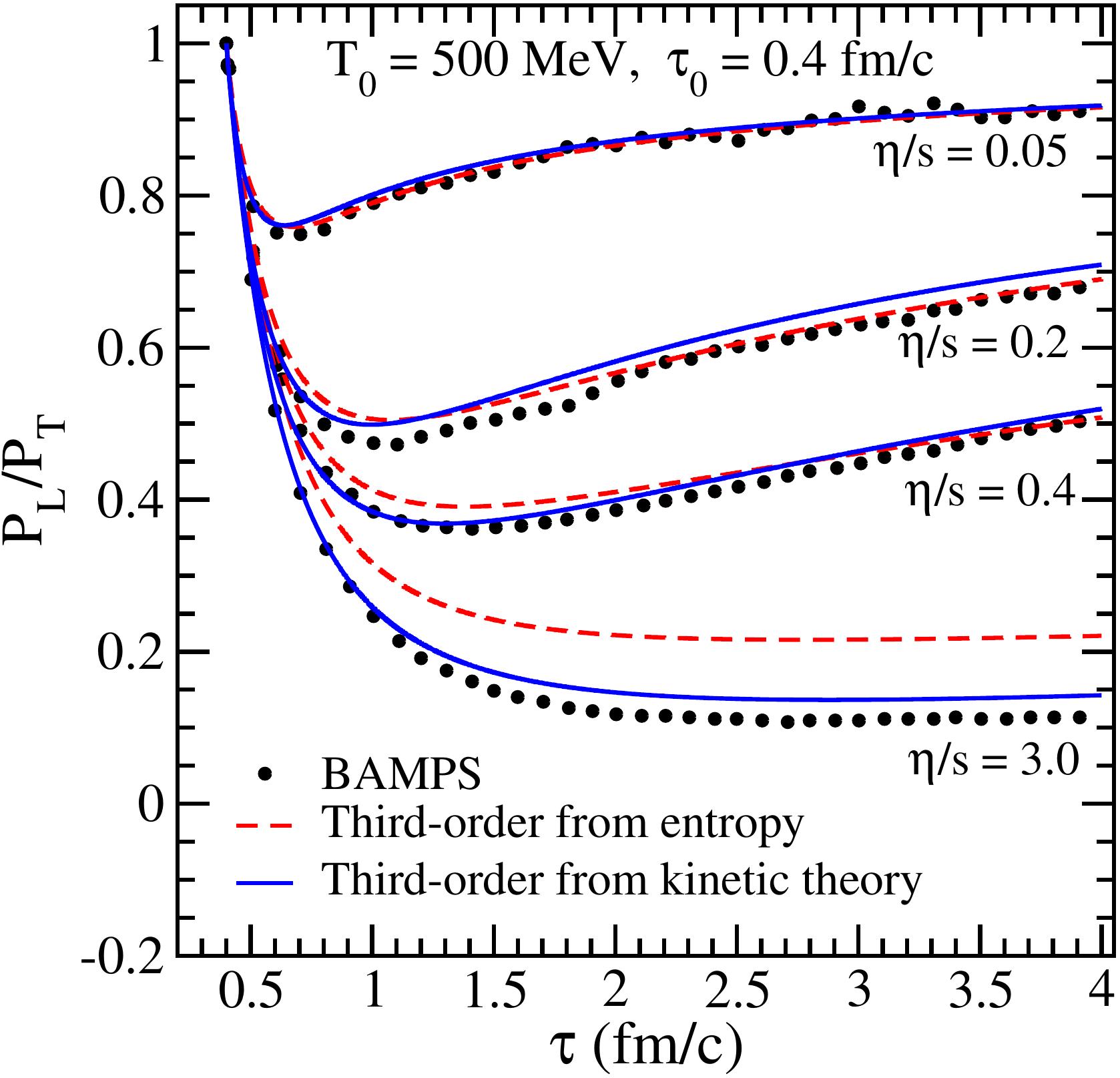}
\end{center}
\vspace{-0.4cm}
\caption[Pressure anisotropy comparison with BAMPS]{Time evolution 
  of $P_L/P_T$ in BAMPS (dots), third-order calculation from entropy 
  method, Eq. (\ref{TOEFC7}) (dashed lines), and the present work 
  (solid lines), for isotropic initial pressure configuration 
  $(\Phi_0=0)$ and various $\eta/s$.} 
\label{PLPTLC7}
\end{figure}

Figure \ref{PLPTLC7}, also shows the time evolution of pressure 
anisotropy for initial temperature $T_0=500$ MeV at initial time 
$\tau_0=0.4$ fm/c which corresponds to Large Hadron Collider initial 
conditions \cite{El:2007vg}. The initial pressure configuration is 
assumed to be isotropic and the evolution is shown for various 
$\eta/s$ values. The solid lines represent the results obtained in 
the present work by solving Eqs. (\ref{BEDC7}) and (\ref{BshearC7}) 
with transport coefficients of Eq. (\ref{BTCC7}). The dashed lines 
corresponds to results of another third-order theory derived based 
on second-law of thermodynamics with transport coefficients given in 
Eq. (\ref{BTCEC7}). The dots represent the results of numerical 
solution of BE using a transport model, the parton cascade BAMPS 
\cite {El:2009vj,Xu:2004mz}. The calculations in BAMPS are performed 
by changing the cross section such that $\eta/s$ remains constant. 
While the results from entropy derivation overestimate the pressure 
anisotropy for $\eta/s>0.2$, those obtained in the present work 
(kinetic theory) are in better agreement with the BAMPS results. 

The RTA for the collision term in BE is based on the assumption that 
the effect of the collisions is to exponentially restore the 
distribution function to its local equilibrium value. Although the 
information about the microscopic interactions of the constituent 
particles is not retained here, it is a reasonably good 
approximation to describe a system which is close to local 
equilibrium. It is important to note that although the third-order 
viscous equations derived here uses BE with RTA for the collision 
term, the evolution shows good quantitative agreement with BAMPS 
results which employs realistic collision kernel \cite{Xu:2004mz}. 
Indeed in Ref. \cite{Dusling:2009df}, it has been shown that for a 
purely gluonic system at weak coupling and hadron gas with large 
momenta, BE in RTA is a fairly accurate description. Furthermore, 
the experimentally observed $1/\sqrt{m_T}$ scaling of the HBT radii, 
which was shown to be broken by including viscous corrections to the 
distribution function \cite{Teaney:2003kp}, can be restored by using 
the form of the non-equilibrium distribution function obtained here 
\cite{Bhalerao:2013pza}. All these factors clearly suggests that the 
BE in RTA can be applied quite successfully in understanding the 
hydrodynamic behaviour of the strongly interacting matter formed in 
heavy-ion collisions.

%%%%%%%%%%%%%%%%%%%%%%%%%%%%%%%%%%%%%%%%%%%%%%%%%%%%%%%%%%%%%%%%%%%%%%%%

\section{Summary and conclusions}

To summarize, we have derived a novel third-order evolution equation 
for the shear stress tensor from kinetic theory within relaxation 
time approximation. Instead of Grad's 14-moment approximation, 
iterative solution of Boltzmann equation was used for the 
nonequilibrium distribution function and the evolution equation for 
shear tensor is derived directly from its definition. Within 
one-dimensional scaling expansion, we have demonstrated that the 
third-order hydrodynamics derived here provides a very good 
approximation to the exact solution of Boltzmann equation in 
relaxation time approximation. Our results also show a better 
agreement with the parton cascade BAMPS for the $P_L/P_T$ evolution 
compared to those obtained from entropy derivation.

As discussed previously, the approach based on the generalized 
second law of thermodynamics fails to capture all the terms in the 
evolution equations of the dissipative quantities when compared with 
similar equations derived from kinetic theory. However, derivation 
of dissipative evolution equations from kinetic theory also fails to 
capture all the terms allowed by symmetry up to a given order in 
derivatives. Starting with the relativistic Boltzmann equation where 
the collision term is generalized to include nonlocal effects via 
gradients of the phase-space distribution function, and using Grad's 
14-moment approximation for the distribution function, we derive 
second-order evolution equations for relativistic dissipative fluids 
in the next chapter. Our method generates all the second-order terms 
that are allowed by symmetry, some of which have been missed by the 
traditional approaches based on the Boltzmann equation with local 
collision term. 

%% file: Chapter8.tex
%########################################################################
\chapter{Nonlocal generalization of collision term and dissipative fluid dynamics}
%########################################################################

%%%%%%%%%%%%%%%%%%%%%%%%%%%%%%%%%%%%%%%%%%%%%%%%%%%%%%%%%%%%%%%%%%%%%%%%

\section{Introduction}

The second-order viscous hydrodynamics has been quite successful in 
explaining the spectra and azimuthal anisotropy of particles 
produced in heavy-ion collisions at the Relativistic Heavy Ion 
Collider (RHIC) \cite{Romatschke:2007mq,Song:2010mg} and recently at 
the Large Hadron Collider (LHC) \cite{Luzum:2010ag,Qiu:2011hf}. 
However, IS theory can lead to unphysical effects such as reheating 
of the expanding medium \cite{Muronga:2003ta} and to a negative 
pressure \cite{Martinez:2009mf} at large viscosity indicating its 
breakdown. Furthermore, from comparison to the transport theory it 
was demonstrated \cite{Huovinen:2008te,El:2008yy} that IS approach 
becomes marginal when the shear viscosity to entropy density ratio 
$\eta/s \gtrsim 1.5/(4\pi)$. With this motivation, the dissipative 
hydrodynamic equations were extended \cite {El:2009vj, 
Jaiswal:2013vta,Jaiswal:2014raa} to third order, which led to an 
improved agreement with the kinetic theory even for moderately large 
values of $\eta/s$.

While it is well known that the approach based on the generalized 
second law of thermodynamics fails to capture several terms in the 
evolution equations of the dissipative quantities when compared with 
similar equations derived from transport theory \cite{Baier:2006um}, 
the derivation from kinetic theory also fails to capture all the 
terms allowed by symmetry. It was pointed out that using directly 
the definitions of the dissipative currents, instead of the second 
moment of the Boltzmann equation as in IS theory, one obtains 
identical equations of motion but with different coefficients \cite
{Denicol:2010xn}. Recently, it has been shown \cite{Denicol:2012cn} 
that a generalization of Grad's 14-moment method \cite{Grad} results 
in additional terms in the dissipative equations.

It is important to note that all formulations that employ the 
Boltzmann equation make a strict assumption of a local collision 
term in the configuration space \cite{Israel:1979wp,Denicol:2010xn}. 
In other words, within an infinitesimal fluid element containing a 
large number of particles and extending over many interparticle 
spacings \cite{Landau}, the different collisions that increase or 
decrease the number of particles with a given momentum $p$ are all 
assumed to occur at the same point $x^\mu$. This makes the collision 
integral a purely local functional of the single-particle 
phase-space distribution function $f(x,p)$ independent of the 
derivatives $\partial_\mu f(x,p)$. In kinetic theory, $f(x,p)$ is 
assumed to vary slowly over space-time, i.e., it changes negligibly 
over the range of interparticle interaction \cite{deGroot}. However, 
its variation over the fluid element may not be insignificant; see 
Fig. \ref{flelC8}. Inclusion of the gradients of $f(x,p)$ in the 
collision term will affect the evolution of dissipative quantities 
and thus the entire dynamics of the system.

In this chapter, we shall provide a new formal derivation of the
dissipative hydrodynamic equations within kinetic theory but using a
nonlocal collision term in the Boltzmann equation. We obtain new
second-order terms and show that the coefficients of the other terms
are altered. These modifications do have a rather strong influence on
the evolution of the viscous medium as we shall demonstrate in the
case of one-dimensional scaling expansion.

%%%%%%%%%%%%%%%%%%%%%%%%%%%%%%%%%%%%%%%%%%%%%%%%%%%%%%%%%%%%%%%%%%%%%%%%

\section{Nonlocal collision term}

Our starting point is the relativistic Boltzmann equation for the
evolution of the phase-space distribution function,
$p^\mu \partial_\mu f =  C[f]$,
where the collision term $C[f]$ is required to be consistent with the
energy-momentum and current conservation.  Traditionally $C[f]$ is
also assumed to be a purely local functional of $f(x,p)$, independent
of $\partial_\mu f$. This locality assumption is a powerful
restriction \cite{Israel:1979wp} which we relax by
including the gradients of $f(x,p)$ in $C[f]$. This
necessarily leads to the modified Boltzmann equation
\begin{equation}\label{MBEC8}
p^\mu \partial_\mu f = C_m[f] 
=  C[f] + \partial_\mu(A^\mu f) + \partial_\mu\partial_\nu(B^{\mu\nu}f) + \cdots,
\end{equation}
where $A^\mu$ and $B^{\mu\nu}$ depend on the type of the
collisions ($2 \leftrightarrow 2,~ 2 \leftrightarrow 3, \ldots$).

\begin{figure}[t]
\begin{center}
\includegraphics[scale=0.5]{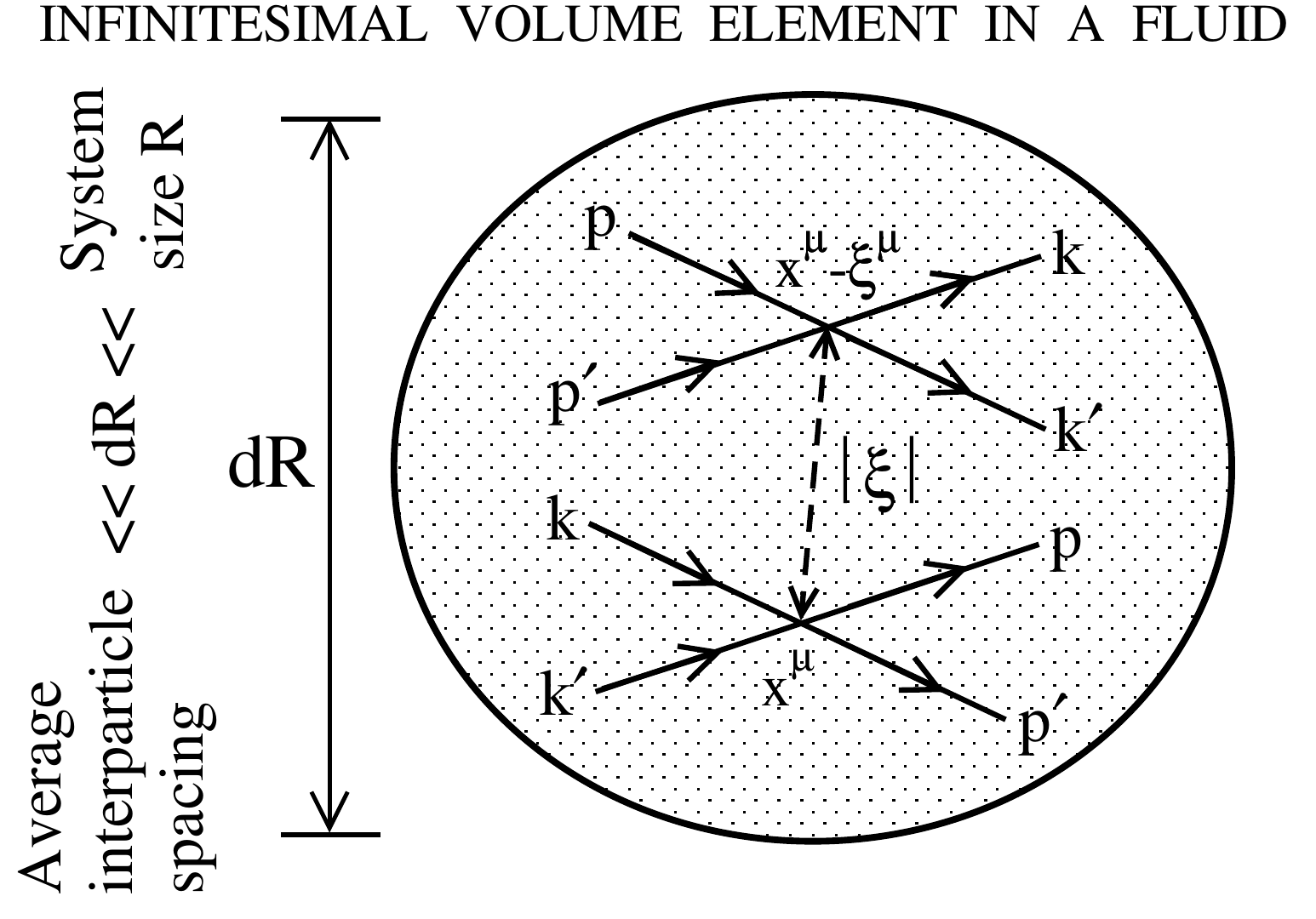}
\end{center}
\vspace{-0.4cm}
\caption[Collisions within an infinitesimal fluid element]{Collisions 
  $kk' \to pp'$ and $pp' \to kk'$ occurring at points $x^\mu$ and 
  $x^\mu-\xi^\mu$ within an infinitesimal fluid element of size $dR$, 
  around $x^\mu$, containing a large number of particles 
  represented by dots.}
\label{flelC8}
\end{figure}

For instance, for $2 \leftrightarrow 2$ 
elastic collisions,
\begin{equation}\label{collC8}
C[f]= \frac{1}{2} \int dp' dk \ dk' \  W_{pp' \to kk'} 
(f_k f_{k'} \tilde f_p \tilde f_{p'} - f_p f_{p'} \tilde f_k \tilde f_{k'}),
\end{equation}
where $ W_{pp' \to kk'}$ is the collisional transition rate, $f_p
\equiv f(x,p)$ and $\tilde f_p \equiv 1-r f(x,p)$ with $r = 1,-1,0$
for Fermi, Bose, and Boltzmann gas, and $dp = g d{\bf p}/[(2 \pi)^3
  \sqrt{{\bf p}^2+m^2}]$, $g$ and $m$ being the degeneracy factor and
particle rest mass. The first and second terms in Eq. (\ref{collC8})
refer to the processes $kk' \to pp'$ and $pp' \to kk'$,
respectively. These processes are traditionally assumed to occur at
the same space-time point $x^\mu$ with an underlying assumption that
$f(x,p)$ is constant not only over the range of interparticle
interaction but also over the entire infinitesimal fluid element of
size $dR$, which is large compared to the average interparticle separation
\cite{Landau}; see Fig. \ref{flelC8}. Equation (1) together with this
crucial assumption has been used to derive the standard second-order
dissipative hydrodynamic equations
\cite{Romatschke:2009im,Israel:1979wp,Denicol:2010xn}. We, however,
emphasize that the space-time points at which the above two kinds
of processes
occur should be separated by an interval $|\xi^\mu| \leq dR$ within
the volume $d^4R$. It may be noted that the large number of particles
within $d^4R$ collide among themselves with various separations
$\xi^\mu$. Further, $\xi^\mu$ is independent of the
arbitrary point $x^\mu$ at which the Boltzmann equation is considered,
and is a function of $(p',k,k')$.
Of course, the points $(x^\mu-\xi^\mu)$ must lie within the past
light-cone of the point $x^\mu$ (i.e., $\xi^2 > 0$ and $\xi^0>0$) to
ensure that the evolution of $f(x,p)$ in Eq. (\ref{MBEC8}) 
does not violate causality.
With this realistic viewpoint, the second term in
Eq. (\ref{collC8}) involves $f(x-\xi,p)f(x-\xi,p')\tilde
f(x-\xi,k)\tilde f(x-\xi,k')$, which on Taylor expansion at $x^\mu$ up
to second order in $\xi^\mu$, results in the modified Boltzmann
equation (\ref{MBEC8}) with
 \begin{eqnarray}\label{coeff1C8}
A^\mu &=& \frac{1}{2} \int dp' dk \ dk' \ \xi^\mu W_{pp' \to kk'}
f_{p'} \tilde f_k \tilde f_{k'}, \nonumber \\
B^{\mu\nu} &=& -\frac{1}{4} \int dp' dk \ dk' \ \xi^\mu \xi^\nu W_{pp' \to kk'}
f_{p'} \tilde f_k \tilde f_{k'}.
\end{eqnarray}

In general, for all collision types ($2 \leftrightarrow 2,~ 2
\leftrightarrow 3, \ldots$), the momentum dependence of the
coefficients $A^\mu$ and $B^{\mu\nu}$ can be made explicit by
expressing them in terms of the available tensors $p^\mu$ and the
metric $g^{\mu\nu} \equiv {\rm diag}(1,-1,-1,-1)$ as $A^\mu = a(x)p^\mu$
and $B^{\mu\nu}= b_1(x) g^{\mu\nu} + b_2(x) p^\mu p^\nu$, in the
spirit of Grad's 14-moment approximation. Equation
(\ref{MBEC8}) forms the basis of our derivation of the second-order
dissipative hydrodynamics.

%%%%%%%%%%%%%%%%%%%%%%%%%%%%%%%%%%%%%%%%%%%%%%%%%%%%%%%%%%%%%%%%%%%%%%%%

\section{Hydrodynamic equations}

As mentioned in Chapter 2, we can express the conserved particle 
current and the energy-momentum tensor as \cite{deGroot}
\begin{equation}\label{NTC8}
N^\mu = \int dp \ p^\mu f, \quad T^{\mu\nu} = \int dp \ p^\mu p^\nu f. 
\end{equation}
The standard tensor decomposition of the above quantities results in
\begin{eqnarray}\label{NTDC8}
N^\mu &=& nu^\mu + n^\mu,  \nonumber\\
T^{\mu\nu} &=& \epsilon u^\mu u^\nu-(P+\Pi)\Delta ^{\mu \nu} 
+ \pi^{\mu\nu},
\end{eqnarray}
where $P, n, \epsilon$ are respectively pressure, number density,
energy density, and $\Delta^{\mu\nu}=g^{\mu\nu}-u^\mu u^\nu$ is the
projection operator on the three-space orthogonal to the hydrodynamic
four-velocity $u^\mu$ defined in the Landau frame: $T^{\mu\nu}
u_\nu=\epsilon u^\mu$. For small departures from equilibrium, $f(x,p)$
can be written as $f = f_0 + \delta f$. The equilibrium distribution
function is defined as $f_0 = [\exp(\beta u\cdot p -\alpha) + r]^{-1}$
where the inverse temperature $\beta=1/T$ and $\alpha=\beta\mu$ ($\mu$
being the chemical potential) are defined by the equilibrium matching
conditions $n\equiv n_0$ and $\epsilon \equiv \epsilon_0$. The scalar 
product is defined as $u.p\equiv u_\mu p^\mu$. The
dissipative quantities, viz., the bulk viscous pressure, the particle
diffusion current and the shear stress tensor are
\begin{eqnarray}\label{DISSC8}
\Pi &=& -\frac{\Delta_{\alpha\beta}}{3}\int dp \ p^
\alpha p^\beta \delta f,  \nonumber\\
n^\mu &=&  \Delta^{\mu\nu} \int dp \ p_\nu \delta f, \nonumber\\
\pi^{\mu\nu} &=& \Delta^{\mu\nu}_{\alpha\beta} \int dp \ p^
\alpha p^\beta \delta f.
\end{eqnarray}
Here
$\Delta^{\mu\nu}_{\alpha\beta} = [\Delta^\mu_\alpha \Delta^\nu_\beta +
  \Delta^\mu_\beta \Delta^\nu_\alpha - (2/3)\Delta^{\mu\nu}
  \Delta_{\alpha\beta}]/2$ is the traceless symmetric projection
operator. Conservation of current, $\partial_\mu N^\mu=0$ and energy-momentum tensor, $\partial_\mu
T^{\mu\nu} =0$, yield the fundamental evolution equations for $n$,
$\epsilon$ and $u^\mu$
\begin{eqnarray}\label{evolC8}
Dn+n\partial_\mu u^\mu + \partial_\mu n^\mu &=& 0, \nonumber \\
D\epsilon + (\epsilon+P+\Pi)\partial_\mu u^\mu - \pi^{\mu\nu}\nabla_{(\mu} u_{\nu)} &=& 0,  \nonumber\\
(\epsilon+P+\Pi)D u^\alpha - \nabla^\alpha (P+\Pi) + \Delta^\alpha_\nu \partial_\mu \pi^{\mu\nu}  &=& 0.
\end{eqnarray}
We use the standard notation $A^{(\alpha}B^{\beta )} = (A^\alpha
B^\beta + A^\beta B^\alpha)/2$, $D=u^\mu\partial_\mu$, and
$\nabla^\alpha = \Delta^{\mu\alpha}\partial_\mu$.
For later use we introduce 
$X^{\langle \mu \rangle}=\Delta^\mu_\nu X^\nu$ and
$X^{\langle \mu \nu \rangle}
=\Delta^{\mu \nu}_{\alpha \beta} X^{\alpha \beta}$.

Conservation of current and energy-momentum 
implies vanishing zeroth and first moments of the
collision term $C_m[f]$ in Eq. (\ref{MBEC8}), i.e., $\int dp \ C_m[f] =
0 = \int dp \ p^\mu C_m[f]$. Moreover, the arbitrariness in $\xi^\mu$
requires that these conditions be satisfied at each order in
$\xi^\mu$. Retaining terms up to second order in derivatives
leads to three constraint equations for the
coefficients ($a, b_1, b_2$), namely $\partial_\mu a = 0$,
\begin{eqnarray}\label{paramC8}
\partial^2\left(b_1 \langle1\rangle_0 \right) 
+ \partial_\mu \partial_\nu\left( b_2 \langle p^\mu p^\nu\rangle_0\right) &=& 0, \nonumber\\
u_\alpha \partial_\mu \partial_\nu \left( b_2 \langle p^\mu p^\nu p^\alpha\rangle_0 \right)  
+ u_\alpha \partial^2 \left(b_1 n u^\alpha \right) &=& 0, 
\end{eqnarray}
where we define $\langle\cdots\rangle_0= \int dp (\cdots) f_0$. It is straightforward to show using
Eq. (\ref{paramC8}) that the validity of the second law of thermodynamics, $\partial_\mu
s^\mu \geq 0$, enforces a further constraint $|a| < 1$,
on the collision term $C_m[f]$.

In order to obtain the evolution equations for the dissipative 
quantities, we follow the approach as described by 
Denicol-Koide-Rischke (DKR) in  Ref. \cite{Denicol:2010xn}. This 
approach employs directly the definitions of the dissipative 
currents in contrast to the IS derivation which uses the second 
moment of the Boltzmann equation. The comoving derivatives of the 
dissipative quantities can be written from their definitions, Eq. 
(\ref{DISSC8}), as
\begin{eqnarray}\label{BE2C8}
\dot\Pi &=& -\frac{\Delta_{\alpha\beta}}{3}\int dp \ 
p^\alpha p^\beta \delta\dot f,  \nonumber\\
\dot n^{\langle\mu\rangle} &=&  \Delta^{\mu\nu} \int dp \ p_\nu \delta\dot f, \nonumber\\
\dot\pi^{\langle\mu\nu\rangle} &=& \Delta^{\mu\nu}_{\alpha\beta} \int dp \ 
p^\alpha p^\beta \delta\dot f,
\end{eqnarray}
where, $\dot X = D X$. Comoving derivative of the nonequilibrium part of 
the distribution function, $\delta\dot f$, can be obtained by writing the 
Boltzmann equation (\ref{MBEC8}) in the form,
\begin{equation}
\delta\dot f = - \dot f_0 - \frac{1}{u.p}p^\mu\nabla_\mu f + \frac{1}{u.p}C_m[f].
\end{equation}

We note that the collision term in the Boltzmann equation, Eq. (\ref 
{collC8}), is not written in the relaxation-time approximation, thus 
making it difficult to solve for the nonequilibrium distribution 
function using Chapman-Enskog like expansion. Therefore to proceed 
further, we take recourse to Grad's 14-moment approximation \cite
{Grad} for the single-particle distribution in orthogonal basis \cite
{Denicol:2010xn}, Eq. (\ref{G14C3}),
\begin{eqnarray}\label{G14C8}
f = f_0 + f_0 \tilde f_0 \left( \lambda_\Pi \Pi + \lambda_n n_\alpha p^\alpha 
+ \lambda_\pi \pi_{\alpha\beta} p^\alpha p^\beta \right).
\end{eqnarray}
The coefficients ($\lambda_\Pi, \lambda_n, \lambda_\pi$) are functions
of ($n,\epsilon,\beta,\alpha$). Using Eqs. (\ref{BE2C8})-(\ref{G14C8}) and 
introducing first-order shear tensor
$\sigma_{\mu\nu}=\nabla_{\langle \mu}u_{\nu \rangle}$, vorticity
$\omega_{\mu \nu}=(\nabla_\mu u_\nu-\nabla_\nu u_\mu)/2$ and expansion
scalar $\theta=\partial \cdot u$, we finally obtain the following
evolution equations for the dissipative fluxes defined in
Eq. (\ref{DISSC8}):
\begin{align}
\dot\Pi = & -\frac{\Pi}{\tau_\Pi'}
- \beta_\Pi' \theta
- \tau_{\Pi n}' n \cdot \dot u 
- l_{\Pi n}' \partial \cdot n
- \delta_{\Pi\Pi}' \Pi\theta 
- \lambda_{\Pi n}' n \cdot \nabla \alpha
+ \lambda_{\Pi\pi}' \pi_{\mu\nu} \sigma^{\mu\nu} \nonumber\\
&+ \Lambda_{\Pi\dot u} \dot u \cdot \dot u
+ \Lambda_{\Pi\omega} \omega_{\mu\nu} \omega^{\nu\mu} 
+ (8 \ {\rm terms}) , \label{bulkC8}\\
\dot n^{\langle\mu\rangle} = & -\frac{n^\mu}{\tau_n'}
+ \beta_n' \nabla^\mu\alpha
- \lambda_{n\omega}' n_\nu \omega^{\nu\mu}
- \delta_{nn}' n^\mu \theta 
- l_{n \Pi}'\nabla^\mu \Pi
+ l_{n \pi}'\Delta^{\mu\nu} \partial_\gamma \pi^\gamma_\nu
+ \tau_{n \Pi}' \Pi \dot u^\mu \nonumber \\
&- \tau_{n \pi}'\pi^{\mu \nu} \dot u_\nu
-\lambda_{n\pi}'n_\nu \pi^{\mu \nu}
+ \lambda_{n \Pi}'\Pi n^\mu 
+  \Lambda_{n \dot u} \omega^{\mu \nu} \dot u_\nu
+ \Lambda_{n \omega} \Delta^\mu_\nu \partial_\gamma \omega^{\gamma \nu}
+ (9 \ {\rm terms}), \label{heatC8}\\
\dot\pi^{\langle\mu\nu\rangle} = & -\frac{\pi^{\mu\nu}}{\tau_\pi'}
+ 2 \beta_\pi' \sigma^{\mu\nu} \!
- \tau_{\pi n}' n^{\langle\mu}\dot u^{\nu\rangle} \!
+ l_{\pi n}' \nabla^{\langle \mu}n^{\nu\rangle} \!
+ 2\lambda_{\pi\pi}' \pi_\rho^{\langle \mu} \omega ^{\nu\rangle \rho} \!
+ \lambda_{\pi n}' n^{\langle\mu} \nabla^{\nu\rangle} \alpha
- \tau_{\pi\pi}' \pi_\rho^{\langle\mu} \sigma^{\nu\rangle\rho} \nonumber\\
&- \delta_{\pi\pi}' \pi^{\mu\nu}\theta
+ \Lambda_{\pi\dot u} \dot u^{\langle \mu} \dot u^{\nu\rangle}
+ \Lambda_{\pi\omega} \omega_\rho^{\langle \mu} \omega^{\nu\rangle\rho} 
+ \chi_1 \dot b_2 \pi^{\mu\nu}
+ \chi_2 \dot u^{\langle \mu} \nabla^{\nu\rangle} b_2
+ \chi_3 \nabla^{\langle \mu} \nabla^{\nu\rangle} b_2. \label{shearC8}
\end{align}
The ``8 terms" (``9 terms'') involve second-order, linear
scalar (vector) combinations of derivatives of $b_1,b_2$.
All the terms in the above equations are inequivalent,
i.e., none can be expressed as a combination of others via equations
of motion \cite{Bhattacharyya:2012ex}. All the coefficients 
in Eqs. (\ref{bulkC8})-(\ref{shearC8})
are obtained as functions of hydrodynamic variables. For example,
some of the transport coefficients related to shear are
\begin{align}\label{coeffC8}
\tau_\pi' = \; & \beta_{\dot\pi}\tau_\pi, \quad
\beta_\pi' = \tilde a \beta_\pi/\beta_{\dot\pi},
\quad
\beta_{\dot\pi} = \tilde a + \frac{b_2}{3\eta\tilde a}
\left[\langle(u.p)^3\rangle_0-m^2n\right], 
\nonumber\\  
\beta_\pi = \; & \frac{4}{5}P + \frac{1}{15}(\epsilon-3P) - \frac{m^4}{15}\left<(u.p)^{-2}\right>_0,
\end{align}
where ${\tilde a}=(1-a)$. The rest of the coefficients can be 
readily calculated by performing the integrations in Eq. (\ref
{BE2C8}), in a way similar to that demonstrated in Appendix B.

Retaining only the first-order terms in
Eqs. (\ref{bulkC8})-(\ref{shearC8}), and using DKR values of bulk 
viscosity $\zeta$, particle diffusion $\kappa$, and shear
viscosity $\eta$, we get the modified first-order
equations for bulk pressure $\Pi = -\tau_{\Pi}' \beta_{\Pi}' \theta 
= - \tilde a \zeta \theta$, heat current $n^\mu = 
\beta_n' \tau_n' \nabla^\mu \alpha $ and shear stress tensor
$\pi^{\mu\nu} = 2 \tau_\pi'
\beta_\pi' \sigma^{\mu\nu} = 2 \tilde a \tau_\pi
\beta_\pi \sigma^{\mu\nu} = 2 \eta {\tilde a} \sigma^{\mu\nu}$. 
Thus the nonlocal collision term modifies even the first-order dissipative
equations. This constitutes one of the main results in the present
study. 

If $a,\ b_1$ and $b_2$ are all set to zero, Eqs. (\ref{bulkC8})-(\ref
{shearC8}) reduce to those obtained by DKR \cite{Denicol:2010xn} 
with the same coefficients. Otherwise coefficients of all the terms 
occurring in the DKR equations get modified. Furthermore, our 
derivation results in new terms, for instance those with 
coefficients $\Lambda_{k \dot u}$, $\Lambda_{k \omega}$, 
($k=\Pi,n,\pi$), which are absent in \cite{Denicol:2010xn} as well as 
in the standard Israel-Stewart approach \cite {Israel:1979wp}. Hence 
these terms have also been missed so far in the numerical studies of 
heavy-ion collisions in the hydrodynamic framework \cite 
{Romatschke:2007mq,Song:2007ux,Luzum:2010ag}. Indeed Eqs. (\ref 
{bulkC8})-(\ref{shearC8}) contain all possible second-order terms 
allowed by symmetry considerations \cite {Bhattacharyya:2012ex}. 
This is a consequence of the nonlocality of the collision term 
$C_m[f]$. However, we note that a generalization of the 14-moment 
approximation is also able to generate all these terms as shown 
recently in Ref. \cite{Denicol:2012cn}.

\begin{figure}[t]
\begin{center}
\includegraphics[scale=0.45]{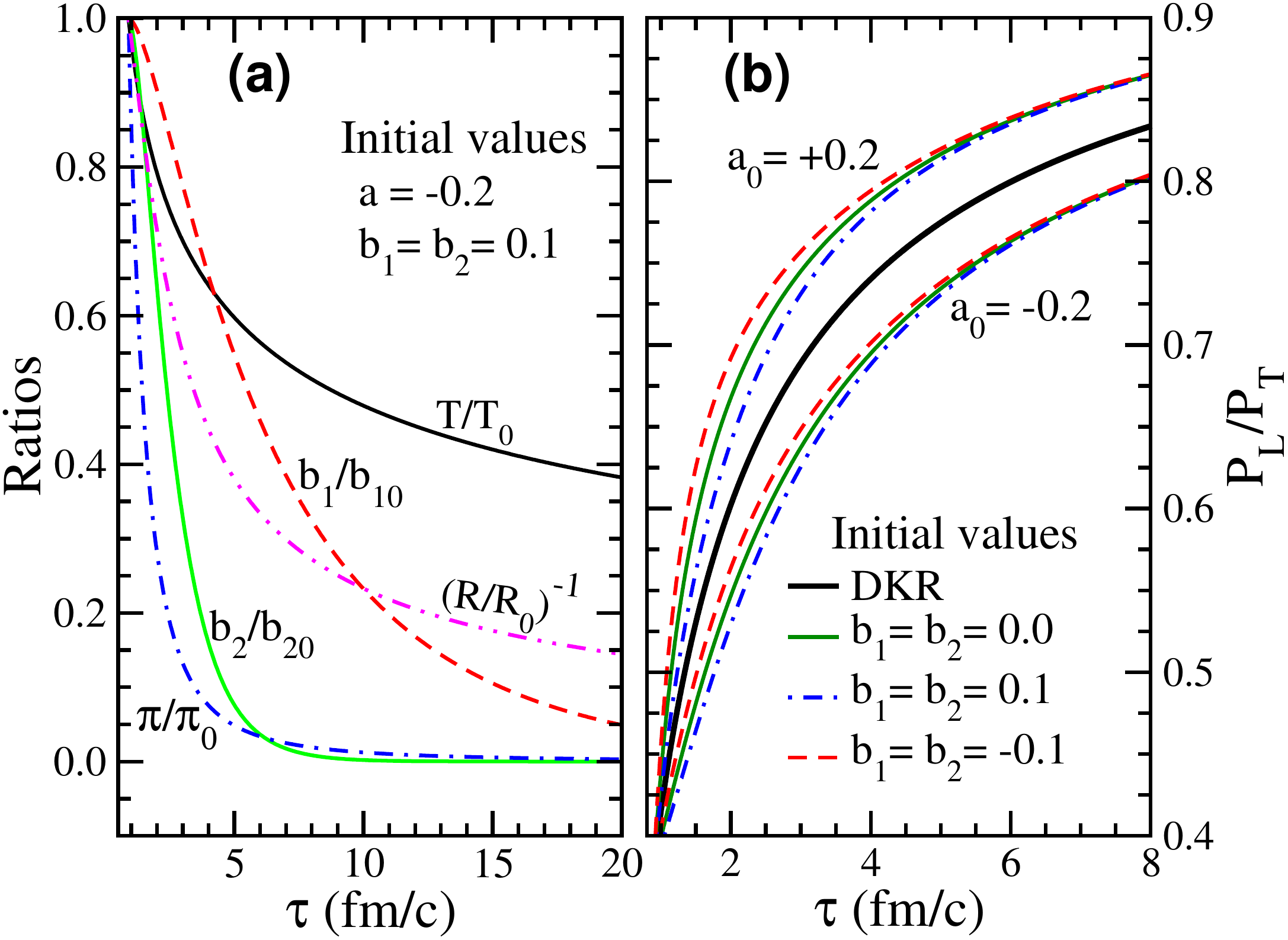}
\end{center}
\vspace{-0.4cm}
\caption[Time evolution of various hydrodynamic quantities]{Time 
  evolution of (a) temperature, shear pressure, inverse Reynolds 
  number and parameters ($b_1, \ b_2$) normalized to their initial 
  values, and (b) anisotropy parameter $P_L/P_T$. Initial values are 
  $\tau_0= 0.9$ fm/c, $T_0=360$ MeV, $\eta/s = 0.16$, $\pi_0 = 
  4\eta/(3\tau_0)$. Units of $b_2$ are GeV$^{-2}$. The curve 
  labelled DKR is obtained by setting $a=b_1=b_2=0$ in Eqs. (\ref 
  {BjpiC8}) and (\ref{coeffuC8}).}
\label{evol1C8}
\end{figure}

%%%%%%%%%%%%%%%%%%%%%%%%%%%%%%%%%%%%%%%%%%%%%%%%%%%%%%%%%%%%%%%%%%%%%%%%

\section{Numerical results}

To demonstrate the numerical significance of the new dissipative
equations derived here, we consider evolution of a massless Boltzmann
gas, with equation of state $\epsilon=3P$, at vanishing net baryon
number density in the Bjorken model \cite{Bjorken:1982qr}. The new
terms, namely $\dot u \cdot \dot u$, $\omega_{\mu\nu}
\omega^{\nu\mu}$, $\omega^{\mu \nu} \dot u_\nu$, $\Delta^\mu_\nu
\partial_\gamma \omega^{\gamma \nu}$, $\dot u^{\langle \mu} \dot
u^{\nu\rangle}$ and $\omega_\rho^{\langle \mu}
\omega^{\nu\rangle\rho}$ containing acceleration and vorticity do not
contribute in this case. However, they are expected to play an
important role in the full 3D viscous hydrodynamics.

In terms of the coordinates ($\tau,x,y,\eta_s$) where $\tau =
\sqrt{t^2-z^2}$ and $\eta_s=\tanh^{-1}(z/t)$, the initial four-velocity
becomes $u^\mu=(1,0,0,0)$. In this scenario $\Pi=0=n^\mu$ and the
equation for $\Phi\equiv -\tau^2\pi^{\eta_s\eta_s}$ reduces to
(see Appendix A for details)
\begin{eqnarray}\label{BjpiC8}
\frac{\Phi}{\tau_\pi} + \beta_{\dot\pi}\frac{d\Phi}{d\tau} = 
\beta_\pi \frac{4}{3\tau} - \lambda \frac{\Phi}{\tau} 
- \psi \Phi \frac{db_2}{d\tau},
\end{eqnarray}
where the coefficients are
\begin{equation}\label{coeffuC8}
\beta_{\dot\pi} = \tilde a + \frac{b_2 (\epsilon + P)}{\tilde a \beta\eta},
\quad \beta_\pi = \frac{4}{5} \tilde a P , 
\quad \psi = \frac{9 (\epsilon + P)}{5 \tilde a \beta \eta},
\quad \lambda = \frac{38}{21}\tilde a 
- \!\left(\! \frac{b_1\beta}{5} - \frac{8b_2}{7\beta} \!\right)\!\frac{\epsilon + P}{\tilde a \eta}.
\end{equation}
For comparison we quote the IS results \cite{Israel:1979wp}: 
$\beta_\pi=2P/3,~ \lambda=2$. The coupled differential equations 
(\ref{evolC8}), (\ref{paramC8}) and (\ref{BjpiC8}) are solved 
simultaneously for a variety of initial conditions: temperature 
$T=360$ or 500 MeV corresponding to typical RHIC and LHC energies, 
and shear pressure $\Phi=0$ or $\Phi=\Phi_{\rm NS}=4\eta/(3\tau_0)$ 
corresponding to isotropic and anisotropic pressure configurations. 
Since the nonlocal effects embodied in the Taylor expansion (\ref
{MBEC8}) are not large, the initial $a,~b_1,~b_2$ are so constrained 
that the corrections to first-order and second-order terms remain 
small; recall also the additional constraints $|a| < 1$ and Eq. (\ref
{paramC8}).

\begin{figure}[t]
\begin{center}
\includegraphics[scale=0.45]{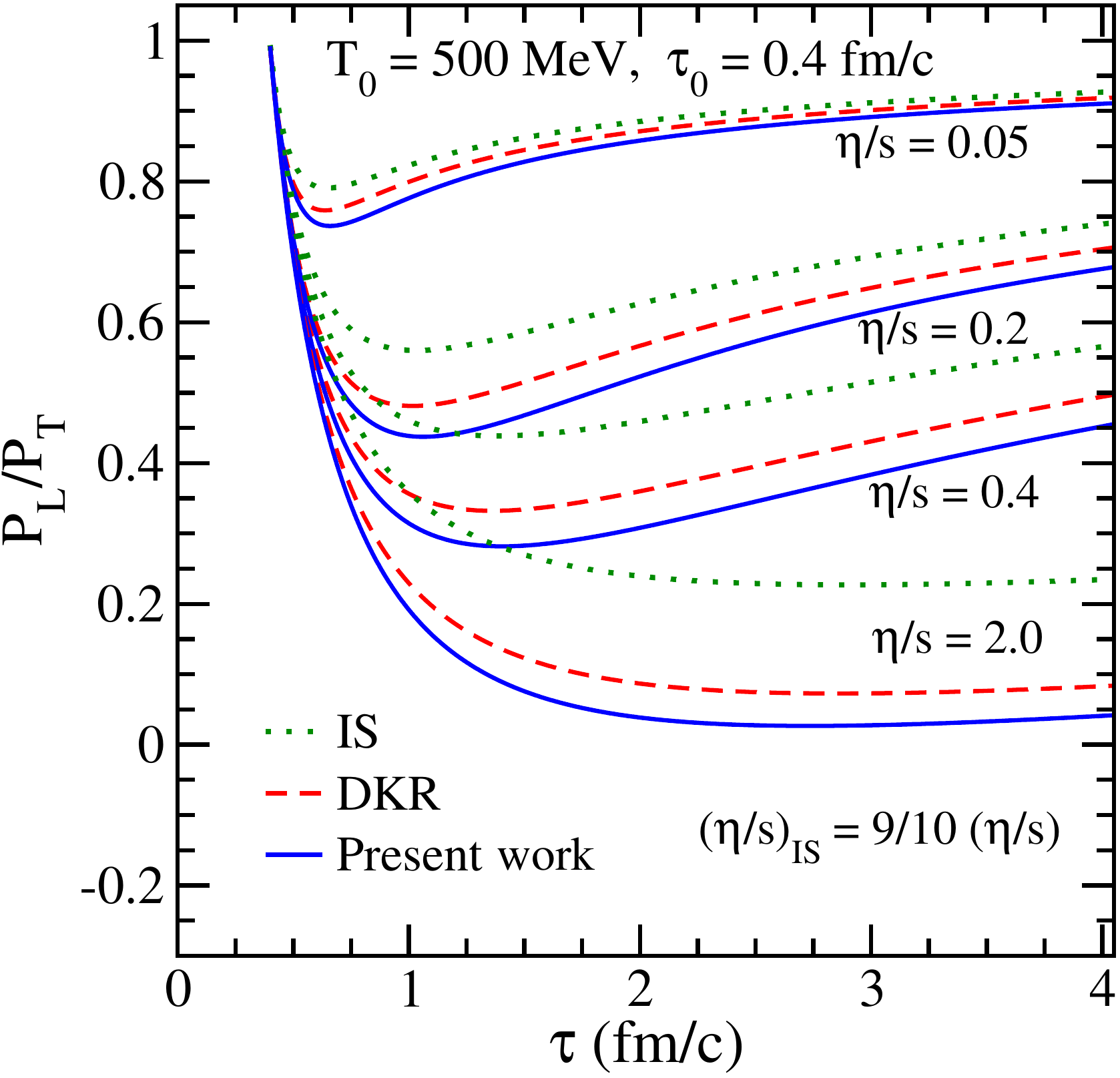}
\end{center}
\vspace{-0.4cm}
\caption[Comparison of pressure anisotropy in various cases]{Time 
  evolution of $P_L/P_T$ in IS \cite{Israel:1979wp}, DKR 
  ($a=b_1=b_2=0$), and the present work, for isotropic initial 
  pressure configuration $(\Phi_0=0)$. The scaling $(\eta/s)_{\rm 
  IS}=9/10(\eta/s)$ ensures that all the results are compared at the 
  same cross section \cite {Denicol:2010xn}.}
\label{evol2C8}
\end{figure}

Figure \ref{evol1C8} (a) illustrates the evolution of these 
quantities for a choice of initial conditions. $T$ decreases 
monotonically to the crossover temperature $~170$ MeV at time $\tau 
\simeq 10$ fm/c, which is consistent with the expected lifetime of 
quark-gluon plasma. Parameter $a$ is constant whereas $b_1$ and $b_2$
vary smoothly and tend to zero at large times indicating reduced 
but still significant presence of nonlocal effects in the collision 
term at late times. This is also evident in Fig. \ref{evol1C8} (b) 
where the pressure anisotropy $P_L/P_T=(P-\Phi)/(P+\Phi/2)$ shows 
marked deviation from IS, controlled mainly by $a$. At late times 
$P_L/P_T$ is largely unaffected by the choice of initial values of 
$b_1,~b_2$. Although the shear pressure $\Phi$ vanishes rapidly 
indicating approach to ideal fluid dynamics, the $P_L/P_T$ is far 
from unity. Faster isotropization for initial $a>0$ may be 
attributed to a smaller effective shear viscosity $(1-a) \eta$ in 
the modified NS equation, and conversely. Figure \ref{evol1C8} (b) 
also indicates the convergence of the Taylor expansion that led to 
Eq. (\ref{MBEC8}).

Figure \ref{evol2C8} shows the evolution of $P_L/P_T$ for isotropic initial
pressure configuration, at various $\eta/s$ for the LHC energy regime.
Compared to IS, DKR leads to larger pressure anisotropy. Further, with
small initial corrections ($ 10$\% to first-order and $\simeq 20$\% to
the second-order terms) due to $a,\ b_1,\ b_2$, the nonlocal
hydrodynamics (solid lines) exhibits appreciable deviation from the
(local) DKR theory. The above results clearly demonstrate the
importance of the nonlocal effects, which should be incorporated in
transport calculations as well. Comparison of nonlocal hydrodynamics
to nonlocal transport would be illuminating.

In a realistic 2+1 or 3+1 D calculation, one has to choose the
thermalization time and the freeze-out temperature together with
suitable initial conditions for hydrodynamic velocity, energy density,
shear pressure as well as for the nonlocal coefficients $a, ~b_1,~
b_2$ to fit $dN/d\eta$ and $p_T$ spectra of hadrons, and then predict,
for example, the anisotropic flow $v_n$ for a given $\eta/s$. Nonlocal
effects (especially via $a$) will affect the extraction of $\eta/s$
from fits to the measured $v_n$. It may also be noted that although
(local) viscous hydrodynamics explains the gross features of $\pi^-$
and $K^-$ spectra for the (0-5)\% most central Pb-Pb collisions at
$\sqrt{s_{NN}}=2.76$ TeV, it strongly disagrees with the measured
$\bar p$ spectrum \cite{Floris:2011ru}. Further the constituent quark
number scaling violation has been observed in the $v_2$ and $v_3$ data
for $\bar p$, at this LHC energy \cite{Krzewicki:2011ee}. The above
discrepancies may be attributed partly to the nonlocal effects which
can have different implications for two- and three-particle
correlations and thus affect the meson and baryon spectra differently.

%%%%%%%%%%%%%%%%%%%%%%%%%%%%%%%%%%%%%%%%%%%%%%%%%%%%%%%%%%%%%%%%%%%%%%%%

\section{Summary and conclusions}

To summarize, we have presented a new derivation of the relativistic 
dissipative hydrodynamic equations by introducing a nonlocal 
generalization of the collision term in the Boltzmann equation. The 
first-order and second-order equations are modified: new terms occur 
and coefficients of others are altered. While it is well known that 
the derivation based on the generalized second law of thermodynamics 
misses some terms in the second-order equations, we have shown that 
the standard derivation based on kinetic theory and 14-moment 
approximation also misses other terms. The method presented in this 
chapter is able to generate all possible terms to a given order that 
are allowed by symmetry. It can also be extended to derive higher 
order fluid dynamic equations. Within one-dimensional scaling 
expansion, we find that nonlocality of the collision term has a 
rather strong influence on the evolution of the viscous medium via 
hydrodynamic equations.

%% file: Chapter9.tex
\chapter{Summary and future outlook}

This thesis presents our work on the theoretical formulation of 
relativistic dissipative fluid dynamics from various approaches 
within the framework of relativistic kinetic theory. Several 
longstanding problems in the formulation as well as in the 
application of relativistic hydrodynamics relevant to heavy-ion 
collisions have been addressed here. The evolution equations for the 
dissipative quantities along with the second-order transport 
coefficients have been derived using the second law of 
thermodynamics within a single theoretical framework. In particular, 
the problem pertaining to the relaxation time for the evolution of 
bulk viscous pressure has been solved here. Subsequently, using the 
same method for two different forms of non-equilibrium 
single-particle distribution functions, viscous evolution equations 
have been derived and applied to study the particle production and 
transverse momentum spectra of hadrons and thermal dileptons. 

An alternate formulation of second-order dissipative hydrodynamics 
has been presented in which iterative solution of the Boltzmann 
equation for non-equilibrium distribution function is employed 
instead of the 14-moment ansatz most commonly used in the 
literature. The evolution equations for the dissipative quantities 
have been obtained directly from their definitions rather than an 
arbitrary moment of Boltzmann equation in the traditional 
Israel-Stewart formulation. Using the iterative solution of 
Boltzmann equation, the form of second-order viscous correction to 
the distribution function has been derived. The effects of these 
corrections on particle spectra and HBT radii are compared to those 
due to 14-moment ansatz. This method has been further extended to 
obtain third-order evolution equation for shear stress tensor. 

Finally, the collision term in the Boltzmann equation corresponding 
to $2\to2$ elastic collisions has been modified to include the 
gradients of the distribution function. This non-local collision 
term has then been used to derive second-order evolution equations 
for the dissipative quantities. The numerical significance of these 
new formulations has been demonstrated within the framework of 
one-dimensional boost-invariant Bjorken expansion of the matter 
formed in relativistic heavy-ion collisions.

Further development of the theory of relativistic dissipative fluid 
dynamics requires to consider the following aspects in the future:
\begin{itemize}
\item {\bf More robust relaxation-time approximation}: The 
relaxation-time approximation for the collision term significantly 
reduces the complexity of solving the Boltzmann equation iteratively 
in contrast to the situation in which the collision term captures 
the microscopic interaction between the constituent particles. The 
relaxation time $\tau_R$ may be either assumed to be independent of 
momenta or parametrized as a simple power law to reflect the 
momentum dependence \cite{Dusling:2009df}. However this 
parametrization, commonly known as the {\it quadratic ansatz}, 
violates the fundamental current and energy-momentum conservation as 
well as the matching conditions \cite{Jaiswal:2013vta}. To remove 
these inconsistencies, the momentum dependence of the relaxation 
time can be incorporated in a parametric form such that the 
conservation equations as well as the matching conditions are not 
violated. Further, causal dissipative relativistic fluid dynamic 
equations can be derived by taking into account the momentum 
dependence of the relaxation time via the nonequilibrium 
distribution function. The momentum dependence of the relaxation 
time may have important bearing on the extraction of $\eta/s$ of the 
QGP from the hydrodynamic analysis of the flow harmonics $v_n(p_T)$.
\item {\bf Fluid dynamics in presence of external forces}: The 
derivation of relativistic dissipative fluid dynamics from kinetic 
theory proceeds by assuming Boltzmann equation in the absence of 
external force fields. However, experiments suggests that strong 
electromagnetic fields are produced in relativistic heavy-ion 
collisions which may have important implications on the 
phenomenology of QGP \cite{Kharzeev:2007jp}. Moreover, the mean 
field effects due to strong interactions are non negligible and may 
affect the dynamics of evolution. Hence it is important to formulate 
fluid dynamic equations in the presence of external forces by 
considering the field term in the Boltzmann equation. For example, 
the derivation of hydrodynamic equation by employing the 
electromagnetic Lorentz force as the field term in the Boltzmann 
equation may change the effective $\eta/s$. The effects due to 
strong interactions can also be incorporated as a mean field term in 
the Boltzmann equation and then derive relativistic fluid dynamic 
equations. 
\item {\bf Complete third-order dissipative fluid dynamics}: In 
order to improve the second-order Israel-Stewart hydrodynamics 
beyond its present scope, third-order evolution equations for the 
shear stress tensor was derived \cite{El:2009vj,Jaiswal:2013vta, 
Jaiswal:2014raa}. However, these formulations are incomplete in the 
sense that they do not take into account the other dissipative 
effects such as bulk viscosity and charge diffusion current and are 
restricted to massless Boltzmann particles. A complete theory of 
relativistic third-order dissipative hydrodynamics from Boltzmann 
equation can be formulated. This involves extending the work done in 
Ref. \cite{Jaiswal:2013vta,Jaiswal:2014raa} to a more general 
system, i.e. for Fermi, Bose and Boltzmann particles, and deriving 
evolution equations for shear stress tensor as well as bulk viscous 
pressure and charge diffusion current.
\item {\bf General relativistic dissipative fluid dynamics}: 
Although the Boltzmann equation can be written for a general 
background spacetime, in the derivation of second-order relativistic 
dissipative fluid dynamic equations from kinetic theory the metric 
is assumed to be Minkowskian \cite{Israel:1979wp}. The resultant 
dissipative equations are therefore valid only in flat spacetime and 
are not applicable to cosmology and astrophysics. It is thus of 
interest to formulate hydrodynamics from kinetic theory in a metric 
independent manner and extend the validity of dissipative equations 
to a general background spacetime. Without assuming a specific form 
of the metric tensor, the nonequilibrium distribution function 
obtained after iteratively solving the general relativistic 
Boltzmann equation in relaxation time approximation can be used to 
formulate a general relativistic theory of second-order dissipative 
hydrodynamics. The resultant second-order coefficients can then be 
compared with those obtained from the strongly coupled ${\cal N}=4$ 
supersymmetric Yang-Mills theory \cite{Baier:2007ix}.
\end{itemize}

In conclusion, the theoretical formulation of relativistic 
dissipative fluid dynamics is experiencing a rapid development, with 
contributions from several different groups. Some major progress 
that we have made in the formulation of relativistic dissipative 
fluid dynamics, within the framework of kinetic theory, has been 
covered in this thesis. The work developed in this thesis has 
several applications to heavy-ion collisions, with implications on 
the evolution of the strongly interacting fluid-like matter created 
at RHIC and LHC. However, as outlined above, there still remains 
numerous interesting aspects in the formulation of relativistic 
dissipative fluid dynamics that need further investigation.

%% file: Appendex.tex
%########################################################################
\chapter{Coordinates and Transformations}
%########################################################################

The three spatial coordinates and  time  form a four dimensional
coordinate system $x^\mu=(t,x,y,z)$, called \emph{Cartesian
coordinates}. Throughout this thesis, we use the metric tensor
$g_{\mu\nu}=g^{\mu\nu}= \mathrm{diag} (1, -1, -1, -1)$, such that
four vectors ($x^\mu$, for example)  transform as follows:
\begin{eqnarray}
x^\mu=(t,x,y,z), \qquad  x_\mu=g_{\mu\nu}x^\nu =(t,-x,-y,-z).
\end{eqnarray}

One generally sets the $z$-axis parallel to the beam direction, and
correspondingly calls the $(x,y)$ plane  the transverse plan (with x
pointing in the direction of the impact parameter). Within the
forward light-cone $|z|<t$, $\eta-\tau$ coordinates $x^\mu=(\tau, x,
y, \eta)$ (with $\tau = \sqrt{ t^2-z^2}$ and $\eta=\frac{1}{2} \ln
\frac{t+z}{t-z}$) prove more useful in high energy particle and
nuclear physics. The metric in this coordinate system reads $g^{\mu\nu}=
\mathrm{diag} (1, -1, -1, -1/\tau^2)$, $g_{\mu\nu}= \mathrm{diag} (1,
-1, -1, -\tau^2)$.

Here we list the transformation between Cartesian and $\eta-\tau$
coordinates:

%
%=========================== Fig. 2 EXEP ===================================
\begin{figure}[h]
  \begin{center}
    \begin{minipage}[b]{0.55\linewidth}
    \includegraphics[width=0.8\linewidth,height=0.5\linewidth,clip=]{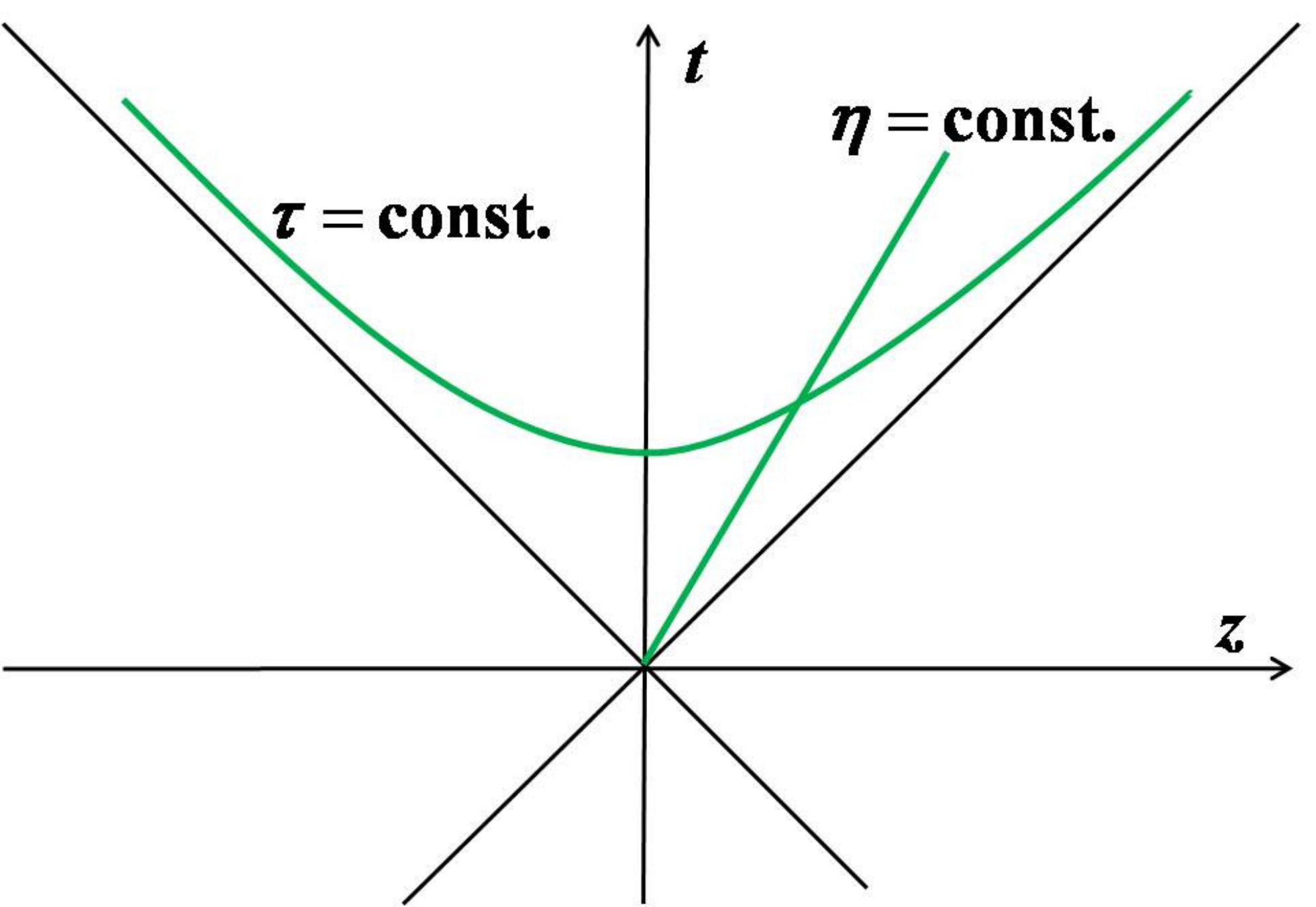}
    \end{minipage}
    \begin{minipage}[b]{0.4\linewidth}
\begin{eqnarray*}
\!\!\!\!\!x^{\mu}=(t,x,y,z) &&  \!\!\!\!\!x^\mu=(\tau, x, y, \eta)         \nonumber \\
\!\!\!\!\!t=\tau\cosh\eta   \ \ \ &\qquad&   \!\!\!\!\!\tau=\sqrt{t^2-z^2} \nonumber \\
\!\!\!\!\!z=\tau\sinh\eta   \ \ \ &\qquad&   \!\!\!\!\!\eta=\arctan (z/t)
\end{eqnarray*}
 \vspace*{0.7cm}
    \end{minipage}
  \end{center}
  \vspace*{+0.3cm}
\end{figure}
%===========================================================================
%

\section{Bjorken Flow}

Bjorken's notion of ``boost invariance" is the statement that at 
longitudinal distance $z$ away from the point of collision and time 
$t$ after the collision, the matter should be moving with a velocity 
$v^z=z/t$. We neglect transverse dynamics $(v^x=v^y=0)$ and 
introduce Milne coordinates proper time $\tau=\sqrt{t^2-z^2}$ and 
spacetime rapidity $\eta=\mathrm{tanh}^{-1}(z/t)$. Hence, 
$t=\tau~\mathrm{cosh}~\eta$ and $z=\tau~\mathrm{sinh}~\eta$. Boost 
invariance for hydrodynamics simply translates into
\begin{equation}
u^t=\frac{t}{\tau} \quad ; \quad u^z=\frac{z}{\tau} \quad ; \quad u^{\eta}=-u^{t}\frac{\mathrm{sinh}~\eta}{\tau}+u^z\frac{\mathrm{cosh}~\eta}{\tau}=0
\label{BI}
\end{equation}
Hence, $u^{\mu}=(1,0,0,0)$ and as a consequence $\epsilon$, $P$, 
$u^{\mu}$ and $\pi^{\mu\nu}$ are all independent of $\eta$ and 
therefore remains unchanged when performing a Lorentz-boost. Even 
though in this highly simplified model the hydrodynamic degrees of 
freedom now only depend on proper time $\tau$, the system dynamics 
is not entirely trivial. The reason for this is that in Milne 
coordinates, the metric is given by $g_{\mu\nu}=$diag$(1, -1, -1, 
-\tau^2)$ and hence is no longer co-ordinate invariant. The 
Christoffel symbols are non-zero and are given by
\begin{equation}
\Gamma^{\eta}_{\eta\tau}=\Gamma^{\eta}_{\tau\eta}=\frac{1}{\tau} \quad ; \quad \Gamma^{\tau}_{\eta\eta}=\tau
\label{CS}
\end{equation}
Since the metric is non-trivial, we need to replace our normal 
derivatives with covariant derivatives. The rule for covariant 
differentiation of a scalar is trivial. If covariant differentiation 
is denoted by $d_{\mu}$ and $A$ is some scalar, then we have
\begin{equation}
d_{\mu}A=\partial_{\mu}A.
\label{CDS}
\end{equation} 
The rule for covariant differentiation of a vector $A^{\alpha}$ and 
$A_{\alpha}$ is
\begin{equation}
d_{\mu}A^{\alpha}=\partial_{\mu}A^{\alpha}+\Gamma^{\alpha}_{\mu\lambda}A^{\lambda} \quad ; \quad 
d_{\mu}A_{\alpha}=\partial_{\mu}A_{\alpha}-\Gamma^{\lambda}_{\mu\alpha}A_{\lambda}.
\label{CDV}
\end{equation} 
For $\theta=\partial_{\mu}u^{\mu}$,
\begin{equation}\label{Bjtheta}
\theta =\nabla_{\mu}u^{\mu}=\Delta^{\rho}_{\mu}d_{\rho}u^{\mu} 
=(g^{\rho}_{\mu}-u^{\rho}u_{\mu})(\partial_{\rho}u^{\mu}+\Gamma^{\mu}_{\rho\lambda}u^{\lambda})
=\Gamma^{\mu}_{\mu\lambda}u^{\lambda}=\Gamma^{\eta}_{\eta\tau}u^{\tau}=\frac{1}{\tau}.
\end{equation}
Lets assume that $\pi^{\mu\nu}$ is diagonal with $\pi^{\tau\tau}=0$ 
and introduce $\Phi=-\tau^2\pi^{\eta\eta}$. For $\pi^{\mu\nu}$ to be 
traceless,
\begin{equation}\label{Bjpietaeta}
\pi^{\eta\eta}=-\Phi/\tau^2~;\quad \pi^{xx}=\pi^{yy}=\Phi/2.
\end{equation}
With these assumption, $\sigma^{\mu\nu}$ becomes
\begin{align}\label{Bjsigma}
\sigma^{\mu\nu}&=\Delta^{\mu\nu}_{\alpha\beta}\nabla^{\alpha}u^{\beta} = \Delta^{\mu\nu}_{\alpha\beta}\Delta^{\alpha\rho}d_{\rho}u^{\beta} 
=\Delta^{\mu\nu}_{\alpha\beta}\Delta^{\alpha\rho}(\partial_{\rho}u^{\beta}+\Gamma^{\beta}_{\rho\gamma}u^{\gamma}) \nonumber\\
&=\Delta^{\mu\nu}_{\eta\eta}g^{\eta\eta}\Gamma^{\eta}_{\eta\tau} \quad\quad [\because \partial_{\rho}u^{\beta} =0 ~~\mathrm{and~only~u^{\tau}}\ne 0] \nonumber\\
&=\frac{1}{\tau}[g^{\mu}_{\eta}g^{\nu\eta}-\frac{1}{3}(g^{\mu\nu}-u^{\mu}u^{\nu})].
\end{align}
Therefore
\begin{equation}\label{Bjsigmaetaeta}
\sigma^{\eta\eta}=\frac{1}{\tau}[g^{\eta}_{\eta}g^{\eta\eta}-\frac{1}{3}(g^{\eta\eta}-u^{\eta}u^{\eta})]
=\frac{1}{\tau}[g^{\eta\eta}-\frac{1}{3}g^{\eta\eta}]=\frac{2}{3\tau}g^{\eta\eta}=-\frac{2}{3\tau^3}.
\end{equation}
Next we calculate the form of the second-order terms in Bjorken case,
\begin{align}\label{Bjpidot}
\dot\pi^{<\eta\eta>}&=\Delta^{\eta\eta}_{\alpha\beta}u^{\mu}d_{\mu}\pi^{\alpha\beta}=[g^{\eta}_{\alpha}g^{\eta}_{\beta}
-\frac{1}{3}g^{\eta\eta}(g_{\alpha\beta}-u_{\alpha}u_{\beta})]u^{\mu}d_{\mu}\pi^{\alpha\beta}=g^{\eta}_{\alpha}g^{\eta}_{\beta}u^{\mu}d_{\mu}\pi^{\alpha\beta} \nonumber\\
&=g^{\eta}_{\alpha}g^{\eta}_{\beta}u^{\mu}(\partial_{\mu}\pi^{\alpha\beta}+2\Gamma^{\alpha}_{\lambda\mu}\pi^{\lambda\beta})
=\partial_\tau \pi^{\eta\eta}+2\frac{\pi^{\eta\eta}}{\tau}=-\frac{1}{\tau^2}\frac{\partial}{\partial\tau}(-\tau^2\pi^{\eta\eta})
=-\frac{1}{\tau^2}\frac{\partial\Phi}{\partial\tau},
\end{align}
and
\begin{align}\label{Bjpisigma}
\pi^{<\eta}_{\gamma}\sigma^{\eta>\gamma}&=\Delta^{\eta\eta}_{\alpha\beta}\pi^{\alpha}_{\gamma}\sigma^{\beta\gamma} 
=[g^{\eta}_{\alpha}g^{\eta}_{\beta}-\frac{1}{3}g^{\eta\eta}(g_{\alpha\beta}-u_{\alpha}u_{\beta})]\pi^{\alpha}_{\gamma}\sigma^{\beta\gamma} \nonumber\\
&=\pi^{\eta}_{\eta}\sigma^{\eta\eta}-\frac{1}{3}g^{\eta\eta}\pi_{\beta\gamma}\sigma^{\beta\gamma} 
=-\frac{2\Phi}{3\tau^3}+\frac{\Phi}{3\tau^3}=-\frac{\Phi}{3\tau^3}.
\end{align}
Hence,
\begin{equation}\label{Bjlambda}
-\tau_{\pi\pi}\,\pi^{<\eta}_{\gamma}\sigma^{\eta>\gamma} - \delta_{\pi\pi}\,\pi^{\eta\eta}\theta = \tau_{\pi\pi}\,\frac{\Phi}{3\tau^3} + \delta_{\pi\pi}\,\frac{\Phi}{\tau^3}
= \left(\frac{1}{3}\tau_{\pi\pi} + \delta_{\pi\pi}\right)\frac{\Phi}{\tau^3}.
\end{equation}
For the scalar $\pi^{\alpha\beta}\sigma_{\alpha\beta}$, we obtain.
\begin{align}\label{BjSpisigma}
\pi^{\alpha\beta}\sigma_{\alpha\beta}&=\pi^{\alpha\beta}(\nabla_{\alpha}u_{\beta}-\frac{1}{3}\Delta_{\alpha\beta}\theta)=\pi^{\alpha\beta}\nabla_{\alpha}u_{\beta}
=\pi^{\alpha\beta}\Delta^{\lambda}_{\alpha}d_{\lambda}u_{\beta}=\pi^{\lambda\beta}(\partial_{\lambda}u_{\beta}-\Gamma^{\alpha}_{\lambda\beta}u_{\alpha}) \nonumber\\
&=-\pi^{\eta\eta}\tau=\frac{\Phi}{\tau}.
\end{align}
Similarly for third-order terms, we find that
\begin{equation}\label{Bjpisigmasigma}
\pi^{\langle\eta}_{\gamma}\pi^{\eta\rangle\gamma}\theta = -\frac{\Phi^2}{2\tau^3}, \quad
\pi^{\rho\langle\eta}\pi^{\eta\rangle\gamma}\sigma_{\rho\gamma} = -\frac{\Phi^2}{2\tau^3},
\end{equation}
and
\begin{align}\label{Bjnablapi}
\nabla_{\gamma}\nabla^{<\eta}\pi^{\eta>\gamma}&=\Delta^{\eta\eta}_{\alpha\beta}\nabla_{\gamma}\nabla^{\alpha}\pi^{\beta\gamma}=\frac{4\Phi}{3\tau^4}, \nonumber\\
\nabla^{<\eta}\nabla_{\gamma}\pi^{\eta>\gamma}&=\Delta^{\eta\eta}_{\alpha\beta}\nabla^{\alpha}\nabla_{\gamma}\pi^{\beta\gamma}=\frac{2\Phi}{3\tau^4}, \nonumber\\
\nabla^2\pi^{<\eta\eta>}&=\Delta^{\eta\eta}_{\alpha\beta}\nabla^{\gamma}\nabla_{\gamma}\pi^{\alpha\beta}=\frac{4\Phi}{3\tau^4}.
\end{align}

%########################################################################
\chapter{Fluid dynamics from Chapman- Enskog expansion: Derivation details}
%########################################################################

%%%%%%%%%%%%%%%%%%%%%%%%%%%%%%%%%%%%%%%%%%%%%%%%%%%%%%%%%%%%%%%%%%%%%%%%

\section{General structure}

The conserved particle current and energy-momentum tensor can be 
expressed in terms of the distribution function as
\begin{align}
N^\mu &= \int dp \ p^\mu f = nu^\mu + n^\mu, \nonumber\\
T^{\mu\nu} &= \int dp \ p^\mu p^\nu f = \epsilon u^\mu u^\nu-(P+\Pi)\Delta ^{\mu \nu} + \pi^{\mu\nu},
\label{NTD}
\end{align}
where $dp = g d{\bf p}/[(2 \pi)^3\sqrt{{\bf p}^2+m^2}]$, $g$ and $m$ 
being the degeneracy factor and particle mass. In the tensor 
decompositions, $\epsilon, P, n$ are respectively energy density, 
pressure, net number density, and $\Delta^{\mu\nu}=g^{\mu\nu}-u^\mu 
u^\nu$ is the projection operator on the three-space orthogonal to 
the hydrodynamic four-velocity $u^\mu$ defined in the Landau frame: 
$T^{\mu\nu} u_\nu=\epsilon u^\mu$. The metric tensor is 
$g^{\mu\nu}\equiv\mathrm{diag}(+,-,-,-)$. The bulk viscous pressure 
$(\Pi)$, shear stress tensor $(\pi^{\mu\nu})$ and particle diffusion 
current $(n^\mu)$ are the dissipative quantities. The net number 
density, energy density and pressure can be expressed as
\begin{align}
n &= u_\mu\int dp \ p^\mu f_0, \nonumber\\ 
\epsilon &= u_\mu u_\nu \int dp \ p^\mu p^\nu f_0, \nonumber\\ 
P &= -\frac{1}{3}\Delta_{\mu\nu} \int dp \ p^\mu p^\nu f_0. \label{NDED}
\end{align}
The equilibrium distribution functions 
$f_0=\frac{1}{\mathrm{exp}(\beta\,u\cdot p-\alpha)+r}$ with 
$r=1,-1,0$ for Fermi, Bose, and Boltzmann particles. Here, 
$\beta=1/T$ is the inverse temperature and $\alpha=\mu/T$ is the 
ratio of chemical potential to temperature.

Current conservation, $\partial_\mu N^{\mu}=0$, and energy-momentum 
conservation, $\partial_\mu T^{\mu\nu} =0$ yields the fundamental 
evolution equations for $n$, $\epsilon$ and $u^\mu$
\begin{align}
\dot n + n\theta + \partial_\mu n^{\mu} &=0~, \label{evol1}\\
\dot\epsilon + (\epsilon+P+\Pi)\theta - \pi^{\mu\nu}\nabla_{(\mu} u_{\nu)} &= 0~,  \label{evol2}\\
(\epsilon+P+\Pi)\dot u^\alpha - \nabla^\alpha (P+\Pi) + \Delta^\alpha_\nu \partial_\mu \pi^{\mu\nu}  &= 0, \label{evol3} 
\end{align}
where, $\theta\equiv\partial_\mu u^\mu$, 
$\nabla^\mu\equiv\Delta^{\mu\nu}\partial_\nu$ and $\dot u^\mu\equiv 
Du^\mu=u^\alpha\partial_\alpha u^\mu$. Obtaining co-moving 
derivatives $\dot n$ and $\dot\epsilon$ from Eq. (\ref{NDED}) and 
substituting in Eqs. (\ref {evol1}) and (\ref {evol2}), we arrive at
\begin{equation}
\dot\beta = \frac{J_{20}^{(0)}n - J_{10}^{(0)} (\epsilon+P)}{J_{20}^{(0)}J_{20}^{(0)}-J_{30}^{(0)}J_{10}^{(0)}}~\theta + \mathcal{O}(\delta^2), \quad 
\dot\alpha = \frac{J_{30}^{(0)}n - J_{20}^{(0)} (\epsilon+P)}{J_{20}^{(0)}J_{20}^{(0)}-J_{30}^{(0)}J_{10}^{(0)}}~\theta + \mathcal{O}(\delta^2),
\label{ABD}
\end{equation}
where $\mathcal{O}(\delta^2)$ means second-order in gradients. We 
define the thermodynamic functions,
\begin{equation}
J_{nq}^{(m)} = \frac{(-1)^q}{(2q+1)!!}\int dp~(u\!\cdot\!p)^{n-2q-m}(\Delta_{\alpha\beta}p^\alpha p^\beta)^q f_0\tilde f_0 ,
\label{JTD}
\end{equation}
where $\tilde f_0 = 1-rf_0$. We similarly define
\begin{equation}
I_{nq}^{(m)} = \frac{(-1)^q}{(2q+1)!!}\int dp~(u\!\cdot\!p)^{n-2q-m}(\Delta_{\alpha\beta}p^\alpha p^\beta)^q f_0,
\label{ITD}
\end{equation}
and state the relations
\begin{equation}
J_{nq}^{(0)} = \frac{1}{\beta}\left[-I_{n-1,q-1}^{(0)} + (n-2q)I_{n-1,q}^{(0)} \right]; \quad
I_{10}^{(0)} = n, \quad I_{20}^{(0)}=\epsilon, \quad I_{21}^{(0)}=-P.
\label{RRIJ}
\end{equation}

Substituting the space-like derivative of pressure ($\nabla^\alpha P$) 
obtained from Eq. (\ref{NDED}) in Eq. (\ref{evol3}) and using Eq. 
(\ref{RRIJ}), we get, 
\begin{equation}
\nabla^\mu\beta = -\beta\dot u^\mu + \frac{n}{\epsilon +P} \nabla^\mu \alpha + \mathcal{O}(\delta^2),
\label{nabb}
\end{equation}

The expressions for the dissipative quantities in terms of away from 
equilibrium part of the distribution functions ($\delta f$), can be 
written using Eq. (\ref{NTD}) as
\begin{align}
\Pi &= -\frac{\Delta_{\alpha\beta}}{3} \int dp \, p^\alpha p^\beta \delta f , \label{FBE}\\
n^\mu &= \Delta^\mu_\alpha \int dp \, p^\alpha \delta f , \label{FCE}\\
\pi^{\mu\nu} &= \Delta^{\mu\nu}_{\alpha\beta} \int dp \, p^\alpha p^\beta \delta f ,\label{FSE}
\end{align}
where
$\Delta^{\mu\nu}_{\alpha\beta} = [\Delta^{\mu}_{\alpha}\Delta^{\nu}_{\beta} + \Delta^{\mu}_{\beta}\Delta^{\nu}_{\alpha} - (2/3)\Delta^{\mu\nu}\Delta_{\alpha\beta}]/2$.

To obtain $\delta f$, we solve Boltzmann equation in 
relaxation time approximation using Chapman-Enskog (CE) expansion.
In the CE expansion, the particle distribution function is expanded 
about its equilibrium value in powers of space-time gradients.
\begin{equation}\label{CEE}
f = f_0 + \delta f, \quad \delta f= \delta f^{(1)} + \delta f^{(2)} + \cdots ,
\end{equation}
where $\delta f^{(1)}$ is first-order in gradients, $\delta f^{(2)}$ 
is second-order, etc. The Boltzmann equation, $p^\mu\partial_\mu f = -(u\!\cdot\!p)\delta f/\tau_R$, in the 
form $f=f_0-(\tau_R/u\!\cdot\! p)\,p^\mu\partial_\mu f$, can be solved 
iteratively as
\begin{equation}\label{F1F2}
f_1 = f_0 -\frac{\tau_R}{u\!\cdot\!p} \, p^\mu \partial_\mu f_0, \quad f_2 = f_0 -\frac{\tau_R}{u\!\cdot\!p} \, p^\mu \partial_\mu f_1, ~~\, \cdots
\end{equation}
where $f_1=f_0+\delta f^{(1)}$ and $f_2=f_0+\delta f^{(1)}+\delta 
f^{(2)}$. To first and second-order in gradients,
\begin{equation}
\delta f^{(1)} = -\frac{\tau_R}{u\!\cdot\!p} \, p^\mu \partial_\mu f_0, \quad
\delta f^{(2)} = \frac{\tau_R}{u\!\cdot\!p}p^\mu p^\nu\partial_\mu\Big(\frac{\tau_R}{u\!\cdot\!p} \partial_\nu f_0\Big). \label{SOC}
\end{equation}

%%%%%%%%%%%%%%%%%%%%%%%%%%%%%%%%%%%%%%%%%%%%%%%%%%%%%%%%%%%%%%%%%%%%%%%%

\section{First-order equations}

The first-order dissipative equations can be obtained from Eqs. 
(\ref{FBE})-(\ref{FSE}) using $\delta f = \delta f^{(1)}$ from Eq. 
(\ref{SOC})
\begin{align}
\Pi &= -\frac{\Delta_{\alpha\beta}}{3}\int dp \, p^\alpha p^\beta \left(-\frac{\tau_R}{u\!\cdot\!p} \, p^\gamma \partial_\gamma f_0\right) , \label{FOBE}\\
n^\mu &= \Delta^\mu_\alpha \int dp \, p^\alpha \left(-\frac{\tau_R}{u\!\cdot\!p} \, p^\gamma \partial_\gamma f_0\right) , \label{FOCE}\\
\pi^{\mu\nu} &= \Delta^{\mu\nu}_{\alpha\beta}\int dp \ p^\alpha p^\beta \left(-\frac{\tau_R}{u\!\cdot\!p} \, p^\gamma \partial_\gamma f_0\right) . \label{FOSE}
\end{align}
Assuming the relaxation time $\tau_R$ to be independent of 
four-momenta, the integrals in Eqs. (\ref {FOBE})-(\ref{FOSE}) 
reduce to (see next section for details)
\begin{equation}\label{FOE}
\Pi = -\tau_R\beta_\Pi\theta, ~~  n^\mu = \tau_R\beta_n\nabla^\mu\alpha, ~~ \pi^{\mu\nu} = 2\tau_R\beta_\pi\sigma^{\mu\nu},
\end{equation}
where 
$\sigma^{\mu\nu}=\Delta^{\mu\nu}_{\alpha\beta}\nabla^{\alpha}u^\beta$. 
The coefficients $\beta_\Pi,~\beta_n$ and $\beta_\pi$ are 
found to be
\begin{align}
\beta_\Pi =&~ \frac{1}{3}\left(1-3c_s^2\right)(\epsilon+P) - \frac{2}{9}(\epsilon-3P)
- \frac{m^4}{9}\left<(u\!\cdot\!p)^{-2}\right>_0 , \label{BB}\\
\beta_n =&~ - \frac{n^2}{\beta(\epsilon+P)} + \frac{2\left<1\right>_0}{3\beta} + \frac{m^2}{3\beta}\left<(u\!\cdot\!p)^{-2}\right>_0 ,\label{BC}\\
\beta_\pi =&~ \frac{4P}{5} + \frac{\epsilon-3P}{15} - \frac{m^4}{15}\left<(u\!\cdot\!p)^{-2}\right>_0 ,\label{BS}
\end{align}
where $\left<\cdots\right>_0=\int dp(\cdots)f_0$, and 
$c_s^2=(dP/d\epsilon)_{s/n}$ is the adiabatic speed of sound squared 
($s$ being the entropy density).

%%%%%%%%%%%%%%%%%%%%%%%%%%%%%%%%%%%%%%%%%%%%%%%%%%%%%%%%%%%%%%%%%%%%%%%%

\section{Second-order evolution equations}

Substituting $\delta f = \delta f^{(1)} +\delta f^{(2)}$ from 
Eq. (\ref{SOC}) in Eq. (\ref{FSE}) and assuming a momentum-independent 
relaxation time, we obtain
\begin{equation}
\frac{\pi^{\mu\nu}}{\tau_R} = 
-\Delta^{\mu\nu}_{\alpha\beta}\int dp \, p^\alpha p^\beta \left[\frac{p^\gamma}{u\!\cdot\! p}\partial_\gamma f_0 
- \frac{p^\gamma p^\rho}{u\!\cdot\! p}\partial_\gamma\left(\frac{\tau_R}{u\!\cdot\! p}\partial_\rho f_0\right) \right] . \label{SOSE}
\end{equation}
The first-order term can be solved as
\begin{align}
\frac{\pi^{\mu\nu}_{(1)}}{\tau_R} &= \Delta^{\mu\nu}_{\alpha\beta}\int dp \, p^\alpha p^\beta \left(-\frac{1}{u\!\cdot\!p} \, 
p^\gamma \partial_\gamma f_0 \right) \nonumber\\
&=\Delta^{\mu\nu}_{\alpha\beta} \int \frac{dp}{u\!\cdot\!p} \, p^\alpha p^\beta p^\gamma 
\left[\{ (u\!\cdot\! p)\partial_\gamma\beta + \beta(\partial_\gamma u_\lambda)p^\lambda -(\partial_\gamma\alpha) \} f_0\tilde f_0  \right]\nonumber\\
&= \Delta^{\mu\nu}_{\alpha\beta} \left[ (\partial_\gamma\beta) J_{(0)}^{\alpha\beta\gamma} + \beta(\partial_\gamma u_\lambda)J_{(1)}^{\alpha\beta\gamma\lambda} 
-(\partial_\gamma\alpha)J_{(1)}^{\alpha\beta\gamma} \right] \nonumber\\
&= 2\, \beta J^{(1)}_{42}\sigma^{\mu\nu}, \label{FOSE1}
\end{align}
where $\sigma^{\mu\nu}=\Delta^{\mu\nu}_{\alpha\beta}(\nabla_\alpha u_\beta)$ 
and $\beta J^{(1)}_{42}$ can be reduced to $\beta_\pi$ after some algebra.

The second-order terms are given by
\begin{align}
\frac{\pi^{\mu\nu}_{(2)}}{\tau_R} &= 
\Delta^{\mu\nu}_{\alpha\beta}\int dp \, p^\alpha p^\beta ~\frac{p^\gamma p^\rho}{u\!\cdot\!p}\partial_\gamma
\left(\frac{\tau_R}{u\!\cdot\!p}\partial_\rho f_0\right) \nonumber\\
&=\Delta^{\mu\nu}_{\alpha\beta}\int dp \, p^\alpha p^\beta \Big[
\underbrace{D\left(\frac{\tau_R}{u\!\cdot\!p}p^\rho\partial_\rho f_0\right) }_{{}(I)}
+ \underbrace{\frac{p^\gamma}{u\!\cdot\!p}\nabla_\gamma(\tau_R\dot f_0) }_{{}(II)}
+ \underbrace{\frac{p^\gamma}{u\!\cdot\!p}\nabla_\gamma\left(\frac{\tau_R}{u\!\cdot\!p}p^\rho\nabla_\rho f_0\right)
}_{{}(III)}
\Big] . \label{SOSE1}
\end{align}
We perform the integrals one-by-one. 
\begin{align}
(I) &= \Delta^{\mu\nu}_{\alpha\beta}D\left(\tau_R \!\int\! \frac{dp}{u\!\cdot\!p}\, p^\alpha p^\beta p^\rho\partial_\rho f_0\right) \nonumber \\
&= -\Delta^{\mu\nu}_{\alpha\beta}D\left[\tau_R \!\int\! \frac{dp}{u\!\cdot\!p}\, p^\alpha p^\beta p^\rho
\left\{(u\!\cdot\!p)\partial_\rho\beta + \beta(\partial_\rho u_\lambda)p^\lambda -(\partial_\rho\alpha) \right\} f_0\tilde f_0\right]\nonumber \\
&=-\Delta^{\mu\nu}_{\alpha\beta}D\left[\tau_R \,\beta\,(\nabla_\rho u_\lambda)J^{\alpha\beta\rho\lambda}_{(1)}\right] \nonumber \\
&= -\Delta^{\mu\nu}_{\alpha\beta}D\left[2\,\tau_R \,\beta\,J^{(1)}_{42}\sigma_{\alpha\beta} \right] \nonumber \\
&=-\dot\pi^{\langle\mu\nu\rangle},
\label{IP}
\end{align}
where 
$A^{\langle\mu\nu\rangle}=\Delta^{\mu\nu}_{\alpha\beta}A_{\alpha\beta}$. 
Eqs. (\ref{RRIJ}) and (\ref{nabb}) has been used to arrive at the 
third step. In the last step, the first-order Eq. (\ref{FOSE1}) has 
been used ($\dot\pi^{\langle\mu\nu\rangle}_{(1)}=\dot\pi^{\langle\mu\nu\rangle}
+\mathcal{O}(\delta^3)$).

Keeping in mind that $\dot f_0=-[(u\cdot p)\dot\beta + \beta p^\lambda\dot u_\lambda - \dot\alpha]f_0\tilde f_0$,
we solve the second term
\begin{align}
(II) &= \Delta^{\mu\nu}_{\alpha\beta}\!\int\! \frac{dp}{u\!\cdot\!p} \, p^\alpha p^\beta p^\gamma \,
\nabla_\gamma(\tau_R\dot f_0) \nonumber \\
&= \Delta^{\mu\nu}_{\alpha\beta}\nabla_\gamma\left(\tau_R\!\int\! \frac{dp}{u\!\cdot\! p} 
\, p^\alpha p^\beta p^\gamma \,\dot f_0\right) 
+ \Delta^{\mu\nu}_{\alpha\beta} (\nabla_\gamma u_\rho) \tau_R\!\int\! \frac{dp}{(u\!\cdot\!p)^2} 
\, p^\alpha p^\beta p^\gamma p^\rho \,\dot f_0 \nonumber \\
&= -2\,\tau_R\left[\left(J^{(0)}_{31}+J^{(1)}_{42}\right)\dot\beta 
- \left(J^{(1)}_{31}+J^{(2)}_{42}\right)\dot\alpha \right]\sigma^{\mu\nu}
-2\nabla^{\langle\mu}\left(\dot u^{\nu\rangle} \tau_R\beta J^{(1)}_{42} \right).
\label{IIP}
\end{align}

Similarly, writing $\nabla_\rho f_0=-[(u\cdot p)\nabla_\rho\beta + \beta(\nabla_\rho u_\lambda)p^\lambda 
-(\nabla_\rho\alpha) ] f_0\tilde f_0$, and using Eq. (\ref{nabb}) the third term becomes
\begin{align}
(III) =&~ 2\nabla^{\langle\mu}\!\left(\dot u^{\nu\rangle} \tau_R\beta J^{(1)}_{42} \right)
\!+2\nabla^{\langle\mu}\!\left[\left(\nabla^{\nu\rangle}\alpha \right)\tau_R \left(J^{(2)}_{42}
\!-\frac{n}{\epsilon+P}J^{(1)}_{42}\right)\! \right] 
\!+ 4\tau_R\beta J^{(1)}_{42}\sigma_\gamma^{\langle\mu}\omega^{\nu\rangle\gamma} \nonumber\\
&- \frac{4}{3}\,\beta\tau_R \left(7J^{(3)}_{63}+5J^{(1)}_{42}\right)\theta\sigma^{\mu\nu}
-4\beta\tau_R\left(2J^{(3)}_{63}+J^{(1)}_{42}\right)\sigma_\gamma^{\langle\mu}\sigma^{\nu\rangle\gamma}.
\label{IIIP}
\end{align}
We observe that the last term in Eq. (\ref{IIP}) and the first term in Eq. (\ref{IIIP}) cancels. 
Adding Eqs. (\ref{IP})-(\ref{IIIP}), we obtain
\begin{align}
\frac{\pi^{\mu\nu}_{(2)}}{\tau_R} =& -\dot\pi^{\langle\mu\nu\rangle}
-2\,\tau_R\left[\left(J^{(0)}_{31}+J^{(1)}_{42}\right)\dot\beta 
- \left(J^{(1)}_{31}+J^{(2)}_{42}\right)\dot\alpha \right]\sigma^{\mu\nu}
+ 4\tau_R\beta J^{(1)}_{42}\sigma_\gamma^{\langle\mu}\omega^{\nu\rangle\gamma} \nonumber\\
&+2\nabla^{\langle\mu}\!\left[\left(\nabla^{\nu\rangle}\alpha \right)\tau_R \left(J^{(2)}_{42}
-\frac{n}{\epsilon+P}J^{(1)}_{42}\right)\! \right] 
- \frac{4}{3}\,\beta\tau_R \left(7J^{(3)}_{63}+5J^{(1)}_{42}\right)\theta\sigma^{\mu\nu}\nonumber\\
&-4\beta\tau_R\left(2J^{(3)}_{63}+J^{(1)}_{42}\right)
\sigma_\gamma^{\langle\mu}\sigma^{\nu\rangle\gamma}.
\label{SOSE2}
\end{align}
Adding Eqs. (\ref{FOSE1}) and (\ref{SOSE2}) and using Eq. (\ref{ABD}) and (\ref{nabb}), 
we get the final evolution equation for shear stress tensor,
\begin{align}
\frac{\pi^{\mu\nu}}{\tau_R} 
=&-\dot{\pi}^{\langle\mu\nu\rangle}
+2\beta_{\pi}\sigma^{\mu\nu}
+2\pi_{\gamma}^{\langle\mu}\omega^{\nu\rangle\gamma}
-\tau_{\pi\pi}\pi_{\gamma}^{\langle\mu}\sigma^{\nu\rangle\gamma}
-\delta_{\pi\pi}\pi^{\mu\nu}\theta 
+\lambda_{\pi\Pi}\Pi\sigma^{\mu\nu}
-\tau_{\pi n}n^{\langle\mu}\dot{u}^{\nu\rangle }  \nonumber \\
&+\lambda_{\pi n}n^{\langle\mu}\nabla ^{\nu\rangle}\alpha
+\ell_{\pi n}\nabla^{\langle\mu}n^{\nu\rangle},
\label{SOSEF}
\end{align}
where,
\begin{align}
\tau_{\pi\pi} =\,& 2\beta\!\left(\!2J^{(3)}_{63}+J^{(1)}_{42}\!\right)\!\!/\beta_\pi, \quad
\delta_{\pi\pi}= \frac{1}{3}\beta \!\left(\!7J^{(3)}_{63}+5J^{(1)}_{42}\!\right)\!\!/\beta_\pi, \quad
\ell_{\pi n} = 2\!\left(\!J^{(2)}_{42} -\frac{n}{\epsilon+P}J^{(1)}_{42}\!\right)\!\!/\beta_n \nonumber \\
\lambda_{\pi\Pi}=\,& 2\bigg[\left(J^{(0)}_{31}+J^{(1)}_{42}\right)\frac{J_{20}^{(0)}n - J_{10}^{(0)} 
(\epsilon+P)}{J_{20}^{(0)}J_{20}^{(0)}-J_{30}^{(0)}J_{10}^{(0)}} - \left(J^{(1)}_{31}+J^{(2)}_{42}\right)
\frac{J_{30}^{(0)}n - J_{20}^{(0)} (\epsilon+P)}{J_{20}^{(0)}J_{20}^{(0)}-J_{30}^{(0)}J_{10}^{(0)}}  \nonumber \\
&\quad + \frac{1}{3}\beta \left(7J^{(3)}_{63}+5J^{(1)}_{42}\right)\bigg]/\beta_\Pi.
\label{coeffs}
\end{align}
The coefficients $\tau_{\pi n}$ and $\lambda_{\pi n}$ contain derivatives of $\ell_{\pi n}$.

Proceeding in a similar way, the evolution equations for bulk pressure and particle diffusion 
current can also be obtained. For bulk pressure, we get the expression
\begin{equation}
\frac{\Pi}{\tau_R} = -\dot{\Pi}
-\beta_{\Pi}\theta 
-\delta_{\Pi\Pi}\Pi\theta
+\lambda_{\Pi\pi}\pi^{\mu\nu}\sigma_{\mu \nu } 
-\tau_{\Pi n}n\cdot\dot{u}
-\lambda_{\Pi n}n\cdot\nabla\alpha
-\ell_{\Pi n}\partial\cdot n, 
\label{BULKF}\\
\end{equation}
where,
\begin{align}
\delta_{\Pi\Pi}=\,& \frac{5}{3}\bigg[\left(J^{(0)}_{31}+J^{(1)}_{42}\right)\frac{J_{20}^{(0)}n - J_{10}^{(0)} 
(\epsilon+P)}{J_{20}^{(0)}J_{20}^{(0)}-J_{30}^{(0)}J_{10}^{(0)}} - \left(J^{(1)}_{31}+J^{(2)}_{42}\right)
\frac{J_{30}^{(0)}n - J_{20}^{(0)} (\epsilon+P)}{J_{20}^{(0)}J_{20}^{(0)}-J_{30}^{(0)}J_{10}^{(0)}}  \nonumber \\
&\quad~ + \frac{1}{3}\,\beta \left(7J^{(3)}_{63}+\frac{23}{5}J^{(1)}_{42}\right)\bigg]/\beta_\Pi, \nonumber \\
\lambda_{\Pi\pi}=\,& \frac{1}{3}\,\beta \left(7J^{(3)}_{63}+J^{(1)}_{42}\right)\!/\beta_\pi, \quad
\ell_{\Pi n} = \frac{5}{3}\left(J^{(2)}_{42} -\frac{n}{\epsilon+P}J^{(1)}_{42}\right)\!/\beta_n.
\label{coeffb}
\end{align}
The coefficients $\tau_{\Pi n}$ and $\lambda_{\Pi n}$ contain derivatives of $\ell_{\Pi n}$.

Finally, the second-order evolution equation for particle diffusion current can also be 
derived by performing a similar kind of calculation. 
\begin{align}
\frac{n^{\mu}}{\tau_R} =& -\dot{n}^{\langle\mu\rangle}
+\beta_{n}\nabla^{\mu}\alpha
-n_{\nu}\omega^{\nu\mu}
-\lambda_{nn}n^{\nu}\sigma_{\nu}^{\mu}
-\delta_{nn}n^{\mu}\theta  +\lambda_{n\Pi}\Pi\nabla^{\mu}\alpha
-\lambda_{n\pi}\pi^{\mu\nu}\nabla_{\nu}\alpha 
-\tau_{n\pi}\pi_{\nu}^{\mu}\dot{u}^{\nu}  \nonumber \\
&+\tau_{n\Pi}\Pi\dot{u}^{\mu}
+\ell_{n\pi}\Delta^{\mu\nu}\partial_{\gamma}\pi_{\nu}^{\gamma}
-\ell_{n\Pi}\nabla^{\mu}\Pi~,  \label{HEATF}
\end{align}
where,
\begin{align}
\lambda_{nn}=\,& 1\!+\!2\!\left(\!\frac{nJ^{(2)}_{42}}{\epsilon+P} - J^{(3)}_{42}\!\right)\!\!/\beta_n, \quad
\delta_{nn}= \frac{4}{3}\!+\!\frac{5}{3}\!\left(\!\frac{nJ^{(2)}_{42}}{\epsilon+P} - J^{(3)}_{42}\!\right)\!\!/\beta_n, \quad
\ell_{n\pi} = -\left(\!\beta J^{(2)}_{42}\!\right)\!\!/\beta_\pi, \nonumber \\
\ell_{n\Pi} =\,& -\left(\!J^{(0)}_{21} 
\frac{J_{20}^{(0)}n - J_{10}^{(0)} (\epsilon+P)}{J_{20}^{(0)}J_{20}^{(0)}-J_{30}^{(0)}J_{10}^{(0)}}
-J^{(1)}_{21}\frac{J_{30}^{(0)}n - J_{20}^{(0)} (\epsilon+P)}{J_{20}^{(0)}J_{20}^{(0)}-J_{30}^{(0)}J_{10}^{(0)}}
+ \frac{5}{3}\beta J^{(1)}_{42}  \!\right)\!\!/\beta_\Pi.
\label{coeffc}
\end{align}
The coefficients $\tau_{n\pi}$ and $\lambda_{n\pi}$ contain 
derivatives of $\ell_{n\pi}$ and the coefficients $\tau_{n\Pi}$ and 
$\lambda_{n\Pi}$ contain derivatives of $\ell_{n\Pi}$.

%########################################################################
\chapter{Effect of second-order viscous correction to the distribution function}
%########################################################################

%%%%%%%%%%%%%%%%%%%%%%%%%%%%%%%%%%%%%%%%%%%%%%%%%%%%%%%%%%%%%%%%%%%%%%%%

\section{Constraints on the viscous correction to the distribution function}

In this appendix, we show that the form of the viscous correction to
the distribution function, $\delta f$, given in Eq. (\ref {SOVCC6})
satisfies the matching condition $\epsilon =\epsilon_0$ and the Landau
frame definition $u_\nu T^{\mu \nu} = \epsilon u^\mu$, at each order
in gradients \cite{deGroot}. We also show that $\delta f$ is
consistent with the definition of the shear stress tensor,
Eq. (\ref{FSEC6}).

The first- and second-order viscous corrections to the distribution
function can be written separately using Eq. (\ref {SOVCC6}). The
first-order correction is given by
\begin{equation}\label{deltaf1}
\delta f_1 =  \frac{f_0\beta}{2\beta_\pi(u\!\cdot\!p)}\, p^\alpha p^\beta \pi_{\alpha\beta},
\end{equation}
whereas the second-order correction is
\begin{align}\label{deltaf2}
\delta f_2 \!= & -\frac{f_0\beta}{\beta_\pi} \bigg[\frac{\tau_\pi}{u\!\cdot\!p}\, p^\alpha p^\beta \pi^\gamma_\alpha\, \omega_{\beta\gamma} 
-\frac{5}{14\beta_\pi (u\!\cdot\!p)}\, p^\alpha p^\beta \pi^\gamma_\alpha\, \pi_{\beta\gamma} \nonumber\\
&+\!\frac{\tau_\pi}{3(u\!\cdot\!p)}p^\alpha p^\beta \pi_{\alpha\beta}\theta  
-\frac{6\tau_\pi}{5} p^\alpha\dot u^\beta\pi_{\alpha\beta}
+\!\frac{(u\!\cdot\!p)}{70\beta_\pi}\pi^{\alpha\beta}\pi_{\alpha\beta}\nonumber\\
&+\frac{\tau_\pi}{5} p^\alpha\! \left(\!\nabla^\beta\pi_{\alpha\beta}\!\right)\!
-\frac{3\tau_\pi}{(u\!\cdot\!p)^2}\, p^\alpha p^\beta p^\gamma \pi_{\alpha\beta}\dot u_\gamma \!+ \frac{\tau_\pi}{2(u\!\cdot\!p)^2}\nonumber\\
&\times\! p^\alpha p^\beta p^\gamma\! \left(\nabla_\gamma\pi_{\alpha\beta}\!\right) 
-\frac{\beta\!+\!(u\!\cdot\!p)^{-1}}{4(u\!\cdot\!p)^2\beta_\pi} \!\left(p^\alpha p^\beta \pi_{\alpha\beta}\!\right)^{\!2}\!\bigg].
\end{align}

In the following, we show that the $\delta f_i$ given in Eqs. (\ref
{deltaf1}) and (\ref{deltaf2}) satisfies the conditions
\begin{equation}\label{Landau1}
L_1[\delta f_i] \equiv \int dp\, (u\cdot p)^2\, \delta f_i = 0,
\end{equation}
corresponding to $\epsilon =\epsilon_0$, and
\begin{equation}\label{Landau2}
L_2[\delta f_i] \equiv \int dp\, \Delta_{\mu\alpha}u_\beta\, p^\alpha p^\beta\, \delta f_i = 0,
\end{equation}
corresponding to $u_\nu T^{\mu \nu} = \epsilon u^\mu$.

At first order, we obtain
\begin{equation}\label{Landaudf1}
L_1[\delta f_1] = \frac{\beta}{2\beta_\pi}\pi_{\alpha\beta}u_\gamma I_{(0)}^{\alpha\beta\gamma},\quad\!
L_2[\delta f_1] = \frac{\beta}{2\beta_\pi}\pi_{\alpha\beta}\Delta_{\mu\gamma} I_{(0)}^{\alpha\beta\gamma},
\end{equation}
where we define the integral
\begin{equation}\label{mom_int}
I^{\mu_1\mu_2\cdots\mu_n}_{(r)} \equiv \int \frac{dp}{(u\!\cdot\! p)^r} p^{\mu_1}p^{\mu_2} \cdots p^{\mu_n} f_0. 
\end{equation}
The above momentum integral can be decomposed into hydrodynamic tensor
degrees of freedom as
\begin{align}\label{tens_decm}
I^{\mu_1\mu_2\cdots\mu_n}_{(r)} =\ & I_{n0}^{(r)} u^{\mu_1}u^{\mu_2}\cdots u^{\mu_n} 
+ I_{n1}^{(r)} \big(\Delta^{\mu_1\mu_2} u^{\mu_3} \cdots u^{\mu_n} \nonumber\\
&+ \mathrm{perms}\big) + \cdots, 
\end{align}
where we readily identify $I_{20}^{(0)}=\epsilon$ and
$I_{21}^{(0)}=-P$. Using the above tensor decomposition for
$I_{(0)}^{\alpha\beta\gamma}$ in Eq. (\ref{Landaudf1}), we obtain
\begin{equation}\label{Landau1df1F}
L_1[\delta f_1] = 0,\quad L_2[\delta f_1] = 0.
\end{equation}

Similarly, for second-order corrections given in Eq. (\ref{deltaf2}),
we obtain
\begin{align}\label{Landau1df2}
L_1[\delta f_2] =&\, 0 + \!\frac{5\beta}{14\beta_\pi^2}\pi_{\alpha\beta}\pi^{\alpha\beta}\!I_{31}^{(0)}\! 
+ 0 + 0 - \!\frac{\beta}{70\beta_\pi^2}\pi_{\alpha\beta}\pi^{\alpha\beta}\!I_{30}^{(0)} \nonumber\\
& - \frac{\beta\tau_\pi}{5\beta_\pi}(\nabla^\alpha\pi_{\alpha\beta})I_{30}^{(0)} u^\beta  + 0
- \frac{\beta\tau_\pi}{\beta_\pi}(\nabla_\gamma\pi_{\alpha\beta})I_{31}^{(0)} \nonumber\\
& \times\! u^{(\alpha}\Delta^{\beta)\gamma} + \frac{\beta}{2\beta_\pi^2}\pi_{\alpha\beta}\pi^{\alpha\beta}
\!\left(\beta I_{42}^{(0)} + I_{42}^{(1)}\right).
\end{align}
Using the identities 
\begin{align}
I_{nq}^{(r)}&=-\frac{1}{2q+1}I_{n-1,q-1}^{(r-1)}, 
\label{prop1} \\
I_{nq}^{(0)}&= \frac{1}{\beta}\left[-I_{n-1,q-1}^{(0)} + (n-2q)I_{n-1,q}^{(0)} \right],
\label{prop2}
\end{align}
and Eq. (\ref{FOEC6}), we obtain
\begin{align}\label{Landau1df2f}
L_1[\delta f_2] =&\, -\frac{25}{14\beta_\pi}\pi_{\alpha\beta}\pi^{\alpha\beta} \!
- \frac{3}{14\beta_\pi}\pi_{\alpha\beta}\pi^{\alpha\beta}\!
+ \frac{12}{8\beta_\pi}\pi_{\alpha\beta}\pi^{\alpha\beta} \nonumber\\
&- \frac{5}{2\beta_\pi}\pi_{\alpha\beta}\pi^{\alpha\beta} 
+ \frac{3}{\beta_\pi}\pi_{\alpha\beta}\pi^{\alpha\beta} \nonumber\\
=&\, 0.
\end{align}

A similar calculation leads to
\begin{align}\label{Landau2df2}
L_2[\delta f_2] =&\, 0 + 0 + 0 + \frac{6\beta\tau_\pi}{5\beta_\pi}I_{31}^{(0)}\Delta_\mu^\alpha\dot u^\beta\pi_{\alpha\beta} + 0 \nonumber\\
&- \frac{\beta\tau_\pi}{5\beta_\pi}I_{31}^{(0)}\Delta_\mu^\alpha \left(\!\nabla^\beta\pi_{\alpha\beta}\!\right)
-\frac{6\beta\tau_\pi}{5\beta_\pi}I_{31}^{(0)}\Delta_\mu^\alpha\dot u^\beta\pi_{\alpha\beta} \nonumber\\
& - \frac{\beta\tau_\pi}{\beta_\pi}I_{42}^{(1)}\Delta_\mu^\alpha \left(\!\nabla^\beta\pi_{\alpha\beta}\!\right) + 0 \nonumber\\
=&\, 0.
\end{align}
To obtain the second equality, we have used Eq. (\ref{prop1}) to
replace $I_{42}^{(1)}=-I_{31}^{(0)}/5$.

Next we show that the form of the viscous correction to the
distribution function, $\delta f=\delta f_1+\delta f_2$ given in Eqs. (\ref
{deltaf1}) and (\ref{deltaf2}),
is consistent with the definition of the shear stress tensor given in
Eq. (\ref {FSEC6}). In other words, we show that $\pi^{\mu\nu}=L_3[\delta
f_1]+L_3[\delta f_2]$, where
\begin{equation}\label{CheckDefn}
L_3[\delta f_i] \equiv \Delta^{\mu\nu}_{\alpha\beta} \int dp \, p^\alpha p^\beta\, \delta f_i.
\end{equation}
 At first order, we get
\begin{equation}\label{CheckDefn1}
L_3[\delta f_1] =\frac{\beta}{2\beta_\pi}\,\Delta^{\mu\nu}_{\alpha\beta}\,\pi_{\gamma\delta}\,I_{(1)}^{\alpha\beta\gamma\delta}.
\end{equation}
Using the tensor decomposition for
$I_{(1)}^{\alpha\beta\gamma\delta}$ in the above equation, we
obtain
\begin{equation}\label{CheckDefn1f}
L_3[\delta f_1] =\frac{\beta}{\beta_\pi}\,I^{(1)}_{42}\,\pi^{\mu\nu} = \pi^{\mu\nu}.
\end{equation}
Here we have used $I^{(1)}_{42}=\beta_\pi/\beta$, obtained 
by employing the recursion relations, Eqs. (\ref{prop1}) and (\ref{prop2}).

Similarly, for the second-order correction $\delta f_2$ given in Eq.
(\ref{deltaf2}), we obtain
\begin{align}\label{CheckDefn2}
L_3[\delta f_2] =&\, -2\tau_\pi \pi_\gamma^{\langle\mu} \omega^{\nu\rangle\gamma}
+ \frac{5}{7\beta_\pi}\pi_\gamma^{\langle\mu} \pi^{\nu\rangle\gamma}
-\frac{2}{3}\tau_\pi \pi^{\mu\nu}\theta + 0 \nonumber\\
&+ 0 + 0 + 0 + \Big( \frac{1}{\beta_\pi}\pi_\gamma^{\langle\mu} \pi^{\nu\rangle\gamma}
+ 2\tau_\pi \pi_\gamma^{\langle\mu} \omega^{\nu\rangle\gamma} \nonumber\\
&+ \frac{2}{3}\tau_\pi \pi^{\mu\nu}\theta \Big)
-\frac{12}{7\beta_\pi}\pi_\gamma^{\langle\mu} \pi^{\nu\rangle\gamma} \nonumber\\
=&\ 0.
\end{align}
Hence $L_3[\delta f]=L_3[\delta f_1]+L_3[\delta f_2]=\pi^{\mu\nu}$. 
This result was expected because no second-order term (e.g., $\pi 
\pi$, $\pi \omega$, etc.) or their linear combinations, when 
substituted in Eq. (\ref{FSEC6}), can result in a first-order term 
($\pi$) which we have on the left-hand side of Eq. (\ref{FSEC6}). In 
fact, each higher-order correction ($\delta f_n$) when substituted 
in Eq. (\ref{FSEC6}) will vanish. The fact that $\delta f$ given in 
Eq. (\ref{SOVCC6}) satisfies the constraints, as demonstrated in 
this Appendix, shows that our method of obtaining the viscous 
corrections to the distribution function is quite robust.

%%%%%%%%%%%%%%%%%%%%%%%%%%%%%%%%%%%%%%%%%%%%%%%%%%%%%%%%%%%%%%%%%%%%%%%%

\section{Second-order viscous corrections to hadron spectra and HBT radii}

Within the one-dimensional scaling expansion, $\dot u = 0 =
\omega^{\mu\nu}$, which reduces the number of terms in
Eq. (\ref{deltaf2}). The non-vanishing terms can be simplified using
Eq. (\ref{visc-fctsC6}) as
\begin{align}
\delta f_2 \!= \frac{f_0 \beta}{\beta_\pi} \Bigg[&-\frac{5\Phi^2 m_T\left\{p_T^2/(4m_T^2) + \sinh^2(y-\eta_s)\right\}}
{14\beta_\pi \cosh(y-\eta_s)}
-\frac{\tau_\pi\Phi\, m_T\left\{p_T^2/(2 m_T^2) - \sinh^2(y-\eta_s)\right\}}{3\tau\cosh(y-\eta_s)} \nonumber \\
&- \frac{3\Phi^2 m_T \cosh(y-\eta_s)}{140\beta_\pi} + \frac{\tau_\pi\Phi\, m_T \cosh(y-\eta_s)}{5 \tau}
-\frac{\tau_\pi  \Phi\, m_T \sinh^2(y-\eta_s)}{\tau \cosh(y-\eta_s)} \nonumber \\
&+ \frac{\Phi^2\beta}{4\beta_\pi \cosh^2(y-\eta_s)} \left\{\! 1 + \frac{(\beta m_T)^{-1}}{ \cosh(y-\eta_s)} \!\right\}
\left\{\!\frac{p_T^2}{2m_T^2} - \sinh^2(y-\eta_s)\!\right\}^{\!2}\Bigg]. 
\label{df2}
\end{align}

The contribution to the hadronic spectra resulting from these
second-order terms is calculated using Eq. (\ref{CFC6}) as
\begin{align}
\frac{\delta dN^{(2)}}{d^2p_Tdy} \equiv &\ \frac{g}{(2\pi)^3} \int m_T \cosh(y-\eta_s) \tau d\eta_s rdr d\varphi\, \delta f_2\nonumber \\
=&\ \frac{g\,\tau\, A_\perp}{4\pi^3 \beta\beta_\pi} \Bigg[\! 
 -\frac{5\Phi^2}{56\beta_\pi} \left( z_p^2\,K_0 + 4z_m\,K_1\right) 
 -\frac{\Phi \tau_\pi}{6\tau}\!\left( z_p^2K_0 \!- 2z_mK_1\right)
 -\frac{3\Phi^2 z_m^2}{280\beta_\pi}\!\left( K_0 \!+\! K_2\right) \nonumber \\
&\qquad\qquad~ +\frac{\Phi \tau_\pi z_m^2}{10\tau}\left( K_0 + K_2 \right)
 -\frac{\Phi\tau_\pi z_m}{\tau} K_1 + \frac{\Phi^2z_m^2}{4\beta_\pi} 
 \Big\{ z_m X^2 {\mathcal I}_1 - 2 z_m X  K_1  \nonumber \\ 
&\qquad\qquad~ + \frac{z_m}{4}\left( K_3 + 3K_1\right)
 + X^2 {\mathcal I}_2 - 2 X K_0 + \frac{1}{2}\left( K_0 + K_2 \right)\Big\}
 \Bigg]\label{SCF},
\end{align}
where $X\equiv z_p^2/(2z_m^2)+1$, $K_n(z_m)$ are the modified Bessel
functions of the second kind
\begin{equation}
  K_n(z) \equiv \int_0^\infty e^{-z\cosh (t)} \cosh (nt)\, dt,
\end{equation}
 and ${\mathcal I}_n$ are the integrals defined as
\begin{equation}
 {\mathcal I}_n(z) \equiv \int_0^\infty e^{-z\cosh (t)}\, {\rm sech}^n(t),
\end{equation}
with the following properties
\begin{equation}
 \frac{d^n {\mathcal I}_n(z)}{dz^n} = (-1)^n\,K_0(z), \quad {\mathcal I}_0(z)=K_0(z).
\end{equation}
The expression for hadron spectra up to second order, by setting 
$f=f_0+\delta f_1+\delta f_2$ in the freezeout prescription, Eq. 
(\ref{CFC6}), becomes
\begin{equation}\label{SCFF}
\frac{d N^{(2)}}{d^2p_Tdy} = \frac{d N^{(1)}}{d^2p_Tdy} + \frac{\delta dN^{(2)}}{d^2p_Tdy}.
\end{equation}

Similarly, within the Bjorken model, one can calculate the
longitudinal HBT radii by including the second-order viscous
corrections in Eq. (\ref{HBTBC6}) using Eq. (\ref{df2}). To this end, we
calculate $N[\delta f_2]$ by setting $f=f_0+\delta f_1+\delta f_2$ in
Eq. (\ref{HBTBC6}) and performing the integrations
\begin{align}
 N[\delta f_2] =&\ \int m_T \cosh^3(y-\eta_s) \tau d\eta_s rdr d\varphi\, \delta f_2 \nonumber \\
=&\ \frac{2 A_\perp \tau}{\beta\beta_\pi} \Bigg[ \!\!
-\frac{5\Phi^2}{112\beta_\pi} \Big\{\!\! \left(z_p^2-z_m^2\right)K_0 + z_p^2\,K_2  
 +z_m^2\,K_4\!\Big\} -\frac{\Phi \tau_\pi}{24\tau}\Big\{\!\! \left(2z_p^2+z_m^2\right)\!K_0 + 2z_p^2\,K_2\nonumber \\
&\qquad\quad~-z_m^2\,K_4\Big\} -\frac{3\Phi^2 z_m^2}{1120\beta_\pi}\left( 3K_0+4 K_2+K_4\right)
 + \frac{\Phi\tau_\pi z_m^2}{40\tau} \left( 3K_0+4 K_2+K_4\right)  \nonumber \\
&\qquad\quad~ -\frac{\Phi\tau_\pi z_m^2}{8\tau}\big( K_4 -K_0 \big) \!+ \frac{\Phi^2z_m^2}{4\beta_\pi} 
\bigg\{\!\!\left(\!\!X^2-X+\frac{3}{8}\right)\!\!K_0 + \!\left(\!\!z_m X^2 \!- \frac{3}{2}z_mX \!+ \frac{5}{8}z_m\!\!\right)\nonumber \\
&\qquad\quad~ \times\! K_1 + \!\left(\frac{1}{2}-X\!\right)\!K_2 + \bigg(\!\frac{5}{16}z_m 
-\frac{1}{2}z_mX \!\bigg)K_3 + \frac{1}{8}K_4 + \frac{1}{16}z_mK_5 \bigg\} \Bigg].
\end{align}

%########################################################################
\chapter{Glossary}
%########################################################################

RHIC: Relativistic Heavy Ion Collider

LHC: Large Hadron Collider

QCD: Quantum Chromo-Dynamics

QGP: Quark Gluon Plasma

LRF: Local Rest Frame

RTA: Relaxation Time Approximation

BAMPS: Boltzmann Approach of MultiParton Scatterings
\\[-0.05in]

NS: Navier-Stokes

IS: Israel-Stewart

BE: Boltzmann Equation

CE: Chapman-Enskog

HBT: Hanburry Brown-Twiss

KSS: Kovtun-Son-Starinets

DKR: Denicol-Koide-Rischke
\\[-0.05in]

$g^{\mu\nu}$: metric tensor

$u^\mu$: fluid four velocity

$\tau$: longitudinal proper time

$\eta_s$: space time rapidity

$\gamma$: Lorentz contraction factor
\\[-0.05in]

$\epsilon$: energy density

$n$: number density

$P$: pressure

$T$: temperature

$s$: entropy density

$c_s$: speed of sound
\\[-0.05in]

$f(x,p)$: distribution function

$f_0(x,p)$: equilibrium distribution function

$\delta f(x,p)$: non-equilibrium part in the distribution function 
$f=f_0+ \delta f$
\\[-0.05in]

$T^{\mu\nu}$: energy momentum tensor

$N^{\mu}$: conserved charge flow

$\pi^{\mu\nu}$: shear pressure tensor

$\Pi$: bulk pressure

$n^\mu$: particle diffusion current

$\sigma^{\mu\nu}$, $\nabla^{\La \mu} u^{\nu \Ra}$ : velocity stress 
tensor

$\omega^{\mu\nu}$: vorticity tensor

$\theta$: expansion scalar
\\[-0.05in]

$\zeta$: bulk viscosity

$\eta$: shear viscosity

$\lambda$: charge conductivity

$\tau_\pi$: relaxation time for shear pressure tensor

$\tau_\Pi$: relaxation time for bulk pressure

$\tau_n$: relaxation time for charge current
\\[-0.05in]

$A$: atomic number

$R_A$: nuclear radius

$A_\perp$: transverse area of the overlap zone of colliding nuclei

$b$: impact parameter
\\[-0.05in]

$\tau_0$: initial time

$T_0$: initial temperature

$T_c$: critical temperature

$T_{fo}$: freeze-out temperature

$y$: momentum rapidity

$p_T$: particle transverse momentum

$m_T$: particle transverse mass